\documentclass[numbib]{comnet}

\jno{iarxxx} 
\received{XX XX XXXX} 
\revised{XX XX XXXX} 
\accepted{XX XX XXXX} 

\usepackage{amsmath,amssymb}
\usepackage{upgreek}
\usepackage{amsfonts}
\usepackage{graphicx}
\usepackage{color}
\usepackage{url}
\usepackage{todonotes}

\definecolor{darkgreen}{rgb}{0,0.5,0}

\graphicspath{{arxiv_v2_figures/}}

\begin{document}

\date{\today}

\title{Network Analysis of Particles and Grains}

\shorttitle{Granular Networks}

\author{%
{\sc Lia Papadopoulos}\\[2pt]
Department of Physics \& Astronomy, University of Pennsylvania, USA\\
{epapad@sas.upenn.edu}\\[6pt]
{\sc Mason A. Porter}\\[2pt]
Department of Mathematics, University of California Los Angeles, USA\\ 
Mathematical Institute, University of Oxford, UK\\
CABDyN Complexity Centre, University of Oxford, UK\\
{mason@math.ucla.edu}\\[6pt]
{\sc Karen E. Daniels}\\[2pt]
Department of Physics, North Carolina State University, USA\\
{kdaniel@ncsu.edu}\\[6pt]
{\sc Danielle S. Bassett}$^*$\\[2pt]
Departments of Bioengineering and Electrical \& Systems Engineering, University of Pennsylvania, USA\\
\email{$^*$Corresponding author: dsb@seas.upenn.edu}}

\maketitle

\begin{abstract}
{The arrangements of particles and forces in granular materials have a complex organization on multiple spatial scales that ranges from local structures to mesoscale and system-wide ones. This multiscale organization can affect how a material responds or reconfigures when exposed to external perturbations or loading. The theoretical study of particle-level, force-chain, domain, and bulk properties requires the development and application of appropriate physical, mathematical, statistical, and computational frameworks. Traditionally, granular materials have been investigated using particulate or continuum models, each of which tends to be implicitly agnostic to multiscale organization. Recently, tools from network science have emerged as powerful approaches for probing and characterizing heterogeneous architectures across different scales in complex systems, and a diverse set of methods have yielded fascinating insights into granular materials. In this paper, we review work on network-based approaches to studying granular matter and explore the potential of such frameworks to provide a useful description of these systems and to enhance understanding of their underlying physics. We also outline a few open questions and highlight particularly promising future directions in the analysis and design of granular matter and other kinds of material networks.} 
{Granular materials, particulate systems, networks, network science, multiscale organization}
\end{abstract}

\newpage

\tableofcontents

\newpage

\section*{Glossary of Terms}

\begin{description}

\item[\emph{Granular materials}:] Granular materials are collections of discrete, macroscopic particles that interact with each other through contact (rather than long-range) forces. Importantly, these systems are non-equilibrium: the particles are large enough to avoid rearrangement under thermal fluctuations, and they lose energy through frictional and inelastic interactions
with neighboring particles. 

\item[\emph{Particulate materials}:] Like granular materials, particulate materials are collections of discrete, macroscopic elements. However, the elements making up the system may be entities --- such as bubbles, foams, colloids, or suspensions --- that include multiple phases of matter.
The term ``particulate material'' is a more general one than ``granular material''. 

\item[\emph{Packing fraction}:] The fraction of a granular material that consists of the particles. One calculates the packing fraction as the ratio of the total volume of all particles to the volume of the region that they occupy. The packing fraction is also sometimes called the ``packing density'' or ``volume fraction''. 

\item[\emph{Force chain}:] Force chains are typically described as the subset of inter-particle contacts in a granular material that carry the largest forces in the system. They often form filamentary networks that align preferentially with the principal stress axes under which a material is loaded.

\item[\emph{Jamming}:] As certain system parameters change, disordered, particulate materials typically undergo a transition from an underconstrained, liquid-like state to a rigid, solid-like state characterized by the onset of mechanical stability. The transition to/from a jammed state may arise through increases/decreases in quantities like packing fraction or contact number, which can occur due to an applied load. A formal theory of jamming exists for idealized situations (with soft, frictionless, and spherical particles).

\item[\emph{Isostatic}:] A jammed packing is isostatic when it has exactly the minimum number of contacts that are required for maintaining mechanical stability through force balance and torque balance. One typically examines isostaticity by calculating the mean number of contacts per particle. ``Hyperstatic'' and ``hypostatic'' packings have more and fewer contacts, respectively. 

\item[\emph{Mono/bi/poly-disperse}:] A particulate material is monodisperse if it is composed of particles of a single species (with the same size, shape, and material properties). Bidisperse materials have particles of two species, and polydisperse materials can either have particles from three or more discrete species or have particles with a continuum of properties.

\item[\emph{Structural rigidity theory}:] For a specified structure composed of fixed-length rods connected to one another by hinges, structural rigidity theory studies the conditions under which the associated structural graph is able to resist deformations and support applied loads. 

\item[\emph{Stress}:] A stress is a force applied to an object's surfaces. (The units measure a force per unit area.) Shear stress arises from the component of the applied force that acts in a direction parallel to the object's cross-section, and normal stress arises from the perpendicular component.

\item[\emph{Strain}:] Strain is the fractional (unitless) deformation of a material that arises due to an applied stress. One calculates strain from the relative displacement of particles in a material, excluding rigid-body motion such as translation or rotation. Like stress, strain has both shear and normal components. 

\item[\emph{Pure shear}:] One can use the term ``pure shear'' to describe either stresses or strains in which an object is elongated along one axis and shortened in the perpendicular direction without inducing a net rotation.

\item[\emph{Axial compression}:] In axial compression, one applies inward forces to an object in one direction (uniaxial), two directions (biaxial), or all directions (isotropic compression). These forces result in uniaxial strain, biaxial strain, or isotropic strain, respectively. 

\item[\emph{Cyclic shear/compression}:] These consist of repeated cycles of shear or compression applied to the same system.

\item[\emph{Shear band}:] A shear band is a narrow region of a particulate material in which most of the strain is localized, whereas other regions remain largely undeformed. A shear band is also sometimes called a region of ``strain localization''. 

\item[\emph{Strain softening/hardening}:] As a material is loaded and undergoes deformation, continuing deformation can become either easier (strain softening) or harder (strain hardening). Eventually, after much deformation, the material can reach a critical state in which there are no further changes in the resistance to deformations. 

\item[\emph{Stress ratio}:] The stress ratio, which is analogous to Coulomb's Law, is the ratio of shear to normal stresses. Frictional failure occurs when the shear force exceeds the product of the normal force and the coefficient of friction.

\item[\emph{Photoelasticity/birefringence}:] Photoelasticity is an optical technique for quantifying internal stresses based on the transmission of polarized light through ``birefringent'' materials, which have preferentially fast and slow directions for the propagation of light.

\item[\emph{DEM} or \emph{MD simulations}:] The Discrete (or Distinct) Element Method and Molecular Dynamics simulations are related numerical techniques that compute the motions of all particles in a system (such as a granular material). In each method, a computer algorithm treats each particle as an object subject to Newton's laws of motion, where forces consist of body forces (e.g., gravity) and those that arise from interactions with the object's neighbors.

\end{description}
\newpage

\section{Introduction}
\label{intro}

Granular materials comprise a subset of the larger set of particulate matter \citep{Jaeger1996,duran1999sands,Mehta2007,Franklin2015,Andreotti2013,Nagel:2017a}. People engage with such materials --- which include sands, beans, grains, powders such as cornstarch, and more --- often in their daily lives. One can define a \emph{granular material} as a large collection of discrete, macroscopic particles that interact only when in contact. Granular materials are inherently non-equilibrium in two distinct ways, characterized by \emph{(1)} the lack of rearrangement under thermal fluctuations and \emph{(2)} the loss of energy through frictional and inelastic dissipation during contact between grains. Nonetheless, they phenomenologically reproduce equilibrium states of matter, exhibiting characteristics of solids (forming rigid materials), liquids (flowing out of a container), or gases (infrequent contacts between grains), depending on the type and amount of driving. In this review, we focus mainly on granular solids and slow (non-inertial) flows \citep{Mort2015}; these are dense materials in which sustained inter-particle contacts provide the dominant contribution to material properties.

\begin{figure}[t]
\centering
\includegraphics[width=0.7\textwidth]{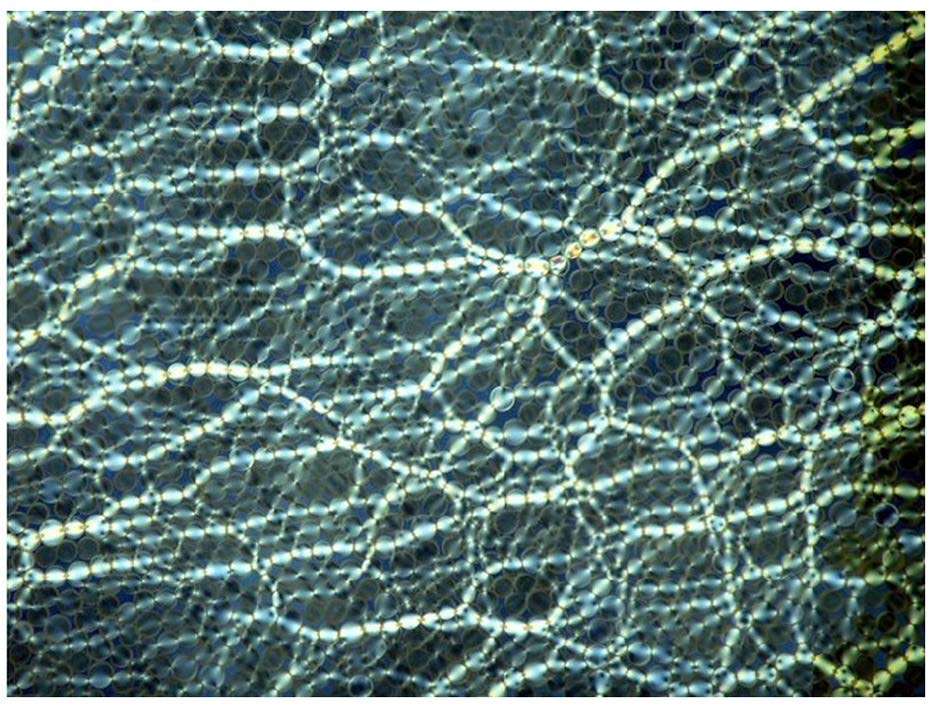}
\caption{\textbf{Force chains in an experimental granular system.} A photoelastic image of a quasi-two-dimensional (quasi-2D) packing of photoelastic disks that were subjected to pure shear. The photoelastic image allows one to visualize the force pattern in a material. Bright particles carry the strongest forces, and one can observe that a network of force chains tends to align along the principal stress axes. Modern techniques allow one to determine vector contact forces at each inter-particle contact. [We adapted this figure, with permission, from \citep{Behringer2014:Statistical}.]}
\label{f:Behringer2014Statistical_Fig3Right}
\end{figure}

The functional properties of granular materials are related in a nontrivial way to the complex manner in which particles interact with one another and to the spatial scales (particle, chain, domain, and bulk) and time scales over which those interactions occur. For example, pairs of particles can exert force on one another in a local neighborhood. However, as particles push on adjacent particles, the combined effect can transmit forces over long distances via important mesoscale structures commonly called \emph{force chains} \citep{Liu:1995aa,Mueth:1998a}.
The idea of networks has been invoked for many years to help provide a quantitative understanding and explanation of force-chain organization \citep{Coppersmith:1996a,Claudin:1998a,Sexton:1999a, Socolar:2002a, Peters:2005aa}.
Broadly speaking, force chains form a network of filamentary-like structures that are visually apparent in images from experiments, like the one shown in Fig.~\ref{f:Behringer2014Statistical_Fig3Right}. In such images, the brighter particles carry larger forces \citep{Liu:1995aa, Howell1999}. Furthermore, force chains tend to align preferentially along the principal stress axes \citep{Majmudar:2005aa}. It can be helpful to think of a force-chain network as the backbone of strong forces that span a system, providing support for both static \citep{Geng:2001a} and dynamic \citep{Howell1999} loading. However, weaker forces can also play a stabilizing role, much as guy-wires do on an aerial tower \citep{Radjai:1998aa, cates1999jamming}.

It is also possible for sets of particles to cluster together into larger geographical domains, with potentially distinct properties, that can have weak structural boundaries between them \citep{bassett2012influence}. At the largest scale, granular materials as a whole exhibit bulk properties, such as mechanical stability or instability in response to sheer or compression \citep{richard2005slow}. All of the aforementioned spatial scales are potentially relevant for understanding phenomena such as transmission of acoustic waves \citep{Owens:2011}, thermal conductivity and heat transfer \citep{Smart:2007a}, electrical properties \cite{Gervois:1989a}, and more. The time scales of interactions in granular materials are also important, and they can vary over many orders of magnitude. For example, in systems under compression, statistical fluctuations of grain displacements depend fundamentally on the length of the strain step (i.e., ``increment'') over which one makes measurements, as fluctuations over short windows are consistent with anomalous diffusion and those over longer windows are consistent with Brownian behavior \citep{combe2015experimental}.

The principled study of such diverse characteristics and organization in a single system can be very challenging, and the development of appropriate physical, mathematical, statistical, and computational models is important to attain a mechanistic understanding of granular materials. Traditionally, it has been common to model granular materials using either particulate-based or continuum-based frameworks \citep{cates1999jamming}. However, both of these approaches are often implicitly agnostic to intermediate-scale organization, which is important for understanding both static granular packings \cite{bassett2015extraction} as well as granular dynamics  \cite{Herrera:2011aa}. Recently, tools from network science \citep{newman2010networks,Bollobas:1998a} and related mathematical subjects --- which include approaches that can account explicitly for mesoscale structures \citep{Fortunato2016,Porter2009,Fortunato2010,csermely2013} --- have been used successfully to study properties of granular materials across multiple spatial and temporal scales. The most common representation of a network, an important idea for the study of complex systems of interacting entities \citep{newman2011complex}, is as a graph \citep{Bollobas:1998a}. A graph consists of a set of nodes (to represent the entities) and a set of edges, each of which represents an interaction between a pair of entities (or between an entity and itself). Increasingly, more complicated network representations (such as multilayer networks \citep{kivela2014multilayer}) are also employed. Moreover, there is also an increasing recognition that it is important to consider the impact of other features, such as spatial embedding and other spatial effects \citep{barth2011}, on network structure and dynamics, rather than taking an approach that promises that ``one size fits all.'' Network science offers methods for quantitatively probing and analyzing large, interacting systems whose associated networks have heterogeneous patterns that defy explanations attained by considering exclusively all-to-all, regular, or lattice-like interactions \citep{newman2010networks}.

There are several open problems in granular physics that may benefit from network-science approaches. In particular, because granular materials have multiple relevant length and time scales \citep{Howell1999,Hidalgo2002,candelier2009building,mehta2008heterogeneities,Keys:2007a}, it can be challenging to model and quantify their structural organization, material properties, and responses to external loads \citep{Digby:1981,Velicky:2002,Goddard:1990,Makse-1999-WEM,Goldenberg-2005-FEE}. However, although complex, the pairwise inter-particle interactions that underlie and govern the structure and behavior of granular systems (and other particulate matter) render them amenable to various network representations and network-based analyses. \citet{Smart:2008b} were among the first to explicitly suggest and formalize the use of ideas from network science to begin to study some of the difficult questions in granular physics. In their paper, they highlighted the ability of a network-based perspective to complement traditional methods for studying granular materials and to open new doors for the analysis of these complex systems. One place in which network analysis may be especially useful is in quantifying how local, pairwise interactions between particles in a granular packing yield organization on larger spatial scales (both mesoscale and system-level). For example, in sheared or compressed granular packings, such organization can manifest as force chains or other intermediate-sized sets of particles that together comprise a collective structure. Network science provides approaches to extract and quantitatively characterize heterogeneous architectures at microscale, mesoscale, and macroscale sizes, and one can use these methods to understand important physical phenomena, including how such multiscale organization relates to bulk material properties or to spatial and temporal patterns of force transmission through a material.

Network-based approaches should also be able to provide new insights into the mechanisms that govern the dynamics of granular materials. For example, as we will discuss, network analysis can helpfully describe certain aspects of complex dynamics (such as granular flows), and can provide quantitative descriptions of how the structure of a dense granular material evolves as a system deforms under external loads (such as those induced by compression, shear, tapping, or impact). It seems sensible to use a network-based approach when a network is changing on temporal scales slower than the time that it takes for information to propagate along it. We also expect ideas from temporal networks \cite{Holme2011} or adaptive networks \cite{thilo-adaptive} to be fruitful for studying faster dynamics, and investigation of granular dynamics in general should benefit from the development of both novel network representations and methods of network analysis that are designed specifically to understand temporally evolving systems. Another important problem in the study of granular materials is to predict when and where a granular system will fail. There has been some progress made in this area using network-based approaches, but it is important to continue to develop and apply tools from network analysis and related areas to gain a deeper understanding of which network features regulate or are most indicative of eventual failure. Another exciting direction for future work is to combine network-based approaches with questions about material design. In particular, can one use network-based approaches to help engineer granular systems --- or other materials that are amenable to network representations --- with desired and specialized properties?

It is also important to note that network-based representations and methods of analysis can provide insightful descriptions of granular materials with various additional complexities, such as systems that are composed of differently-shaped particles, 3-dimensional (3D) materials, and so on. This flexibility makes the application of tools from network science a powerful approach for studying the structural properties and dynamics of granular networks. Such a framework also allows one to compare network architectures in diverse situations, such as between simulations and experiments, across systems that are composed of different types of particles or exposed to different loading conditions, and more. Exploiting these capabilities will yield improved understanding of which properties and behaviors of granular materials are general, versus which are specific to various details of a system. With the continued development of physically-informed network-analysis tools, network-based approaches show considerable promise for further development of both qualitative and quantitative descriptions of the organization and complex behavior of granular materials.

The purpose of our paper is to review the nascent application of network theory (and related topics) to the study of granular materials. We begin in Sec.~\ref{s:networks} with a mathematical description of networks. In Sec.~\ref{s:network_measures}, we briefly review a set of measures that one can calculate on graphs and which have been useful in past investigations of granular materials. In Sec.~\ref{s:granular_networks}, we review several different ways in which granular materials have been represented as networks, and we discuss investigations of such networks to quantify heterogeneous, multiscale organization in granular materials and to understand how these systems evolve when exposed to external perturbations. We also point out insights into the underlying physics that have resulted from network-based investigations of granular matter. We close in Sec.~\ref{open_problems} with some thoughts on the many remaining open questions, and we describe a few specific future directions that we feel are important to pursue. We hope that our review will be helpful for those interested in using tools from network science to better understand the physics of granular systems, and that it will spur interest in using these techniques to inform material design.  

\section{Network construction and characterization}

\subsection{What is a network?}
\label{s:networks}
\begin{figure}[t]
\centering
\includegraphics[width = \textwidth]{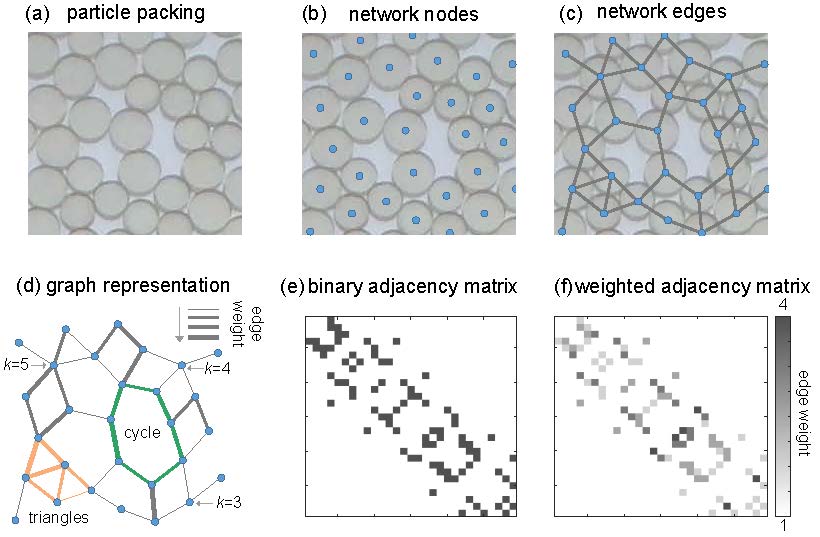}
\caption{\textbf{From a packing to a network.} \emph{(a)} A sample packing of grains. \emph{(b)} Representation of particles as network nodes. \emph{(c)} Representation of contacts as network edges. \emph{(d)} Graph representation of nodes and edges. The edges highlighted in green illustrate a cycle, which is a loop of physical contacts. The edges highlighted in peach illustrate a set of triangles, which are minimally-rigid structures in the context of structural rigidity. Edge weights often represent contact forces, which we illustrate here with different line widths. The degree $k$ of a node is equal to the number of edges attached (i.e., \emph{incident}) to that node, and the strength (i.e., weighted degree) $s$ of a node is given by the sum of weights of the edges attached to that node. One can \emph{(e)} encode an unweighted (i.e., binary) graph as an unweighted adjacency matrix and \emph{(f)} encode a weighted graph as a weighted adjacency matrix.
}
\label{f:packing}
\end{figure}

It is often useful to model a complex system as a network, the simplest type of which is a graph \citep{Newman:2006aa, newman2010networks}. For most of our article, we will use the terms \emph{network} and \emph{graph} synonymously, but the former concept is more general than the latter.\footnote{Indeed, it is increasingly important to examine network representations that are more complicated than graphs (see Secs.~\ref{alt_methods} and \ref{s:homology}) --- such as multilayer networks \citep{kivela2014multilayer}, simplicial complexes \citep{giusti2016twos}, and others --- and it is also essential to study dynamical processes on networks, rather than focusing exclusively on structural characteristics \citep{porter2016dynamical}.}  A graph $G$ consists of \emph{nodes} (i.e., vertices), where pairs of nodes are adjacent to each other via \emph{edges} (i.e., links). 
We denote the set of nodes by $\mathcal{V}$ and the set of edges by $\mathcal{E}$. 
A node can also be adjacent to itself via a \emph{self-edge} (which is also sometimes called a \emph{self-loop}), and a \emph{multi-edge} describes the presence of two or more edges that are attached to (i.e., \emph{incident} to) the same pair of nodes. (Unless we state otherwise, we henceforth assume that our networks have neither self-edges nor multi-edges.) The number of nodes in a graph is the \emph{size} of the graph, and we also use the word ``size'' in the same way for other sets of nodes. A \emph{subgraph} of a graph $G$ is a graph constructed using a subset of $G$'s nodes and edges.

An edge between two nodes represents some sort of relationship between them. For example, edges can represent information flow between different parts of the internet \citep{liben2008tracing}, friendship or other social interactions between people \citep{Scott:2012a}, trading between banks \citep{hurd2017framework}, anatomical or functional connections between large-scale brain regions \citep{Sporns:2013aa,Bassett:2017a}, physical connections between particles in contact \citep{Walker:2010aa, bassett2012influence}, and so on. Edges can be either unweighted or weighted, and they can be either undirected or directed \citep{newman2010networks}. In an unweighted (i.e., \emph{binary}) network, an edge between two nodes is assigned a binary value (traditionally $0$ or $1$) to encode the absence or presence of a connection. In a weighted network, edges can take a variety of different values to convey varying strengths of relationships between nodes. In an undirected network, all edges are bidirectional, so one assumes that all relationships are reciprocal. In a directed network, however, edges have a direction that encodes a connection from one node to another.

An \emph{adjacency matrix} is a useful way to represent the information in a graph. For an unweighted and undirected graph, an element of the adjacency matrix $\mathbf{A}$ of an $N$-node network is 
\begin{equation}
	A_{ij} = \left\{\begin{array}{ll}
	         1\,, \text{ if there is an edge between nodes $i$ and $j$\,,} \\
	       	0\,, \text{ otherwise\,,}
	        \end{array} \right.
\end{equation}
where $i,j \in \{1, \dots, N\}$. For a network in which nodes do not have labels, one can apply a permutation to ${\bf A}$'s rows and columns --- one uses the same permutation for each --- to obtain another adjacency matrix that represents the same network. In this paper, when we refer to ``the'' adjacency matrix ${\bf A}$ of a graph with unlabeled nodes, we mean any one of these matrices, which have the same properties (spectra, etc.).

For a weighted graph, if nodes $i$ and $j$ are adjacent via an edge, we denote the corresponding edge weight by $w_{ij}$ (which is usually given by a nonnegative real number (e.g., the value of the normal or tangential component of the force between two contacting particles)\footnote{We do not consider edges with negative weights, although it may be interesting to do so in future work if there is an appropriate physical reason.}). An element in the associated weighted adjacency matrix (which is sometimes called a \emph{weight matrix}) $\mathbf{W}$ is
\begin{equation}
	W_{ij} = \left\{\begin{array}{ll}
	         w_{ij}\,, \text{ if there is an edge between nodes $i$ and $j$\,,} \\
	       	0\,, \text{ otherwise\,.}
	        \end{array} \right.\
\end{equation}

For the more general case of a weighted, directed graph, if there is an edge from node $j$ to node $i$, then we let $w_{ij}$ represent the weight of that edge \citep{newman2010networks}. The associated weighted and directed adjacency matrix $\mathbf{W}$ is 
\begin{equation}
	W_{ij} = \left\{\begin{array}{ll}
	         w_{ij}\,, \text{ if there is an edge from node $j$ to node $i$\,,} \\
	       	0\,, \text{ otherwise\,.}
	        \end{array} \right.\
\end{equation}

An adjacency matrix associated with an undirected network is symmetric, but an adjacency matrix for a directed network need not be (and is symmetric if and only if all directed edges are reciprocated). In the present review, we primarily consider undirected networks, although we will occasionally make remarks about directed situations.

For weighted graphs, it is often also important to consider a binary adjacency matrix $\mathbf{A}$ associated with a weight matrix $\mathbf{W}$. Note that $\mathbf{A}$ captures only the \textit{connectivity} of nodes (i.e., their adjacencies), irrespective of how strongly they interact with each other. In terms of $\mathbf{W}$, the corresponding binary network (which can be either directed or undirected) is
\begin{equation}
	A_{ij} = \left\{\begin{array}{ll}
	         1\,, \text{ if } W_{ij} \neq 0\,, \\
	       	0\,, \text{ otherwise\,.}
	        \end{array} \right.\,
\end{equation}
It is common to use terms like \emph{network topology} when discussing structural properties of $\mathbf{A}$, and sometimes one uses terms like \emph{network geometry} when discussing properties that also depend on edge weights. Because we will also employ ideas from subjects like algebraic topology (see Sec.~\ref{s:homology}), we will need to be very careful with such terminology.

The network representations that have been used to study granular matter (and other kinds of materials) employ diverse definitions of edges (both weighted and unweighted, and both directed and undirected), and some generalizations of graphs have also been considered. See Fig.~\ref{f:packing} for a schematic showing possible choices of nodes and edges for a network representation of a granular packing. A variety of tools and measures from network analysis have been used to study granular networks. We discuss some of these ideas in Sec.~\ref{s:network_measures}.

\subsection{Some tools for characterizing granular networks}
\label{s:network_measures}

Network theory \citep{newman2010networks} provides myriad ways to characterize and quantify the topological and geometrical organization of complex networks. Thus, different network methods can reveal different important features of the underlying system, and these features, in turn, can help explain how a system behaves in certain situations. In the context of granular matter, for example, it is often desirable to understand the stability of a material, mechanical responses to external stresses, or wave propagation through a system. Recent investigations have demonstrated that network analysis can inform understanding of the mechanisms that underlie these phenomena. In this section, we discuss several network concepts; and in Sec.~\ref{s:granular_networks}, we describe how they have been used for the study of granular materials. We are, of course, not presenting anything close to an exhaustive list of tools from network science. See \citep{newman2010networks} and other books and reviews (and references therein) for discussions of other tools from network science. For simplicity, we primarily give definitions for undirected networks, though many of the ideas that we present also have directed counterparts. We start with basic network diagnostics, and also discuss some more complicated methods.

\subsubsection{Degree.}
\label{degree}

One local property of a network is node \emph{degree}. In an undirected network, a node's degree is equal to the number of edges that are attached to it (see Fig.~\ref{f:packing}\emph{d}). We denote the degree of node $i$ as $k_{i}$, and we recall that $N$ denotes the total number of nodes. For an unweighted graph with adjacency matrix ${\bf A}$, one can calculate $k_{i}$ with the formula
\begin{equation}\label{eq:node_degree}
	k_{i} = \sum_{j=1}^{N} A_{ij}\,.
\end{equation} 

One can generalize the idea of degree to \emph{strength} (i.e., weighted degree) using a weight matrix $\mathbf{W}$ \citep{Barrat:2004aa,Newman:2004aa}. The strength $s_i$ of node $i$ is equal to the sum of the weights of the edges that are attached to that node: 
\begin{equation}\label{eq:node_strength}
	s_{i} = \sum_{j=1}^{N} W_{ij}\,.
\end{equation} 
Its associated degree $k_i$ is still given by Eq.~\eqref{eq:node_degree}.

A common network representation of granular materials is to treat particles as nodes and physical contacts between particles as either unweighted or weighted edges (see Fig.~\ref{f:packing}\emph{a}--{c}). In this representation, node degree and node strength quantify information at the scale of single particles. 

One can compute the mean degree (a global property) of a network by calculating
\begin{equation}\label{eq:avg_degree}
	\langle{k}\rangle = \frac{1}{N}\sum_{i} k_{i}\,,
\end{equation}
and one can similarly compute the mean strength with the formula
\begin{equation}\label{eq:avg_strength}
	\langle{s}\rangle = \frac{1}{N}\sum_{i} s_{i}\,.
\end{equation}
In an undirected network, mean degree is related to $N$ and the total number $m$ of edges through the relation $\langle{k}\rangle = \frac{2m}{N}$. It is sometimes useful to characterize a network using its degree distribution $P(k)$ \citep{newman2010networks}, which gives the probability that a node has degree $k$.

When one represents a granular packing as a contact network (see Fig.~\ref{f:packing} and Sec.~\ref{s:contact_network}), which is a binary network (i.e., unweighted network), the degree $k_i$ of node $i$ is known more commonly in the physics literature as its \emph{contact number} or \emph{coordination number} $Z_i$. If every node has the same degree (or even if most nodes have this degree), such as in a regular lattice, one often refers to the mean coordination number $Z$ of the lattice or packing. It is well-known that coordination number is related to the stability of granular packings \citep{Alexander:1998aa} and plays a critical role in the jamming transition \citep{Wyart:2005a,Liu:2010aa,VanHecke2010}, a change of phase from an underconstrained state to a rigid state that is characterized by the onset of mechanical stability.\footnote{Unless we note otherwise, we use the phrase \emph{jamming} in the formal sense of the jamming transition as defined by \citep{Liu:2010aa,VanHecke2010}. Packings of particles above the jamming point (a critical point related to the jamming transition) are rigid and overconstrained (i.e., ``hyperstatic''), those at this point are marginally stable and exactly constrained (i.e., ``isostatic''), and those below this point are underconstrained (i.e., ``hypostatic''). Additionally, packings below the jamming point are sometimes called ``unjammed'', and those above the jamming point are called ``jammed''.} We discuss these ideas further throughout the review. 

\subsubsection{Walks and paths.} 
\label{paths}

In a network, a \emph{walk} is an alternating sequence of nodes and edges that starts and ends at a node, such that consecutive edges are both incident to a common node. A walk thus describes a traversal from one node to another node (or to itself) along edges of a network. A \emph{path} is a walk that does not intersect itself or visit the same node (or the same edge) more than once, except for a \emph{closed path}, which starts and ends at the same node (see Sec.~\ref{cycles}). One can compute the number of (unweighted) walks of a given length from a binary, undirected adjacency matrix ${\bf A}$ \citep{newman2010networks}. The length $l$ of an unweighted walk is defined as the number of edges in the associated sequence (counting repeated edges the number of times that they appear). Letting $\Xi^{l}_{ij}$ denote the number of walks of length $l$ between nodes $i$ and $j$, one calculates
\begin{equation}\label{eq:num_paths}
	\Xi^{l}_{ij} = [\mathbf{A}^{l}]_{ij}\,.
\end{equation}
Various types of random walks yield short paths between nodes in a network, and such ideas (and their relation to topics such as spectral graph theory) are very insightful for studying networks \cite{masuda2016}.

In an undirected network, a path from node $i$ to node $j$ is necessarily also a path from node $j$ to node $i$. However, this is not typically true in directed networks, which have sometimes been utilized in studies of granular force chains (see, e.g., \citep{Socolar:2002a}). Depending on a network's structure, there may be several or no paths between a given pair of nodes. An undirected network is called \emph{connected} if there exists a path between each pair of nodes, and a directed network is called \emph{strongly connected} if there is a path between each pair of nodes \citep{newman2010networks}. (A directed network is called \emph{weakly connected} if its associated undirected network is connected, and a strongly connected network is necessarily also weakly connected.) In networks of granular packings, both the existence and lengths of paths can impact system behavior. A connected network consists of a single \emph{component}. When a network has multiple components, it is common to study one component at a time (e.g., focusing on a \emph{largest connected component} (LCC), which is one that has the largest number of nodes).

The length of an unweighted path is the number of edges in the associated sequence, and it is sometimes also called the \emph{hop distance} (or occasionally, unfortunately, the \emph{topological distance}). Paths in a network can also be weighted by defining some (possibly abstract) notion of distance associated with the edges of the network. For example, in a spatially-embedded network \citep{barth2011}, distance may refer to actual physical distance along an edge, in which case the length of a weighted path in the network is given by the sum of the physical distances along the sequence of edges in the path. However, one can also consider ``distance" more abstractly. For example, in a transportation or flow network, one can define a distance between two adjacent nodes to be some measure of resistance between those nodes, and then the length of a weighted path in such a network is given by the sum of the resistances along the sequence of edges in the path.\footnote{One can also calculate distances between nodes if they occupy positions in a metric space (such as a latent one that determines the probability that a pair of nodes is adjacent to each other) \cite{boguna2010}, and the properties of that metric space can influence the distances (namely, the ones along a network) that concern us.}

We use the term \emph{network distance} to indicate a distance between two nodes (which can be either unweighted or weighted) that is computed by summing along edges in a path. A \textit{geodesic path} --- i.e., a shortest path (which need not be unique) --- between two nodes can be particularly relevant (though other short paths are often also relevant), and a \textit{breadth-first search} (BFS) algorithm \citep{skiena2008algorithm} is commonly employed to find geodesic paths in a network. The \textit{diameter} of a graph is the maximum geodesic distance between any pair of nodes. 

Denoting the shortest, unweighted network distance (i.e., shortest-path distance) between nodes $i$ and $j$ as $d_{ij}$, the mean shortest-path distance $L$ between pairs of nodes in a graph is \citep{Watts:1998aa}
\begin{equation}\label{eq:shortest_path}
	L = \frac{1}{N(N-1)}\sum_{i,j \,\, (i \neq j)} d_{ij} \,.
\end{equation}
Note that one must be cautious when computing the mean shortest-path distance on disconnected networks (i.e., on networks that have more than one component), because the usual convention is to set the distance between two nodes in different components to be infinite \citep{newman2010networks}. Therefore, in a network with multiple components, the distance $L$ from Eq.~\eqref{eq:shortest_path} is infinite. One solution to this problem is to compute $L$ for each component separately. Another network notion that relies on paths is \emph{network efficiency} \citep{latora2001efficient,Rubinov2009}
\begin{equation}
	E = \frac{1}{N(N-1)}\sum_{i,j\,\, (i \neq j)}\frac{1}{d_{ij}}\,.
\label{eq:E}
\end{equation}

One can generalize measures based on walks and paths to incorporate edge weights \citep{Latora2003,Rubinov2009}. Letting $d^{w}_{ij}$ denote the shortest, weighted network distance between nodes $i$ and $j$, one can define a weighted mean shortest-path distance $L^{w}$ and weighted efficiency $E^{w}$ as in Eqs.~\eqref{eq:shortest_path} and ~\eqref{eq:E}, respectively, but now one uses $d_{ij}^{w}$ instead of $d_{ij}$. The network efficiency $E$ is a normalized version of the Harary index of a graph \cite{estrada2016}. Additionally, the convention $d_{ij} = \infty$ (or $d^{w}_{ij} = \infty$) if there is no path from $i$ to $j$ allows one to use Eq.~\eqref{eq:E} (or its weighted counterpart $E^{w}$) on connected graphs or on graphs with more than one component. For both unweighted and weighted scenarios, large values of network efficiency tend to correspond to small values of mean shortest-path length, and vice versa. One can also readily generalize notions of paths, distances, and efficiency to directed networks \citep{latora2001efficient,Latora2003,Rubinov2009}.

In later sections, we will describe the use of paths, walks, and related ideas for investigating the structure of granular materials and their response to perturbations --- including, but not limited to, how these quantities change as a granular packing is compressed and goes through the jamming transition \citep{Arevalo:2010aa} --- and we will also describe their use in specific applications, such as in understanding heat transfer through a granular material \cite{Smart:2007a}.

\subsubsection{Cycles.} 
\label{cycles}

A \textit{cycle} (i.e., a \emph{closed walk}) in a network is a walk that begins and ends at the same node \cite{Bollobas:1998a}. As with other walks, one way to characterize a cycle is by calculating its length. An \emph{$l$-cycle} is a cycle in which $l$ edges are traversed (counting repeated edges the number of times that they appear in the cycle). A \emph{simple cycle} is a cycle that does not include repeated nodes or edges, aside from one repetition of the origin node at the termination of a closed cycle. Thus, for example, a simple 3-cycle in an undirected network is a triangle. For the remainder of this review, we assume that cycles are simple cycles, unless we explicitly state otherwise. In the context of a granular packing, one can directly map particle \textit{contact loops} --- sets of physically-connected grains arranged in a circuit --- to cycles in a corresponding graphical representation (see Fig.~\ref{f:packing}d). The length $l$ is odd for an \emph{odd cycle} and even for an \emph{even cycle}.

We briefly note a few related concepts that are used to examine cycles in graphs because of their relevance to several network-based studies of granular materials. These are the notions of \textit{cycle space}, \textit{cycle basis}, and \textit{minimum cycle basis} \cite{Bollobas:1998a,Gross:2005a}. The \textit{cycle space} of an undirected graph is the set of all simple cycles in a graph along with all subgraphs that consist of unions of edge-disjoint simple cycles (i.e., they can share nodes but not edges) \cite{Kavitha:2009a,Griffin:2017a}. A \textit{cycle basis} is a minimal set of simple cycles such that any element of the cycle space can be written as a symmetric difference of cycles in the cycle basis \cite{Kavitha:2009a}. Finally, for unweighted networks, a \textit{minimum cycle basis} is a basis in which the total length of all cycles in the basis is minimal. For weighted networks, it is a basis in which the sum of the weights of all cycles in the basis is minimal.

Minimum cycle bases can provide useful information about the structure and organization of cycles in a network, so several algorithms have been developed to extract them (see, for example, \cite{Horton:1987a,Mehlhorn:2006a}). Once one has determined a minimum cycle basis, one can examine the distribution of cycle lengths or define measures to quantify the participation of different nodes in cycles of different lengths. For example, \citet{Tordesillas:2014a,Walker:2015b} defined the concept of a \emph{cycle-participation vector} $X_{i}^{\mathrm{cycle}} = [x_{i}^{0},x_{i}^{3},\dots,x_{i}^{l}]$ for each node $i$. The elements of this vector count the number of cycles of each length in which node $i$ participates. In this definition, $x_{i}^{3}$ is the number of 3-cycles in which node $i$ participates, $x_{i}^{4}$ is the number of 4-cycles in which node $i$ participates, and so on (up to cycles of length $l$). If a node is not part of any cycle, then $x_{i}^{0} = 1$ and $x_{i}^{j} = 0$ for all $j \geq 3$; otherwise, $x_{i}^{0} = 0$.

One reason to examine cycles in granular networks \citep{Smart:2008aa,Arevalo:2009aa,Arevalo:2010ba,Arevalo:2010aa,Tordesillas:2010aa,Walker:2010aa} is that they can help characterize mesoscale structural features of a network. Cycles (that are nontrivial) involve more than a single node, but they do not typically embody global structures of a large network. This makes them appealing for studying network representations of granular materials, because mesoscale features seem to play a role in the behavior of these systems \citep{bassett2012influence}. Perhaps the most important motivation, however, is that cycles appear to be relevant for stability and rigidity of a system. Specifically, in the context of structural rigidity theory, 3-cycles tend to be stabilizing structures that can maintain rigidity under applied forces \citep{rivier2006extended}, whereas 4-cycles can bend or deform (see Sec.~\ref{rigidity}). In Sec.~\ref{s:granular_networks}, we discuss in more detail how cycles can help characterize granular systems.

\subsubsection{Clustering coefficients.}
\label{clustering_coefficient}

Clustering coefficients are commonly-used diagnostics to measure the density of triangles either locally or globally in a network \citep{newman2010networks}. For an unweighted, undirected network, the local clustering coefficient $C_{i}$ is usually defined as the number of triangles involving node $i$ divided by the number of triples centered at node $i$ \citep{Watts:1998aa,Newman:2003aa}. A triple is a set of three nodes that can include either three edges (to form a 3-cycle) or just two of them. In terms of the adjacency matrix and node degree, the local clustering coefficient is
\begin{equation}\label{eq:local_cluster}
	C_{i} = \frac{\sum_{hj} A_{hj} A_{ih} A_{ij}}{k_{i}(k_{i} - 1)}
\end{equation}
for $k_i \geq 2$ (and $C_{i} = 0$ if $k_i \in \{0,1\}$). One can then calculate a global clustering coefficient of a network as the mean of $C_{i}$ over all nodes:
\begin{equation}
\label{eq:global_cluster}
	C = \frac{1}{N} \sum_{i}^{N} C_{i}\,.
\end{equation}
There is also another (and simpler) common way of defining a global clustering coefficient in a network that is particularly useful when trying to determine analytical approximations of expectations over ensembles of random graphs \citep{Barrat:2000aa, newman2010networks}.
\begin{figure}[t]
\centering
\includegraphics[width=0.75\textwidth]{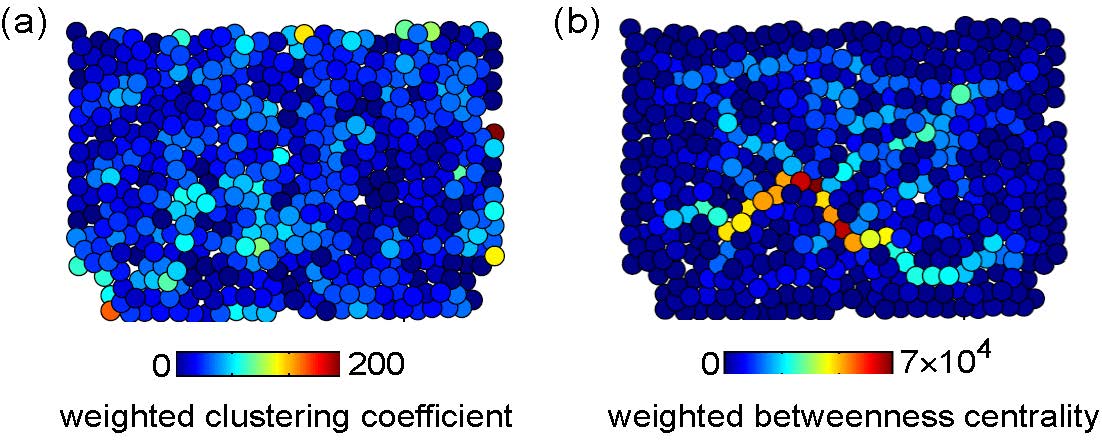}
\caption{\textbf{Network diagnostics reveal different types of structure in granular packings.} The two panels show the same compressed collection of photoelastic disks. One can measure the forces between particles and use the magnitude of the force between particle $i$ and particle $j$ to weight the edge between node $i$ and node $j$ in a weighted, undirected graph representation of the packing. \emph{(a)} One can compute a weighted clustering coefficient at each network node (i.e., particle) to probe local structure in the packing. \emph{(b)} One can compute a weighted betweenness centrality at each network node to probe how often its associated particle lies on a weighted shortest path between any two particles in the packing. [We adapted this figure, with permission, from \citep{bassett2012influence}.]
}
\label{f:diagnostics}
\end{figure}

The notion of a local clustering coefficient has also been extended to weighted networks in several ways \citep{jari2007,Barrat:2004aa,onnela2005,zhang2005general}. In one formulation \citep{Barrat:2004aa}, a local, weighted clustering coefficient $C^{w}_{i}$ is defined as
\begin{equation}
	C^{w}_{i} = \frac{1}{s_{i}(k_{i} - 1)} \sum_{j,h} \frac{(W_{ij} + W_{ih})} {2} A_{ij} A_{ih} A_{jh}
\end{equation}
for strength $s_i > 0$ and degree $k_i \geq 2$. The quantity $C^{w}_{i} = 0$ if either $s_i = 0$ (so that $k_i = 0$) or $k_i = 1$. 
Recall that $\mathbf{W}$ and $\mathbf{A}$ are, respectively, associated weighted and unweighted adjacency matrices. The mean of $C^{w}_{i}$ over all nodes gives a weighted clustering coefficient $C^{w}$ of a network. As we will discuss later (see Secs.~\ref{s:role_of_cycles} and \ref{s:force_threshold}), clustering coefficients have been employed in several studies of granular materials. For example, they have been used to examine stability in granular packings \citep{Arevalo:2009aa,Arevalo:2010ba,Arevalo:2010aa,Tordesillas:2010aa,Walker:2010aa}. See Fig.~\ref{f:diagnostics}\emph{a} for an example of the spatial distribution of a clustering coefficient in a granular packing.

\subsubsection{Centrality measures.}
\label{s:centrality}

In network analysis, one calculates centrality measures to attempt to quantify the importance of particular nodes, edges, or other structures in a network \citep{newman2010networks}. Different types of centralities characterize importance in different ways. The \emph{degree centrality} (i.e., degree) of a node, for example, is simply the number of edges attached to it (see Sec.~\ref{degree}). A few other types of centrality that have been used to study granular materials are closeness centrality, node betweenness centrality, edge betweenness centrality, and subgraph centrality.

Notions of \emph{closeness centrality} of a node measure how close that node is to other nodes in a network \citep{Freeman:1977aa}. For a given node $i$, the most standard notion of closeness is defined as the inverse of the sum over the shortest-path lengths from node $i$ to all other nodes $j$ in a network. That is, node $i$'s closeness centrality is
\begin{equation}
\label{eq:closeness}
	H_{i} = \frac{N-1}{\sum_{j \neq i} d_{ij}}\,.
\end{equation}
Note that if we use the convention that the distance between two nodes in different components is infinite, then Eq.~\eqref{eq:closeness} only makes sense for connected networks. For any network with more than one component, Eq.~\eqref{eq:closeness} yields a closeness centrality of $0$.

The \emph{geodesic node betweenness centrality} of node $i$ is the fraction of geodesic paths (either unweighted or weighted) between distinct nodes (not including $i$) that traverse node $i$ \citep{Freeman:1977aa}. Let $\psi_{gh}(i)$ denote the number of geodesic paths from node $g$ to node $h$ that traverse node $i$ (with $i \not \in \{g,h\}$), and let $\psi_{gh}$ denote the total number of geodesic paths from node $g$ to node $h$. The geodesic node betweenness centrality of node $i$ is then
\begin{equation}
\label{eq:betweenness}
	B_{i} = \sum_{g,h;\,g \neq h}\frac{\psi_{gh}(i)}{\psi_{gh}}\,, \quad i \not\in \{g,h\}\,.
\end{equation}

Geodesic node betweenness can probe the heterogeneity of force patterns in granular networks. See Fig.~\ref{f:diagnostics}\emph{b} for an example spatial distribution of a geodesic node betweenness centrality in an experimental granular packing. One can also compute a \emph{geodesic edge betweenness centrality} of an edge by calculating the fraction of shortest paths (either unweighted or weighted) that traverse it \cite{Girvan2002}. Let $\psi_{gh}(i,j)$ denote the number of geodesic paths from node $g$ to node $h$ that traverse the edge that is attached to nodes $i$ and $j$, and let $\psi_{gh}$ denote the total number of geodesic paths from node $g$ to node $h$. The geodesic edge betweenness centrality of this edge is then
\begin{equation}
\label{eq:edge_betweenness}
	B^{e}_{ij} = \sum_{g,h;\, g \neq h} \frac{\psi_{gh}(i,j)}{\psi_{gh}}\,.
\end{equation}

Another measure of node importance is \emph{subgraph centrality} $Y$ \citep{Estrada:2005aa,estrada2012}, which quantifies a node's participation in closed walks of all lengths. Recall from Sec.~\ref{paths} that one can write the number of length-$l$ walks from node $i$ to node $j$ in terms of powers of the adjacency matrix $\mathbf{A}$. To calculate closed walks of length $l$ that begin and end at node $i$, we take $i = j$ in Eq.~\eqref{eq:num_paths}. The subgraph centrality of node $i$, with a specific choice for how much we downweight longer paths, is then given by
\begin{equation}
\label{eq:subgraph_centrality_1}
	Y_{i} = \sum_{l=0}^{\infty} \frac{[\mathbf{A}^{l}]_{ii}}{l!}\,.
\end{equation}

Because shorter walks are weighted more strongly than longer walks in Eq.~\eqref{eq:subgraph_centrality_1}, they contribute more to the value of subgraph centrality. (In other contexts, centrality measures based on walks have also been used to compare the spatial efficiencies of different networks, and such ideas are worth exploring in granular materials \cite{estrada2016}.) One can also express subgraph centrality in terms of the eigenvalues and eigenvectors of the adjacency matrix \citep{Estrada:2005aa}. Let $v^{i}_{\alpha}$ denote the $i{\mathrm{th}}$ component of the $\alpha{\mathrm{th}}$ eigenvector $\bm{v}_{\alpha}$ of $\mathbf{A}$, and let $\lambda_\alpha$ denote the corresponding $\alpha{\mathrm{th}}$ eigenvalue. 

One can then write
\begin{equation}\label{eq:subgraph_centrality}
	Y_{i} = \sum_{\alpha=1}^{n} (v^{i}_{\alpha})^{2} e^{\lambda_\alpha}\,.
\end{equation}

One can then calculate a mean subgraph centrality $Y$ by averaging $Y_i$ over the nodes in a network. In one study of granular materials \citep{Walker:2010aa}, a subgraph centrality was examined for weighted networks by considering the eigenvalues and eigenvectors of the weight matrix $\mathbf{W}$ in Eq.~\eqref{eq:subgraph_centrality}. One can also compute \emph{network bipartivity} $R$ \citep{Estrada:2005ab} to quantify the contribution to mean subgraph centrality $Y$ from closed walks of even length. In particular, the network bipartivity $R_{i}$ of node $i$ is 
\begin{equation}
	R_{i} = \frac{{Y}^{\mathrm{even}}_{i}}{Y_{i}}\,,
\label{eq:bipartivity}
\end{equation} 
where $Y^{\mathrm{even}}_{i}$ is the contribution to the sum in Eq.~\eqref{eq:subgraph_centrality_1} from even values of $l$ (i.e., even-length closed walks). As with other node diagnostics, one can average bipartivity over all nodes in a network to obtain a global measure, which we denote by $R$. 

In Sec.~\ref{s:granular_networks}, we will discuss calculations of closeness, betweenness, and subgraph centralities in granular packings. Obviously, our discussion above does not give an exhaustive presentation of centrality measures, and other types of centralities have also been used in studies of granular materials (see, for example, \citep{bassett2012influence}).
	
\subsubsection{Subgraphs, motifs, and superfamilies.} 
\label{s:motifs}

\begin{figure}[t]
\centering
\includegraphics[width=0.75\textwidth]{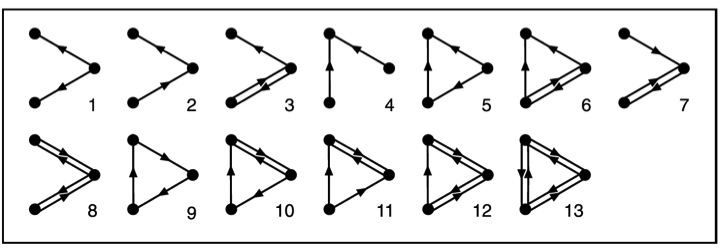}
\caption{\textbf{Subgraphs in networks.} We show all 13 different 3-node subgraphs that can occur in a directed, unweighted graph. A \emph{motif} is a subgraph that occurs often (typically relative to some null model) in a particular network or set of networks \citep{shenorr2002network}. [We reproduced this figure, with permission, from \citep{Milo:2002a}.]} 
\label{f:subgraphs}
\end{figure}

One can interpret the local clustering coefficient in Eq.~\eqref{eq:local_cluster} as a relationship between two small subgraphs: a triangle and a connected triple. Recall that a subgraph of a graph $G$ is a graph constructed using a subset of $G$'s nodes and edges. Conceptually, one can interpret small subgraphs as building blocks or subunits that together can be used to construct a network. For example, in a directed network, there exist three possible 2-node subgraphs (i.e., \emph{dyads}): the dyad in which node $i$ is adjacent to node $j$ by a directed edge, the dyad in which node $j$ is adjacent to node $i$ by a directed edge, and the dyad in which both of these adjacencies exist. In a directed, unweighted graph, there are 13 different connected 3-node subgraphs \citep{Milo:2002a} (see Fig.~\ref{f:subgraphs}). 

The term \emph{motif} is sometimes used for a small subgraph that occurs often in a particular network or set of networks (typically relative to some null model, such as a randomly rewired network that preserves the original degree distribution) \citep{shenorr2002network, Milo:2002a, milo2004superfamilies,alon2007network}. Borrowing terminology from genetics, these motifs appear to be \emph{overexpressed} in a network (or set of networks). Unsurprisingly, the number of $n$-node subgraphs increases very steeply with $n$, so identifying subgraphs in large networks is computationally expensive, and many algorithms have been developed to estimate the number of subgraphs in an efficient (though approximate) way. See, for example, \citep{Milo:2002a,schreiber2005frequency,wernicke2006efficient,grochow2007network,omidi2009moda,kashani2009kavosh}. In applying algorithms for motif counting to data, one seeks to identify subgraphs that are present more often than expected in some appropriate random-network null model. 

The over-representation of a motif in a network is often interpreted as indicative of its playing a role in the function of that network (though one has to be cautious about drawing such conclusions). For example, 3-node motifs can form feedforward loops in which there are directed edges from node $i_1$ to node $i_2$, from node $i_2$ to node $i_3$, and from node $i_3$ to node $i_1$. The identification and characterization of motifs has yielded insights into the structure and function of a variety of systems, including food webs \cite{paulau2015motif}, gene-regulation networks of yeast \citep{Milo:2002a}, neuronal networks of the macaque monkey \citep{sporns2004motifs}, and others. For different types of networks, one can also identify so-called \emph{superfamilies}, which are sets of networks that have similar motif-frequency distributions \citep{milo2004superfamilies}. There also exists a less-stringent definition of a superfamily in which one disregards whether a subgraph is a motif in the sense of it being more abundant than expected from some random-graph null model and instead considers a superfamily to be a set of networks that have the same rank-ordering of the number of $n$-node subgraphs for some fixed value of $n$ \cite{Xu:2008aa}. In either case, one can examine different superfamilies to help understand the role that specific motifs (or subgraphs) or sets of motifs (or subgraphs) may have in potentially similar functions of networks in a given superfamily. 

Subgraphs, motifs, and superfamilies have been examined in several studies that applied network analysis to granular materials \cite{Walker:2014aa,Walker:2015b,Walker:2015c,Tordesillas:2012aa}. They have revealed interesting insights into the deformation and reconfiguration that occurs in granular systems for different types of loading conditions and external perturbations. We discuss these ideas further in Secs.~\ref{subgraph} and \ref{kinematic}.

\subsubsection{Community structure.} 
\label{s:comm_structure}

Many real-world networks also have structure on intermediate scales (\emph{mesoscales}) that can arise from particular organizations of nodes and edges \citep{Fortunato2016,peixoto2017,Porter2009,Fortunato2010,csermely2013}. The most commonly-studied mesoscale network property is \textit{community structure} \citep{Fortunato2016,Porter2009}, which describes sets of nodes, called \textit{communities}, that are densely (or strongly) interconnected to each other but only weakly connected to other dense sets of nodes. In other words, a community has many edges (or large total edge weight, in the case of weighted networks) between its own nodes, but the number and/or weight of edges between nodes in different communities is supposed to be small. Once one has detected communities in a network, one way to quantify their organization is to compute and/or average various network quantities over the nodes (or edges) within each community separately, rather than over an entire network. For example, one can compute the size (i.e., number of nodes) of a community, mean path lengths between nodes in a given community, or some other quantity to help characterize the architecture of different communities in a network. Studying community structure can reveal useful insights about granular systems, whose behavior appears to be influenced by mesoscale network features \cite{bassett2012influence,bassett2015extraction,Giusti:2016a,papadopoulos2016evolution,Walker:2012aa,Tordesillas:2013a,Walker:2014ba,Walker:2012a}. 

Community structure and methods for detecting communities have been studied very extensively \cite{Fortunato2016}. We will briefly discuss the method of \textit{modularity maximization} \citep{NG2004,Newman2006b,newman2010networks}, in which one optimizes an (occasionally infamous) objective function known as \emph{modularity}, as this approach has been employed previously in several studies of granular materials (see, e.g., Sec.~\ref{s:comm_detect}). However, myriad other approaches exist for studying community structure in networks. These include stochastic block models (SBMs) and other methods for statistical inference (which are increasingly favored by many scholars) \cite{peixoto2017}, approaches based on random walks (e.g., InfoMap \citep{rosvall2008maps}), various methods for detecting \emph{local community structure} (see, e.g., \cite{clauset2005finding,jeub2015}), edge-based communities \citep{ahn2010link}, and many others.

The goal of modularity maximization is to identify communities of nodes that are more densely (or more strongly) interconnected with other nodes in the same community than expected with respect to some null model. To do this, one maximizes a modularity objective function
\begin{equation}\label{eq:one}
	Q = \sum_{i,j} [W_{ij} - \gamma P_{ij}] \delta(g_{i},g_{j})\,,
\end{equation}
where $g_{i}$ is the community assignment of node $i$ and $g_{j}$ is the community assignment of node $j$, and where the Kronecker delta $\delta(g_{i},g_{j})=1$ if $g_{i} = g_{j}$ and $\delta(g_{i},g_{j})=0$ otherwise. The quantity $\gamma$ is a resolution parameter that adjusts the relative average sizes of communities \citep{Good2010,Fortunato2007}, where smaller values of $\gamma$ favor larger communities and larger values of $\gamma$ favor smaller communities \citep{bassett2013robust}. The element $P_{ij}$ is the expected weight of the edge between node $i$ and node $j$ under a specified null model. In many contexts, the most common choice is to determine the null-model matrix elements $P_{ij}$ from the Newman--Girvan (NG) null model \citep{NG2004,Newman2006,Bazzi2016}, for which
\begin{equation}
	P^{\mathrm{NG}}_{ij} = \frac{s_{i} s_{j}}{2m}\,,
	\label{eq:ng_null}
\end{equation}	
where $s_{i}=\sum_j W_{ij}$ is the strength (and, for unweighted networks, the degree $k_i$) of node $i$ and $m=\frac{1}{2}\sum_{i,j} W_{ij}$ is the total edge weight (and, for unweighted networks, the total number of edges) in the network. There are several other null models, which are usually based on a random-graph model, and they can incorporate system features (such as spatial information) in various ways \cite{Sarzynska:2015aa}. In the next part of this subsubsection, we discuss a physically-motivated null model that is useful for studying granular force networks.
\begin{figure}[t]
\centering
\includegraphics[width=0.75\textwidth]{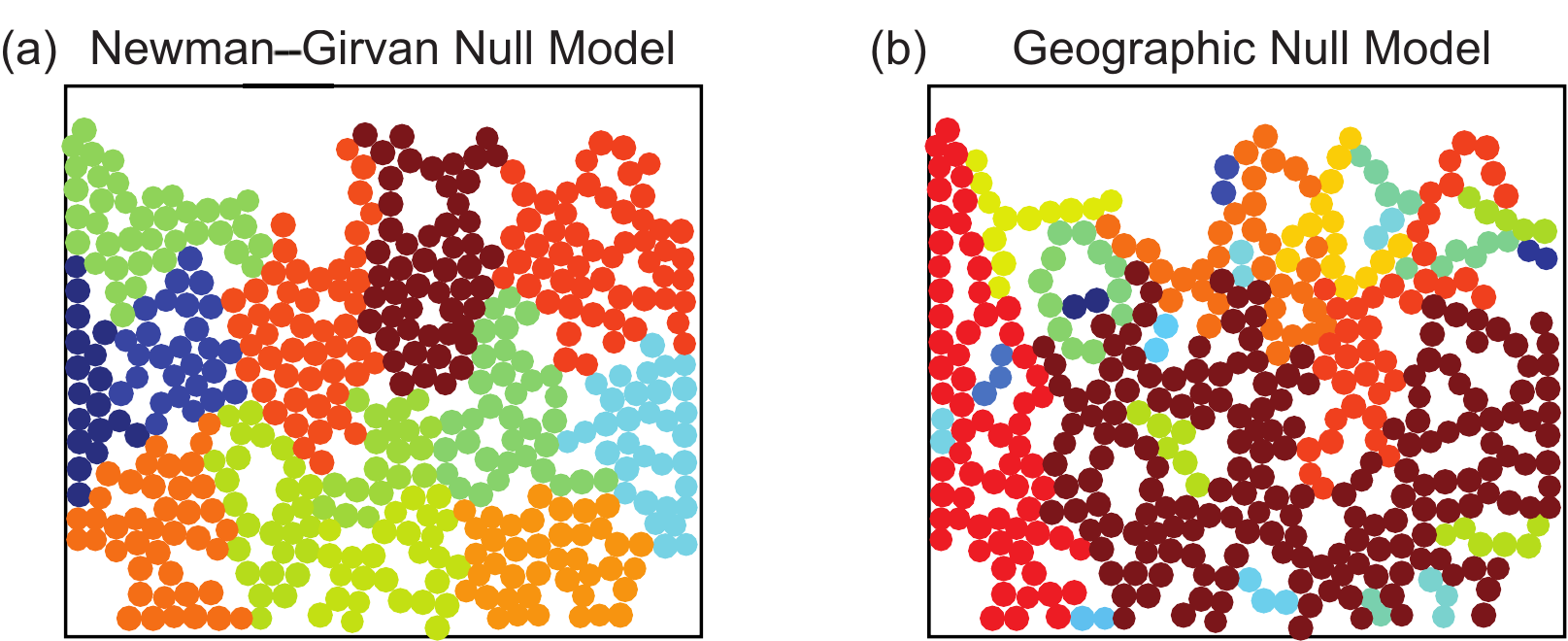}
\caption{\textbf{Modularity maximization with different null models can reveal distinct types of community structures in granular networks.} In a network representation in which particles are represented by nodes and contact forces are represented by weighted edges, \emph{(a)} using the Newman--Girvan null model helps uncover contiguous domains in a granular system, and \emph{(b)} using a geographical null model on the same network helps detect chain-like structures that are reminiscent of force chains. In both panels, nodes (i.e., particles) of the same color are assigned to the same community.}
\label{f:nullmodels}
\end{figure}

Maximizing $Q$ is NP-hard \citep{Brandes2008}, so it is necessary to use computational heuristics to identify near-optimal partitions of a network into communities of nodes \citep{Good2010}. Two well-known choices are the Louvain \citep{Blondel2008} and Louvain-like \cite{genlouvain2016} locally greedy algorithms, which begin by placing all nodes in their own community, and they then iteratively agglomerate nodes when the resulting partition increases modularity $Q$. Because of the extreme near degeneracy of the modularity landscape (a very large number of different partitions can have rather similar values of the scalar $Q$), it is often useful to apply such an algorithm many times to construct an ensemble of partitions, over which one can average various properties to yield a consensus partition \citep{bassett2013robust,Lancichinetti2012,Sarzynska:2015aa,jeub-santo2017}.

~\\
\noindent \emph{Physical considerations.} Community-detection tools, such as modularity maximization, have often been applied to social, biological, and other networks \cite{Porter2009,Fortunato2016}. In applying these techniques to granular materials, however, it is important to keep in mind that the organization of particulate systems (such as the arrangements of particles and forces in a material) is subject to significant spatial and physical constraints, which can severely impact the types of organization that can arise in a corresponding network representation of the material. When studying networks that are embedded in real space or constructed via some kind of physical relationship between elements, it is often crucial to consider the spatial constraints --- and, more generally, a system's underlying physics --- and their effects on network architecture \cite{barth2011}. Such considerations also impact how one should interpret network diagnostics such as path lengths and centrality measures, the null models that one uses in procedures such as modularity maximization, and so on. The NG null model was constructed to be appropriate for networks in which a connection between any pair of nodes is possible. Clearly, in granular materials --- as in other spatially-embedded systems \citep{barth2011} --- this assumption is unphysical and therefore problematic. 

Bassett et al. \citep{bassett2013robust} defined a null model that accounts explicitly for geographical (and hence spatial) constraints in granular materials, in which each particle can contact only its nearest neighbors \citep{bassett2015extraction}. In the context of granular networks with nodes representing particles and edges representing forces between those particles, the \emph{geographical null model} ${\bf P}$ in \citep{bassett2013robust} has matrix elements
\begin{equation}
\label{geog}
	P_{ij} = \rho A_{ij} \,,
\end{equation}
where $\rho$ is the mean edge weight in the network and ${\bf A}$ is the binary adjacency matrix of the network. In this particular application, $\rho = \overline{f} := \langle f_{ij} \rangle$ is the mean inter-particle force. As we illustrate in Fig.~\ref{f:nullmodels}, modularity maximization with the geographical null model [Eq.~\eqref{geog}] produces different communities than modularity maximization with the NG null model [Eq.~\eqref{eq:ng_null}] \citep{bassett2015extraction,Giusti:2016a,papadopoulos2016evolution}.  
\begin{figure}[t]
\centering
\includegraphics[width=0.75\textwidth]{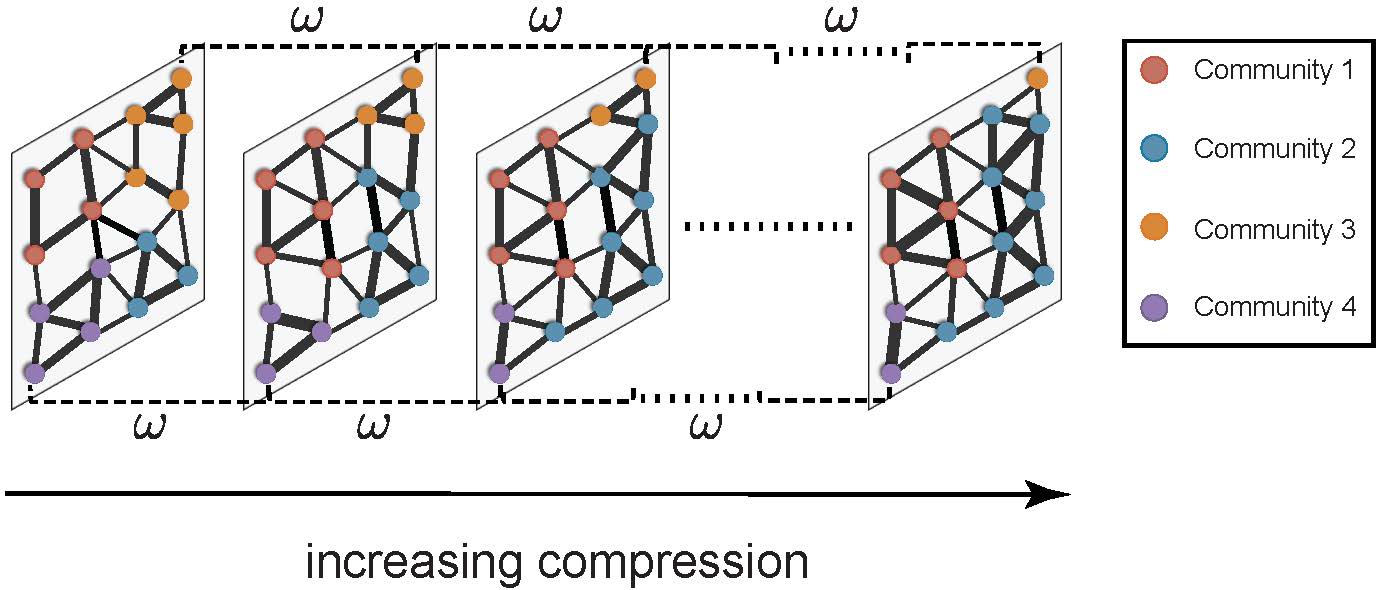}
\caption{\textbf{A schematic of a multilayer network with layer-dependent community structure.} In this example, each layer represents a static granular force network in which nodes (i.e., particles) are adjacent to one another via intralayer weighted edges (e.g., representing contact forces). Additionally, the same particle in consecutive layers is adjacent to itself via an interlayer edge of uniform weight $\omega$. For clarity, we only show two such couplings, but these interlayer edges exist for all particles (and between all consecutive layers). One can extract communities that change across layers --- e.g., if layers represent time, these are time-evolving communities --- to study mesoscale organization in a granular system and to help understand how it reconfigures due to external loading (such as compression). In this schematic, we use different colors to label the particles that belong to different communities. Note that the same community can persist across several (or all) layers and reconfigure in terms of its constituent particles and the mean strength of its nodes. [We reproduced this figure, with permission, from \citep{papadopoulos2016evolution}.]}
\label{f:multilayer_schematic}
\end{figure}

~\\
\noindent \emph{Generalization of modularity maximization to multilayer networks.} Although studying community structure in a given granular packing can provide important insights, one is also typically interested in how such mesoscale structures reconfigure as a material experiences external perturbations, such as those from applied compression or sheer. To examine these types of questions, one can optimize a \emph{multilayer} generalization of modularity to study multilayer granular force networks in which each layer represents a network at a different step in the evolution of the system (for example, at different time steps or for different packing fractions) \citep{papadopoulos2016evolution}. In Fig.~\ref{f:multilayer_schematic}, we show a schematic of a multilayer construction that has been employed in such investigations. See \cite{kivela2014multilayer,Boccaletti2014} for reviews of multilayer networks (including generalizations of this construction).

One way to detect multilayer communities in a network is to use a generalization of modularity maximization \citep{Mucha2010}, which was derived for multilayer networks with interlayer edges between counterpart nodes in different layers. For simplicity, we suppose that all edges are bidirectional. One maximizes
\begin{equation} \label{eq:dynamic_modularity}
	Q_{\mathrm{multi}} = \frac{1}{2 \eta} \sum_{ijqr} [(\mathcal{W}_{ijq} - \gamma_{q}\mathcal{P}_{ijq}) \delta_{qr} + \omega_{jqr}\delta_{ij}] \delta(g_{iq},g_{jr})\,,
\end{equation}

where $\mathcal{W}_{ijq}$ is the $(i,j){\mathrm{th}}$ component of the $q{\mathrm{th}}$ layer of the adjacency tensor $\mathcal{W}$ \cite{domenico2013mathematical} associated with the multilayer network, $\mathcal{P}_{ijq}$ is the $(i,j){\mathrm{th}}$ component of the $q{\mathrm{th}}$ layer of the null-model tensor, $\gamma_{q}$ is a resolution parameter (sometimes called a \emph{structural resolution parameter}) for layer $q$, and $\omega_{jqr}$ is the \textit{interlayer coupling} between layers $q$ and $r$. (In the context of multilayer representations of temporal networks, if $\mathcal{\omega}_{jqr} = \omega$ for all $j$, $q$, and $r$, one can interpret $\omega$ as a \textit{temporal resolution parameter}.)
More specifically, $\mathcal{\omega}_{jqr}$ is the strength of the coupling that links node $j$ in layer $q$ to itself in layer $r$. (This type of interlayer edge, which occurs between counterpart nodes in different layers, is called a \emph{diagonal} edge \cite{kivela2014multilayer}.) The quantities $g_{iq}$ and $g_{jr}$, respectively, are the community assignments of node $i$ in layer $q$ and node $j$ in layer $r$. The intralayer strength of node $j$ in layer $q$ is $s_{jq} = \sum_{i} \mathcal{W}_{ijq}$, and the interlayer strength of node $j$ in layer $q$ is $\zeta_{jq} = \sum_{r}{\mathcal{\omega}_{jqr}}$, so the multilayer strength of node $j$ in layer $q$ is given by $\kappa_{jq} = s_{jq} + \zeta_{jq}$. Finally, the normalization factor $\eta = \frac{1}{2} \sum_{jq} \kappa_{jq}$ is the total strength of the adjacency tensor.\footnote{In the study of multilayer networks, it is common to use the term ``tensor'' to refer to a multidimensional array \cite{kivela2014multilayer} (as is common in some disciplines \cite{Kolda2009Tensor}), and proper tensorial structures have been explored briefly in adjacency tensors \cite{domenico2013mathematical}.} 

Maximizing multilayer modularity [Eq.~\eqref{eq:dynamic_modularity}] allows one to examine phenomena such as evolving communities in time-dependent networks or communities that evolve with respect to some other parameter, and communities in networks with multiple types of edges. Capturing such behavior has been useful in many applications, including financial markets \citep{Bazzi2016}, voting patterns \cite{Mucha2010}, international relations \cite{Cranmer:2014ut}, international migration \cite{danchev2016}, disease spreading \cite{Sarzynska:2015aa}, human brain dynamics \citep{Bassett2011b,bassett2015learning,braun2015dynamic}, and more. In the context of granular matter, multilayer community detection allows one to examine changes in community structure of a force network, in which communities can either persist or reconfigure, with respect to both particle content and the mean strength of nodes inside a community, due to applied loads on a system.  

\subsubsection{Flow networks.} 
\label{s:max_flow}

One can examine many natural and engineered systems --- such as animal and plant vasculature, fungi, and urban transportation networks \citep{Blinder:2013aa, Katifori:2010a, shl2017, Bebber:2007a, Banavar:2000a, Newman:2006a, Kurant:2006a} --- from the perspective of \emph{flow networks} (which are often directed) that transport a load (of fluids, vehicles, and so on) along their edges. It is of considerable interest to examine how to optimize flow through a network \citep{Bertsekas:1998, newman2010networks}. A well-known result from optimization theory is the \emph{maximum-flow--minimum-cut theorem} \citep{Ahuja:flow,Bertsekas:1998,newman2010networks}: for a suitable notion of flow and under suitable assumptions, the maximum flow that can pass from a source node to a sink node is given by the total weight of the edges in the \emph{minimum cut}, which is the set of edges with smallest total weight that, when removed, disconnect the source and the sink. A related notion, which applies to networks in which there is some cost associated with transport along network edges, is that of \emph{maximum-flow--minimum-cost}. In this context, one attempts to find a route through a network that maximizes flow transmission from source to sink, while minimizing the cost of flow along network edges \citep{Ahuja:flow,Bertsekas:1998}. The maximum-flow--minimum-cut and maximum-flow--minimum-cost problems are usually examined under certain constraints, such as flow conservation at each node and an upper bound (e.g., limited by a capacitance) on flow through any edge. One can examine granular networks from such a perspective by considering a flow of force along a network formed by contacting grains. We discuss relevant studies in Sec.~\ref{s:flow_networks}. 
	
\subsubsection{Connected components and percolation.}
\label{s:largest_comp}

Sometimes it is possible to break a network into connected subgraphs called components (which we introduced briefly in Sec.~\ref{paths}). A \emph{component}, which is sometimes called a \textit{cluster}, is a subgraph $G_C$ of a graph $G$ such that at least one path exists between each pair of nodes in $G_C$ \citep{newman2010networks}. Components are maximal subsets in the sense that the addition of another node of $G$ to it destroys the property of connectedness. An undirected graph is connected when it consists of a single component. Networks with more than one component often have one component that has many more nodes than the other components, so there can be one large component and many small components. One can find the components of a graph using a breadth-first search (BFS) algorithm \citep{skiena2008algorithm}, and one can determine the number of components by counting the number of $0$ eigenvalues of a graph's combinatorial Laplacian matrix \citep{newman2010networks}. To study graph components, one can also use methods from computational algebraic topology. Specifically, the zeroth Betti number $\beta_{0}$ indicates the number of connected components in a graph \citep{Kaczynski:2004aa} (see Sec.~\ref{s:homology}). 

Percolation theory \citep{newman2010networks,kesten-whatis,Stauffer:1994a,saberi2015}, which builds on ideas from subjects such as statistical physics and probability theory, is often used to understand the emergence and behavior of connected components in a graph \citep{porter2016dynamical}. For example, in the traditional version of what is known as \emph{bond percolation} (which is also traditionally studied on a lattice rather than on a more general network) \citep{broadbent1957percolation}, edges are occupied with probability $p$, and one examines quantities such as the size distributions of connected components as a function of the parameter $p$, called the \textit{bond occupation probability}. It is especially interesting to determine a critical value $p_c$, called the \textit{percolation threshold}, at which there is a phase transition: below $p_{c}$, there is no \emph{percolating component} (or cluster), which spans the system and connects opposite sides; above $p_{c}$, there is such a cluster \citep{Albert2002,broadbent1957percolation}. In the latter scenario, it is common to say that there is a ``percolating network". In percolation on more general networks, one can study how the size of the largest component, as a fraction of the size of a network, changes with $p$. Related ideas also arise in the study of components in Erd\H{o}s--R\'{e}nyi random graphs $G(N,p)$, in which one considers an ensemble of $N$-node graphs and $p$ is the independent probability that an edge exists between any two nodes \citep{Bollobas:1998a,newman2010networks,Erdos:1959aa, Erdos:1960aa}. In the limit $N \rightarrow \infty$, the size of the largest connected component (LCC) undergoes a phase transition at a critical probability $p_{c} = 1/N$. When $p < p_{c}$, the ER graph in expectation does not have a \textit{giant connected component} (GCC); at $p = p_{c}$, a GCC emerges whose size scales linearly with $N$ for $p > p_{c}$. Similarly, for bond percolation on networks, a transition occurs at a critical threshold $p_c$, such that for $p > p_c$, there is a GCC (sometimes also called a ``giant cluster'' or ``percolating cluster'') whose size is a finite fraction of the total number $N$ of nodes as $N \rightarrow \infty$ \citep{newman2010networks,porter2016dynamical}. When studying percolation on networks, quantities of interest include the fraction of nodes in the LCC, the mean component size, the component-size distribution, and critical exponents that govern how these quantities behave just above the percolation threshold \citep{Albert2002,Stauffer1979a,newman2010networks}.

We will see in Sec.~\ref{s:granular_networks} that it can be informative to use ideas from percolation theory to study the organization of granular networks. For example, it is particularly interesting to examine how quantities such as the number and size of connected components evolve as a function of packing density (or another experimental parameter). \citep{Slotterback:2012aa,Herrera:2011aa,Arevalo:2010aa,Kondic:2012aa,Kramar:2013aa,Kramar:2014b,Kramar:2014aa,Ardanza-Trevijano:2014aa,Kondic:2016a,Pugnaloni:2016a}. Some studies have considered \emph{connectivity percolation} transitions, which are characterized by the appearance of a connected component that spans a system (i.e., a percolating cluster, as reflected by an associated GCC in the infinite-size limit of a network); or \emph{rigidity percolation transitions}, which can be used to examine the transition to jamming \citep{Feng1985,Moukarzel:1995a,Jacobs1995,Aharonov:1999a,Lois:2008a,Shen:2012a,Kovalcinova:2015a,Henkes2016}. Rigidity percolation is similar to ordinary bond percolation (which is sometimes used to study connectivity percolation), except that edges represent the presence of rigid bonds between network nodes \cite{Thorpe1985a,Thorpe:1999aa} and one examines the emergence of rigid clusters in the system as a function of the fraction of occupied bonds.  One can also study percolation in force networks by investigating the formation of connected components and the emergence of a percolating cluster of contacts as a function of a force threshold, which is a threshold applied to a \emph{force-weighted adjacency matrix} (representing contact forces between particles) to convert it to a binary adjacency matrix  \citep{Arevalo:2010aa,Kondic:2012aa,Kramar:2013aa,Kramar:2014b,Kramar:2014aa,Ardanza-Trevijano:2014aa,Kondic:2016a,Pugnaloni:2016a,Kovalcinova:2016a,Kovalcinova:2015a,PastorSatorras2012,pathak2017force}. (See Sec.~\ref{s:force_threshold} for additional discussion.) However, it is important to note that when studying networks of finite size, one needs to be careful with claims about GCCs and percolation phase transitions, which are defined mathematically only in the limit of infinite system size.

\subsubsection{Methods from algebraic topology and computational topology.} 
\label{s:homology}

\begin{figure}[t]
\centering
\includegraphics[width=0.75\textwidth]{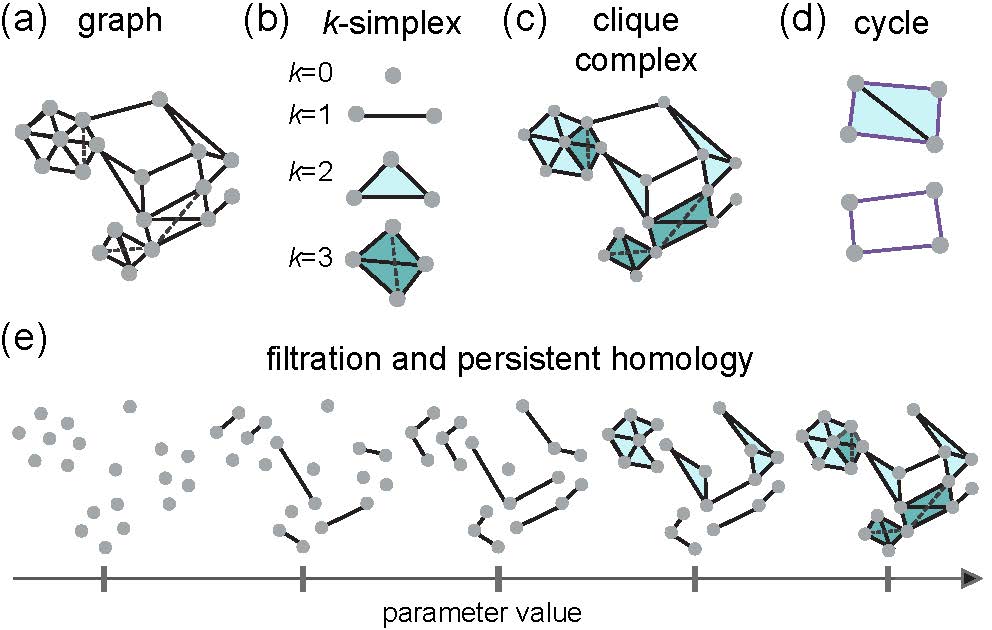}
\caption{\label{f:algebraic_topology} \textbf{Algebraic topology and clique complexes.} \emph{(a)} An interaction in a graph is a dyad (i.e., a 2-node subgraph) or a node and a self-edge. \emph{(b)} An alternative fundamental unit is a $k$-simplex. A 0-simplex is a node, a 1-simplex is an edge, a 2-simplex is a filled triangle, and so on. \emph{(c)} A collection of $k$-simplices is called a simplicial complex, and one type of simplicial complex that can be used to encode the information in a graph is a clique complex (sometimes also called a \emph{flag complex}). One constructs a clique complex by taking every $k$-clique (a complete subgraph of $k$ nodes) in a graph $G$ to be a simplex of the same number of nodes. \emph{(d)} An interesting feature that can occur in a simplicial complex is a cycle, which is a closed arrangement of a collection of $k$-simplices. The purple edges in the upper object indicate a 1-dimensional cycle that encloses a region filled in by simplices, whereas the purple edges in the lower object indicate a 1-dimensional cycle that encloses a hole. \emph{(e)} One can use a filtration to decompose a weighted graph into a sequence of binary graphs. For example, if one uses edge weight as a filtration parameter, one can represent a weighted graph as a sequence of unweighted graphs, which in turn yields a sequence of unweighted clique complexes. [We adapted this figure, with permission, from \citep{Bassett:2017a}.]}
\end{figure}

The tools that we have described thus far rely on the notion of a dyad (i.e., a 2-node subgraph) as the fundamental unit of interest (see Fig.~\ref{f:algebraic_topology}\emph{a}). However, recent work in algebraic topology and computational topology \citep{Edelsbrunner:2010a,Carlsson:2009a,ghrist2014elementary,Kaczynski:2004aa} offers a complementary view, in which the fundamental building blocks that encode relationships between elements of a system are \emph{$k$-simplices} (each composed of $k+1$ nodes), rather than simply nodes and dyadic relations between them (see Fig.~\ref{f:algebraic_topology}\emph{b}). These structures can encode ``higher-order'' interactions and can be very useful for understanding the architecture and function of real-world networks (e.g., they yield a complementary way to examine mesoscale network features), and they have been insightful in studies of sensor networks \citep{dlotko2012distributed}, contagion spreading \citep{taylor2015topological}, protein interactions \citep{sizemore2016classification}, neuronal networks \citep{giusti2016twos,sizemore2016cliques}, and many other problems. See \cite{otter2015,Patania2017} for further discussion and pointers to additional applications. The discussion in \cite{Stolz2016} is also useful.

\begin{figure}
\centering
\includegraphics[width=0.95\textwidth]{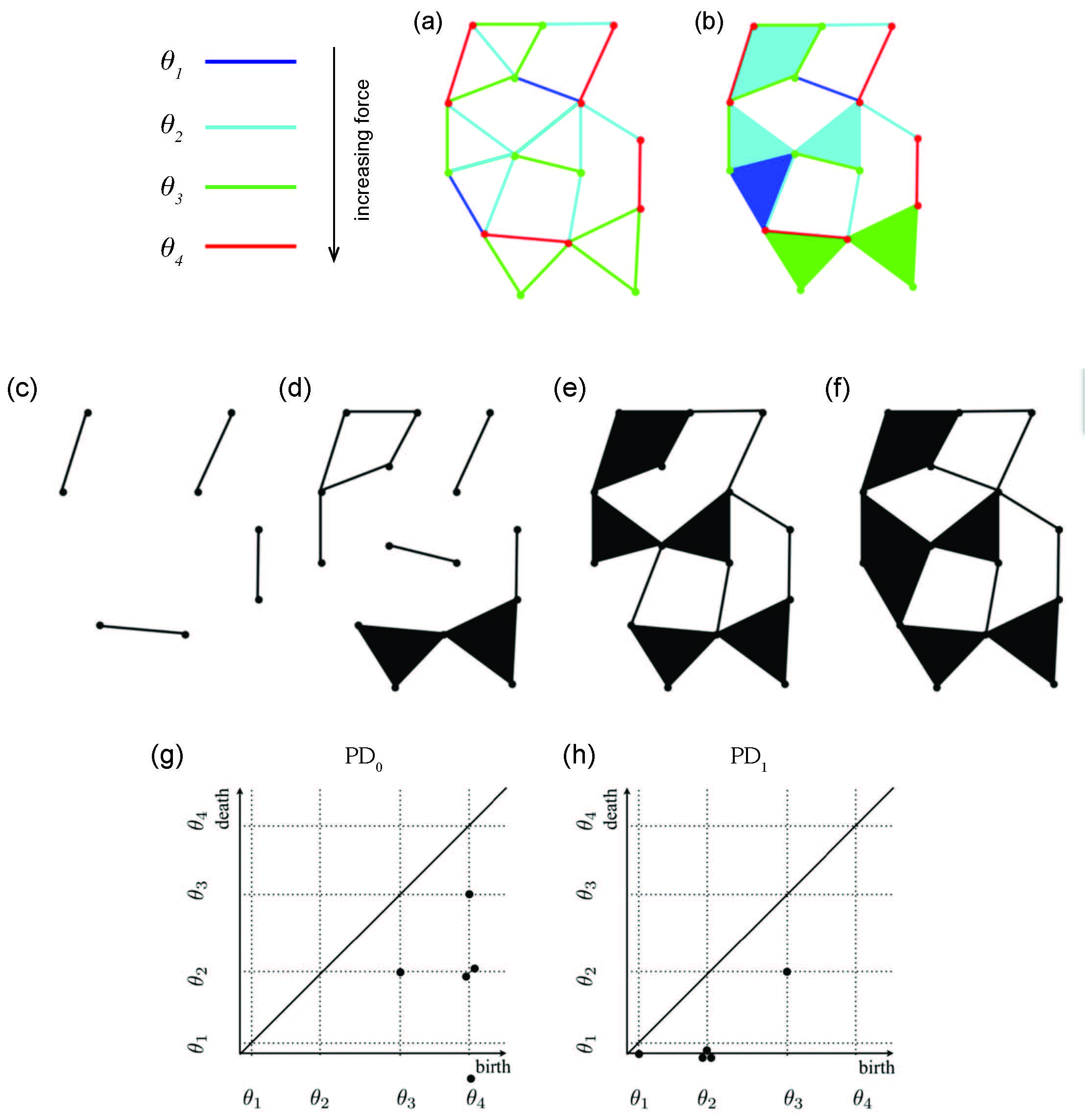}
\caption{\label{f:PH_schematic} \textbf{An example force network, an associated filtration over the flag complex (i.e., clique complex), and persistence diagrams.} \emph{(a)} In the force network, colored edges represent the magnitude of the force between contacting particles, which are represented as nodes in the network. In order from smallest to largest, the four values of the force are $\theta_{1}$ (dark blue), $\theta_{2}$, (cyan), $\theta_{3}$ (green), and $\theta_{4}$ (red). \emph{(b)} The flag complex is formed by filling in all 3-particle loops (i.e., triangular loops) with the smallest value of the force along any of its edges. Defining a filtration over the flag complex avoids counting these 3-particle loops. \emph{(c)--(f)} The sequence of complexes corresponding to the filtration over the flag complex; one obtains the sequence by descending the four levels of the force threshold $\theta$. \emph{(g)} The $\beta_{0}$ persistence diagram $\mathrm{PD}_{0}$. \emph{(h)} The $\beta_{1}$ persistence diagram $\mathrm{PD}_{1}$. [We adapted this figure, with permission, from \citep{Kramar:2014b}.]}
\end{figure}

A collection of simplices that are joined in a compatible way is called a \emph{simplicial complex}, which is a generalization of a graph that can encode non-dyadic relations \citep{ghrist2014elementary}. More precisely, and following \citep{giusti2016twos}, we define an \emph{(abstract) simplicial complex} $\mathcal{X}$ as a pair of sets: $V_{\mathcal{X}}$, called the \emph{vertices} (or \emph{nodes}); and $S_{\mathcal{X}}$, called the \emph{simplices}, each of which is a finite subset of $V_{\mathcal{X}}$, subject to the requirement that if $\sigma \in S_{\mathcal{X}}$, then every subset $\tau$ of $\sigma$ is also an element of $S_{\mathcal{X}}$. A simplex with $k$ elements is called a \emph{$(k-1)$-simplex}, and subsets $\tau \subset \sigma$ are called \emph{faces} of $\sigma$. Using this notation, a $0$-simplex is a node, a $1$-simplex is an edge and its two incident nodes (i.e., a dyad), a 2-simplex is a filled triangle, and so on (see Fig.~\ref{f:algebraic_topology}\emph{b}). One type of simplicial complex that can be used to encode the information in a graph is a \emph{clique complex} (sometimes also called a \emph{flag complex}); we show an example in Fig.~\ref{f:PH_schematic}. To construct the clique complex of a graph $G$, one associates every $k$-clique (a complete --- i.e., fully connected --- subgraph of $k$ nodes) in $G$ with a $(k-1)$-simplex. One can thus think of building the clique complex of a graph $G$ as ``filling in'' all of the $k$-cliques in $G$ (see Fig.~\ref{f:algebraic_topology}c). Note that we use the terms $k$-simplex and $k$-clique because they are standard, but it is important not to confuse the use of $k$ in this context with the use of $k$ as the (also standard) notation for node degree.

One important feature of a simplicial complex is the potential presence of \emph{cycles}.\footnote{Although we use the term \emph{cycle}, which is standard in algebraic topology, note that this concept of a cycle is distinct from (though related to) the standard network-science use of the word ``cycle'' (see Sec.~\ref{cycles}). The latter is sometimes called a \emph{circuit}, a term that we will use occasionally for clarity (especially given our focus on connected graphs).} A cycle can consist of any number of nodes, and a $k$-dimensional cycle is defined as a closed arrangement of $k$-simplices, such that a cycle has an empty boundary\footnote{The precise mathematical definition of a cycle requires a more detailed presentation than what we include in our present discussion. For more information and further details from a variety of perspectives, see \cite{ghrist2014elementary,otter2015,kozlov2007combinatorial,nanda2014simplicial,Kaczynski:2004aa,Edelsbrunner:2010a,sizemore2016classification,Stolz2016}.}.
For example, Fig.~\ref{f:algebraic_topology}\emph{d} illustrates a closed arrangement of $1$-simplices (i.e., edges) that forms a 1-dimensional cycle. It important to distinguish between cycles that encircle a region that is filled by simplices with cycles that enclose a void (which is often called a ``hole'' for the case of 1-dimensional cycles). For example, the set of purple edges in the object in the upper portion of Fig.~\ref{f:algebraic_topology}\emph{d} constitute a 1-dimensional cycle that surrounds a region filled by 2-simplices (i.e., filled triangles), whereas the purple edges in the object in the lower portion of Fig.~\ref{f:algebraic_topology}\emph{d} constitute a $1$-dimensional cycle that encloses a hole.
 
Characterizing the location and prevalence of void-enclosing cycles in the clique complex of a network representation of a granular packing can offer fascinating insights into the packing's structure \citep{Kondic:2012aa}. One way to do this is by computing topological invariants such as \emph{Betti numbers} \citep{Kaczynski:2004aa,Edelsbrunner:2010a,Kramar:2014b}. 
The $k{\mathrm{th}}$ Betti number $\beta_k$ counts the number of inequivalent $k$-dimensional cycles that enclose a void, where two $k$-dimensional cycles are \emph{equivalent} if they differ by a boundary of a collection of $(k+1)$-simplices. In other words, the $k{\mathrm{th}}$ Betti number $\beta_k$ counts the number of nontrivial \emph{equivalence classes} of $k$-dimensional cycles and can thus also be interpreted as counting the number of voids (i.e., ``holes" of dimension $k$).\footnote{In the literature, it is common to abuse terminology and refer to an equivalence class of $k$-dimensional cycles simply as a $k$-dimensional cycle.} The zeroth Betti number $\beta_{0}$ gives the number of connected components in a network, the first Betti number $\beta_{1}$ gives the number of inequivalent 1-dimensional cycles that enclose a void (i.e., it indicates loops), the second Betti number $\beta_{2}$ gives the number of inequivalent 2-dimensional cycles that enclose a void (i.e., it indicates cavities), and so on. 

Another useful way to examine the topological features that are determined by equivalence classes of $k$-dimensional cycles (i.e., components, loops, cavities, and so on) is to compute \emph{persistent homology} (PH) of a network. For example, to compute PH for a weighted graph, one can first decompose it into a sequence of binary graphs. One way to do this is to begin with the empty graph and add one edge at a time in order of decreasing edge weights (see Fig.~\ref{f:algebraic_topology}\emph{e}). More formally and following \citep{sizemore2016cliques}, this process can translate information about edge weights into a sequence of binary graphs as an example of what is called a \emph{filtration} \cite{ghrist2014elementary,otter2015}. The sequence $G_0 \subset G_1 \subset \dots \subset G_{|\mathcal{E}|}$ of unweighted graphs begins with the empty graph $G_0$, and one adds one edge at a time (or multiple edges, if some edges have the same weight) in order from largest edge weight to smallest edge weight. (One can also construct filtrations in other ways). Constructing a sequence of unweighted graphs in turn yields a sequence of clique complexes \citep{Petri2013a}, allowing one to examine equivalence classes of cycles as a function of the edge weight $\theta$ (or another filtration parameter). Important values of $\theta$ include the weight $\theta_{\mathrm{birth}}$ associated with the first graph in which an equivalence class (i.e., a topological feature) occurs (i.e., its \emph{birth} coordinate) and the edge weight $\theta_{\mathrm{death}}$ associated with the first graph in which the feature disappears (i.e., its \emph{death} coordinate), such as by being filled in with higher-dimensional simplices or by merging with an older feature. One potential marker of the relative importance of a particular feature (a component, a loop, and so on) in the clique complex is how long it persists, as quantified by its \emph{lifetime} $\theta_{\mathrm{birth}} - \theta_{\mathrm{death}}$ (although short-lived features can also be meaningful \cite{Stolz2016,otter2015}). A large lifetime indicates robust features that persist over many values of a filtration parameter. \emph{Persistence diagrams} (PDs) are one useful way to visualize the evolution of $k$-dimensional cycles with respect to a filtration parameter. PDs encode birth and death coordinates of features as a collection of \emph{persistence points} $(\theta_{\mathrm{birth}},\theta_{\mathrm{death}})$ in a planar region. One can construct a PD for each Betti number: a $\beta_{0}$ PD (denoted by $\mathrm{PD}_{0}$) encodes the birth and death of components in a network, a $\beta_{1}$ PD (denoted by $\mathrm{PD}_{1}$) encodes the birth and death of loops, and so on.

To demonstrate some key aspects of a filtration, the birth and death of topological features, and PDs, we borrow and adapt an example from \citet{Kramar:2014b}. Consider the small granular force network in Fig.~\ref{f:PH_schematic}\emph{a}; the nodes represent particles in a 2D granular packing, and the colored edges represent the magnitude of the inter-particle forces (of which there are four distinct values) between contacting particles. In a 2D system like this one, the only relevant Betti numbers are $\beta_{0}$ and $\beta_{1}$, as all others are $0$. In Fig.~\ref{f:PH_schematic}\emph{b}, we show the \emph{flag complex} (which is essentially the same as a \emph{clique complex} \cite{otter2015}) of the granular network, where the color of a triangle indicates the value corresponding to the minimum force along any of its edges. Computing PH on a flag complex (which has been done in several studies of PH in granular force networks \citep{Kramar:2013aa,Kramar:2014b,Kramar:2014aa,Pugnaloni:2016a,Kondic:2016a,Ardanza-Trevijano:2014aa}) only counts loops that include 4 or more particles. That is, it does not count 3-particle loops (which are sometimes called ``triangular loops''). Loops with 4 or more particles are associated with \emph{defects}, because they would not exist in a collection of monosized disks that are packed perfectly and as densely as possible into a ``crystalline" structure (which has only triangular loops) \citep{Kramar:2014b}. 

In Fig.~\ref{f:PH_schematic}\emph{c--f}, we show the sequence of complexes that correspond to the filtration over the flag complex. One descends the four threshold levels (i.e., edge weights), beginning with the largest ($\theta_{4}$) and ending with the smallest ($\theta_{1}$). In Fig.~\ref{f:PH_schematic}\emph{g,h}, we show the corresponding PDs for $\beta_{0}$ and $\beta_{1}$. It is helpful to discuss a few features of these diagrams. In $\mathrm{PD}_{0}$, we observe four points that are born at $\theta_{4}$; these points correspond to the four connected components that emerge at the first level of the filtration in Fig.~\ref{f:PH_schematic}\emph{c}. Two of the components merge into one component at $\theta_{3}$ (see Fig.~\ref{f:PH_schematic}\emph{d}); this corresponds to the point at $(\theta_{4},\theta_{3})$. A new component forms at $\theta_{3}$ and dies at $\theta_{2}$; this is represented by the point at $(\theta_{3},\theta_{2})$ (see Fig.~\ref{f:PH_schematic}\emph{d}). Additionally, two components born at $\theta_4$ die at $theta_2$, corresponding to the two points at $(\theta_{4},\theta_{2})$. One can continue this process until the end of the filtration, where there is just a single connected component (see Fig.~\ref{f:PH_schematic}\emph{f}). This component is born at $\theta_{4}$; it persists for all thresholds, and we use \citet{Kramar:2014b}'s convention to give it a death coordinate of $-1$; this yields the persistence point at $(\theta_{4},-1)$. In $\mathrm{PD}_{1}$, we observe that a loop emerges at $\theta_{3}$ (see Fig.~\ref{f:PH_schematic}\emph{d}), and it is then filled by triangles at $\theta_{2}$ (see Fig.~\ref{f:PH_schematic}\emph{e}), leading to the point at $(\theta_{3},\theta_{2})$. Three more loops are born at $\theta_{2}$ and never die (see Fig.~\ref{f:PH_schematic}\emph{e}); using the convention in \citep{Kramar:2014b}, we assign these features a death coordinate of $0$, so there are three persistence points at $(\theta_{2},0)$. Finally, one more loop appears at $\theta_{1}$ and does not die (see Fig.~\ref{f:PH_schematic}\emph{e}); this is represented by a point at $(\theta_{1},0)$.

\citet{Kramar:2014b} gave an in-depth exposition of how to apply PH to granular networks, and we refer interested readers to this paper for more information. Because studying PH is a general mathematical approach, it can be applied to different variations of force networks and can also be used on networks constructed from different types of experimental data (e.g., digital image data, particle-position data, or particle-interaction data). \citet{Kramar:2014b} also discussed a set of measures that can be used to compare and contrast the homology of force networks both within a single system (e.g., at two different packing fractions) and across different systems (e.g., if one uses particles of different sizes or shapes), and they explored the robustness of PH computations to noise and numerical errors. In Sec.~\ref{s:granular_alg_topology}, we further discuss applications of methods from algebraic and computational topology to granular materials.

\subsection{Some considerations when using network-based methods}
\label{s:relationships_measures}

Because there are many methods that one can use to analyze granular networks and many quantities that one can compute to measure properties of these networks, it is useful to discuss some relationships, similarities, and distinctions between them. Naturally, the meaning of any given network feature depends on how the network itself is defined, so we focus the present discussion on the most common representation of a granular system as a network. (See Sec.~\ref{s:other_representations} for discussions of other representations.) In this representation (see Fig.~\ref{f:packing}), nodes correspond to particles and edges correspond to contacts between particles. Edge weights can represent quantities such as normal or tangential forces between particles. In this type of granular network, it is important to be aware of which network quantities explicitly take into account spatial information or physical constraints in a system, which consider only network topology, and which consider only network geometry (i.e., both topology and edge weights, but not other information). Granular materials have a physical nature and are embedded in real space, so such considerations are extremely important. For discussions of how such issues manifest in spatial networks more generally, see \cite{barth2011}.

One way to explicitly include spatial or physical information into network analysis is to calculate quantities that are defined from some kind of distance (e.g., a Euclidean distance between nodes), whether directly or through a latent metric space, rather than a hop distance.
For example, as discussed in Sec.~\ref{paths}, one can define the edge length between two adjacent nodes from the physical distance between them, which allows quantities such as mean shortest path length, efficiency, and some centrality measures to directly incorporate spatial information. However, traditional network features such as degree and clustering coefficient depend only on network connectivity, although their values are influenced by spatial effects. In Sec.~\ref{s:comm_structure}, we also saw that one can incorporate physical constraints from granular networks into community-detection methods by using a geographical null model, rather than the traditional NG null model, in modularity maximization.

Different computations in network analysis can also probe different spatial, topological, or geometrical scales. For example, measures such as degree, strength, and clustering coefficients are local measures that quantify information about the immediate neighborhood of a node. However, measures such as the mean shortest path length and global efficiency are global in nature, as they probe large-scale network organization. In between these extremes are mesoscale structures. A network-based framework can be very helpful for probing various types of intermediate-scale structures, ranging from very small ones (e.g., motifs, such as small cycles) to larger ones (e.g., communities), and tools such as PH were designed to reveal robust structural features across multiple scales. Crucially, although there are some clear qualitative similarities and differences between various network-analysis tools (and there are some known quantitative relationships between some of them \cite{newman2010networks}), it is a major open issue to achieve a precise understanding of the relationships between different network computations. Moreover, in spatially-embedded systems (as in any situation where there are additional constraints) one can also expect some ordinarily distinct quantities to become more closely related to each other \cite{barth2011}. Furthermore, the fact that a granular particle occupies a volume in space (volume exclusion) gives constraints beyond what arises from embeddedness in a low-dimensional space.

\section{Granular materials as networks}
\label{s:granular_networks}

We now review network-based models and approaches for studying granular materials. Over the past decade, network analysis has provided a novel view of the structure and dynamics of granular systems, insightfully complementing and extending traditional perspectives. See \citep{Jaeger1996,duran1999sands,Mehta2007,Franklin2015,Andreotti2013} for reviews of non-network approaches.

Perhaps the greatest advantages of using network representations and associated tools are their natural ability to \emph{(1)} capture and quantify the complex and intrinsic heterogeneity that manifests in granular materials (e.g., in the form of force chains), and to \emph{(2)} systematically and quantitatively investigate how the structure and organization of a granular system changes when subjected to external loads or perturbations (such as compression, shear, or tapping). In particular, network science and related subjects provide a set of tools that help quantify structure (and changes in structure) over a range of scales --- including local, direct interactions between neighboring particles; larger, mesoscale collections of particles that can interact and reconfigure via more complicated patterns; and system-wide measurements of material (re)organization. It is thought that local, intermediate, and system-wide scales are all important for regulating emergent, bulk properties of granular systems. Because structure at each of these scales can play a role in processes such as acoustic transmission and heat transfer, it can be difficult to obtain a holistic, multiscale understanding of granular materials. For example, microscale particle-level approaches may not take into account collective organization that occurs on slightly larger scales, and continuum models and approaches that rely on averaging techniques may be insensitive to interesting and important material inhomogeneities \citep{Digby:1981,Velicky:2002,Goddard:1990,Goldenberg-2005-FEE}. 

Network representations also provide a flexible medium for modeling different types of granular materials (and other particulate matter). For example, network analysis is useful for both simulation and experimental data of granular materials, and methods from complex systems and network science can help improve understanding of both dense, quasistatically-deforming materials as well as granular flows. In any of these cases, one often seeks to understand how a system evolves over the course of an experiment or simulation. To study such dynamics, one can examine a network representation of a system as a function of a relevant physical quantity that parameterizes the system evolution. For example, for a granular system in which the packing fraction increases in small steps as the material is compressed quasistatically, one can extract a network representation of the system at each packing fraction during the compression process and then study how various features of that network representation change as the packing fraction increases. Even a particular type of granular system is amenable to multiple types of network representations, which can probe different aspects of the material and how it evolves under externally applied loads. For instance, one can build networks based only on knowledge of the locations of particles (which, in some cases, may be the only information available) or by considering the presence or absence of physical contacts between particles. 

If one knows additional information about the elements in a system or about their interactions, one can construct more complicated network representations of it. For example, it has long been known that granular materials exhibit highly heterogeneous patterns of force transmission, with a small subset of the particles carrying a majority of the force along force chains \citep{Dantu1957, Drescher1972}. Recall from Sec.~\ref{intro} that, broadly speaking, a force chain (which is also sometimes called a \emph{force network}) is a set of contacts that carry a load that is larger than the mean load \citep{Liu:1995aa, Howell1999}, and the mean orientation of a force chain often encodes the direction of the applied stress \citep{Majmudar:2005aa}. We illustrated an example of force chain structure in Fig.~\ref{f:Behringer2014Statistical_Fig3Right}, and we further discuss force-chain organization in Sec.~\ref{s:force_weighted}. Because of the nature of the distribution of force values and the interesting way in which forces are spatially distributed in a material, it is often very useful to consider network representations of granular materials that take into account information about inter-particle forces (see Sec.~\ref{s:force_weighted}) and to use network-based methods that allow one to quantitatively investigate how the structure of a force network changes when one includes only contacts that carry at least some threshold force. See Sec.~\ref{s:force_threshold} and Sec.~\ref{s:granular_alg_topology}.

In our ensuing discussion, we describe several network constructions that have been used to study granular materials, discuss how they have been investigated using many of the concepts and diagnostics introduced in Sec.~\ref{s:network_measures}, and review how these studies have improved scientific understanding of the underlying, complex physics of granular systems.

\subsection{Contact networks}
\label{s:contact_network}

A \emph{contact network} is perhaps the simplest way to represent a granular system. Such networks (as well as the term ``contact network'') were used to describe granular packings long before explicitly network science-based approaches were employed to study granular materials; see, for example, \citep{luding1997stress,silbert2002statistics}. The structure of a contact network encodes important information about a material's mechanical properties. As its name suggests, a contact network embodies the physical connectivity and contact structure of the particles in a packing (see Fig.~\ref{f:packing}). In graph-theoretic terms, each particle in the packing is represented as a node, and an edge exists between any two particles that are in physical contact with one another. Note that it may not always be possible to experimentally determine which particles are in physical contact, and one may need to approximate contacts between particles using information about particle positions, radii, and inter-particle distances. (See Sec.~\ref{limit} for details.) By definition (and however it is constructed), a contact network is unweighted and undirected, and it can thus be described with an unweighted and undirected adjacency matrix (see Sec.~\ref{s:networks}):
\begin{equation}
	A_{ij} = \left\{\begin{array}{ll}
	         1\,, \text{ if particle \textit{i} and \textit{j} are in contact\,,} \\
	       	 0\,, \text{ otherwise\,.}
	        \end{array} \right.
\end{equation}

\begin{figure}[t!]
\centering
  \includegraphics[width=0.65\textwidth]{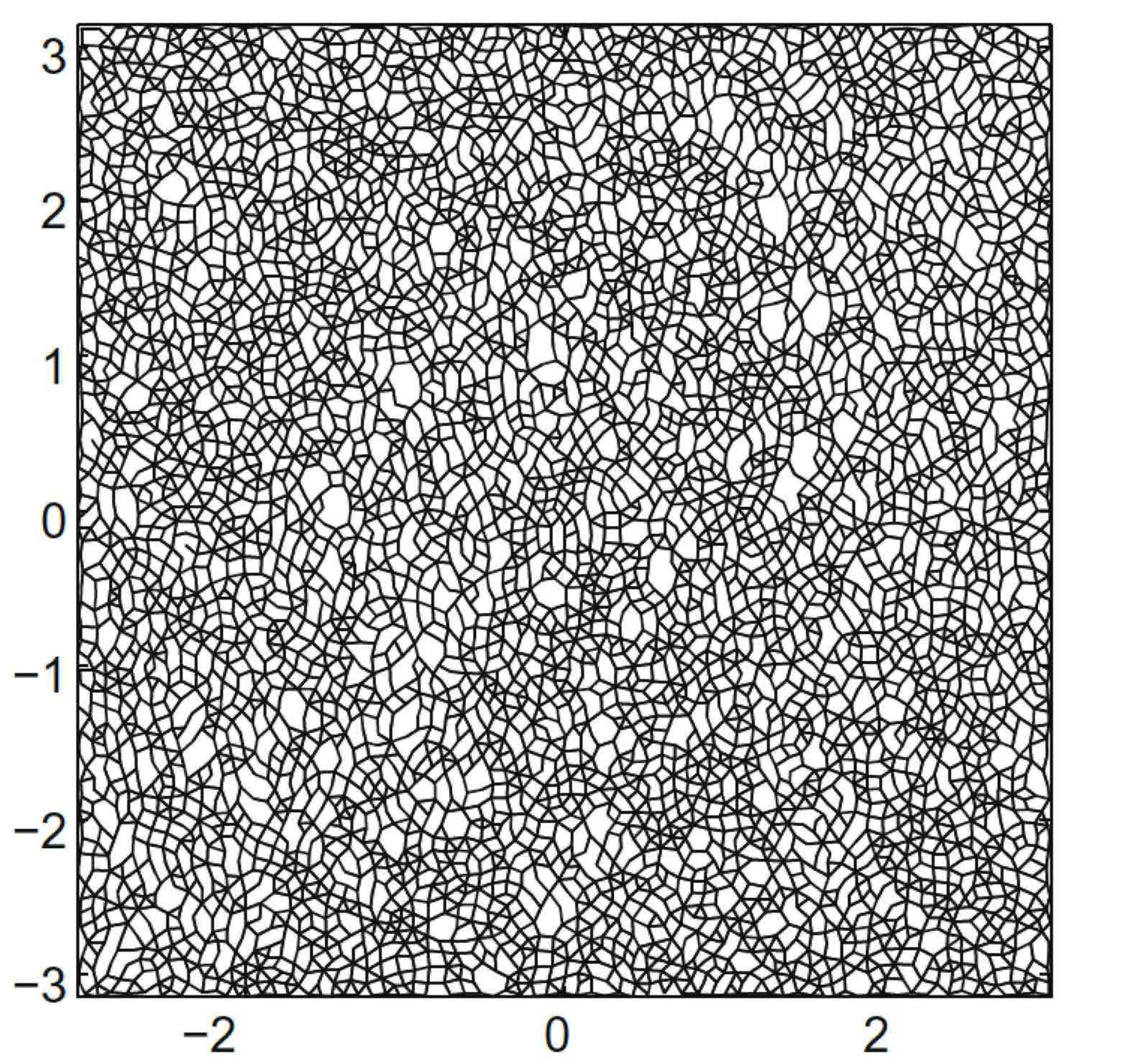}
\caption{\textbf{An example of a contact network from a discrete-element-method (DEM) simulation of a densely-packed 2D system of polydisperse, disk-shaped particles.} In this case, the granular material was subjected to biaxial compression and constrained to move along a plane. This snapshot corresponds to a network at an axial strain before full shear-band formation. [We adapted this figure, with permission, from \citep{Walker:2010aa}.]}
\label{f:contact_network}
\end{figure}

Because the organization of a contact network depends on and is constrained by the radii of the particles and their locations in Euclidean space, a contact network is a \emph{spatially-embedded} graph \citep{barth2011}. In Sec.~\ref{s:comm_detect}, we will see that this embedding in physical space has important consequences for the extraction of force-chain structures via community-detection techniques (see Sec.~\ref{s:comm_structure}). In Fig.~\ref{f:contact_network}, we show an example of a contact network generated from a discrete-element-method (DEM) simulation (see Sec.~\ref{limit}) of biaxial compression \citep{Tordesillas:2007aa}. The granular system in this figure is \emph{polydisperse}, as it has more than two types of particles. (In this case, the particles have different sizes.) If all particles are identical in a granular system, it is called \emph{monodisperse}; if there are two types of particles in a system, it is called \emph{bidisperse}. In practice, although the presence or absence of a contact is definitive only in computer simulations, one can set reasonable thresholds and perform similar measurements in experiments \citep{Majmudar:2007aa} (see Sec.~\ref{limit}). 
It is also important to note that packing geometry and the resulting contact network do not completely define a granular system on their own. In particular, one can associate a given geometrical arrangement of particles with several configurations of inter-particle forces that satisfy force and torque balance constraints and the boundary conditions of a system \citep{Snoeijer-2004-FNE,Snoeijer-2004-ETF,Tighe2010a,Kollmer2017}. This is a crucial concept to keep in mind when conducting investigations based only on contact networks, and it also motivates the inclusion of contact forces to construct more complete network representations of granular systems (see Sec.~\ref{s:force_weighted}).

In the remainder of this subsection, we review some of the network-based approaches for characterizing contact networks of granular materials and how these approaches have been used to help understand the physical behavior of granular matter. We primarily label the following subsubsections according to the type of employed methodology. However, we also include some subsubsections about specific applications to certain systems.

\subsubsection{Coordination number and node degree.} 
\label{s:kz}

One can study a contact network in several ways to investigate different features of a granular system. We begin our discussion by associating the mean node degree of a contact network with the familiar and well-studied \emph{coordination number} (i.e., \emph{contact number}) $Z$. Although early investigations of granular materials did not consciously make this connection, the mean degree and coordination number are synonymous quantities. The contact degree $k_i$ of particle $i$ is the number of particles with which $i$ is directly in contact, and one can calculate it easily from an adjacency matrix [see Eq.~\eqref{eq:node_degree}]. A contact network is undirected, so its adjacency matrix ${\bf A}$ is symmetric, and its row sum and column sum each yield a vector of node degrees. The mean degree $\langle k \rangle$ of a contact network [ Eq.~\eqref{eq:avg_degree}] is then what is usually known as the mean coordination number (i.e., contact number) $Z$, and it gives the mean number of contacts per particle. As we noted previously (see Sec.~\ref{degree}), $Z$ is an important quantity in granular systems because of its connection with mechanical stability and rigidity --- which, loosely speaking, is the ability of a system to withstand deformations --- in these systems and its characterization of the jamming transition \citep{Liu:2010aa,Liu:2011aa} and other mechanical properties. In particular, the condition for mechanical stability --- i.e., the condition to have all translational and rotational degrees of freedom constrained so that there is force and torque balance --- in a packing of frictionless spheres in $d$ dimensions \citep{Alexander:1998aa,VanHecke2010,Henkes:2010aa,Liu:2010aa,Liu:2011aa} is 
\begin{equation}
	Z \geq 2d \equiv Z_{\mathrm{iso}}\,. 
\end{equation}
The isostatic number $Z_{\mathrm{iso}}$ indicates the condition for \emph{isostaticity}, which is defined as the minimum contact number that is needed for mechanical stability. One can use the coordination number (which is often tuned by changing the packing fraction $\phi$) as an order parameter to describe the jamming transition for frictionless spheres in two and three dimensions \citep{Liu:2010aa, VanHecke2010,Henkes:2010aa}. Specifically, there is a critical packing fraction $\phi_{c}$ such that below $\phi_{c}$, the contact number for these systems is $Z = 0$ (i.e. there are no load-bearing contacts), and at the critical packing fraction $\phi_{c}$, the contact number jumps to the critical value $Z_c = Z_{{\mathrm iso}} = 2d$. One can also generalize the use of the coordination number in order to examine mechanical stability and jamming in granular systems of frictional spheres. In these systems, the condition for stability is
\begin{align}
	Z &\geq Z^{m}_{{\mathrm{iso}}}\,, \notag \\
	Z^{m}_{{\mathrm{iso}}} &\equiv (d+1) + \frac{2N_{m}}{d}\,,
\end{align}
where $N_{m}$ is the mean number of contacts that have tangential forces $f_{t}$ equal to the so-called \emph{Coulomb threshold} --- i.e., $N_{m}$ is the mean number of contacts with $f_{t} = \mu f_{n}$, where $\mu$ is the coefficient of friction and $f_{n}$ is the normal force \citep{Shundyak:2007aa,Henkes:2010aa,VanHecke2010} --- and $Z^{m}_{{\mathrm{iso}}}$ again designates the condition for isostaticity. Results from experimental systems have demonstrated that contact number also characterizes the jamming transition in frictional, photoelastic disks \citep{Majmudar:2007aa}. 

The coordination number has been studied for several years in the context of granular materials and jamming, and it is fruitful to connect it directly with ideas from network science. Several recent studies have formalized the notion of a \emph{contact network}, and they deliberately modeled granular systems as such networks to take advantage of tools like those described in Sec.~\ref{s:network_measures}. Such investigations of contact networks allow one to go beyond the coordination number and further probe the rich behavior and properties of granular materials --- including stability and the jamming transition \citep{Liu:2010aa}, force chains \citep{Liu:1995aa,Mueth:1998a,Howell1999}, and acoustic propagation \citep{Digby:1981,Makse-1999-WEM,Owens:2011,bassett2012influence}.

Perhaps the simplest expansion of investigations into the role of the coordination number is the study of the degree distribution $P(k)$ of the contact network of a packing. Calculating degree distributions can provide potential insights into possible generative mechanisms of a graph \citep{Albert2002,newman2010networks}, although one has to be very careful to avoid over-interpreting the results of such calculations \citep{stumpf2012critical}. In granular physics, it has been observed that the degree distribution of a contact network can track changes in network topology past the jamming transition in isotropically compressed simulations of a 2D granular system \citep{Arevalo:2010aa}. Specifically, the peak of $P(k)$ shifts from a lower value of $k$ to a higher value of $k$ near the transition. Moreover, changes in the mean degree $\langle k \rangle$ and its standard deviation can anticipate the onset of different stages of deformation in DEM simulations (i.e., molecular-dynamics simulations) of granular systems under various biaxial compression tests \citep{Walker:2010aa, Tordesillas:2010a}. 

\subsubsection{Investigating rigidity of a granular system using a contact network.} 
\label{rigidity}

An important area of research in granular materials revolves around attempts to \emph{(1)} understand how different types of systems respond when perturbed, and \emph{(2)} determine what features of a system improve its integrity under such perturbations. As we noted in Sec.~\ref{s:kz}, it is well-known that coordination number (and hence node degree) is a key quantity for determining mechanical stability and understanding jamming in granular materials. However, contact networks obviously have many other structural features, and examining them can be very helpful for providing a more complete picture of these systems.

\begin{figure}[t]
	\centering
	\includegraphics[width=0.75\textwidth]{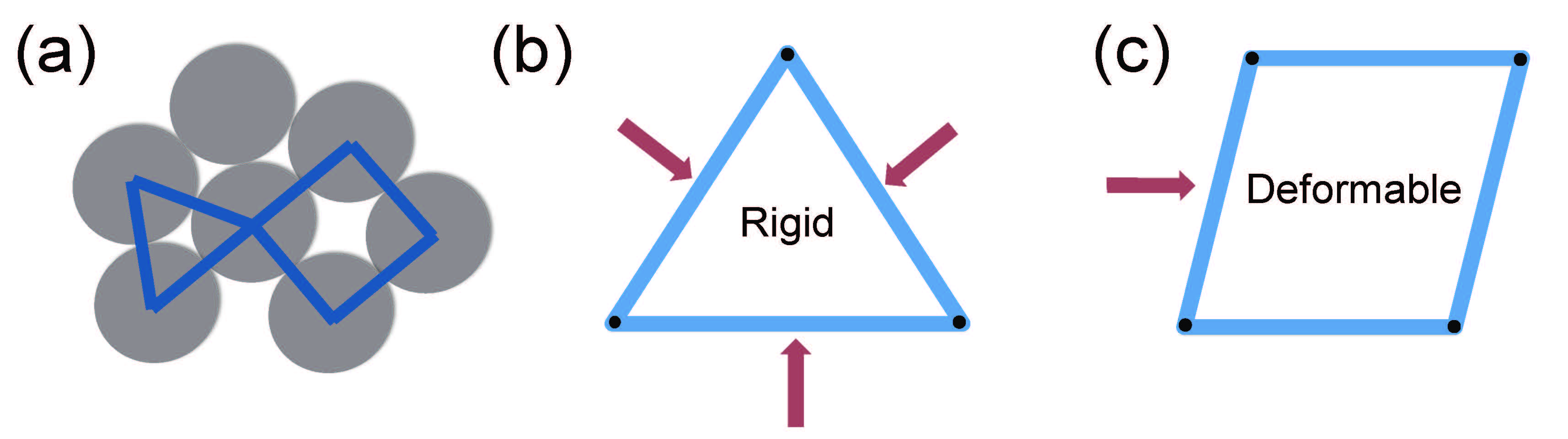}
	\caption{\textbf{Some ideas from structural rigidity.}  \emph{(a)} An example of a 3-cycle and a 4-cycle that one can examine using concepts from structural rigidity theory by considering the edges to be rods of fixed length that are connected to one another by rotating hinges. \emph{(b)} Triangular structures are rigid under a variety of applied forces (represented as red arrows), whereas \emph{(c)} squares can deform under such perturbations. By Laman's theorem, a 2D network with $N$ nodes is minimally rigid if it has exactly $2N-3$ edges, and each of its subgraphs satisfies the analogous constraint (so an $\tilde{N}$-node subgraph has no more than $2\tilde{N}-3$ edges). The network in panel (b) satisfies this criterion, but the network in panel (c) does not.
	}
	\label{f:structural_rigidity}
\end{figure}

To the best of our knowledge, the stability of granular materials was first studied from a graph-theoretic standpoint in the context of structural rigidity \citep{Jacobs1995,Duxbury1999,Thorpe:1999aa}, and it has since been applied to amorphous solids more generally \citep{Wyart:2005a}. In structural rigidity theory, thought to have been studied originally by James Clerk Maxwell \citep{Maxwell:1864aa}, rods of fixed length are connected to one another by hinges, and one considers the conditions under which the associated structural graphs are able to resist deformations and support applied loads (see Fig.~\ref{f:structural_rigidity}). A network is said to be \emph{minimally rigid} (or \emph{isostatic}) when it has exactly the number of bars needed for rigidity. This occurs when the number of constraints is equal to the number of degrees of freedom in the system (i.e., when \emph{Laman's minimal-rigidity criterion} is satisfied). The network is flexible if there are too few rods, and it is overconstrained (i.e., self-stressed) if there are more rods than needed for minimal rigidity. Triangles are the smallest isostatic structures in two dimensions \citep{Laman:1970aa, Asimow:1978aa, Crapo:1979aa}; there are no allowed motions of the structure that preserve the lengths and connectivity of the bars, so triangles (i.e., 3-cycles) do not continuously deform due to an applied force. In comparison, a 4-cycle is structurally flexible and can continuously deform from one configuration to another while preserving the lengths and connectivity of the rods (see Fig.~\ref{f:structural_rigidity}). 

Extending a traditional network of rods and hinges, concepts from structural rigidity yield interesting insights into contact networks of particulate matter. See \citep{Guyon:1990aa} for a discussion of some of the earliest applications of such ideas to disordered systems and granular materials. \citet{Moukarzel:1998aa} used structural rigidity theory to derive conditions for the isostaticity of a granular packing; and he tied the fact that random packings of particles tend to be isostatic to the origin of instabilities in granular piles. Later, similar concepts were used to show that in granular networks, cycles with an even number of edges allow contacting grains to roll without slipping when subject to shear; however, these relative rotations are ``frustrated" in cycles with an odd number of edges, so such cycles can act as stabilizing structures in a network \citep{rivier2006extended}. Several later studies (such as \cite{Smart:2008aa,Tordesillas:2010aa,Walker:2010aa,Tordesillas:2010a,Tordesillas:2011aa,Tordesillas:2014a,Walker:2015b,Tordesillas:2010b,Arevalo:2010aa,Arevalo:2010ba,Walker:2014c}) have confirmed that contact loops are often stabilizing mesoscale features in a contact network of a granular material. We specifically consider the role of cycles in granular contact networks in Sec.~\ref{s:role_of_cycles}. 

Another type of network approach for understanding rigidity in granular systems is rigidity percolation \cite{Thorpe1985a,Duxbury1999,Thorpe:1999aa} (see Sec.~\ref{s:largest_comp}). \citet{Feng1985} conducted an early investigation of an idealized version of bond percolation in a granular context. It is now known that hallmarks of this bond-percolation transition occur below isostaticity: \citep{Shen:2012a} identified that a percolating (i.e., system-spanning) cluster of non-load-bearing contacts forms at a packing density below the jamming point. In modern contexts, the rigidity-percolation approach can be used to determine if a network is both percolating and rigid (see Sec.~\ref{s:largest_comp}). Note that a rigid granular network is also percolating, but a percolating network need not be rigid. Rigidity percolation relies on tabulating local constraints via a \emph{pebble game} \citep{Jacobs1995}, which reveals connected, rigid regions (sometimes called ``clusters'') in a network. In a series of papers \citep{Jeng2008,Jeng2010,Cao2012,Lopez:2013aa} on simulated packings, Schwarz and coworkers went beyond Laman's minimal-rigidity criterion to investigate local versus global rigidity in a network, the size distribution of rigid clusters, the important role of spatial correlations, and the necessity of force balance. Building on the above work, \citep{Henkes2016} recently utilized a rigidity-percolation approach to identify floppy versus rigid regions in slowly-sheared granular materials and to characterize the nature of the phase transition from an underconstrained system to a rigid network. See also the recent paper \cite{heroy2017}.

\subsubsection{Exploring the role of cycles.}
\label{s:role_of_cycles}

We now consider the role of circuits (i.e., the conventional network notion of cycles, which we discussed in Sec.~\ref{cycles}) in granular contact networks. Cycles in a contact network can play crucial stabilizing roles in several situations. Specifically, as we will discuss in detail in this section, simulations (and some experiments) suggest that \emph{(1)} odd cycles (especially 3-cycles) can provide stability to granular materials by frustrating rotation among grains and by providing lateral support to surrounding particles, and that \emph{(2)} a contact network loses these stabilizing structures as the corresponding granular system approaches failure. 

Noting that 3-cycles are the smallest arrangement of particles that can support (via force balance) a variety of 2D perturbations to a compressive load without deforming the contact structure, 
\citet{Smart:2008aa} studied the effects of friction and tilting on the evolution of contact-loop organization in a granular bed. In their simulations, they implemented tilting by incrementally increasing the angle of a gravity vector with respect to the vertical direction, while preserving the orientation of the granular bed and maintaining quasistatic conditions. In untilted granular packings, they observed that lowering inter-particle friction yields networks with a higher density of 3-cycles and 4-cycles, where they defined the ``density'' of an $l$-cycle to be the number of $l$-cycles divided by the total number of particles. By examining the contact network as a function of tilting angle, \citet{Smart:2008aa} also observed that the density of 4-cycles increases prior to failure --- likely due to the fracture of stabilizing 3-cycles --- and that this trend was distinguishable from changes in coordination number alone. 

Cycles have also been studied in the context of DEM simulations of dense, 2D granular assemblies subject to quasistatic, biaxial compression tests \citep{Tordesillas:2010aa, Walker:2010aa, Tordesillas:2011aa,Tordesillas:2010a}. In many of these studies, the setup consists of a collection of disks in 2D that are compressed slowly at a constant strain rate in the vertical direction, while allowed to expand under constant confining pressure in the horizontal direction \citep{Tordesillas:2010aa, Walker:2010aa, Tordesillas:2011aa}. In another variation of boundary-driven biaxial compression, a sample can be compressed with a constant volume but a varying confining pressure \citep{Tordesillas:2010a}). Before describing specifics of the network analysis for these systems, it is important to note that for the previously described conditions, the axial strain increases in small increments (i.e., ``steps'') as compression proceeds, and one can extract the inter-particle contacts and forces at each strain value during loading to examine the evolution of a system as a function of strain. Additionally, these systems undergo a change in behavior from a solid-like state to a liquid-like state and are characterized by different regimes of deformation as a function of increasing axial strain \citep{Tordesillas:2007aa}. In particular, the granular material first undergoes a period of \emph{strain hardening}, followed by \emph{strain softening}, after which it enters a \emph{critical state}. In the strain-hardening regime, the system is stable and the shear stress increases monotonically with axial strain up to a peak value. After the peak shear stress, strain softening sets in; this state is marked by a series of steep drops in the shear stress that indicate reduced load-carrying capacity. Finally, in the critical state, a persistent shear band has fully formed, and the shear stress fluctuates around a steady-state value. The shear band is a region of localized deformation and gives one signature of material failure \cite{Oda:1998a}. Inside the shear band, force chains can both form and buckle \citep{Tordesillas:2009b}. One can also associate increases in the energy dissipation rate of the system with particle rearrangements (such as those that occur during force-chain buckling) and loss of stability in the material.
\begin{figure}[t!]
\centering
  \includegraphics[width=0.75\textwidth]{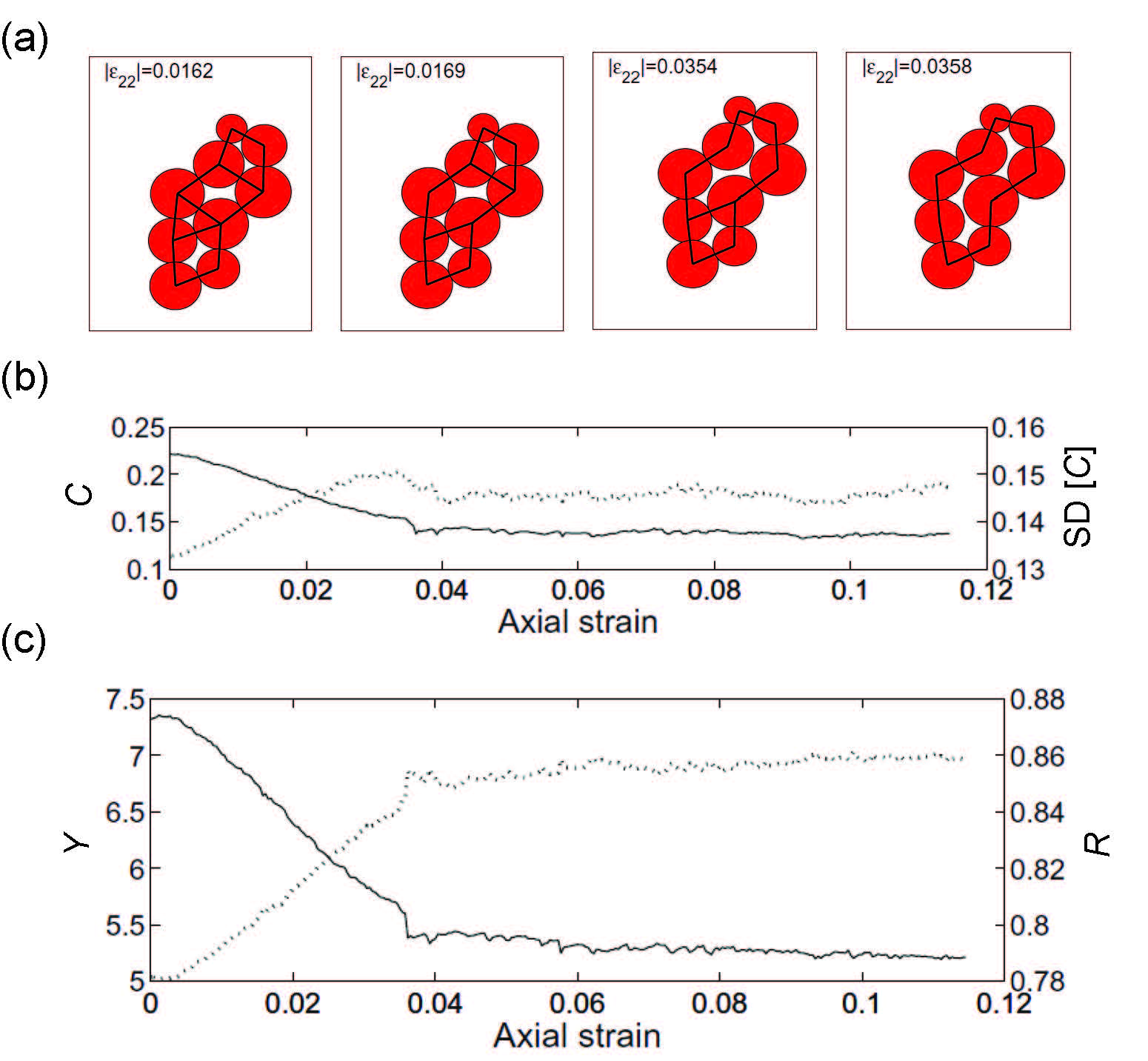}
\caption{\textbf{Evolution of cycles in a deforming granular material.} \emph{(a)} One can track a subset of particles and their corresponding contact network from a DEM simulation for increasing axial strain values $|\varepsilon_{22}|$. The system consists of 5098 spherical, polydisperse particles that were subjected to a quasistatic, biaxial compression test. 
At the smallest displayed axial strain, the set of particles in this figure yields a network that is composed of 3-cycles and 4-cycles. During loading, contacts are lost and longer cycles arise until only a single 9-cycle remains. \emph{(b)} One way to quantify these structural changes is by calculating the global clustering coefficient $C$ (solid curve), which undergoes a sharp drop at peak stress, signifying the onset of material failure. (The dashed curve shows the standard deviation of the distribution of local clustering coefficients $C_{i}$). \emph{(c)} A decrease in mean subgraph centrality $Y$ (solid curve) also illustrates the loss of short cycles during deformation. More specifically, the mean network bipartivity $R$ (dashed curve) increases with axial strain, highlighting that, during loading, closed walks of even (respectively, odd) length contribute more (respectively, less) to the mean subgraph centrality of the contact network.
[We adapted this figure, with permission, from \citep{Walker:2010aa}.]
}
\label{f:cycle_evolution}
\end{figure}

Examining the temporal evolution of cycles in an evolving granular contact network can reveal important information about changes that occur in a material during deformation. Using DEM simulations (see the previous paragraph), \citet{Tordesillas:2010aa} computed the total number of cycles of different lengths in a minimal cycle basis (see Sec.~\ref{cycles}) of a contact network at each strain state during loading, and they observed that there are many more 3-cycles and 4-cycles than longer cycles in the initial, solid-like state of the system. However, as axial strain increases and one approaches the maximum shear stress, the total number of 3-cycles falls off steeply. (The same is true for 4-cycles, though it is less dramatic.) Additionally, during axial-strain steps (i.e., axial-strain ``intervals'') corresponding to drops in shear stress, \citet{Tordesillas:2010aa} observed large increases in the number of 3-cycles and 4-cycles that open up to become longer cycles. In Fig.~\ref{f:cycle_evolution}\emph{a}, we show an example of the evolution of cycle organization with increasing axial strain for a subset of particles from a DEM simulation of a granular material under biaxial compression, carried out by \citet{Walker:2010aa}. The authors observed that in this system, both the global clustering coefficient $C$ [Eq.~\eqref{eq:global_cluster}] and the mean subgraph centrality $Y$ decrease with increasing axial strain, drop sharply at peak shear stress, and then level out (see Fig.~\ref{f:cycle_evolution}\emph{b,c}). Recalling that $C$ is a measure of triangle density in a graph and that subgraph centrality measures participation of nodes in closed walks (with more weight given to shorter paths), these results also imply that the loss of small cycles co-occurs with the deformation and failure of a system due to increasing load. \citet{Walker:2010aa} also computed the network bipartivity $R$ \citep{Estrada:2005ab} of the contact network to quantify the contribution to mean subgraph centrality $Y$ from closed walks of even length [see Eq.~\ref{eq:bipartivity}]. They observed that $R$ increases with increasing axial strain, revealing that closed walks of even length become more prevalent during loading (see Fig.~\ref{f:cycle_evolution}c). The authors suggested that this trend may be due to a decrease in the prevalence of 3-cycles (which are stabilizing, as discussed in Sec.~\ref{cycles} and elsewhere). \citet{Tordesillas:2011aa} also examined the stability of cycles of various lengths in both DEM simulations and experimental data, and they observed that, during loading, 3-cycles tend to be more stable (as quantified by a measure of stability based on a structural-mechanics framework \citep{Bagi:2007a}) than cycles of other lengths in a minimal cycle basis of the network. 

Minimal cycle bases and the easier-to-compute subgraph centrality have also been used to examine fluctuations in kinetic energy in simulations of deforming sand. \citet{Tordesillas:2014a} computed a minimal cycle basis and then constructed cycle-participation vectors (see Sec.~\ref{cycles}) from a contact network after each strain step (i.e., at each strain state) during loading. They observed that temporal changes in the cycle-participation vectors of the particles between consecutive strain steps are correlated positively with temporal changes in kinetic energy over those steps.
They also observed that large values in the temporal changes of particle cycle-participation vectors and particle subgraph centrality occur in the shear-band region. \citet{Walker:2015b} also studied a minimal cycle basis and corresponding cycle-participation vectors to examine structural transitions in a 3D experimental granular system of hydrogel spheres under uniaxial compression. As pointed out in \citep{Tordesillas:2014a}, developing quantitative predictors that are based on topological information alone is extremely important for furthering understanding of how failure and rearrangements occur in systems in which energy or force measurements are not possible. 

Examining cycles in contact networks can also shed light on the behavior of force chains. The stability, load-bearing capacity, and buckling of force chains depend on neighboring particles (so-called \emph{spectator grains}) to provide lateral support \citep{Radjai:1998aa, cates1999jamming,Tordesillas:2009aa}. Because 3-cycles appear to be stabilizing features, it is interesting to consider the co-evolution of force chains and 3-cycles in a contact network. Such an investigation requires a precise definition of what constitutes a force chain, so that it is possible to \emph{(1)} extract these structures from a given packing of particles and \emph{(2)} characterize and quantify force-chain properties. Several definitions of force chains have been proposed; see, e.g., \citep{Cates:1998aa, Howell1999, Peters:2005aa,bassett2015extraction}. The studies that we describe in the next three paragraphs used a notion of ``force chains'' from \citep{Peters:2005aa,Muthuswamy2006a}, in which force chain particles are identified based on their particle-load vectors (where each particle is associated with a single particle-load vector that indicates the main direction of force transmission). More specifically, a single chain is a set of three or more particles for which the magnitude of each of their particle-load vectors is larger than the mean particle-load vector magnitude over all particles, and for which the directions of the particle load vectors are, within some tolerance, aligned with one another (i.e., they are ``quasilinear''.) We note that an important point for future work is to conduct network-based studies of force-chain structure for different definitions of force chains, and to investigate if there are qualitative differences in their associated network properties.

Using DEM simulations of a densely packed system of polydisperse disks under biaxial loading --- i.e., compressed quasistatically at a constant strain rate in the vertical direction, while allowed to expand under constant confining pressure in the horizontal direction --- \citet{Tordesillas:2010aa} quantified the co-evolution of force chains and 3-cycles in several ways. For example, they computed a minimal cycle basis (see Sec.~\ref{cycles}) of a contact network and then examined \emph{(1)} the ratio of 3-cycles to the total number of cycles in which particles from a force chain participate and \emph{(2)} the force chain's 3-cycle \emph{concentration}, which is defined as the ratio of 3-cycles involving force-chain particles to the total number of particles in the force chain. When averaged over all force chains, the above two measures decrease rapidly with increased loading. Additionally, \citet{Tordesillas:2010aa} observed that force chains that do not fail by buckling (see \citep{Tordesillas:2007aa} for how ``buckling" was defined) have a larger ratio of 3-cycle participation to total cycle participation than force chains that do buckle. \citet{Tordesillas:2011aa} observed, in both DEM simulations of biaxial loading (see above) and 2D photoelastic disk experiments under pure shear, that a particular measure (developed by \citet{Bagi:2007a}) of structural stability of force chains is correlated positively with the mean of the local clustering coefficient [Eq.~\eqref{eq:local_cluster}] over force-chain particles. Their results also suggest that 3-cycles are more stable structures than cycles of longer length during loading and that force chains with larger 3-cycle participation tend to be more structurally stable. These observations suggest that cycles --- and especially 3-cycles --- in contact networks are stabilizing structures that can provide lateral support to force chains. It would be interesting to study these ideas further, and to relate them to structural rigidity theory (see Fig.~\ref{f:structural_rigidity} and Sec.~\ref{rigidity}), especially in light of the difference between 3-cycles (which are rigid) and deformable cycles (e.g., 4-cycles). 
	
DEM simulations of 3D, ellipsoidal-particle systems subject to triaxial compression also suggest that 3-cycles are important features in granular contact networks \citep{Tordesillas:2010b}. 
Similar to the aforementioned results from simulations of 2D systems with disk-shaped particles, the number of 3-cycles in a minimal cycle basis of the contact networks (and the global clustering coefficient [Eq.~\eqref{eq:global_cluster}]) initially decrease and then saturate with increasing load, and particles in force chains have a larger number of 3-cycles per particle than particles that are not in force chains. \citet{Tordesillas:2010b} also observed that the set of 3-cycles that survive throughout loading tend to lie outside the strain-localization region (where force chains buckle). The dearth of 3-cycles in certain regions in a material may thus be a signature of strain-localization zones. 
Another paper to note is \citep{Tordesillas:2010a}, which examined and compared the temporal evolution of cycles (and several other contact-network diagnostics) in a set of DEM simulations using a variety of different material properties and boundary conditions.

In another interesting study, \citet{Walker:2014c} examined the phenomenon of \emph{aging} \citep{kob1997aging,kabla2004contact} --- a process in which the shear strength and stiffness of a granular material increase with time --- in collections of photoelastic disks subject to multiple cycles of pure shear under constant volume. Because aging is a slow process, it can be difficult both to uncover meaningful temporal changes in dynamics and to characterize important features in packing structure that accompany aging. To overcome these challenges, \citet{Walker:2014c} first analyzed the time series of the stress ratio (using techniques from dynamical-systems theory) to uncover distinct temporal changes in the dynamics of the system. (See \citep{Walker:2014c} for details.) After each small, quasistatic strain step, they also extracted the contact network of the packing at that time to relate aging to changes in topological features of the network structure. As one approaches the shear-jammed regime during prolonged cyclic shear, they observed on average that force chains are associated with more 3-cycles and 4-cycles from the minimal cycle basis. 
\begin{figure}
\centering
\includegraphics[width=0.50\textwidth]{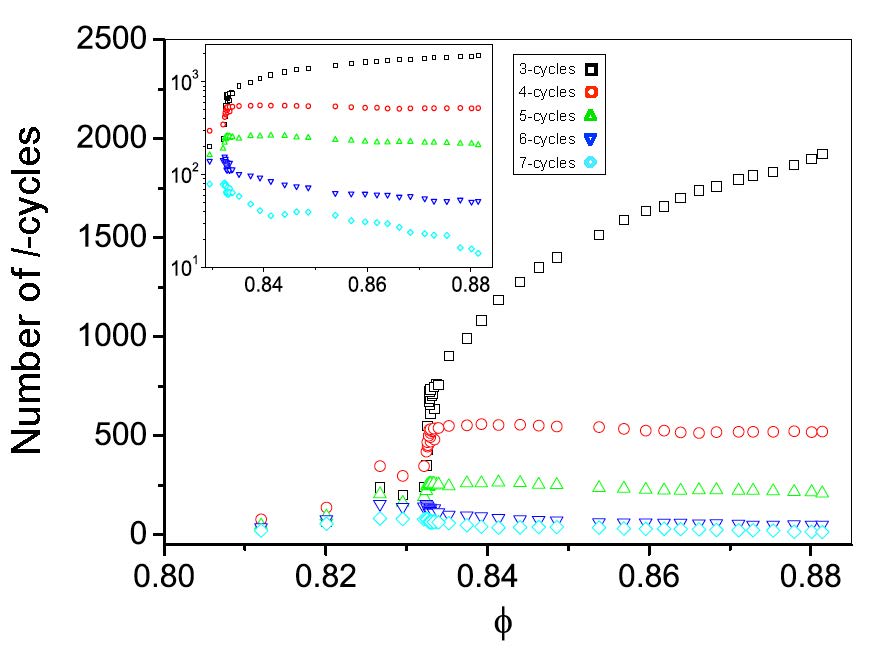}
\caption{\textbf{The number of $l$-cycles in a contact network versus the packing fraction $\phi$ from a DEM simulation of isotropic compression of a granular system.} Each color represents a cycle of a different length $l$: 3-cycles (black squares), 4-cycles (red circles), 5-cycles (green triangles), 6-cycles (dark-blue inverted triangles), and 7-cycles (cyan diamonds). As the packing transitions from a fluid-like state to a solid-like one, there is an increase in the number of cycles in the contact network near the critical packing fraction $\phi_{c}$. In addition, while 3-cycles continue to grow in number after the transition, cycles of longer lengths slowly decrease. The inset shows the same data on a semi-logarithmic plot. [We adapted this figure, with permission, from \citep{Arevalo:2010aa}.]}
\label{f:jamming_triangles}
\end{figure}

We have just discussed many papers that concern transitions in granular matter from a solid-like regime to a liquid-like regime. One can also use changes in the loop structure of a contact network to describe the opposite transition, in which a granular material goes from an underconstrained, flowing state to a solid-like state during the process known as jamming (see Sec.~\ref{s:kz}). Studying 2D frictional simulations of isotropically compressed granular packings, \citet{Arevalo:2010aa} examined a granular contact network as a function of packing fraction. They observed that the number of cycles (which were called \textit{polygons} in \citep{Arevalo:2010aa}) in the contact network grows suddenly when the packing fraction approaches the critical value $\phi_{c}$ that marks the transition to a rigid state (see Fig.~\ref{f:jamming_triangles}). They also observed that 3-cycles appear to be special: they continue to grow in number above the jamming point, whereas longer cycles slowly decrease in number after jamming. Although they observed a nonlinear relationship near the jamming point between $Z$ (the contact number, which is the usual order parameter for the jamming transition) and the number of 3-cycles \citep{Arevalo:2010ba}, these quantities appear to depend linearly on each other after the transition. These results suggest that one can use the evolution of contact loops to understand the transition to a rigid state and to characterize subsequent changes in the system.
\begin{figure}
\centering
\includegraphics[width=0.75\textwidth]{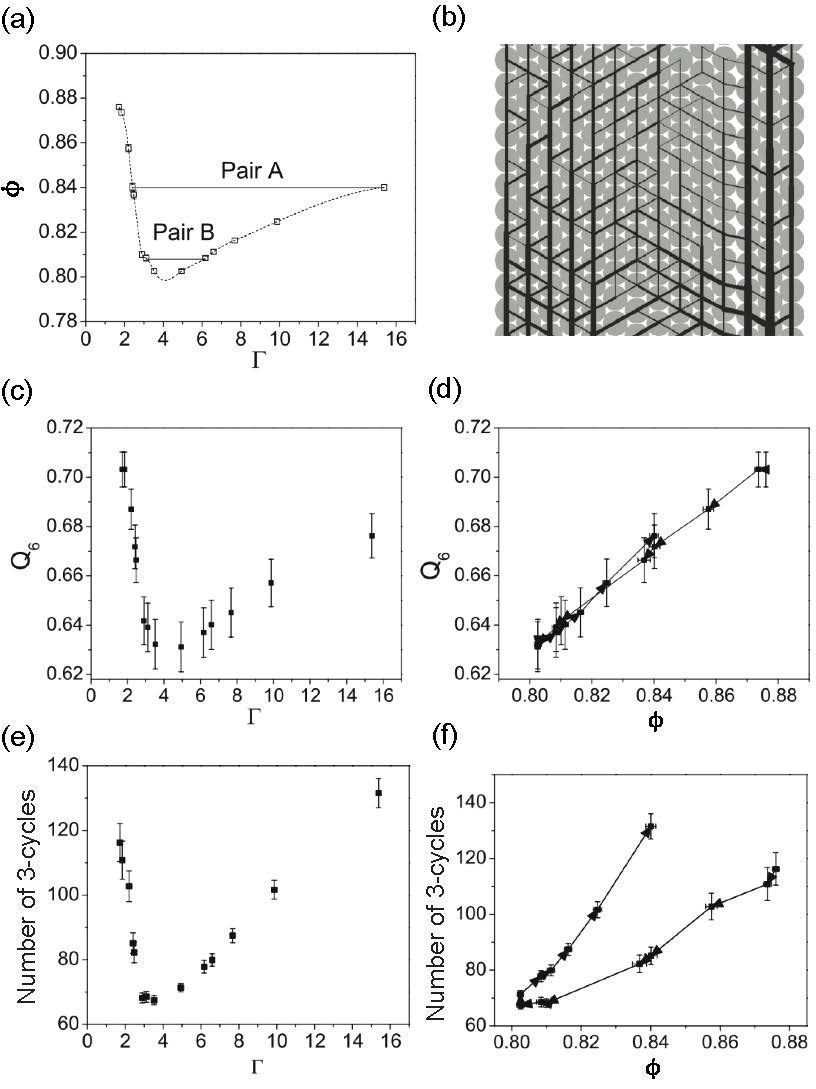}
\caption{\textbf{Using a contact network to distinguish states of the same packing fraction in simulations of tapped granular materials.} \emph{(a)} Packing fraction $\phi$ versus tap intensity $\Gamma$. The horizontal lines connect states at the same packing fraction that were obtained using different tap intensities. \emph{(b)} A section of one of the packings. \emph{(c)} The mean value of the ``bond orientational order parameter'' (or simply ``bond order parameter'') as a function of tap intensity. The bond order parameter, which is often used to quantify local and long-range crystalline structure in a material \citep{Steinhardt:1983a}, was computed on each subgraph of particles defined by a central particle and the set of its neighbors within a distance of 1.2 particle diameters, and it was then averaged over all such subgraphs to obtain a mean value. \emph{(d)} The mean value of the bond order parameter as a function of packing fraction, where the arrows indicate the direction of increasing $\Gamma$. It is difficult to differentiate between different states at the same packing fraction using this quantity.
\emph{(e, f)} The same as panels (\emph{c,d}), but now the vertical axis is the number of 3-cycles in the contact network. Calculating the number of 3-cycles successfully separates different states of the system with the same packing fraction. [We adapted this figure, with permission, from \citep{Arevalo:2013aa}.]} 
\label{f:contact_tapping}
\end{figure}

~\\
\noindent \emph{Application to tapped granular materials.} 
\label{s:tapping_contact}

Properties of contact networks have also been used to study \emph{tapped granular materials}, in which a packing of grains is subject to external pulses of excitation against it. In most studies of tapped granular materials, the packing and pulses are both vertical. The intensity $\Gamma$ of these mechanical perturbations (so-called ``taps") is usually quantified as a dimensionless ratio of accelerations, such as the ratio of the peak acceleration of the excitation to the acceleration of gravity \cite{Nowak:1998a,Pugnaloni:2010aa}. Tapped granular materials are interesting, because the packing fraction $\phi$ is not a monotonic function of the tapping intensity $\Gamma$ \citep{Pugnaloni:2008aa, Gago:2009aa,Carlevaro:2011a}. It reaches a minimum value $\phi_{\mathrm{min}}$ at an intensity of $\Gamma_{\mathrm{min}}$, and it then increases as the tap intensity increases (see Fig.~\ref{f:contact_tapping}\emph{a}). Consequently, one can achieve steady states with the same packing fraction by using different tap intensities (i.e., both a ``low" tap intensity, which is smaller than $\Gamma_{\mathrm{min}}$, and a ``high" tap intensity, which is larger than $\Gamma_{\mathrm{min}}$). These steady states are not equivalent to each other, as they have different force-moment tensors \citep{Pugnaloni:2010aa}, for example. An interesting question is thus the following: What features of a granular packing distinguish between states at the same packing fraction that are reached by using different tap intensities? 

Recent work has suggested that properties of contact networks --- especially cycles (which, in this case, are particle contact loops) --- can distinguish between steady-state configurations at the same packing fraction but that are generated from different tap intensities in simulated 2D granular packings subjected to tapping \citep{Arevalo:2013aa, Arevalo:2013ba} (see Fig.~\ref{f:contact_tapping}\emph{b}). For example, as $\Gamma$ is increased in the regime $\Gamma < \Gamma_{\mathrm{min}}$, the number of 3-cycles (i.e., triangles) and the number of 4-cycles (i.e., squares) both decrease. As $\Gamma$ is increased in the regime $\Gamma > \Gamma_{\mathrm{min}}$, the opposite trend occurs, so the numbers of 3-cycles and 4-cycles increase. This makes it possible to differentiate configurations at the same $\phi$ obtained from low and high tap intensities. (See Fig.~\ref{f:contact_tapping}\emph{e,f} for a plot of the number of triangles versus $\Gamma$ and $\phi$.) However, geometrical measures like the pair-correlation function, distributions of Voronoi tessellation areas, or bond orientational order parameters do not seem to be as sensitive to differences in these two different states of the system (see Fig.~\ref{f:contact_tapping}\emph{c,d}), perhaps because they quantify only local proximity rather than directly examining contacts. (See \cite{Arevalo:2013aa} and references therein for details of these descriptors.) These results suggest that topological features (e.g., mesoscale features) of a contact network can capture valuable information about the organization of granular packings.

\subsubsection{Other subgraphs in contact networks.}
\label{subgraph}

When studying contact networks, it can also be helpful to explore network motifs other than cycles. Recall from Sec.~\ref{s:motifs} that motifs are subgraphs that appear often in a network (e.g., in comparison to a null model) and are thus often construed as being related to system function \citep{Milo:2002a,itzkovitz2005subgraphs,alon2007network,shoval2010snap}. Network motifs, which traditionally refer to small subgraphs, are a type of mesoscale feature, and it can be insightful to examine how their prevalences change in a granular material as it deforms.

One system in which motifs and their dynamics have been studied is frictional, bidisperse, photoelastic disks subject to quasistatic cyclic shear \citep{Tordesillas:2012aa}. After each small strain increment (i.e., strain step) in a shear cycle, the authors considered the contact network of the granular packing. For each particle $i$ in the contact network, they extracted the subgraph of particles (nodes) and contacts (edges) formed by the central particle $i$ and particle $i$'s contacting neighbors. This process results in a set of $N$ subgraphs\footnote{In some degenerate cases (e.g., when there is a network component that consists of a clique), this includes duplications of subgraphs, even when the nodes of a network are labeled.} (which, borrowing terminology from \citep{Tordesillas:2012aa}, we call \textit{conformation subgraphs}), where $N$ is the number of particles in the network.

To examine packing rearrangements as a system is sheared, \citep{Tordesillas:2012aa} represented each distinct conformation subgraph present at any time during loading as one ``state" in a Markov transition matrix, and they studied transitions between the conformation subgraphs as a discrete-time Markov process. More specifically, each element in the $n_{c} \times n_{c}$ transition matrix (where $n_{c}$ is the total number of unique conformation subgraphs and hence the number of unique states) captured the fraction of times that each conformation subgraph transformed to each other conformation subgraph (including itself) after four quasistatic steps of the shearing process. \citet{Tordesillas:2012aa} reported that force-chain particles typically occur in network regions with high mean degree, high mean local clustering coefficients, and many 3-cycles. (Note that this study, as well as the others that we describe in the present and the following paragraph, define force-chains as in \citep{Peters:2005aa,Muthuswamy2006a}.) Furthermore, when considering the conformation subgraphs of particles in force chains that fail by buckling (see \citep{Tordesillas:2009aa,Tordesillas:2009b} for details on the definition of ``buckling"), the most likely transformations to occur tend either to maintain the topology of those conformation subgraphs or to involve transitions from conformation subgraphs in which the central particle has a larger degree or is part of more 3-cycles to conformation subgraphs in which the degree of the central particle is smaller or in which it participates in fewer 3-cycles. \citet{Tordesillas:2012aa} also used force information to compute a measure of structural stability (based on a structural-mechanics framework \citep{Bagi:2007a,Tordesillas:2011aa} and summarized in a single number) for each conformation subgraph. They then split the full range of the stability values into several smaller ``stability intervals'' (i.e., small ranges of contiguous structural-stability values) and modeled transitions between stability intervals as a Markov chain. They examined the number of conformation subgraphs that occupy each stability interval and observed pronounced peaks in some intervals that persist during loading. They also reported that conformation subgraphs whose central particles belong to force chains tend to be more stable and that conformation subgraphs whose central particles are part of buckling force chains have a higher probability of transitioning from high-stability states to low-stability states than vice versa. (For details, see Fig.~7 of \citep{Tordesillas:2012aa} and the corresponding discussions.) 
\begin{figure}[t!]
\centering
 \includegraphics[width=\textwidth]{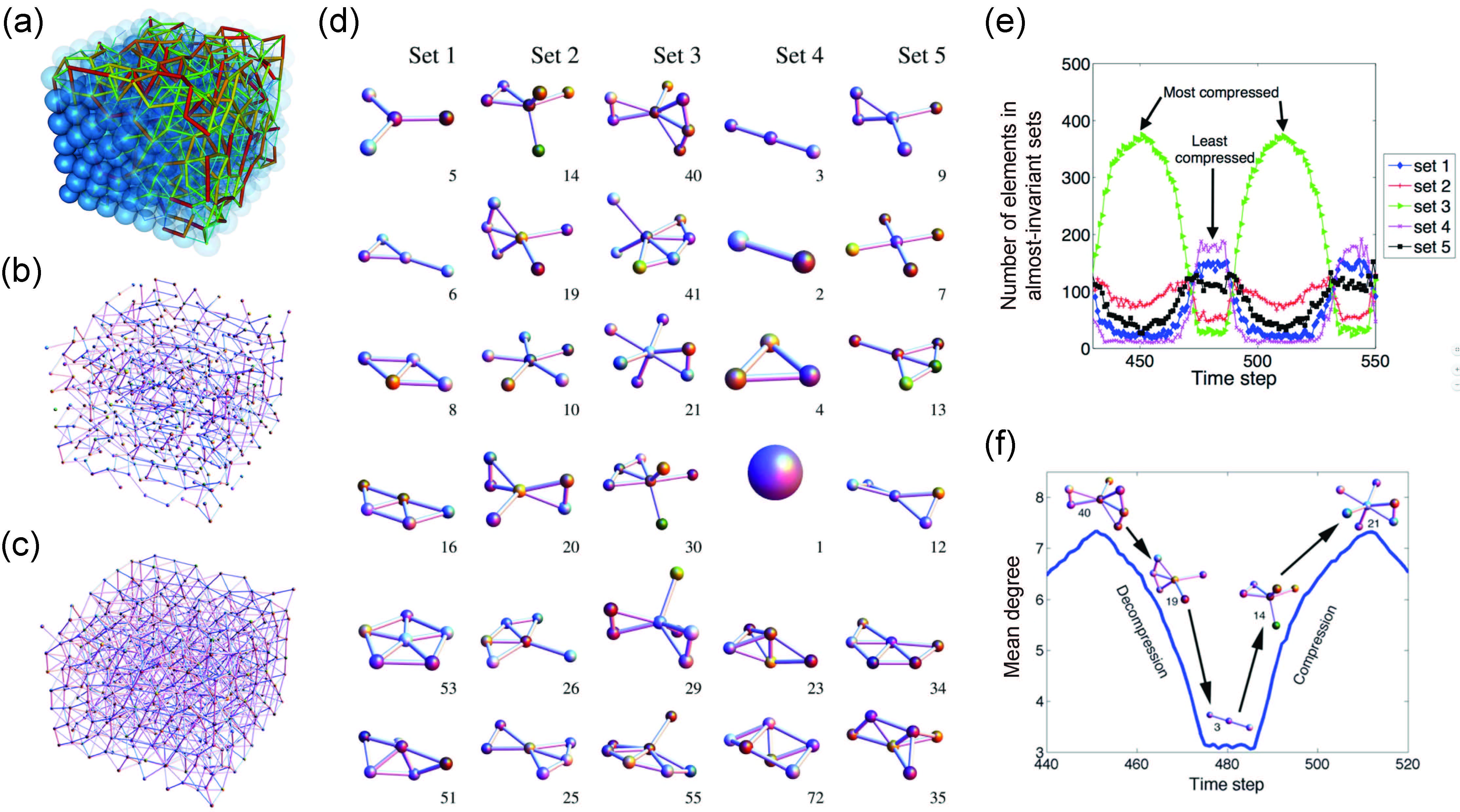}
\caption{\textbf{Conformation subgraphs can help quantify structural transitions in a 3D network of hydrogels.} \emph{(a)} A rendering of a granular system composed of spherical, hydrogel beads \citep{Brodu2015, Dijksman2017} that are subject to multiple cycles of compression along a single axis. Redder colors indicate stronger contact forces. \emph{(b)} An example of a contact network in a decompressed state and \emph{(c)} an example of a contact network in a compressed state. \emph{(d)} Pictorial representations of the most common conformation subgraphs in each almost-invariant set. The almost-invariant sets are collections of conformation subgraphs that, during loading, tend to transition amongst themselves more than they transition to conformation subgraphs in another almost-invariant set. (See the main text and \citep{Walker:2015b} for details.) \emph{(e)} The evolving 
cardinalities of almost-invariant sets can track structural transitions during cycles of compression and decompression. (The number of elements in an almost-invariant set at a given time step is the number of particles in the system whose induced conformation subgraph belongs to that set.) In highly-compressed states, most conformation subgraphs in a packing are members of Set 3, whereas most conformation subgraphs belong to Set 4 for the least-compressed states. \emph{(f)} An example of a possible \emph{transition pathway}, which consists of a sequence of conformation subgraphs in a given almost-invariant set and the transitions between them that occur during loading. In this example, the pathway corresponds to the following sequence of transitions between sets: Set 3 $\rightarrow$ Set 2 $\rightarrow$ Set 4 $\rightarrow$ Set 2 $\rightarrow$ Set 3. The first conformation subgraph (40) corresponds to the most prevalent conformation subgraph in Set 3, and the conformation subgraphs for the subsequent transitions between sets are those with the largest transition probabilities. For example, of all conformation subgraphs in Set 2, subgraph 19 is the one to which conformation subgraph 40 (of Set 3) is most likely to transition. These transition pathways can potentially inform constitutive-modeling efforts. [We adapted this figure, with permission, from \citep{Walker:2015b}.]
}
\label{f:3D_subgraphs}
\end{figure}

\citet{Walker:2015b} used similar methods to study self-assembly in an almost-frictionless, 3D system of hydrogel spheres under quasistatic, cyclic uniaxial compression (see Fig.~\ref{f:3D_subgraphs}\emph{a}--\emph{c}). After every compression step, they constructed the contact network for the system and examined two types of subgraphs for each particle: \emph{(1)} conformation subgraphs, which (as discussed earlier) consist of a single central particle $i$ and that particle's contacts; and \emph{(2)} the cycle-participation vector of each particle (see Sec.~\ref{cycles}). \citet{Walker:2015b} determined the set of all unique conformation subgraphs that exist during the above compression process. They then used each of those conformation subgraphs as one state in a transition matrix, the elements of which give the fraction of times (across the whole experiment) that a particle in one state transitions to any other state in consecutive compression steps. 
To focus on the presence or absence of a particle in an $l$-cycle (using cycle lengths up to $l = 10$), they binarized each element of the cycle-participation vectors. (The new vectors thus indicate, for each particle, whether a particle is part of at least one $l$-cycle.) They then constructed a transition matrix in which each state is a unique binarized cycle-participation vector that occurs during the experiment. The two transition matrices capture useful information about the most likely transformations that occur between different conformation subgraphs and cycle-participation vectors as one compresses or decompresses the granular system. For both types of mesoscale structures, \citet{Walker:2015b} used their transition matrices to extract \textit{almost-invariant sets}, which indicate sets of conformation subgraphs or cycle-participation vectors (i.e., states) that tend to transition among themselves more than to states in another almost-invariant set. (See \citep{Walker:2015b} for details.) In Fig.~\ref{f:3D_subgraphs}\emph{d}, we show the most common conformation subgraphs in each almost-invariant set of the 
subgraphs. The conformation subgraphs formed by force-chain particles belong mostly to Set 3 (see Fig.~\ref{f:3D_subgraphs}\emph{d}), which consists of densely-connected conformation subgraphs in which there are many contacts between particles. To characterize structural changes that occur in a packing as it moves towards or away from a jammed configuration, \citet{Walker:2015b} tracked the number of conformation subgraphs (and cycle-participation vectors) in each almost-invariant set across time. In Fig.~\ref{f:3D_subgraphs}\emph{e}, we show the temporal evolution of the numbers of elements in the almost-invariant sets of the conformation subgraphs. \citet{Walker:2015b} also proposed transition pathways (see Fig.~\ref{f:3D_subgraphs}\emph{f}) that may be useful for thermo-micro-mechanical constitutive-modeling efforts \citep{sepiani2009thermo}. (A \emph{transition pathway} consists of a sequence of conformation subgraphs in different almost-invariant sets, and transitions between them.) 

\begin{figure}
\centering
\includegraphics[width=\textwidth]{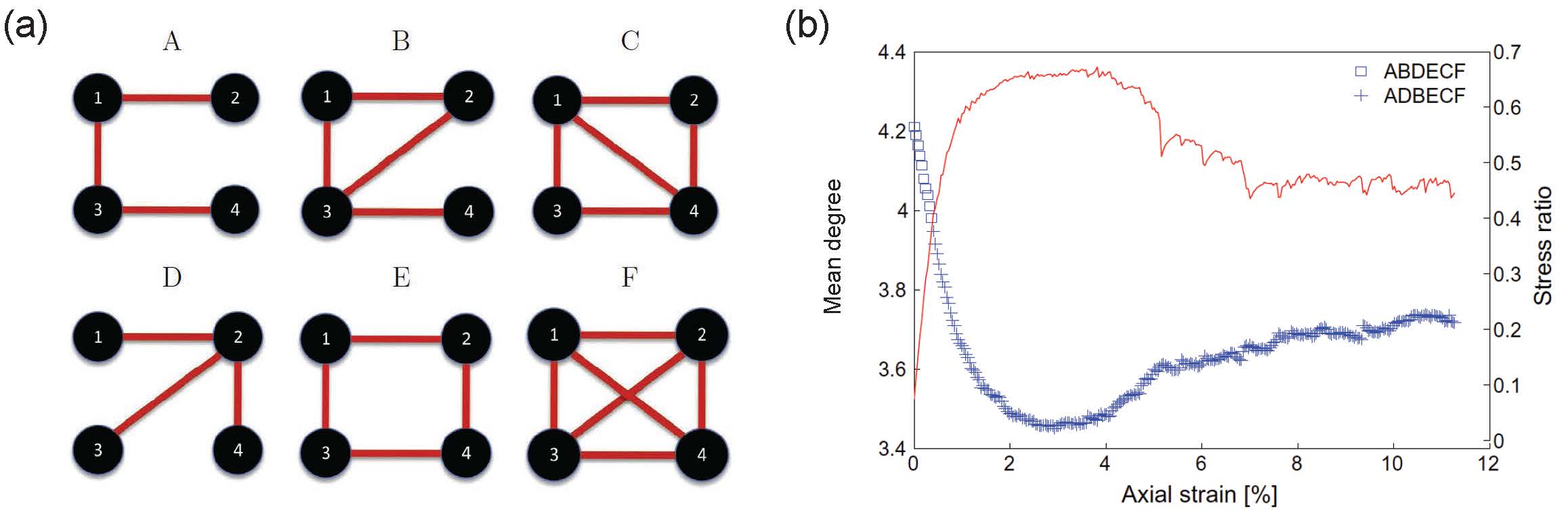}
\caption{\textbf{Examination of superfamilies in a granular material.}
\emph{(a)} One can use a rank-ordering of relative frequencies of sets of 4-node subgraphs to define a superfamily in a contact network. \emph{(b)} Mean degree (blue) and stress ratio (red) versus axial strain in a DEM simulation of a deforming granular material. (The stress ratio is the minimum stress divided by the maximum stress.) The symbols at each strain increment correspond to the superfamily of the associated contact network at that point. During loading, the system is characterized by a superfamily transition (specifically, ABDECF $\longrightarrow$ ADBECF) that corresponds to a shift from one with a higher prevalence of a subgraph that includes a triangle (B) to one with a higher prevalence of a subgraph without this stabilizing feature (D). [We adapted this figure, with permission, from \citep{Walker:2015c}.]
}
\label{f:superfamilies}
\end{figure}

Another way to study various types of subgraphs in granular materials is through the classification of superfamilies \citep{milo2004superfamilies,Xu:2008aa} (see Sec.~\ref{s:motifs}). A recent investigation by \citet{Walker:2015c} considered superfamilies that result from examining 4-particle subgraphs (see Fig.~\ref{f:superfamilies}\emph{a}) in a variety of different granular systems, including experimental packings of sand and photoelastic disks, and DEM simulations for different types of loading and in different dimensions. In their study, the authors defined a superfamily as a set of networks in which the prevalence of the different 4-node subgraphs have the same rank-ordering. (They did not consider whether the subgraph was a motif in the sense of occurring more frequently than in a random-graph null model.) Despite the diversity of system types, they observed several trends in the transitions between superfamilies that occur as a system transitions from a pre-failure regime to a failure regime. The most important change in the superfamilies appears to be a switch in relative prevalence of 4-edge motifs with 3 edges arranged as a triangle to acyclic 3-edge motifs (see Fig.~\ref{f:superfamilies}). This observation highlights the important role that small mesoscale structures can play as building blocks in granular systems. It also suggests that examining the prevalence and temporal evolution of such motifs can \emph{(1)} help characterize the macroscopic states of a granular system and \emph{(2)} help quantify what structural changes occur as a system transitions between different states. 

Notably, although calculating the prevalence of cycles and small motifs can be useful for gaining insights into contact-network structure, it is also important to employ other types of network analysis that examine structure on larger scales. For example, in simulations of 2D packings of disks under isotropic compression, \citet{Arevalo:2010aa} observed that the mean shortest-path length (\ref{eq:shortest_path}) of a contact network reflects changes in the organization of a packing as one approaches the jamming point and changes that occur after the jamming transition takes place. The path length appears to reach a maximum value at a packing fraction below $\phi_{c}$. With further increases in $\phi$ below $\phi_{c}$, the path length then decreases rapidly, likely due to the formation of new contacts that shorten the distance between particles as the system nears the jamming point. After the jamming transition, the path length decreases further.

Before moving on, we note that because it can be difficult to measure force information accurately in many experimental granular systems, continuing to develop relevant measures for studying contact topology (i.e., without incorporating weights) remains an important area of investigation.

\subsection{Force-weighted networks}
\label{s:force_weighted}

Although studying only a contact network can be rather informative (see Sec.~\ref{s:contact_network}), it is important to incorporate more information into network representations to more fully capture the rich behavior of granular materials. Many of the approaches for quantifying unweighted networks can be generalized to weighted networks (see Sec.~\ref{s:network_measures}), although significant complications often arise (e.g., because typically there are numerous choices for precisely how one should do the generalizing). From both physics and network-science perspectives, it is sensible to construct weighted networks that incorporate information about the forces between particles. This can shed considerable light on phenomena that have eluded understanding from studying only unweighted contact networks.
\begin{figure}
\centering
\includegraphics[width=0.75\textwidth]{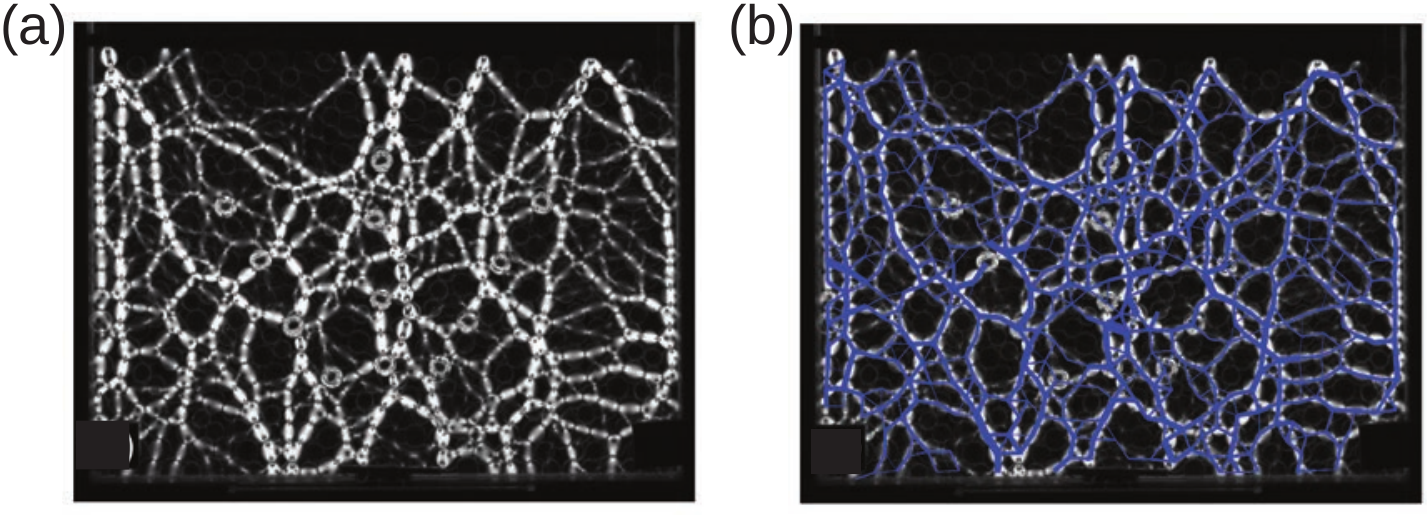}
\caption{\textbf{Photoelastic techniques allow the extraction of force-weighted networks from experimental granular packings.} \emph{(a)} An example of the photoelastic stress pattern from a vertical, 2D collection of bidisperse disks that are compressed from the top. \emph{(b)} Corresponding structure of the force network overlaid on the image. Each particle represents a node in the network, and line segments represent edges. Line thicknesses indicate edge weights and are proportional to the normal forces between contacting particles. [We adapted this figure, with permission, from \citep{bassett2015extraction}.]}
\end{figure}

One important physical motivation for incorporating information about inter-particle forces is that photoelastic-disk experiments and numerical simulations have both highlighted that, particularly just above isostaticity (see the bottom of Sec.~\ref{degree}), loads placed on granular systems are not shared evenly across all particles. Instead, forces are carried primarily by a backbone of force chains. It has often been claimed that the statistical distribution of the forces are approximately exponential \citep{Liu:1995aa}, but in fact, it depends on whether forces are recorded at a granular system's boundary or in its bulk \citep{Tighe:2011a}, as well as on the loading history \citep{Majmudar:2005aa}. Illuminating how force-chain structures arise provides crucial information for understanding how one can control the elastic modulus and mechanical stability \citep{Aharonov:1999a} and acoustic transmission \citep{Makse-1999-WEM} in granular materials.

However, despite the ability of humans to see force chains easily in photoelastic images, it is difficult to characterize quantitatively what is or is not a force chain, and it can also be difficult to quantify how force chains evolve under compression or shear. Part of the challenge lies in the fact that force chains are spatially anisotropic, can exhibit long-range spatial correlations \citep{Majmudar:2005aa}, and can have complex temporal fluctuations \citep{Howell1999,Hidalgo2002}. Consequently, understanding emergent phenomena in granular systems is typically difficult using continuum theories or approaches based only on local structure. On the other hand, a network-theoretic perspective provides a fruitful way to explore interesting material properties and organization that arise from inter-particle contact forces in granular materials. Importantly, in addition to data from simulations, multiple techniques now exist for measuring inter-particle forces between grains in experiments; these include photoelasticity \citep{Majmudar:2005aa,Daniels2017}, x-ray diffraction measurements of microscopic \citep{Hurley:2016a} or macroscopic \citep{Mukhopadhyay2011,Saadatfar2012} deformations, and fluorescence with light sheets \citep{Brodu2015,Dijksman2017}. As we will see, incorporating information about inter-particle forces into network-based investigations has yielded fascinating insights into the organization and collective structure in granular packings (and other particulate materials) for both numerically-simulated and experimental systems.

The most common method for constructing a network that captures the structure of forces in a granular system is to let a node represent a particle and then weight the edge between two particles that are in contact according to the value of the force between them. One can describe such a network with a weighted adjacency matrix (see Sec.~\ref{s:networks}) ${\bf W}$ with elements
\begin{equation}
	W_{ij} = \left\{\begin{array}{ll}
	         f_{ij}\,, \text{ if particles $i$ and $j$ are in contact\,,} \\
	       	 0\,, \text{ otherwise\,,}
	        \end{array} \right.
\end{equation}
where $f_{ij}$ is the inter-particle force between particles $i$ and $j$. Such a force network also encodes the information in the associated contact network, and one can recover a contact network from a force-weighted network by setting all non-zero weights in the latter to $1$. Although most work has determined edge weights using the normal component of the inter-particle force, one could alternatively weight the edges by inter-particle tangential forces. With the advent of high-performance computational capabilities, one can determine inter-particle forces from DEM simulations \citep{Poschel2005} of hundreds to millions of particles. In experiments, it is possible to determine inter-particle forces using photoelastic disks (in 2D) \citep{Daniels2017} or x-ray tomography (in 3D) \citep{weis:17}, although typically these techniques are limited to systems of hundreds to thousands of particles.

We now review network-based approaches for investigating force-weighted networks constructed from granular materials and discuss the resulting insights that these approaches have provided into the physical behavior of such systems. We label most of the following subsubsections according to the type of employed methodology, although we also include a subsubsection about some specific applications to different systems.

\subsubsection{Examining weighted cycles and other structural features.}
\label{s:weighted_cycles}
\begin{figure}
	\centering
	\includegraphics[width = \textwidth]{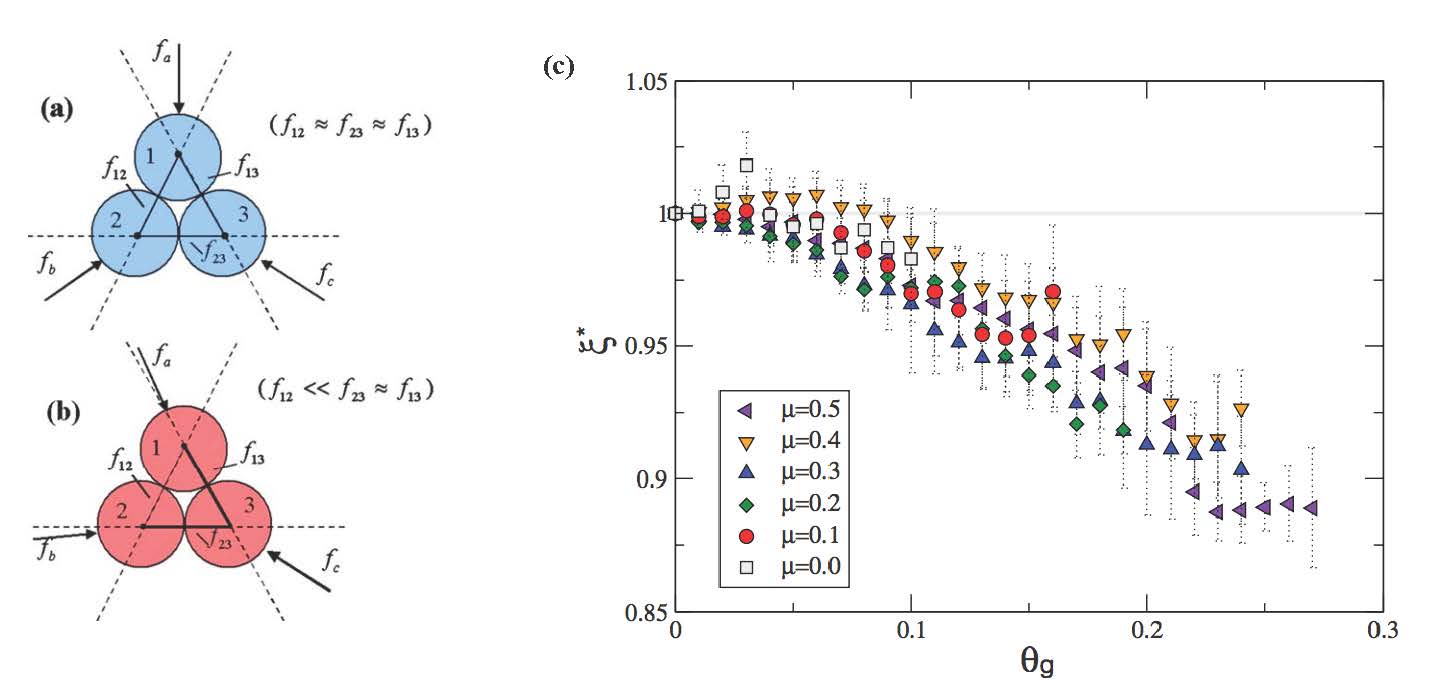}
	\caption{\textbf{A measure of the stability of 3-cycles quantifies the effects of tilting a granular packing.} The \emph{loop stability} $\xi_{3}$ of a 3-cycle is determined from the contact forces along each edge of the loop. (See the main text for details.) \emph{(a)} A contact loop with approximately equal forces on all edges is very stable to perturbations in the direction of the compressive force vectors on the loop, and has a loop stability that is close to $1$. \emph{(b)} A contact loop in which one of the edges has a much smaller force than the others is very unstable to perturbations; it has a loop stability near $0$. \emph{(c)} The normalized mean loop stability $\xi^{*}_{3}$ tends to decrease with increasing tilting angle $\theta_{\text{g}}$. [We adapted this figure, with permission, from \citep{Smart:2008aa}.]} 
	\label{f:loop_stability}
\end{figure}

In Sec.~\ref{s:contact_network}, we discussed why examining cycles can be useful for studying granular contact networks. It is also useful to examine cycles when investigating force networks, which are weighted. For example, \citet{Smart:2008aa} studied the evolution of weighted contact loops in a simulation of a quasistatically tilted granular packing. They used topological information (i.e., which particles are in contact) to define the presence of a cycle, and they defined a notion of \textit{loop stability}, 
\begin{equation}
 	\xi_{l} = \frac{1}{\overline{f}^{l}}\prod_{i=1}^{l}f^{\text{edge}}_{i}\,,
\label{eq:loop_stability}
\end{equation}	
to quantify the range of compressive loads that a given loop can support. In Eq.~\eqref{eq:loop_stability}, $l$ is the number of edges in the loop (i.e., its \emph{length}), $f^{\text{edge}}_{i}$ is the contact force of the $i^{\mathrm{th}}$ edge, and $\overline{f}$ is the mean edge weight (i.e., mean force) over all of the edges in the loop. See Fig.~\ref{f:loop_stability} for a schematic of this stability measure for a $3$-cycle. For $l = 3$, the quantity $\xi_{3} \approx 1$ corresponds to having approximately equal contact forces on all edges and is the most stable configuration (see Fig.~\ref{f:loop_stability}\emph{a}). The value of $\xi_{3}$ approaches $0$ as the contact force on one edge becomes much smaller than those on the other two edges. As illustrated in Fig.~\ref{f:loop_stability}\emph{b}, this situation is rather unstable. Both the density of $3$-cycles (specifically, the number of $3$-cycles in the system divided by the total number of particles) and a normalized $3$-cycle loop stability 
\begin{equation*}
	\xi^{*}_{3} = \frac{\langle \xi_{3}(\theta_{g}) \rangle}{\langle \xi_{3}(\theta_{g} = 0) \rangle}
\end{equation*}	
(where the brackets denote means over all $3$-cycles in a network) tend to decrease with increasing tilting angle $\theta_g$ (see Fig.~\ref{f:loop_stability}\emph{c}). \citet{Smart:2008aa} also reported that the effect of tilting on loop stability is largely independent from the effect of tilting on mean coordination number (i.e., mean degree).

\begin{figure}[t!]
\centering
\includegraphics[width=0.5\textwidth]{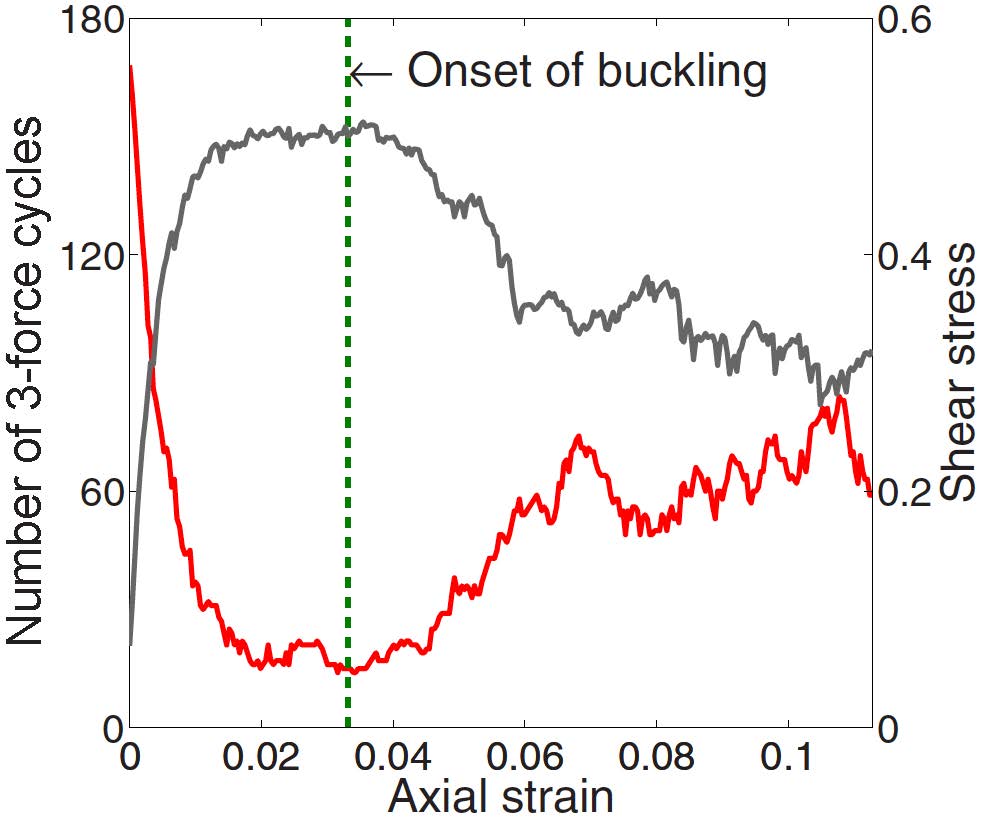}
\caption{\textbf{Evolution of shear stress (gray curve) and the number of $3$-force cycles in a minimal cycle basis of a contact network (red curve) as a function of axial strain in a DEM simulation of a granular material under quasistatic, biaxial loading.} The number of 3-force cycles decreases rapidly during the initial stages of strain hardening, but it begins to increase at the onset of force-chain buckling (dashed line). [We adapted this figure, with permission, from \citep{Tordesillas:2010aa}.]
}
\label{f:Tordesillas2010aa_Fig10b}
\end{figure}

\citet{Tordesillas:2010aa} examined what they called a \emph{force cycle}, which is a cycle of physically-connected particles in which each contact carries a force above the global mean. Using DEM simulations of a biaxially compressed, dense granular system --- the sample was compressed quasistatically at a constant strain rate in the vertical direction, while allowed to expand under constant confining pressure in the horizontal direction --- they studied the evolution of $3$-force cycles (i.e., force cycles with $3$ particles) in a minimal cycle basis of the contact network with respect to axial strain. They observed that $3$-force cycles initially decrease in number during strain hardening, before increasing in number at the onset of force-chain buckling \citep{Tordesillas:2007aa}, and finally leveling out in number in the critical-state regime. (See the third paragraph of Sec.~\ref{s:role_of_cycles} for a brief description of these different regimes of deformation.) 
In Fig.~\ref{f:Tordesillas2010aa_Fig10b}, we show a plot of the number of $3$-force cycles and the shear stress versus axial strain. The $3$-force cycles that arise at the onset of buckling are often part of force chains (using the definition from \citep{Peters:2005aa,Muthuswamy2006a}). Additionally, these $3$-force cycles tend to concentrate in the region of the shear band, where they may act as stabilizing structures both by frustrating relative rotations and by providing strong lateral support to force chains. However, with increased loading, the system eventually fails, and \citet{Tordesillas:2010aa} suggested that the increase in the number of $3$-force cycles may be an indicator of failure onset. Qualitatively similar results have been observed when examining the evolution of $3$-force cycles in three DEM simulations (each with slightly different material properties and boundary conditions) \citep{Tordesillas:2010a} and in DEM simulations of 3D ellipsoidal-particle packings subject to triaxial compression \citep{Tordesillas:2010b}. 

Using similar DEM simulations for biaxial compression as those described in the previous paragraph, \citet{Walker:2010aa} examined the evolution of force-weighted networks with axial strain using several of the network concepts that we discussed in Sec.~\ref{s:network_measures}. Unsurprisingly, they found that including contact forces in their network analysis yields a more complete characterization of a granular system than ignoring them.

One measure that they felt was particularly useful is a weighted version of subgraph centrality (see Sec.~\ref{s:centrality}). From a contact network, \citet{Walker:2010aa} first extracted all conformation subgraphs. As described in Sec.~\ref{subgraph}, a conformation subgraph is a subgraph that consists of a given particle $i$ and that particle's immediate contacts. (Each particle in a network thus yields one conformation subgraph.) To incorporate inter-particle force information, \citet{Walker:2010aa} generated \textit{force-weighted conformation subgraphs} by weighting each edge in the conformation subgraphs by the magnitude of the normal force component along that contact. They then computed a weighted subgraph centrality $Y^{w}_{i}$ for each force-weighted conformation subgraph, and they computed changes $|\upDelta \tilde{Y}^{w}|$ in magnitude between consecutive strain steps of a suitably averaged version ($\tilde{Y}^{w}$) of this quantity (see \citep{Walker:2010aa} for details). They observed that the temporal evolution of $|\upDelta \tilde{Y}^{w}|$ with strain step is effective at tracking large changes in energy dissipation rate that can occur due to rearrangement events (e.g., force-chain buckling) associated with the loss of inter-particle contacts. They also observed that the central particle in the conformation subgraphs that undergo the largest changes in weighted subgraph centrality seems to be associated with locations with much dissipation, such as in the shear band and in buckling force chains or the neighboring particles of those force chains.

See \cite{Peters:2005aa,Muthuswamy2006a} for the employed definition of force chains and \cite{Tordesillas:2007aa} for the employed specification of force-chain buckling. \citet{Walker:2010aa} highlighted that network analysis --- and especially examination of mesoscale features --- can be helpful for gaining insights into mechanisms that regulate deformation in granular materials. Such studies can perhaps also help guide efforts in thermo-mechanical constitutive modeling \citep{Tordesillas:2008a}.

\subsubsection{Extracting multiscale architectures from a force network using community detection.}
\label{s:comm_detect}

A major benefit of studying a network representation of a granular system (and using associated computational tools) is that it provides a natural framework in which to probe structure and dynamics across several spatial scales. One can examine different spatial scales in multiple ways, including both physically (e.g., using distance in Euclidean space or in some other metric space) or topologically (e.g., using the hop distance along edges in a network). In these studies, one can use network diagnostics and approaches like the ones discussed in Sec.~\ref{s:network_measures}. The ability to successfully study mesoscale architecture, which manifests at intermediate scales between particle-level and system-wide organization, is an especially important contribution of network analysis. One of the most common ways to examine mesoscale structures is with community detection (see Sec.~\ref{s:comm_structure}), which one can use to extract sets of densely-connected nodes in a network \citep{Fortunato2016,Porter2009}. One can also tune community-detection methods to examine sets of nodes of a range of sizes, from very small sets (with a lower limit of one node per set) to large sets (with an upper limit of all nodes in a single set).

By applying multiscale community-detection methods to force-weighted contact networks of photoelastic disks, \citet{bassett2015extraction} identified chain-like structures that are visually reminiscent of force chains in granular packings. Notably, the algorithmic extraction of these ``distributed'' mesoscale structures \citep{bassett2015extraction,Giusti:2016a,Huang:2016a,papadopoulos2016evolution}, in contrast to ``compact'', densely-connected geographical domains in a material \citep{bassett2012influence}, required the development of a geographical null model, which can be used in modularity maximization and which encodes the fact that a given particle can exert force only on other particles with which it is in direct contact \citep{bassett2013robust} (see Sec.~\ref{s:comm_structure}). The different type of mesoscale organization extracted by this geographical null model highlights the fact that using physically motivated network-based approaches, which incorporate spatial and/or physical constraints, may give different information about a granular system than what one obtains when doing network calculations based only on network structure (i.e., without considering known context about a network, so that ``one size fits all''). In a modularity-maximization approach to community detection, one can also tune a resolution parameter of a modularity objective function to identify and characterize network communities of different sizes. This can also be combined with inference procedures to determine particularly important scales.
\begin{figure}
	\centering
	\includegraphics[width=\textwidth]{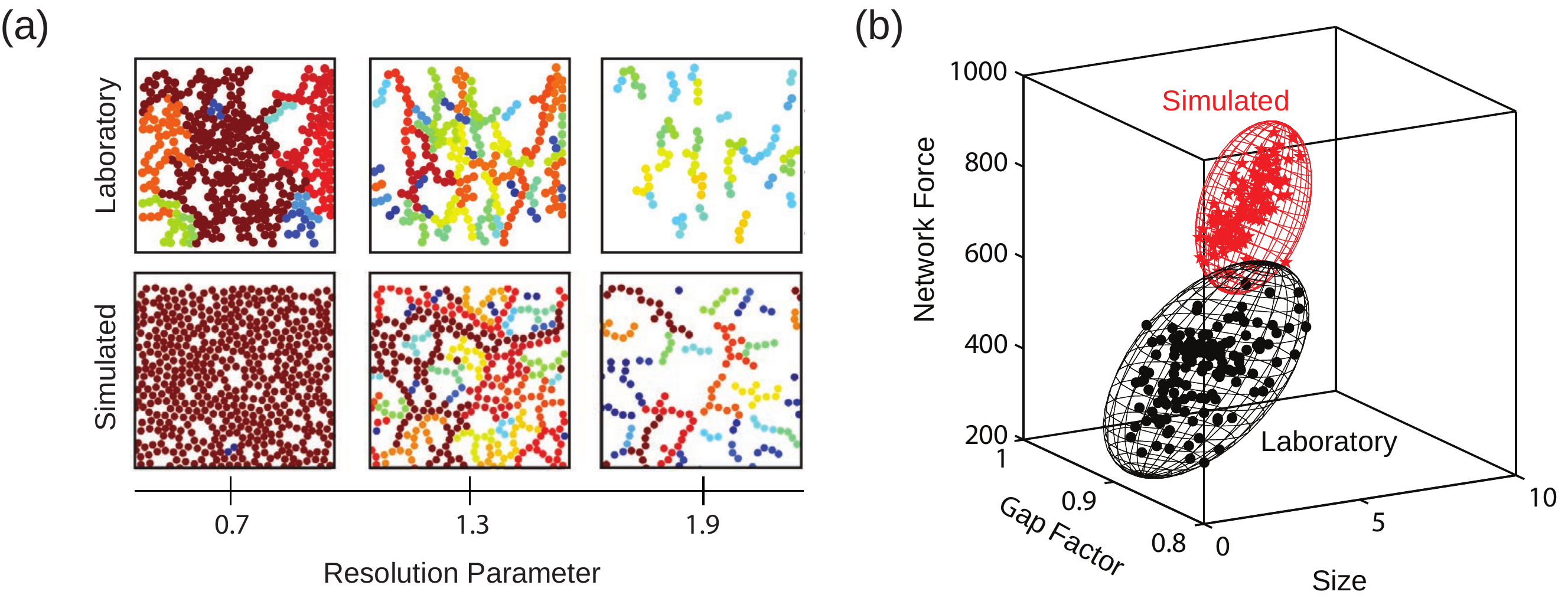}
	\caption{\textbf{Community-detection techniques can uncover multiscale force-chain structures in laboratory and simulated granular packings.} \emph{(a)} Examples of force-chain communities extracted using modularity maximization for networks constructed from 2D laboratory and simulated granular packings. Tuning a resolution parameter allows one to detect mesoscale features at multiple scales in a packing: smaller resolution-parameter values yield larger and more compact communities, and larger values yield smaller and chain-like communities. \emph{(b)} One can characterize community structure using different measures, which are able to differentiate between laboratory and simulated packings in a quantitative manner. In this figure, the \emph{size} of a community is the number of particles in that community, the \emph{network force} of a community is the contribution of that community to modularity, and the \emph{gap factor} is a diagnostic that measures the presence of gaps and the extent of branching in a community. (The gap factor is small for very linear or compact communities, and it is large for communities with curves and/or branching.) [We adapted this figure, with permission, from \citep{bassett2015extraction}.]
	}
\end{figure}

One interesting result from \citep{bassett2015extraction} is that properties of force chain-like communities can distinguish frictional, laboratory packings from frictionless, simulated ones, allowing a quantification of structural differences between these systems. In later work, \citet{Huang:2016a} used similar techniques to examine friction-dependence and pressure-dependence of community structure in 3D simulations of compressed granular materials. To further quantify such mesoscale organization and examine how it changes with compression, \citet{Giusti:2016a} extracted communities using the geographical null model, and used ideas from algebraic topology to define a \emph{topological compactness factor} that quantifies the amount of branching --- versus compact, densely-interconnected regions --- in communities of a force network from 2D granular systems. The approach from \citep{bassett2015extraction} was extended to multilayer networks (see Sec.~\ref{s:comm_structure}) in \citep{papadopoulos2016evolution}, providing a way to link particulate communities across compression steps (rather than extracting new ones at each step) when examining how such communities reconfigure. These studies helped lay groundwork to improve understanding of how the multiscale nature of force-chain architecture impacts bulk material properties. 
Various community-detection approaches have also been used for identifying other types of inhomogeneities in granular matter \citep{Navakas:2014aa}.

Before moving on, it is important to note that --- although related --- the definition of force-chain structure using the community-detection approaches that we described above \cite{bassett2012influence,bassett2015extraction,Huang:2016a,Giusti:2016a,papadopoulos2016evolution} differs from the definitions of force chains that have been used in some other studies (e.g., see \cite{Peters:2005aa}). In future work, it is important to examine how the properties of force chains differ when they are defined in different ways. 

\subsubsection{Some applications.}

\noindent \emph{Comminution processes.} A network-based approach can give fascinating insights into \emph{comminution}, the fragmentation of a material into smaller pieces. \citet{Walker:2011a} used DEM simulations to study comminution in a granular material under uniaxial compression and reported that the degree distribution of the system's contact network (which we recall is unweighted) evolves towards a power law during this process. This is consistent with the development of a power-law grain-size distribution, in which large particles are hubs that have many smaller, neighboring particles, which make up the majority of a packing. \citet{Walker:2011a} also examined several other features (such as measures of network efficiency, node betweenness, and cycle populations) of both contact networks and networks weighted by the normal force between particles as a function of increasing strain to examine what changes occur in a granular system during comminution. 

~\\
\noindent \emph{Heat transfer.} Another problem that has been examined using network-based methods is heat transfer in granular matter. Using a heat-transport model on simulations of a compressed granular material, \citet{Smart:2007a} probed the effects of heterogeneity in the force distribution and the spatial arrangements of forces in a system on heat transfer through the material. Specifically, they compared measures of transport in the (normal) force-weighted network of a granular system to two null-model networks with the same contact topology but with either \emph{(1)} homogenous, uniform edge weights that were equal to the mean force of the packing; or \emph{(2)} the same set of heterogeneous edge weights from the actual granular network, but assigned uniformly at random to the edges. \citet{Smart:2007a} estimated the thermal diffusivity and effective conductivity from simulations on each network, and they observed that the real granular system has significantly higher diffusivity and effective conductivity than the homogenous null model. Additionally, comparing the results from the real material to the null model with randomly reassigned edge weights demonstrated that the qualitative differences between the real granular network and the homogenous null model could not be explained by the heterogeneity in the force distribution alone, as the authors observed that this second null model (with randomly reassigned edge weights) was also not a good medium for heat transfer. 

To investigate what features of a granular network facilitate efficient heat transfer, \citet{Smart:2007a} defined a weighted network distance (see Sec.~\ref{paths}) between particles $i$ and $j$ as $d^{w}_{ij} = 1/H_{ij}$, where $H_{ij}$ is the \emph{local heat-transfer coefficient}, such that the network distance between two particles in contact is proportional to that contact's resistance to heat transfer. Note that $H_{ij} \propto f_{ij}^{\nu}$, where $f_{ij}$ is the magnitude of the normal force between $i$ and $j$, and $\nu \geq 0$ is a constant. They then defined a network-based (but physically-motivated) measure of heat transport efficiency as the weighted efficiency $E^{w}$ (see Sec.~\ref{paths}) computed using the distances $d^{w}_{ij}$. In a comparison between the real granular system and the two null models, $E^{w}$ gave the same quantitative results as the effective conductivity. In particular, the calculations in \citep{Smart:2007a} revealed that the real granular system has a larger efficiency than that of either null model, suggesting that the spatial distribution of force-chain structure in the granular network appears to facilitate heat transport. Finally, iterative edge removals in decreasing order of geodesic edge betweenness centrality [Eq.~\eqref{eq:edge_betweenness}] yield a faster decrease in effective conductivity than either edge removals done uniformly at random or edge removals in decreasing order of the local heat-transport coefficient, further illustrating the utility of network-theoretic measures for examining transport phenomena in granular systems.

~\\
\noindent \emph{Acoustic transmission.} One can also examine the effect of network structure on properties such as electrical conductivity in systems composed of metallic particles or on other types of transport (such as sound propagation) through a particulate material. The transmission of acoustic signals through granular materials is poorly understood \citep{Owens:2011}, and it is particularly challenging to explain using continuum or particulate models \citep{Digby:1981,Velicky:2002,Goddard:1990}. A few years ago, \citet{bassett2012influence} represented compressed, 2D packings of bidisperse, photoelastic disks as force-weighted contact networks, and found that some network diagnostics are able to identify injection versus scattering phases of acoustic signals transmitted through a granular material. Among the diagnostics that they computed, the authors observed that network efficiency (see Eq.~\eqref{eq:E} in Sec.~\ref{paths}) is correlated positively with acoustic transmission during the signal-injection phase, suggesting that high-amplitude and high-energy signals are transmitted preferentially along short paths (and perhaps even shortest paths) through a force-weighted contact network. In contrast, low-amplitude and low-energy signals that reverberate through a packing during the subsequent signal-scattering phase correlate positively with the intra-community strength $z$-score, which characterizes how strongly a node connects to other nodes in its own community. These results suggest that one can use network diagnostics that probe diverse spatial scales in a system to describe different bulk properties. Because \citep{bassett2012influence} did not use community-detection approaches informed by a geographical null model (see Sec.~\ref{s:comm_structure}), it did not address (and it is not yet fully understood) how acoustic transmission depends on the multiscale architecture of chain-like structures reminiscent of force chains. This remains an open issue, and network-based approaches --- e.g., using geographical null models and other ideas that pay attention to the role of space --- are likely to be important in future work on this topic.

\subsubsection{Thresholded force networks.}
\label{s:force_threshold}

Before researchers started using network-based methods as a common perspective for studying granular materials, \citet{Radjai:1998aa,Radjai:1999aa} reported that under stress, the force network in simulations of granular matter organizes into two subsets: one set with ``strong" contacts and another set with ``weak" contacts. The \emph{strong subnetwork} of forces forms a backbone of chain-like structures that carry most of a system's load and which tend to align approximately with the direction of compression. Between these strong force chains, there is a \emph{weak subnetwork} of contacts that carry forces that are less than the mean. This weak subnetwork tends to have an anisotropy that is orthogonal to the compression, and it may provide support to the backbone of strong forces. Such heterogeneity in a force network is an interesting feature of granular materials, and network-based approaches provide a direct way to examine how strong and weak contacts can play important roles in material properties and stability.
\begin{figure}
\centering
\includegraphics[width=0.75\columnwidth]{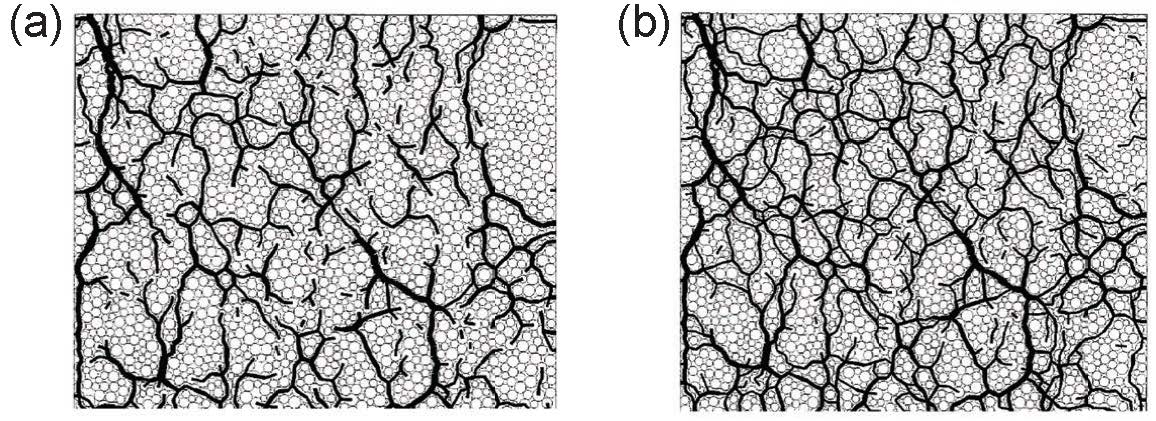}
\caption{\textbf{The contacts and forces in a granular network at two different force thresholds.} \emph{(a)} When only contacts with forces $f > 1.3 \langle f \rangle$ (where $\langle f \rangle$ is the mean force) are included, the resulting network is composed of many disconnected components. \emph{(b)} As one decreases the threshold, the components grow, and if only contacts with $f > \langle f \rangle$ are included, the network is connected and ``percolates" along the axis of compression. The system is a numerical simulation of a dense, 2D collection of hard spheres under biaxial compression. [We adapted this figure, with permission, from \citep{Radjai:1998aa}.]
}
\label{f:Radjai1997a_Fig5}
\end{figure}

Theses ideas have been explored using \emph{force-thresholded} networks \citep{Arevalo:2009aa,Arevalo:2010aa,Arevalo:2010ba,Pena:2009aa}, in which one retains only contacts that carry a force of at least some threshold $f_{\mathrm{th}}$. That is,
\begin{equation}
	A^{\mathrm{th}}_{ij} = \left\{\begin{array}{ll}
	         1\,, \text{ if particles $i$ and $j$ are in contact and } f_{ij} \geq f_{\mathrm{th}}\,, \\
	       	0\,, \text{ otherwise.}
	        \end{array} \right.
\label{eq:force_threshold}
\end{equation}

The threshold $f_\mathrm{th}$ should be lower-bounded by the smallest non-zero contact force $f_{\mathrm{min}}$ in the system being considered. (Note that when $f_{\mathrm{th}} = f_{\mathrm{min}}$, one includes all contacts in the force-thresholded network.) It is common to use a system's mean force $\langle f \rangle$ as a reference point and to systematically vary $f_{\mathrm{th}}$ to be different fractions of the mean force. Whether one uses the normal or the tangential component of the force, varying $f_{\mathrm{th}}$ results in a series of thresholded networks that one can subsequently characterize using traditional network or percolation-based analyses (which we discuss in the following three paragraphs of this subsubsection) or using methods from computational algebraic topology (which we discuss in Sec.~\ref{s:granular_alg_topology}). See Fig.~\ref{f:Radjai1997a_Fig5} for an example depicting the contacts and forces in a simulation of a 2D granular material at two different values of the force threshold. There is also a connection between the idea of force-thresholded networks and carrying out modularity maximization with the geographical null model [Eq.~\eqref{geog}] and a resolution parameter $\gamma$ (see Sec.~\ref{s:comm_structure}). Similar to how increasing the threshold $f_{\mathrm{th}}$ selects the subset of particles in a network for which inter-particle forces exceed or equal the threshold, in modularity maximization, increasing the resolution-parameter value $\gamma$ tends to yield communities of particles such that within a community, the inter-particle forces are at least as large as the threshold $\gamma \langle f \rangle$.

\citet{Arevalo:2009aa, Arevalo:2010aa,Arevalo:2010ba} examined several network diagnostics --- e.g., mean degree, shortest-path length, diameter, LCC size, and component-size distributions --- as a function of the force threshold $f_{\mathrm{th}}$ (and at different values of the packing fraction) in DEM simulations of 2D granular packings under isotropic compression. The computations in \citep{Arevalo:2009aa, Arevalo:2010aa} suggest that the way in which many of these measures change as a function of $f_{\mathrm{th}}$ depends on packing fraction, and many of the measures can thus potentially be used to help understand differences in the organization of force networks above versus below the jamming point. For example, for packing fractions above the jamming point, the LCC size and the shortest-path length undergo qualitative changes in behavior near a threshold $f_{\mathrm{th}} \approx \langle f \rangle$, signifying interesting structural changes in the organization of the force networks at that threshold. The relationship between the number of 3-particle contact cycles (i.e., triangles) and the force threshold was examined in \citet{Arevalo:2010ba,Arevalo:2010aa}. In the jammed state, they observed a steep decline in the number of 3-cycles in the networks as they increased the threshold and considered only progressively larger forces. (As we discuss in Sec.~\ref{s:granular_alg_topology}, one can also use methods from computational algebraic topology to examine the evolution of cycle organization in dense granular materials.) This observation suggests that triangles (or at least one of their contacts) belong primarily to the weak subnetwork of forces that help support strong, filamentary force-chain structures. See Fig.~\ref{f:Behringer2014Statistical_Fig3Right} and Secs.~\ref{s:role_of_cycles} and ~\ref{s:weighted_cycles} for other discussions of the roles of cycles and their relationship to force chains.

Another way to study force-thresholded granular networks is using a percolation-like approach (see Sec.~\ref{s:largest_comp}). For example, one can examine the sizes and number of connected components in a thresholded network as a function of $f_{\mathrm{th}}$ \citep{Ostojic:2007a,PastorSatorras2012,Kovalcinova:2015a,Kovalcinova:2016a,pathak2017force}. For dense packings, the intuition is that when $f_{\mathrm{th}}$ is very large, a force-thresholded granular network splits up into several disconnected components; and as one decreases $f_{\mathrm{th}}$, these components begin to merge until eventually all contacts are in a single component. In this type of bond percolation, which edges are included in the network thus depends on the force that the edges carry. One can look for a critical threshold $f^{\mathrm{c}}_{\mathrm{th}}$ such that for $f_{\mathrm{th}} > f_{\mathrm{th}}^{\mathrm{c}}$, the network fragments into many small components, but as one lowers the threshold towards $f^{\mathrm{c}}_{\mathrm{th}}$, a large, percolating cluster forms in the system (see Sec.~\ref{s:largest_comp}). Quantities that are often investigated when studying this type of force-percolation transition include $f^{\mathrm{c}}_{\mathrm{th}}$ and ``critical exponents'' for the transition (see Sec.~\ref{s:largest_comp}). Because (both experimental and computational) granular systems are finite in practice, such investigations often use finite-size scaling techniques.

Several simulation-based studies of granular systems have deployed percolation analyses based on force-thresholded networks to quantify the organization of granular force networks, how such organization changes with increasing compression, and other phenomena \citep{Ostojic:2007a,PastorSatorras2012,Kovalcinova:2016a,pathak2017force}. For example, \citet{PastorSatorras2012} studied force percolation in the \emph{q-model} \citep{Coppersmith:1996a} of anisotropic granular force networks, in which there is a preferred direction of force propagation. They concluded that the asymmetry in the model has a significant effect on the percolation transition, and they found that the critical exponents differ from those of isotropically compressed granular force networks. \citet{Kovalcinova:2016a} investigated force percolation in a variety of simulations of slowly compressed, 2D granular systems. They examined the effects of polydispersity and friction, finding that these factors can qualitatively influence various features of the percolation transition. Very recently, \citet{pathak2017force} also investigated the force-percolation transition in simulations of jammed granular packings at fixed pressures.

\subsubsection{Methods from computational algebraic topology.} 
\label{s:granular_alg_topology}

In addition to traditional approaches for network analysis, one can also study the architecture of granular networks using ideas from algebraic topology. Persistent homology (PH) \citep{Kaczynski:2004aa,Edelsbrunner:2010a,otter2015,Kramar:2014b} (see Sec.~\ref{s:homology}) seems especially appropriate, and over the past several years, it has provided a fruitful perspective on the structure of compressed \citep{Kondic:2012aa,Kramar:2013aa,Kramar:2014b,Kramar:2014aa,Kondic:2017a}, and tapped \citep{Pugnaloni:2016a,Kondic:2016a,Ardanza-Trevijano:2014aa} granular materials. Very recently, it has also been used to study responses of granular materials to impact \cite{kondic-impact2017,lim-impact2017}. One way to characterize the organization and evolution of granular force networks is to examine how Betti numbers (see Sec.~\ref{s:homology})) change as a function of a force threshold (and also as a function of packing fraction) in compressed granular systems. This includes studying the birth and death of components (determined by $\beta_{0}$) and loops (determined by $\beta_{1}$) as a function of a force threshold by computing and analyzing persistence diagrams (see Fig.~\ref{f:PH_schematic} of Sec.~\ref{s:homology}). Examining when and how long different features persist in a network provides a detailed characterization of the structure of granular force networks, and one can quantify differences between two networks by defining measures of ``distance" between their associated persistence diagrams. These capabilities allow a PH framework to provide a distinct set of contributions that complement more traditional network-based analyses of components, cycle structure, and other features.

\citet{Kondic:2012aa} investigated how simulated 2D granular force networks evolve under slow compression as they cross the jamming point. They first demonstrated that one can identify the jamming transition by a significant change in behavior of $\beta_{0}$ (specifically, there is an increase in the number of components at a force threshold approximately equal to the mean force $\langle f \rangle$), and that structural properties of the network --- such as the size of the connected components --- continue to change above jamming. \citet{Kondic:2012aa} also demonstrated that $\beta_0$ and $\beta_1$ can quantitatively describe the effects of friction and polydispersity on the organization of force networks (and can distinguish how friction and polydispersity alter the structure of a force network). This work was extended in \citet{Kramar:2013aa}, who examined numerical simulations of 2D, slowly compressed, dense granular materials using PH. In addition to examining the values of the Betti numbers, they also computed $\beta_{0}$ and $\beta_{1}$ persistence diagrams ($\mathrm{PD}_{0}$ and $\mathrm{PD}_{1}$, respectively) as the system was compressed to quantify the appearance, disappearance, and lifetimes of components and loops. In \citep{Kramar:2013aa}, they defined a filtration over the clique complex of the networks (see Sec.~\ref{s:homology}), so only loops with four or more particles were counted. To extract useful information from the PDs, they binned the persistence points in each diagram into different regions corresponding to features that \emph{(1)} are born at any force threshold but have relatively short lifetimes compared to those that are born at either \emph{(2)} strong, \emph{(3)} medium, or \emph{(4)} weak forces and that persist for a large range of thresholds. Their persistence analysis led to several insights into the structure of a normal-force network as a granular system is compressed, as well as insights into differences in the structure of a normal-force network for systems with different amounts of friction and different polydispersities. For example, \citet{Kramar:2013aa} observed that, near the jamming point, frictionless packings appear to have more ``extreme'' features than frictional packings, in the sense that frictionless packings have many more $\beta_{0}$ persistence points that are born at either weak or strong forces and that are relatively long-lived. Observations on the effects of polydispersity and friction may be difficult to observe using traditional measures such as the probability density function of the normal forces. 
\begin{figure}[t!]
\centering
\includegraphics[width=0.9\textwidth]{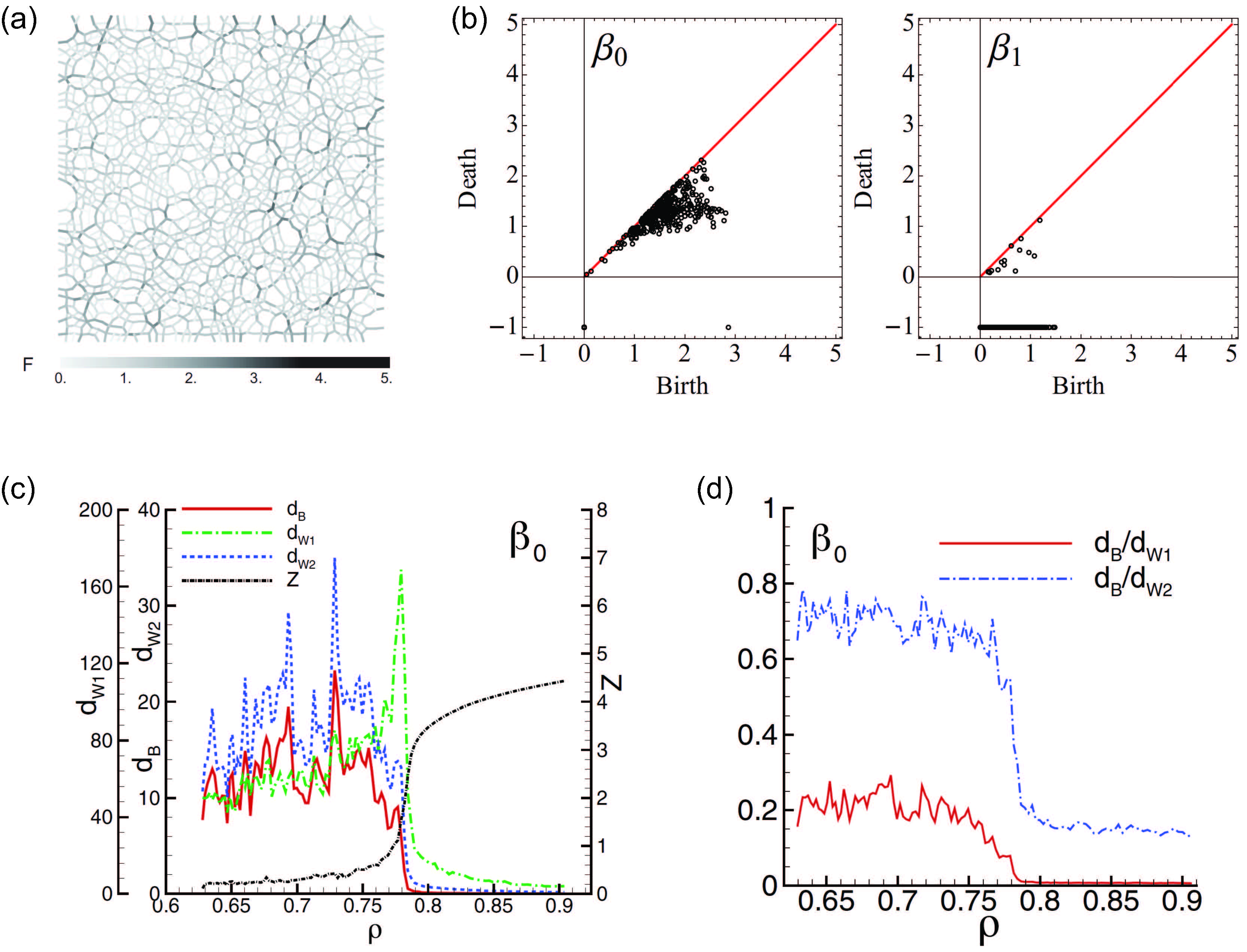}
\caption{\textbf{Examination of granular force networks using persistent homology.} \emph{(a)} An example of a force network from a simulation of a dense, 2D, frictional, polydisperse granular material under compression. The grayscale bar indicates the magnitude of the normal force (normalized by the mean force in the packing) between contacting particles. \emph{(b)} The left panel shows the $\beta_{0}$ persistence diagram ($\mathrm{PD}_{0}$) of the force network, and the right panel shows the $\beta_{1}$ persistence diagram ($\mathrm{PD}_{1}$) of the force network. These PDs indicate the appearance and disappearance of connected components and loops, respectively. Because the filtration is defined over the clique complex of the force network, one counts loops only when they include four or more particles. \emph{(c)} One can examine the evolution of the force network as a function of packing fraction $\rho$ by using various distances (left axis) to quantify differences between the persistence diagrams computed from force information from consecutive samples (see the main text for details) of the system during compression. These distances (the \emph{bottleneck distance} $d_B$ and two variants, $d_{W1}$ and $d_{W2}$, of the \emph{Wasserstein distance} \citep{otter2015}) capture both local and global changes in force geometry as one compresses the system. The curve for the mean coordination number $Z$ (right axis) gives the mean number of contacts per particle, and the steep rise in $Z$ signifies the onset of jamming. The values of the distances (which we show for consecutive $\mathrm{PD}_{0}$s) change dramatically as the system goes through the jamming point, after which they vary much more smoothly. See \citep{Kramar:2014aa} for details about these measures and for an example computation of the distances between consecutive $\mathrm{PD_1}$s as a function of packing fraction. \emph{(d)} The ratios of some of the distance measures during compression. Considered in conjunction with the behavior of the distance measures in panel \emph{(c)}, the dramatic drop in these ratios near the jamming point signifies a rapid global restructuring of the force network that occurs via many small changes in the organization of the network's connected components. [We adapted this figure, with permission, from \citep{Kramar:2014aa}.]
}
\label{f:Kramar2014aa_LP}
\end{figure}

\citet{Kramar:2014aa} also used PH to study force networks from simulations of slowly compressed, polydisperse packings of disks in 2D as they traverse the jamming transition (see Fig.~\ref{f:Kramar2014aa_LP}\emph{a}). As they compressed the system 
through a range of different packing fractions $\rho \in [0.63,0.9]$, they extracted force information at approximately fixed time intervals during the simulation, and they then computed PDs of components and loops (i.e., $\mathrm{PD}_{0}$ and $\mathrm{PD}_{1}$, respectively) for each force network sampled during compression (see Fig.~\ref{f:Kramar2014aa_LP}\emph{b}). As in other studies, \citep{Kramar:2014aa} used the clique complex of the force networks to avoid counting 3-particle loops. They then used the bottleneck distance $d_{B}$ and two variants, $d_{W1}$ and $d_{W2}$, of the Wasserstein distance \citep{otter2015} to quantify differences between two PDs and thereby help quantify differences in local and global features between two granular force networks. As described in \citet{Kramar:2014aa}, the bottleneck distance captures only the largest difference between two PDs, whereas the Wasserstein distances include all differences between two PDs, with $d_{W1}$ more sensitive than $d_{W2}$ to small changes. (See \citep{Kramar:2014aa} for more details.) Calculating the various types of distance between the two $\beta_0$ PDs and the two $\beta_1$ PDs for consecutive samples (i.e., consecutive states) of a force network allows one to characterize different kinds of variations in the geometry of force networks as a function of packing fraction (see Fig.~\ref{f:Kramar2014aa_LP}\emph{c}). Using the employed distance measures, \citet{Kramar:2014aa} observed that in the unjammed state, there can be significant (but localized) reorganization in force geometry as a packing is compressed. They also concluded that the jamming transition is characterized by rapid and dramatic global rearrangements of a force network, followed by smoother and less dramatic reconfiguration in the system above jamming, where the distances between consecutive states of a packing are much smaller than in the unjammed state. \citet{Kramar:2014aa} also found that tangential-force networks seem to exhibit similar behavior (in most respects) to that of normal-force networks. They also observed that friction can have a significant impact on how the geometry of the forces reconfigures with compression. For example, they showed that the rate of change of loop features (as measured by the distances between consecutive $\mathrm{PD}_{1}$s) is larger for a frictional system than for a frictionless one below the jamming point, but just before jamming and thereafter (i.e., during the entire jammed state) differences in loop structure between consecutive packing fractions are larger for frictionless systems than for frictional ones. In very recent work, \citet{Kondic:2017a} used PH to examine the temporal scales on which granular force networks evolve during slow compression. They simulated dense 2D granular materials and studied the influence of the externally-imposed time scale --- set by the rate of compression --- on how frequently one should sample a system during compression to be able to evaluate its dynamics near the jamming transition. By varying the sampling rate and carrying out a persistence analysis to quantify the distance between consecutive sampled states of the system, their results indicate that close to jamming, a force network evolves on a time scale that is much faster than the one imposed by the external compression. See \citep{Kondic:2017a} for further details.

One can also use PH to study force networks in tapped granular systems. \citet{Pugnaloni:2016a} examined DEM simulations of two different 2D systems exposed to tapping. One type of packing consisted of monosized disk-shaped particles, and the other type consisted of monosized pentagon-shaped particles. \citet{Pugnaloni:2016a}'s investigation suggested that particle shape can play an important role in mechanical responses, which is consistent with observations from classical investigations of granular materials \cite{Geng:2001a}. More specifically, \citet{Pugnaloni:2016a} computed $\beta_{0}$ and $\beta_{1}$ as a function of force threshold in both normal-force networks and tangential-force networks. They observed for both types of force-weighted networks (but particularly for the tangential one) that the first two Betti numbers are able to clearly distinguish between disks and pentagons, where $\beta_{0}$ (respectively, $\beta_{1}$) is consistently larger (respectively, smaller) for pentagons across a wide range of force thresholds. However, using only $\beta_{0}$ and $\beta_{1}$, \citep{Pugnaloni:2016a} were unable to clearly differentiate states with similar packing fractions but that result from different tap intensities. 

In a follow-up investigation, \citet{Kondic:2016a} simulated a series of several taps to granular packings and used PH to examine how normal and tangential force-weighted networks vary between individual tap realizations. Specifically, they computed distances between $\mathrm{PD}_{0}$s and between $\mathrm{PD}_{1}$s of force networks associated either with individual realizations of tapping to the same system or with individual realizations of tapping to two different systems. In one part of their study, they examined systems of disks exposed to a series of taps at two different tap intensities. (See Sec.~\ref{s:tapping_contact} for a rough delineation of ``low" tapping intensity versus ``high" tapping intensity.) They observed that in terms of loop structure, the set of networks generated from a series of taps at low intensity differ far more substantially from each other --- as quantified by the distribution of distances between the $\mathrm{PD}_{1}$s for different realizations of the tapping --- than do the set of force networks from a series of taps at high intensity. They also observed that the distances between different realizations of low-intensity tapping are as large as the distances between low-intensity tapping and high-intensity tapping realizations. Therefore, although the high-intensity tapping and low-intensity tapping regimes yield networks with approximately the same packing fraction (see Sec.~\ref{s:tapping_contact}), one can use methods from PH to help explain some of the differences between the packing structure in the two regimes. In another part of their study, \citet{Kondic:2016a} carried out a persistence analysis of tapped packings of disks and tapped packings of pentagons, and they observed clear distinctions between the two systems based on calculations of $\beta_1$ PDs. For example, for each system, they computed the $\mathrm{PD}_{1}$s for a set of networks associated with several individual realizations of the same tapping intensity, and they then computed a distance between each pair of $\mathrm{PD}_{1}$s for realizations within the same packing and across the two types of packings. They observed that the distribution of distances between the $\mathrm{PD}_{1}$s of individual tapping realizations to the packing of pentagons is narrower and centered at a smaller value than the distribution of distances between individual realizations of taps for the packing of disks. They also observed that the distances between the disk and pentagon systems are much larger than those between different realizations of the disk system. Thus, \citet{Kondic:2016a} were able to distinguish clearly between tapped disk packings and tapped pentagon packings using PH, especially when considering properties of loop structures. Past work using 2D experiments has also been able to distinguish between the dynamics of disk and pentagon packings using conventional approaches \cite{Geng:2001a}.

One can also use methods from computational algebraic topology to study granular networks in which one uses edge weights from something other than a force. For example, \citet{Ardanza-Trevijano:2014aa} used only particle positions (in the form of point clouds) and computed Betti numbers to distinguish states at the same density but that are at different mechanical equilibria. Using both experimental and simulated 2D granular packings of monodisperse particles, they constructed networks by locating the center of each particle and then introducing a filtration parameter $\delta$, such that any two particles separated by a Euclidean distance less than or equal to $\delta$ are adjacent to each other in a graph. They considered $\delta \in [d,1.12d]$, where $d$ is the particle diameter and the domain for $\delta$ resembles the choices that are used for determining if particles are in physical contact with each other. The authors computed, as a function of $\delta$, the first Betti number $\beta_{1}$ on the whole network to count the total number of loops at a given $\delta$. They also computed $\beta_{1}$ on the flag
complex (see Sec.~\ref{s:homology}), thus counting the number of loops with four or more nodes at a given value of $\delta$. For values of $\delta$ that are slightly larger than $d$, \citet{Ardanza-Trevijano:2014aa} were able to separate states at the same packing fraction that are generated by tapping at different intensities. They observed for a fixed packing fraction that states that arise from lower-intensity tapping have a larger value of $\beta_{1}$ when computed on the whole network and a lower value of $\beta_{1}$ when computed on the flag complex. Their results were robust to both noise and errors in particle-position data.

\subsection{Other network representations and approaches}
\label{s:other_representations}

\subsubsection{Network-flow models of force transmission.}
\label{s:flow_networks}

Another technique for gaining insight into the organization of forces in deforming granular systems, and how microscale aspects of a force network lead to macroscale phenomena such as shear bands and material failure, is to view force transmission from the perspective of maximum-flow--minimum-cut and maximum-flow--minimum-cost problems (see Sec.~\ref{s:max_flow}). To examine a granular system using such a perspective, one can consider the ``flow" of force through a contact network (with some contacts able to transmit more force than others), which in turn yields a ``cost'' to the system in terms of energy dissipation at the transmitting contacts \citep{Tordesillas:2015a, Tordesillas:2015b, Tordesillas:2013b, Lin:2013a, Lin:2014a}. One can calculate flow and costs in routes through a network (and hence determine bottlenecks in force transmission) to gain understanding of how contact structure relates to and constrains a system's ability to transmit forces in a material. For example, \citet{Lin:2013a, Lin:2014a} constructed flow networks from DEM simulations of a system of polydisperse particles compressed quasistatically under a constant strain rate in the vertical direction and allowed to expand under constant confining pressure in the horizontal direction. At a given axial-strain value (i.e., ``strain state''), they assigned uniform capacities $u_{ij}$ to each edge of the contact network to reflect the maximum flow that can be transmitted through each contact, and they assigned costs $c_{ij}$ to each edge to model dissipation of energy at each contact. After each axial-strain increment during loading, they then solved the maximum-flow--minimum-cost problem for the network at the given strain state, finding that edges in the minimum cut (yielding \emph{bottlenecks} in the force-transmission networks) localize in the material's shear band. By using costs $c_{ij}$ that reflected the type of inter-particle contact (specifically, elastic contacts versus various types of plastic contacts) \citet{Lin:2013a, Lin:2014a} were able to track different stages of deformation (i.e., strain-hardening, strain-softening, and the critical-state regime). They also computed a minimal cycle basis and observed that a large majority of force-chain particles and particles in 3-cycles are involved in the set of contacts that comprise the maximum-flow--minimum-cost routes.

One can use the above approach with various definitions of force capacity and cost functions. Using simulations of the same type of system as that in the previous paragraph, \citet{Tordesillas:2013b} constructed networks --- one for each strain state as a system was loaded until it failed --- that incorporated information about both the inter-particle contacts at a given strain state and the particle displacements that occur between consecutive strain steps. Specifically, if nodes $i$ and $j$ are in contact at a given strain state, the weight of the edge between them is the inverse of the absolute value of the magnitude of the relative displacement vector of particles $i$ and $j$, where one computes the displacement for each particle from the previous strain state to the current one. The distance between nodes $i$ and $j$ is $0$ if the associated particles are not in contact. The intuition behind this capacity function is that, when there is more relative motion between a pair of particles, one expects those particles to have less capacity to transmit force to each other. \citet{Tordesillas:2015b} used capacities that incorporate 3-cycle memberships of edges in a study of minimum cuts of a flow network in two samples of 3D sand under triaxial compression and in a 3D DEM simulation of simple shear. Grains in the bottlenecks localize early during loading, and are indicative of subsequent shear-band formation. Other work \citep{Tordesillas:2015a} studied DEM simulations of compressed, 3D bonded granular materials (where \emph{bonded} signifies that the grains are connected via solid bonds of some strength) and a system of 2D photoelastic disks under shear stress with the goal of testing the hypothesis that an appropriate maximum-flow--minimum-cost approach can identify experimentally-determined load-bearing particles and force-chain particles without relying on knowledge of contact forces. \citet{Tordesillas:2015a} examined different combinations of force-transmission capacity and cost functions, and they examined the fraction of force-chain particles that are part of the associated maximum-flow--minimum-cost network for a given capacity and cost function. In both cases, costs based on 3-cycle membership of edges seem to yield large values of these fractions, and \citet{Tordesillas:2015a} were able to successfully forecast most of the particles that eventually become part of force chains without using information about contact forces.

\subsubsection{Broken-link networks.}
\label{s:broken_link}

\begin{figure}
\centering
\includegraphics[width=\textwidth]{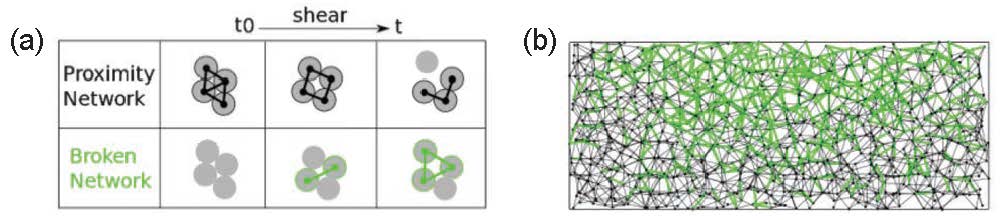}
\caption{\textbf{A broken-link network can yield insights into dynamic reconfiguration in granular shear flows.} \emph{(a)} A schematic of the construction of a broken-link network based on changes in a proximity network that occur as a system is sheared. \emph{(b)} An example of a broken-link network (with edges in green) at one instant in time. [We adapted this figure, with permission, from \citep{Herrera:2011aa}.]
}
\label{f:Herrera2011aa_Fig3}
\end{figure}

In previous discussions (see Secs.~\ref{s:contact_network},~\ref{s:force_weighted}), we have seen that one way to investigate the evolution of a granular system under an applied load is \emph{(1)} to compute contact networks or force-weighted networks for the system as a function of packing fraction, strain, or some other control parameter, and then \emph{(2)} to study how different features and properties of the networks emerge and change as one varies that parameter. One can also use other network constructions to explore different mesoscale features and examine system dynamics. For example, \citet{Herrera:2011aa} designed a \emph{broken-link network} (see Fig.~\ref{f:Herrera2011aa_Fig3}) to study the dynamics of 3D granular flows. They conducted an experiment on a collection of acrylic beads immersed in a box of liquid medium, shearing the system at a constant rate ($\Omega \approx 1.05 \times 10^{-3}$ rad/s) by a rotating circular disk at the bottom of the box. (See \citep{Herrera:2011aa} for details about their experiments.) First, they constructed \textit{proximity networks} (a variant of a contact network) as a function of time. (Specifically, they constructed one network every 3-degree increment of rotation.) In their proximity network, they assigned a ``contact" (edge) to each pair of particles whose distance from one another in the given frame was within a specified distance threshold, which they chose to be a conservative upper bound for particle contact. They then defined a \emph{broken link} (relative to some reference frame) as an existing edge between two particles in the reference frame that was subsequently absent in at least two later time frames (due to the particles having moved apart). A broken-link network for the frame in which a pair of particles moved apart had an edge between the two particles that were initially in contact, and broken links were not allowed to reform later. In Fig.~\ref{f:Herrera2011aa_Fig3}\emph{a}, we illustrate this procedure for constructing a broken-link network. 

Studying the temporal evolution of a broken-link network provides a quantitative approach for examining particle rearrangement events in granular matter. To examine the temporal evolution of a granular system, \citet{Herrera:2011aa} examined the size of the LCC in a sequence of broken-link networks as a function of applied shear, drawing an analogy between the fraction $\chi_{b}$ of broken links and the occupation probability in traditional percolation problems (see Sec.~\ref{s:largest_comp}).  
They observed that the fraction $s_{g}$ of nodes in the LCC of the broken-link network grows with $\chi_b$ in a way that suggests that there is a continuous phase transition in $s_{g}$, and they approximated the value of $\chi_{b}$ at which this transition occurs. (However, as we noted in Sec.~\ref{s:largest_comp}, because these networks have finite sizes, one needs to be cautious regarding statements about percolation and phase transitions.) 
From a physical standpoint, this transition region corresponds to a characteristic deformation at which broken links --- which are due to particle rearrangements --- start to become globally connected, relative to a reference proximity network. By examining $\chi_{b}$ as they applied shear to the system, \citet{Herrera:2011aa} approximated a characteristic amount of strain associated with this transition region (by mapping the value of $\chi_b$ associated with the transition region to a corresponding value of strain), and they suggested that the determined strain scale may be useful for identifying the onset of global reorganization in the system. 

In a later study, \citet{Slotterback:2012aa} used a similar approach to examine progression from reversible to irreversible dynamics in granular suspensions under oscillatory shear strain. In experiments similar to the one described above \citep{Herrera:2011aa}, the authors considered a series of 20 shear cycles of the following form: for each cycle in one experiment, a suspension was sheared with a rotating disk up to a strain amplitude of $\theta_{r}$, after which the direction of rotation was reversed and the disk was rotated by $-\theta_{r}$ back to its original location. The authors performed experiments for $\theta_r$ values of $2^\circ$, $4^\circ$, $10^\circ$, $20^\circ$, and $40^\circ$. In a qualitative sense, measuring ``reversibility" of the dynamics in this setup entails determining the extent to which particles return to their original positions after some number of shear cycles. To quantify this idea, one can compute the mean square displacement (MSD) of the particles after each cycle, where smaller MSD values correspond to more reversible dynamics. To study this system, \citet{Slotterback:2012aa} adjusted the idea of a broken-link network to include ``healing", so that broken links that are repaired in later frames do not contribute edges in the broken-link network, and they studied the temporal evolution of this broken-link network as a function of the cyclic shear and for different amplitudes $\theta_r$. In addition to the extent of \textit{spatial (ir)reversibility} measured by calculating the MSD, \citet{Slotterback:2012aa} also proposed a notion of \emph{topological (ir)reversibility} by examining the temporal evolution of the size of the LCC in their broken-link networks. For low values of $\theta_r$ (specifically, for $\theta_r \leq 20^\circ$), the system appears to be almost reversible: proximity-based contacts break, but they reform after shear reversal, and the fraction of particles in the LCC of the broken-link network thus grows before subsequently shrinking to almost $0$ after reversal. However, for a higher shearing amplitude (specifically, for $\theta_r = 40^\circ$), the system shows signatures of irreversibility. Many broken links do not reform after a shear cycle, and after reversal, the LCC of the broken-link network remains at a value that constitutes a substantial fraction of the total system size.

\subsubsection{Constructing networks from time series of node properties or from kinematic data.}
\label{kinematic}

Another way to examine the organization of deforming granular materials is to construct networks based on the temporal evolution of particle properties \citep{Walker:2012aa,Walker:2013b,Walker:2014ba,Navakas:2014aa}, an idea that draws from earlier work in complex systems on constructing networks from different kinds of time-series data. (See, for example, \citep{Zhang:2006aa, Xu:2008aa, Yang:2008a, Lacasa:2008a, Gao:2009a, Marwan:2009a}.) In one type of construction, which yields what are sometimes called \emph{functional networks} \cite{Stolz2016}, one records the time series of some property of each particle and places an edge (which can potentially be weighted and/or directed) between two particles according to some relationship (e.g., some type of distance or other measure of similarity) between the particle-property time series. When using a different property, the nodes are the same, but the edges and especially edge weights will in general be different. Once networks have been constructed, one can examine them using techniques such as those in Sec.~\ref{s:network_measures}. Some authors have also used the generic term \emph{particle-property networks} to describe networks that they constructed in some way from particle properties.

\citet{Walker:2012aa} used particle-property networks to study the evolution of DEM simulations of a quasistatically deforming, 2D granular material under biaxial compression. They constructed time series for two features (as well as their coevolution) for each particle --- membership in force chains (determined as in \cite{Peters:2005aa,Muthuswamy2006a}) and membership in 3-cycles of a minimal cycle basis --- by recording a $1$ if a particle has the given property and a $0$ if it does not. They then quantified the similarity of particle-property evolution using the Hamming distance \citep{MacKay:2003}, and they added (undirected and unweighted) edges between each particle and its $k$ closest (i.e., most similar) particles until eventually obtaining a single network with one connected component. \citet{Walker:2012aa} then extracted sets of particles that exhibit similar dynamic behavior by detecting communities (see Sec.~\ref{s:comm_structure}) in the particle-property networks. This uncovered distinct regions in the material --- including the shear band and different subnetworks composed of primarily force chain or primarily non-force chain particles --- as well as interlaced regions in the shear band that alternate between jammed and unjammed configurations. See \citep{Walker:2013b,Walker:2014ba} for additional studies that used membership in cycles of length up to $l = 7$ for the construction of particle-property networks. 

One can also construct particle-property networks from data that do not rely on knowledge of contact or force networks. For example, \citet{Walker:2012a} studied deformation in sand subject to plane-strain compression using measurements of grain-scale displacements from digital image correlations (DICs) \citep{Rechenmacher:2006a,Rechenmacher:2010a}. From these data, they constructed \textit{kinematic networks}, a type of network that arises from some measurement of motion (such as displacement or a rotation) over a small time increment. \citet{Walker:2012a} considered each observation grid point of digital images of a sample to be a node, and they placed edges between nodes with similar displacement vectors during a small axial-strain increment. This yields a collection of (undirected and unweighted) time-ordered kinematic networks. In another study, \citet{Tordesillas:2013a} calculated particle rotations and displacements for triaxial compression tests on sand using x-ray micro-tomography scanning \citep{Ando:2012a}. They generated an ordered set of networks from these data for several strain steps by treating each grain as a node and linking nodes with similar kinematics during the specified interval. More specifically, they represented the displacements and rotations of each particle as points in a state space, and they connected particles that are nearby in that state space according to Euclidean distance. In the network that they constructed, each particle is adjacent to $k$ nearest neighbors in state space, where $k$ is as small as possible so that the (unweighted and undirected) network is connected. Various notions of what it means to be ``similar'', and thus how to quantitatively define edges, are discussed in \citep{Walker:2012a,Walker:2014ba}. To probe the collective dynamics of interacting groups of particles for the network in each strain step, \citet{Tordesillas:2013a} detected communities corresponding to mesoscale regions in the material that exhibit similar dynamic behavior. Calculating the mean shortest-path length between pairs of particles in the same community yields a potentially important intermediate spatial scale (that they concluded is consistent with the shear-band diameter) of a granular system. For each strain step, they computed a variant of closeness centrality --- it is similar to the one in Sec.~\ref{s:centrality}, but it is not exactly the same --- of each particle in the corresponding network, and observed that particles with large closeness centrality localize in the region of the shear band early in loading (and, in particular, before the shear band develops). This study highlights the potential of network analysis to provide early warning to detect regions of failure in particulate systems. 

Methods from nonlinear time-series analysis have also been used in network-based studies of stick-slip dynamics in a granular packing sheared by a slider \citep{Walker:2014aa}. Using so-called \emph{phase-space networks} (see \citep{Xu:2008aa} for a description of them) to construct networks from measurements of a slider time series, \citet{Walker:2014aa} associated network communities with slip events.

\subsection{Comparing and contrasting different network representations and approaches}

Network representations of granular and other particulate systems, in combination with methods from network science and related disciplines, provide a plethora of ways to analyze granular materials. For these tools to be optimally useful in furthering understanding of the physics of these systems, it is also important to draw connections between the many existing approaches. The application of network-based frameworks to the study of granular matter is relatively young, and little work thus far has focused specifically on exploring such connections, but it is crucial to determine \emph{(1)} how conclusions from different approaches relate to one another and \emph{(2)} how similarities and differences in these conclusions depend on the system being studied. In this short subsection, we point out a few relationships and places to begin thinking about such questions.

First, it is important to consider the network representation itself. We have discussed several different representations in this review, and some are more related to each other than others. Broadly speaking, one class of granular networks tries to encode the physical structure of a material --- in the sense that edges exist only when there is an inter-particle contact --- at a given point in an experiment or simulation. Such networks include contact networks (see Sec.~\ref{s:contact_network}) and force-weighted networks (see Sec.~\ref{s:force_weighted}). One can obtain a contact network from an associated force-weighted network by discarding the edge weights and keeping information only about connectivity. Force-weighted networks thus contain much more information, and they allow one to investigate phenomena --- such as force-chain organization --- that may not be possible to probe with the contact network alone. However, it is still important to develop tools for the analysis of contact networks and understand what phenomena arises from features of the connectivity alone (or what one can learn and explain from connectivity alone), as force information may not always be available or may not be necessary to understand certain behaviors of a granular system. Generalizing some quantities (e.g., betweenness centralities) from unweighted networks to weighted networks also involves choices, and it is often desirable to conduct investigations that have as few of these potentially confounding factors as possible. Other types of granular networks do not encode physical connectivity, but instead directly represent something about the dynamics or changes that occur in a system during an experiment or simulation. Examples of such networks include broken-link networks (see Sec.~\ref{s:broken_link}) and particle-property networks (see Sec.~\ref{kinematic}). These different classes of network representations offer distinct ways of studying granular materials, and utilizing each of them should improve understanding of these systems. 

For a given network representation, it is reasonable to expect that conclusions that arise from similar network quantities or methods of analysis are related to one another, whereas conclusions that result from tools designed to probe very different kinds of organization in a network provide rather different information about the underlying granular system \cite{bassett2012influence, Giusti:2016a}. For example, some studies have suggested that in deforming granular materials, results based on calculations of clustering coefficients are similar to those from studying 3-cycles. This is intuitively reasonable, given that calculating a local clustering coefficient yields one type of local 3-cycle density. We have also observed that conclusions drawn from examinations of small subgraphs in a deforming granular system may also be related to whether or not those subgraphs contain cycles of certain lengths. As we discussed in Sec.~\ref{s:relationships_measures}, another way to draw connections between different approaches is to consider the spatial, topological, or other scales that are probed by the different approaches. For instance, in a force-weighted network, node strength is a particle-scale property, and it encompasses only very local information about a granular material. However, granular systems exhibit collective organization and dynamics on several larger scales that may be difficult to understand by exclusively computing local measures and distributions of such measures. To obtain an understanding of larger-scale structures, it is necessary to also employ different methods, such as community detection, persistent homology, and the examination of conformation subgraphs that are composed of more than just a single particle. Utilizing such approaches has provided insights into force-chain structure, shear band formation, and reconfiguration in granular systems under load that one may not be able to obtain by considering only local network quantities. It is thus important to continue to use and develop methods to analyze granular networks across multiple scales, as doing so can provide important and new information about a system. Finally, we note that even a single approach, depending on how it is used, can provide multiple types of information about a granular network. A good illustration of this is community detection. In Sec.~\ref{s:comm_structure}, for example, we saw that using different null models in modularity maximization allows one to probe rather different types of mesoscale architecture in granular force networks.

We look forward to forthcoming studies that directly compare the results and assumptions in different approaches (both network-based and traditional ones) and different network representations of granular and other particulate systems. Conducting principled investigations into how conclusions from various network-based approaches are related is indeed an important direction for future work.

\subsection{Limitations and practicalities of simulations and experiments} 
\label{limit}

Network-based studies of granular materials have examined inter-particle contact and force data (and associated dynamics, such as in the presence of external loading) from both experiments and simulations. A recent summary of the many experimental techniques available for obtaining data about inter-particle contacts is available in the focus issue \citep{amon_focus_2017}. Such techniques include laser-sheet scanning \citep{Dijksman2017}, photoelasticity \citep{Daniels2017}, x-ray tomography \citep{weis:17}, and nuclear magnetic resonance \citep{stannarius:17}. Using each of the first three approaches, it is possible to measure both particle positions and inter-particle forces. If one is careful, it is sometimes possible to measure the forces as vectors (i.e., including both the normal and tangential components), but some techniques or systems do not have sufficient resolution to allow more than scalar or coarse-grained values. Determining the forces also helps experimentalists to confidently construct contact (i.e., unweighted) networks of particulate materials. In deciding whether or not two particles are actually in contact, rather than merely being adjacent in space, it is necessary to perform a detailed study of the effects of thresholding the data \citep{Majmudar:2007aa}. Any experimental technique will imperfectly report the presence versus absence of the weakest contacts in a system. Additionally, because of the difficulty of accessing the interior of granular materials, much more data is available for 2D force networks than for 3D force networks \citep{amon_focus_2017}.

The most widely-used simulation techniques are discrete element methods (DEMs) \citep{Poschel2005}, in which the dynamics of individual particles (usually spheres) are determined by their pairwise interactions under Newton's laws of motion. The normal forces are typically determined from a Hertzian-like contact law (see, e.g., the sidebar in \cite{PT2015} for an introduction to Hertzian contacts) via an energy penalty for the overlap of two particles. The tangential (frictional) forces are most commonly modeled using the Cundall--Strack method \citep{Cundall1979} of a spring and a dashpot, but they have also been modeled via the surface roughness created by a connected set of smaller spheres \citep{Papanikolaou2013}. For a given application, it is not known whether these simplified models capture all of the salient features of inter-grain contacts, and the situation likely differs for different applications. For example, experimental measurements of sound propagation in photoelastic disks \citep{Owens:2011} suggest that the amplitude of sounds waves may be largest along a force chain network, an effect not observed in DEM simulations \citep{Somfai2005}. This is likely a consequence of real particles physically deforming their shape to create an increased contact area through which sound can be transmitted; existing DEM simulations do not account for this effect. Another important use of particle simulations is to provide a means to investigate the robustness of network-based analyses to various amounts of experimental error \citep{Kramar:2014b}. Simulations provide an important check on experimental uncertainties in the determination of force-weighted networks and other network representations of granular materials. Conversely, network-based approaches provide a means to compare how faithfully simulations are able to reproduce experimental results.

\section{Open problems and future directions}
\label{open_problems}

We now discuss a few open problems and research directions for which we anticipate important progress in the near future. We divide our comments into three main areas: the construction of different types of networks that encode various physical relationships or other properties (see Sec.~\ref{alt_methods}), the application of network analysis to additional types of materials (see Sec.~\ref{extensions}), and the application of network-based approaches to the design of materials (see Sec.~\ref{net_design}). Network tools can provide valuable insights --- both explanatory and predictive --- into particulate materials and their dynamics, and a lot of fascinating research is on the horizon.

\subsection{Network representations and computations}
\label{alt_methods}

To briefly explore the potential of different approaches for constructing granular (and other particulate) networks for providing insights into the physics of granular materials (and particulate matter more generally), we discuss choices of nodes, choices of edges, edge-to-node dual networks, multilayer networks, and annotated networks.\footnote{Other ideas that are worth considering include memory networks \cite{Rosvall2014NatComm}, adaptive networks \cite{thilo-adaptive}, and various representations of temporal networks \cite{Holme2015EurPhysJB}.} It is also worth thinking about what calculations to do once one has constructed a network representation of a particulate system, so we also briefly consider the important issue of developing physically-informed methods and diagnostics for network analysis.

\subsubsection{Definitions of nodes and edges.} 

There are many choices --- both explicit and implicit --- for constructing a network \citep{butts2009,newman2010networks}, and these choices can impact the physics that one can probe in granular networks \citep{bassett2012influence}. Perhaps the most obvious choices lie in how one defines nodes and edges. 

In the study of granular materials, a common definition is to treat individual particles as nodes and to treat contacts as edges (often with weights from the inter-particle forces). A natural set of open questions lies in how contact network architectures depend on different features of the grains in a system. For example, there have been several recent studies on systems composed of particles that are neither spheres nor disks --- including ones with U-shaped particles \citep{gravish2012entangled}, Z-shaped particles \citep{murphy2015freestanding}, squares and rods \citep{Hidalgo:2009aa,trepanier2010column}, dimers and ellipses \citep{Schreck:2010a}, and others \citep{Athanassiadis:2014a,durian2017}. It would be interesting to build network representations of these systems, examine how different grain geometries affect network organization, and investigate how that organization relates to the mechanical properties of a system \citep{Azema:2013aa}. It seems particularly important to develop an understanding of which (quantitative and qualitative) aspects of network structure depend on features of grains (such as shape, polydispersity, friction, cohesiveness, and so on \citep{Kovalcinova:2015a,Kondic:2016a,Pugnaloni:2016a,bassett2015extraction,Kovalcinova:2016a,Kramar:2013aa,Tordesillas:2010a}) and which are more universal.

\begin{figure}[h]
\centering
  \includegraphics[width=\textwidth]{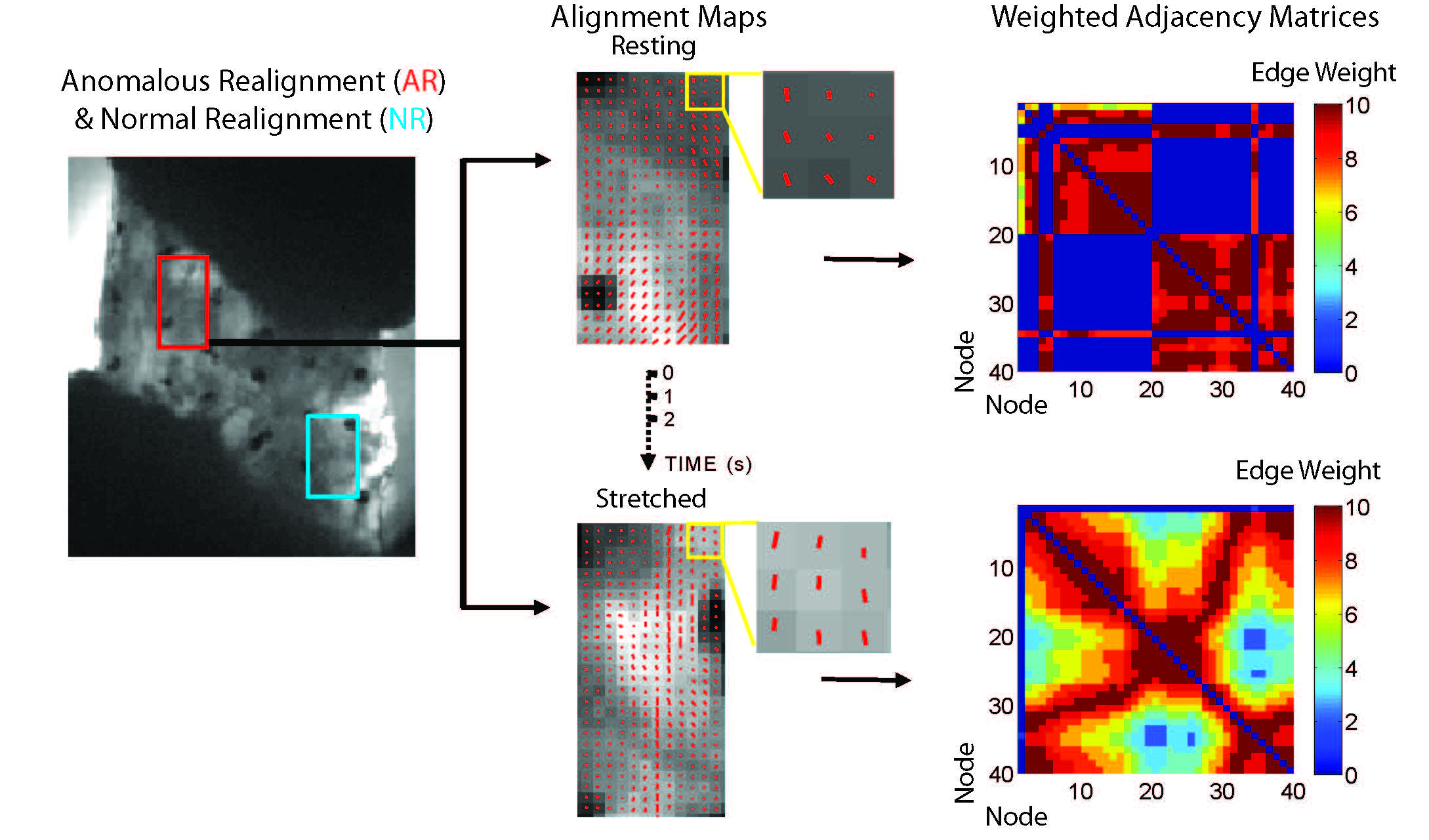}
\caption{\textbf{Constructing a network representation of collagen-fiber alignment.} We show quantitative polarized light images (QPLIs) of the human cervical facet capsular ligament taken both before and during loading. These data were collected as part of a study of collagen-fiber realignment following a mechanical stretching of the material. Such realignment is a mechanical process that commonly leads to both acute and chronic neck pain, particularly following car accidents. QPLIs were used to generate pixel-wise collagen alignment maps in the selected regions of interest (ROIs) of anomalous realignment (AR; red box) and normal realignment (NR; blue box) at rest (time point $t = 1$), at the onset of anomalous realignment (time point $t = T$), and in between these two points in 1-second increments. Rectangular ROIs were defined using the upper-left and lower-right fiduciary markers as the common information between different time points. Each $3 \times 3$ pixel window in an ROI is a node in a network. We show representative $3 \times 3$ pixel windows in the resting and stretched states, along with a corresponding demonstration of alignments. For each node, we calculate the alignment angle, and we use a measure of similarity between alignment angles to weight the edges in the network. In this way, weighted adjacency matrices, with nodes numbered spatially, illustrate a pairwise similarity in alignment angles between ROIs. [We reproduced this figure, with permission, from \citep{Zhang:2015}.]
}
\label{f:collagen}
\end{figure}

One can also consider defining particulate networks in a variety of other ways. For example, when determining edges and edge weights, one can examine the tangential (rather than, or in addition to, the usual normal) component of the force between two grains. Such extensions may facilitate increasingly detailed investigations into a packing's organization \citep{Kramar:2014aa}. It may also be useful to retain information about both the magnitude and direction of forces when defining edges. One may even wish to construct signed networks, for which edges can take either positive or negative values, thereby conveying further information about the relationship between nodes. In such studies, one can perhaps take advantage of advancements in community-detection techniques, such as by using signed null models \citep{Gomez:2009a,Traag:2009a}. Additionally, as we discussed in Sec.~\ref{s:other_representations}, particle-property networks \citep{Walker:2012aa,Walker:2013b,Walker:2014ba} and networks constructed from particle-displacement information \citep{Herrera:2011aa, Slotterback:2012aa,Tordesillas:2013a,Walker:2012a} are other informative ways to build networks for particulate systems. One can also construct edges (and determine edge weights) by incorporating information about inter-grain relationships based on similarities in particle properties such as orientation \citep{Zhang:2015} (see Fig.~\ref{f:collagen}), coefficient of friction \cite{Puckett2013}, or size \citep{shaebani2012influence}. Constructing networks whose edges are determined or weighted by inter-particle similarities may be particularly useful for achieving a better understanding of mesoscale physics in polydisperse packings, which are thought to depend on the spatial distributions of particles of different types \citep{Kumar:2016a}. A perhaps nonintuitive choice is to use a bipartite representation of a granular network, such as the approach used in \cite{slanina2017}.
	
The above choices for network construction give a grain-centric view of the physics of particulate materials. One can also consider edge-to-node ``duals'' of such networks to provide a contact-centric perspective. In a contact-centric approach, one treats contacts between physical objects as nodes and the objects themselves as edges between contacts. (Compare this idea to the notion of a \emph{line graph} \cite{harary1972} of a network $G$.)
A contact-centric network approach was used recently in the study of nanorod dispersions \citep{Shi:2013aa, Shi:2014aa}. Contacts between rods were treated as nodes, and the effective conductance of the rod was treated as a weighted edge. The treatment of a grain or other physical object as an edge rather than as a node is also a particularly appealing way to describe networks of fibers in both human-made and natural systems. Recent examples of such fiber networks that have benefited from network analysis include collagen networks \citep{abhilash2014remodeling,Zhang:2015} (see Fig.~\ref{f:collagen}), fibrin networks \citep{purohit2011protein}, and axonal-fiber networks \citep{Bullmore2009,Bassett:2017a}.

Another way to study granular systems (especially porous ones) is to consider a network constructed from the \emph{pore space} \citep{Blunt:2001a}, in which \emph{pores} (i.e., empty volumes between contacting grains) are nodes and \emph{throats} (i.e., flow pathways that connect pores) are edges (which can be weighted in various ways). Conveniently, there are several methods to precisely determine pores and grains \citep{Al-Raoush:2003a}. Studying pore networks is a common way to examine flow through porous materials or to understand the responses of granular materials to external stresses, but only more recently have such networks been studied explicitly from a network-science perspective. See \citep{Vo:2013,Walker:2013a,Linden:2016a,Russell:2016a,Martinez:2017a} for some recent examples, and see \cite{porous2017} for a study of force chains in porous media. Given how pore networks are formulated, we expect that they can also be studied using PH (see Sec.~\ref{s:homology}).

\subsubsection{Multilayer networks.} 

When considering different ways to construct a network, it is also natural to ask whether it is beneficial to combine two or more approaches in an integrated way to more accurately (and elaborately) represent a particulate system as a network. Ideally, doing network analysis using a more complicated representation will also lead to improved physical understanding. One way to incorporate heterogeneous node types, multiple types of relationships between nodes, and time-dependence into network representations is with \emph{multilayer networks} \citep{kivela2014multilayer,Boccaletti2014,domenico2013mathematical}. Recall that \citep{papadopoulos2016evolution} (see Sec.~\ref{s:comm_structure}) studied one type of multilayer community structure in a compressed granular material. Another type of multilayer network that may be useful for the study of particulate systems are \emph{multiplex networks}, in which one can encode different types of relationships between nodes in different \emph{layers} of a network. For example, one can construct a multiplex network in which two particles are linked both by normal force-weighted contacts and also according to another relationship (such as the tangential force between them, their contact angle, or by a measure of similarity of one or several particle-properties, as discussed in Sec.~\ref{kinematic}). One can also envision using multilayer networks to study particulate systems with multiple types of particles. For example, if a system consists of particles of different shapes or sizes, one possibility is that each particle of a given shape (or size) is in a different layer, intralayer edges represent interactions between particles of the same type, and interlayer edges represent interactions between particles of different types. Another possibility is to let each layer represent a time window, perhaps with intralayer edges representing a mean interaction strength during that window. The study of multilayer networks is one of the most popular areas of network science, and we expect that they will be very illuminating for studies of particulate matter.

\subsubsection{Annotated graphs.}
\label{annotated}

The network-construction techniques that we have been discussing throughout this review represent data in the form of an adjacency matrix (for an ordinary graph) or an adjacency tensor (for a multilayer network) \citep{domenico2013mathematical,kivela2014multilayer,Mucha2010}. However, it may be desirable to encode not only relationships between grains, but also features of the grains and/or their interactions. One option is to use \emph{annotated graphs} (one can also use multilayer networks) to encode inter-node relationships, individual-node properties, and individual-edge properties \citep{newman2016structure,tiago2016}. One can use annotated graphs --- also sometimes called \emph{labeled graphs} or \emph{tagged graphs} \citep{palla2008fundamental} --- to study properties of interaction patterns (such as force-weighted contacts) that may be driven by local particle properties (e.g., size, shape, spatial position in a material, membership in cycles or force chains, and so on). Available tools for annotated graphs include clustering techniques in the form of stochastic block models (SBMs) that combine information from both connectivity and annotations to infer densely-connected sets of nodes (e.g., including when there is sparse or missing data) \citep{newman2016structure,tiago2016}.

\subsubsection{Beyond pairwise interactions.}
\label{hyper}

It is also desirable to explicitly examine interactions between three or more particles, rather than restricting one's study to representations of pairwise interactions. We discussed this idea briefly in Sec.~\ref{s:homology} in the context of simplicial complexes \cite{edels2010}, and we note that one can also encode edges among three or more nodes by using hypergraphs \cite{newman2010networks}.

\subsubsection{Physically-informed network calculations.}

In addition to exploring different choices of how to construct particulate networks, it is also important to consider ways to generalize the tools of network science to incorporate physical constraints \citep{barth2011,bassett2015extraction}. A natural way to begin is to build null models of spatially-embedded graphs that obey simple mechanical constraints like force balance \cite{Ramola:2017a}. One can also develop network diagnostics that incorporate physical geometry, spatial embedding, and latent spatial structure of networks \citep{barth2011}. See \cite{jkt2017} for an example of a kinetic approach. It should also be useful to incorporate ideas of flow and other forms of dynamics into community detection \cite{beguerisse2014interest,jeub2015}. Additionally, it is desirable to develop methods to more explicitly characterize the relationship between a network's structure and the geometry of a space in which it is embedded \citep{Bassett2010,Modes:2016a,barth2011} (as well as latent geometry) and to use such techniques to better understand the plurality of packings that are consistent with a single force distribution via a \emph{force-network ensemble}, which is the space of allowed force configurations for a fixed contact geometry \citep{Snoeijer-2004-ETF,Snoeijer-2004-FNE,Tighe2008,Tighe2010a,Kollmer2017}. We also point out that a spatial embedding can induce correlations between network diagnostics \cite{bassett2012influence}, and it is therefore important to develop a better understanding of the extent to which this occurs in networks that arise from particulate systems.

\subsection{Beyond granular materials}
\label{extensions}

Although we have focused primarily on network analysis of the canonical type of granular systems, which consist of discrete, macroscopic particles that interact via contact forces, one can potentially also use network-based approaches to characterize granular materials with more complex interactions as well as soft materials more broadly \cite{ronhovde2011detection}. As reviewed recently in \citep{Nagel:2017a}, these materials include colloids, glasses, polymers, and gels. In these systems (and others) the particles (or other entities) can interact via various (and often complicated) means.\footnote{One can also examine particulate networks of hard materials that admit Hamiltonian descriptions \cite{amorphous2017}.} For example, system components can have attractive and/or repulsive long-range interactions \citep{muller2009homogeneous}, can be cohesive \citep{Mitarai:2006a}, and/or can interact with one another via chemical gradients \citep{wrobel2014directed,huttenlocher2008reverse}, electric charges \citep{hartveit2012electrical,nualart2013biochim,Pahtz:2010a}, or through other molecular or mechanical processes \citep{ladoux2015mechanotransduction}. In each of these cases, interaction strengths can yield edge weights, either as the sole relationship studied between entities in a graph or as one of several inter-entity relationships in a multilayer network. 

One particularly interesting avenue for future work may be to study polymer and fiber networks \citep{Bausch:2006aa,Broedersz:2014aa}, which are important in biological systems. For example, in biopolymer assemblies, cross-linking can glue filaments together into large-scale web-like structures (e.g., as in the cytoskeleton of a cell) \citep{Lieleg:2009a}. Such cross-linked actin filaments are critical for cellular function and integrity \citep{Fletcher:2010aa}, and it is thus important to understand the structural organization of these kinds of networks, their mechanical properties, and how force transmission is regulated in them \citep{Mizuno2007,Gardel:2008aa,actin2017}. Indeed, there has already been some work examining network representations of gels and polymers, and employing graph-theoretic analyses to quantify the structural properties of these systems (e.g.\citep{Billen2009,Kim2014,Gavrilov2015,Liang:2016,samal2017,Bouzid2017}). One can also use network analysis to study systems at smaller spatial scales. Because traditional network-based approaches are agnostic to the physical scale of a system, off-the-shelf calculations and algorithms are directly portable to microscale, nanoscale, and even smaller-scale systems \citep{ronhovde2011detection,Ahnert:2017aa}. However, despite the technical portability of common tools, the investigation of much smaller-scale systems should benefit from the extension of classical network-based tools in ways that incorporate additional underlying physics. For example, we have described extensions of network tools to assimilate ideas and constraints from classical physics (e.g., spatial embeddedness) \citep{bassett2015extraction,Sarzynska:2015aa}. For such investigations, we expect that ideas from the study of random geometric graphs (RGGs) and their extensions will be helpful \citep{bassett2012influence,setford2014}. One can also consider employing ideas that take into account principles from quantum physics \citep{Mulken:2011aa,Bianconi:2015a} or other areas. The study of quantum networks is a new and exciting area of network science, and there are many opportunities for progress \citep{Biamonte:2017a}.

\subsection{Implications for material design}
\label{net_design}

As is the case with mathematics more generally, network analysis gives a flexible approach for studying many systems, because it is agnostic to many details of their physical (or other) nature \citep{newman2010networks,Newman:2003aa,bocca2006}. Such versatility supports the application of network-science techniques to both living and non-living materials to examine the architectures and dynamics of these systems and to gain insights into relationships between structure and function. The tools and approaches of network science also have the potential to inform the design of new materials. For example, it should be possible to use network theory (e.g., via the tuning of a system's network architecture) to provide guidance for how to engineer a material to exhibit specific mechanical, electrical, or other properties. Material design has become increasingly popular with recent advances in the study and development of \emph{metamaterials} \citep{Liu:2011, Turpin:2014, Lee2012}. Metamaterials can take advantage of precisely-defined component shapes, geometries, orientations, arrangements, and connectivity patterns (rather than specific material and physical characteristics of individual units) to produce tailored mechanical \citep{Greaves:2011a}, acoustic \citep{Lee2012, Rocklin:2015, Fang:2006, Nicolaou:2012aa,PT2015}, and electromagnetic \citep{Liu:2011, Simovski2012, Smith:2004aa} properties. The control of a single unit or component is relatively straightforward, but the question of how to link many components in ways that yield complex material properties is a very challenging one. Approaches that use ideas from network science have the potential to offer guidance for construction patterns of material units that support desired bulk properties. 

There are likely many ways to use network-based approaches to inform the design of new materials. One reasonable possibility is to employ evolutionary and genetic computer algorithms \citep{eiben2015from,diaz2016review} and other tools from algorithmic game theory \citep{papadimitriou2014algorithms}. For example, the combination of multi-objective functions and Pareto optimality \citep{Goldberg:1989} can offer a targeted path through the space of possible network architectures (i.e., through a \emph{network morphospace} \citep{McGhee:1997}). If exact simulations are not computationally tractable, one can use machine-learning techniques to offer fast estimates of material properties and behavior \cite{valera2017machine}. One can perhaps begin with a single material network structure that is physically realizable and then rewire the initial network architecture with a cost function that drives the system towards an arrangement that enhances a desired property. One can perform such a rewiring explicitly along Pareto-optimal fronts using a set of rewiring rules that preserve physical constraints and laws. This approach, which builds on prior applications to other types of spatially-embedded networks \citep{Avena-Koenigsberger:2014, Avena-Koenigsberger:2013, Goni:2013}, selects for physically-feasible network designs that purposely maximize specific properties. A network analysis of these ``evolved'' (and evolving) networks may help elucidate relationships between structural features of system architecture (e.g., clustering-coefficient values) and material properties (e.g., stability under load), providing a link between structure and function. Such an evolutionary-design approach can complement recent efforts to identify optimal shapes with which to construct a packing \citep{jaeger2016evolutionary,miskin2013adapting}, to design rules with which to pack those shapes \citep{miskin2014evolving,roth2016optimizing}, and to construct ``allosteric materials'' with specific functionalities via evolution according to fitness functions \citep{Yan:2017a}.

One set of problems for which network-based tools may be useful are those related to rigidity and mechanical responses of disordered material networks, which have been studied previously using a rigidity-percolation framework \citep{Feng1985,Jacobs1995,Ellenbroek:2015aa,Henkes2016}. In terms of material design, a particularly interesting line of future work may be to use network analysis in conjunction with methods such as \emph{tuning-by-pruning}, in which individual bonds are selectively removed from a disordered material network to engineer a specific property into a system \citep{Goodrich:2015a}. For example, beginning with simulated, disordered spring networks (derived from jammed particle packings), \citet{Goodrich:2015a} used tuning-by-pruning to design networks with different ratios of the shear modulus to the bulk modulus. Motivated by allosteric responses in proteins, \citep{Rocks:2017a} developed an approach that allows careful control of local mechanical responses in disordered elastic networks. In particular, with the removal of very few bonds, one can tune the local strain in one region of a network via a response to an applied strain in a distant region of a system. \citet{Driscoll2016} studied three different spring networks and continuously tuned their mechanical rigidity (measured by different parameters, such as the distance above isostaticity) to examine the effect of such tuning on material failure. They observed that for a fixed amount of disorder (which the authors measured using a scalar, following the approach in \citep{shekhawat2013damage}), the width of the failure zone in a stressed material increases as the rigidity decreases. Recently, \citep{ReidAuxeticMetamaterials:2017} studied a model of amorphous networks that incorporates angle-bending forces, and employed pruning-based methods to design auxetic metamaterials. In light of these findings, it is natural to ask what network quantities and methods may be related to the above types of global or local mechanical properties, and whether they can inform which bonds to remove (or rearrange \citep{Lopez:2013aa}) to invoke particular types of functional responses.

In developing a network-based approach for designing and building new materials, it is desirable to capitalize on the ability of network analysis to quantify multiscale structures (with respect to space, time, and network architectures) in a wide variety of systems (regardless of the exact details of their composition). For example, network analysis has revealed mesoscale architectures that are often crucial for determining material properties in disordered media, but such heterogeneities also appear to be important for biological materials, including networks of collagen fibers \citep{Quinn:2011,Zhang:2015}, tissues and tendons \citep{Zhao:2014,Han:2013}, muscle fibers \citep{Pong:2011}, and axonal fibers \citep{Sporns:2013aa, Sporns:2014aa}. Tools from network science should be useful (and, in principle, flexible) for designing nontrivial structural organizations that yield desired material functions. One can imagine using a network-theoretic framework to design localized, mesoscale, and/or system-level properties, to design and manipulate human-made and natural (including biological) materials, and to precisely control both static and dynamic material properties. 

\section{Conclusions} 
\label{conclusion}

Network science is an interdisciplinary subject --- drawing on methods from physics, mathematics, statistics, computer science, social science, and many other disciplines --- that has been used successfully to help understand the structure and function of many complex systems. Much recent work on networks has yielded fascinating insights into granular materials, which consist of collections of discrete, macroscopic particles whose contact interactions give rise to many interesting behaviors and intricate organization on multiple spatial and temporal scales. These insights have increased scientific understanding of the structure and dynamics of the heterogeneous material architecture that manifests in granular matter (as well as network-based approaches to quantify such architecture) and the response of granular systems to external perturbations such as compression, shear, tapping, and tilting. In this paper, we have reviewed the increasingly fertile intersection of network science and granular materials. Future efforts should help provide a better understanding of the physics of particulate matter more generally, elucidate the roles of mesoscale interaction patterns in mechanical failure, inform the design of new materials with desired properties, and further scientific understanding of numerous important problems and applications in granular physics, soft-matter physics more generally, and even biophysics. 

\section*{Acknowledgements}

We thank Alejandro J. Mart\'inez, Ann E. Sizemore, Konstantin Mischaikow, Jen Schwarz, and an anonymous referee for helpful comments. We also thank our collaborators, students, and mentors, who have shaped our views on the subjects of this review. KD is grateful for support from the National Science Foundation (DMR-0644743, DMR-1206808) and the James S. McDonnell Foundation. LP is grateful to the National Science Foundation for a Graduate Research Fellowship. DSB is grateful to the Alfred P. Sloan Foundation, the John D. and Catherine T. MacArthur Foundation, the Paul G. Allen Foundation, and to the National Science Foundation (PHY-1554488). The content is solely the responsibility of the authors and does not necessarily represent the official views of any of the funding agencies.


\newpage

\begin{thebibliography}{389}
	\providecommand{\natexlab}[1]{#1}
	\providecommand{\url}[1]{\texttt{#1}}
	\expandafter\ifx\csname urlstyle\endcsname\relax
	\providecommand{\doi}[1]{doi: #1}\else
	\providecommand{\doi}{doi: \begingroup \urlstyle{rm}\Url}\fi
	
	\bibitem[Jaeger et~al.(1996)Jaeger, Nagel, and Behringer]{Jaeger1996}
	H~M Jaeger, S~R Nagel, and R~P Behringer.
	\newblock Granular solids, liquids, and gases.
	\newblock \emph{Rev Mod Phys}, 68\penalty0 (4):\penalty0 1259--1273, 1996.
	
	\bibitem[Duran(1999)]{duran1999sands}
	J~Duran.
	\newblock \emph{Sands, Powders, and Grains: {A}n Introduction to the Physics of
		Granular Materials}.
	\newblock Springer-Verlag, 1999.
	
	\bibitem[Mehta(2007)]{Mehta2007}
	A~Mehta.
	\newblock \emph{Granular Physics}.
	\newblock Cambridge University Press, 2007.
	
	\bibitem[Franklin and Shattuck(2015)]{Franklin2015}
	S~V Franklin and M~D Shattuck, editors.
	\newblock \emph{Handbook of Granular Materials}.
	\newblock CRC Press, 2015.
	
	\bibitem[Andreotti et~al.(2013)Andreotti, Forterre, and
	Pouliquen]{Andreotti2013}
	B~Andreotti, Y~Forterre, and O~Pouliquen.
	\newblock \emph{Granular Media: {B}etween Solid and Fluid}.
	\newblock Cambridge University Press, 2013.
	
	\bibitem[Nagel(2017)]{Nagel:2017a}
	S~R Nagel.
	\newblock Experimental soft-matter science.
	\newblock \emph{Rev Mod Phys}, 89\penalty0 (2):\penalty0 025002, 2017.
	
	\bibitem[Mort et~al.(2015)Mort, Michaels, Behringer, Campbell, Kondic,
	Kheiripour~Langroudi, Shattuck, Tang, Tardos, and Wassgren]{Mort2015}
	P~Mort, J~N Michaels, R~P Behringer, C~S Campbell, L~Kondic,
	M~Kheiripour~Langroudi, M~Shattuck, J~Tang, G~I Tardos, and C~Wassgren.
	\newblock Dense granular flow --- {A} collaborative study.
	\newblock \emph{Powder Technol}, 284:\penalty0 571--584, 2015.
	
	\bibitem[Liu et~al.(1995)Liu, Nagel, Schecter, Coppersmith, Majumdar, Narayan,
	and Witten]{Liu:1995aa}
	C-H Liu, S~R Nagel, D~A Schecter, S~N Coppersmith, S~Majumdar, O~Narayan, and
	T~A Witten.
	\newblock Force fluctuations in bead packs.
	\newblock \emph{Science}, 269\penalty0 (5223):\penalty0 513--515, 1995.
	
	\bibitem[Mueth et~al.(1998)Mueth, Jaeger, and Nagel]{Mueth:1998a}
	D~M Mueth, H~M Jaeger, and S~R Nagel.
	\newblock Force distribution in a granular medium.
	\newblock \emph{Phys Rev E}, 57\penalty0 (3):\penalty0 3164--3169, 1998.
	
	\bibitem[Coppersmith et~al.(1996)Coppersmith, Liu, Majumdar, Narayan, and
	Witten]{Coppersmith:1996a}
	S~N Coppersmith, C-h Liu, S~Majumdar, O~Narayan, and T~A Witten.
	\newblock Model for force fluctuations in bead packs.
	\newblock \emph{Phys Rev E}, 53\penalty0 (5):\penalty0 4673--4685, 1996.
	
	\bibitem[Claudin et~al.(1998)Claudin, Bouchaud, Cates, and
	Wittmer]{Claudin:1998a}
	P~Claudin, J-P Bouchaud, M~E Cates, and J~P Wittmer.
	\newblock Models of stress fluctuations in granular media.
	\newblock \emph{Phys Rev E}, 57\penalty0 (4):\penalty0 4441--4457, 1998.
	
	\bibitem[Sexton et~al.(1999)Sexton, Socolar, and Schaeffer]{Sexton:1999a}
	M~G Sexton, J~E~S Socolar, and D~G Schaeffer.
	\newblock Force distribution in a scalar model for noncohesive granular
	material.
	\newblock \emph{Phys Rev E}, 60\penalty0 (2):\penalty0 1999--2008, 1999.
	
	\bibitem[Socolar et~al.(2002)Socolar, Schaeffer, and Claudin]{Socolar:2002a}
	J~E~S Socolar, D~G Schaeffer, and P~Claudin.
	\newblock Directed force chain networks and stress response in static granular
	materials.
	\newblock \emph{Eur Phys J E}, 7\penalty0 (4):\penalty0 353--370, 2002.
	
	\bibitem[Peters et~al.(2005)Peters, Muthuswamy, Wibowo, and
	Tordesillas]{Peters:2005aa}
	J~F Peters, M~Muthuswamy, J~Wibowo, and A~Tordesillas.
	\newblock Characterization of force chains in granular material.
	\newblock \emph{Phys Rev E}, 72\penalty0 (4):\penalty0 041307, 2005.
	
	\bibitem[Howell et~al.(1999)Howell, Behringer, and Veje]{Howell1999}
	D~Howell, R~P Behringer, and C~Veje.
	\newblock Stress fluctuations in a {2D} granular {C}ouette experiment: {A}
	continuous transition.
	\newblock \emph{Phys Rev Lett}, 82\penalty0 (26):\penalty0 5241--5244, 1999.
	
	\bibitem[Majmudar and Behringer(2005)]{Majmudar:2005aa}
	T~S Majmudar and R~P Behringer.
	\newblock Contact force measurements and stress-induced anisotropy in granular
	materials.
	\newblock \emph{Nature}, 435\penalty0 (1079):\penalty0 1079--1082, 2005.
	
	\bibitem[Geng et~al.(2001)Geng, Howell, Longhi, Behringer, Reydellet, Vanel,
	Cl\'ement, and Luding]{Geng:2001a}
	J~Geng, D~Howell, E~Longhi, R~P Behringer, G~Reydellet, L~Vanel, E~Cl\'ement,
	and S~Luding.
	\newblock Footprints in sand: {T}he response of a granular material to local
	perturbations.
	\newblock \emph{Phys Rev Lett}, 87\penalty0 (3):\penalty0 035506, 2001.
	
	\bibitem[Radjai et~al.(1998)Radjai, Wolf, Jean, and Moreau]{Radjai:1998aa}
	F~Radjai, D~E Wolf, M~Jean, and J-J Moreau.
	\newblock Bimodal character of stress transmission in granular packings.
	\newblock \emph{Phys Rev Lett}, 80\penalty0 (1):\penalty0 61--64, 1998.
	
	\bibitem[Cates et~al.(1999)Cates, Wittmer, Bouchaud, and
	Claudin]{cates1999jamming}
	M~E Cates, J~P Wittmer, J-P Bouchaud, and P~Claudin.
	\newblock Jamming and static stress transmission in granular materials.
	\newblock \emph{Chaos}, 9\penalty0 (3):\penalty0 511--522, 1999.
	
	\bibitem[Bassett et~al.(2012)Bassett, Owens, Daniels, and
	Porter]{bassett2012influence}
	D~S Bassett, E~T Owens, K~E Daniels, and M~A Porter.
	\newblock Influence of network topology on sound propagation in granular
	materials.
	\newblock \emph{Phys Rev E}, 86\penalty0 (4):\penalty0 041306, 2012.
	
	\bibitem[Richard et~al.(2005)Richard, Nicodemi, Delannay, Ribi{\`e}re, and
	Bideau]{richard2005slow}
	P~Richard, M~Nicodemi, R~Delannay, P~Ribi{\`e}re, and D~Bideau.
	\newblock Slow relaxation and compaction of granular systems.
	\newblock \emph{Nat Mater}, 4\penalty0 (2):\penalty0 121--128, 2005.
	
	\bibitem[Owens and Daniels(2011)]{Owens:2011}
	E~T Owens and K~E Daniels.
	\newblock Sound propagation and force chains in granular materials.
	\newblock \emph{EPL (Europhysics Letters)}, 94\penalty0 (5):\penalty0 54005,
	2011.
	
	\bibitem[Smart et~al.(2007)Smart, Umbanhowar, and Ottino]{Smart:2007a}
	A~Smart, P~Umbanhowar, and J~Ottino.
	\newblock Effects of self-organization on transport in granular matter: {A}
	network-based approach.
	\newblock \emph{EPL (Europhysics Letters)}, 79\penalty0 (2):\penalty0 24002,
	2007.
	
	\bibitem[Gervois et~al.(1989)Gervois, Ammi, Travers, Bideau, Messager, and
	Troadec]{Gervois:1989a}
	A~Gervois, M~Ammi, T~Travers, D~Bideau, J-C Messager, and J-P Troadec.
	\newblock Importance of disorder in the conductivity of packings under
	compression.
	\newblock \emph{Physica A}, 157\penalty0 (1):\penalty0 565--569, 1989.
	
	\bibitem[Combe et~al.(2015)Combe, Richefeu, Stasiak, and
	Atman]{combe2015experimental}
	G~Combe, V~Richefeu, M~Stasiak, and A~P~F Atman.
	\newblock Experimental validation of a nonextensive scaling law in confined
	granular media.
	\newblock \emph{Phys Rev Lett}, 115\penalty0 (23):\penalty0 238301, 2015.
	
	\bibitem[Behringer et~al.(2014)Behringer, Bi, Chakraborty, Clark, Dijksman,
	Ren, and Zhang]{Behringer2014:Statistical}
	R~P Behringer, D~Bi, B~Chakraborty, A~Clark, J~Dijksman, J~Ren, and J~Zhang.
	\newblock Statistical properties of granular materials near jamming.
	\newblock \emph{Journal of Statistical Mechanics: Theory and Experiment},
	2014\penalty0 (6):\penalty0 P06004, 2014.
	
	\bibitem[Bassett et~al.(2015{\natexlab{a}})Bassett, Owens, Porter, Manning, and
	Daniels]{bassett2015extraction}
	D~S Bassett, E~T Owens, M~A Porter, M~L Manning, and K~E Daniels.
	\newblock Extraction of force-chain network architecture in granular materials
	using community detection.
	\newblock \emph{Soft Matter}, 11\penalty0 (14):\penalty0 2731--2744,
	2015{\natexlab{a}}.
	
	\bibitem[Herrera et~al.(2011)Herrera, McCarthy, Slotterback, Cephas, Losert,
	and Girvan]{Herrera:2011aa}
	M~Herrera, S~McCarthy, S~Slotterback, E~Cephas, W~Losert, and M~Girvan.
	\newblock Path to fracture in granular flows: {D}ynamics of contact networks.
	\newblock \emph{Phys Rev E}, 83\penalty0 (6):\penalty0 061303, 2011.
	
	\bibitem[Newman(2010)]{newman2010networks}
	M~E~J Newman.
	\newblock \emph{Networks: {A}n Introduction}.
	\newblock Oxford University Press, 2010.
	
	\bibitem[Bollob{\'a}s(1998)]{Bollobas:1998a}
	B~Bollob{\'a}s.
	\newblock \emph{Modern Graph Theory}.
	\newblock Springer-Verlag, 1998.
	
	\bibitem[Fortunato and Hric(2016)]{Fortunato2016}
	S~Fortunato and D~Hric.
	\newblock Community detection in networks: {A} user guide.
	\newblock \emph{Phys Rep}, 659:\penalty0 1--44, 2016.
	
	\bibitem[Porter et~al.(2009)Porter, Onnela, and Mucha]{Porter2009}
	M~A Porter, J-P Onnela, and P~J Mucha.
	\newblock Communities in networks.
	\newblock \emph{Not Amer Math Soc}, 56\penalty0 (9):\penalty0 1082--1097,
	1164--1166, 2009.
	
	\bibitem[Fortunato(2010)]{Fortunato2010}
	S~Fortunato.
	\newblock Community detection in graphs.
	\newblock \emph{Phys Rep}, 486\penalty0 (3--5):\penalty0 75--174, 2010.
	
	\bibitem[Csermely et~al.(2013)Csermely, London, Wu, and Uzzi]{csermely2013}
	P~Csermely, A~London, L-Y Wu, and B~Uzzi.
	\newblock Structure and dynamics of core--periphery networks.
	\newblock \emph{Journal of Complex Networks Complex Networks}, 1\penalty0
	(2):\penalty0 93--123, 2013.
	
	\bibitem[Newman(2011)]{newman2011complex}
	M~E~J Newman.
	\newblock Complex systems: {A} survey.
	\newblock \emph{Am J Phys}, 79:\penalty0 800--810, 2011.
	
	\bibitem[Kivel\"a et~al.(2014)Kivel\"a, Arenas, Barthelemy, Gleeson, Moreno,
	and Porter]{kivela2014multilayer}
	M~Kivel\"a, A~Arenas, M~Barthelemy, J~P Gleeson, Y~Moreno, and M~A Porter.
	\newblock Multilayer networks.
	\newblock \emph{Journal of Complex Networks}, 2\penalty0 (3):\penalty0
	203--271, 2014.
	
	\bibitem[Barth{\'e}lemy(2011)]{barth2011}
	M~Barth{\'e}lemy.
	\newblock Spatial networks.
	\newblock \emph{Phys Rep}, 499\penalty0 (1):\penalty0 1--101, 2011.
	
	\bibitem[{Cruz Hidalgo} et~al.(2002){Cruz Hidalgo}, Grosse, Kun, Reinhardt, and
	Herrmann]{Hidalgo2002}
	R~{Cruz Hidalgo}, C~U Grosse, F~Kun, H~W Reinhardt, and H~J Herrmann.
	\newblock {Evolution of Percolating Force Chains in Compressed Granular Media}.
	\newblock \emph{Phys Rev Lett}, 89\penalty0 (20):\penalty0 205501, 2002.
	
	\bibitem[Candelier et~al.(2009)Candelier, Dauchot, and
	Biroli]{candelier2009building}
	R~Candelier, O~Dauchot, and G~Biroli.
	\newblock Building blocks of dynamical heterogeneities in dense granular media.
	\newblock \emph{Phys Rev Lett}, 102\penalty0 (8):\penalty0 088001, 2009.
	
	\bibitem[Mehta et~al.(2008)Mehta, Barker, and Luck]{mehta2008heterogeneities}
	A~Mehta, G~C Barker, and J~M Luck.
	\newblock Heterogeneities in granular dynamics.
	\newblock \emph{Proc Natl Acad Sci}, 105\penalty0 (24):\penalty0 8244--8249,
	2008.
	
	\bibitem[Keys et~al.(2007)Keys, Abate, Glotzer, and Durian]{Keys:2007a}
	A~S Keys, A~R Abate, S~C Glotzer, and D~J Durian.
	\newblock Measurement of growing dynamical length scales and prediction of the
	jamming transition in a granular material.
	\newblock \emph{Nat Phys}, 3\penalty0 (4):\penalty0 260--264, 2007.
	
	\bibitem[Digby(1981)]{Digby:1981}
	PJ~Digby.
	\newblock The effective elastic moduli of porous granular rocks.
	\newblock \emph{J Appl Mech}, 48\penalty0 (4):\penalty0 803--808, 1981.
	
	\bibitem[Velick\'y and Caroli(2002)]{Velicky:2002}
	B~Velick\'y and C~Caroli.
	\newblock Pressure dependence of the sound velocity in a two-dimensional
	lattice of {H}ertz--{M}indlin balls: {M}ean-field description.
	\newblock \emph{Phys Rev E}, 65\penalty0 (2):\penalty0 021307, 2002.
	
	\bibitem[Goddard(1990)]{Goddard:1990}
	J~D Goddard.
	\newblock Nonlinear elasticity and pressure-dependent wave speeds in granular
	media.
	\newblock \emph{Proc R Soc A}, 430\penalty0 (1878):\penalty0 105--131, 1990.
	
	\bibitem[Makse et~al.(1999)Makse, Gland, Johnson, and Schwartz]{Makse-1999-WEM}
	H~A Makse, N~Gland, D~L Johnson, and L~M Schwartz.
	\newblock Why effective medium theory fails in granular materials.
	\newblock \emph{Phys Rev Lett}, 83\penalty0 (24):\penalty0 5070--5073, 1999.
	
	\bibitem[Goldenberg and Goldhirsch(2005)]{Goldenberg-2005-FEE}
	C~Goldenberg and I~Goldhirsch.
	\newblock {Friction Enhances Elasticity In Granular Solids}.
	\newblock \emph{Nature}, 435:\penalty0 188--191, 2005.
	
	\bibitem[Smart and Ottino(2008{\natexlab{a}})]{Smart:2008b}
	A~Smart and J~M Ottino.
	\newblock Granular matter and networks: {T}hree related examples.
	\newblock \emph{Soft Matter}, 4\penalty0 (11):\penalty0 2125--2131,
	2008{\natexlab{a}}.
	
	\bibitem[Holme and Saram{\"a}ki(2012)]{Holme2011}
	P~Holme and J~Saram{\"a}ki.
	\newblock Temporal networks.
	\newblock \emph{Phys Rep}, 519\penalty0 (3):\penalty0 97--125, 2012.
	
	\bibitem[Sayama et~al.(2013)Sayama, Pestov, Schmidt, Bush, Wong, Yamanoi, and
	Gross]{thilo-adaptive}
	H~Sayama, I~Pestov, J~Schmidt, B~J Bush, C~Wong, J~Yamanoi, and T~Gross.
	\newblock Modeling complex systems with adaptive networks.
	\newblock \emph{Computers \& Mathematics with Applications}, 65\penalty0
	(10):\penalty0 1645--1664, 2013.
	
	\bibitem[Newman et~al.(2006)Newman, Barab{\'a}si, and Watts]{Newman:2006aa}
	M~E~J Newman, A-L Barab{\'a}si, and D~J Watts.
	\newblock \emph{The Structure and Dynamics of Networks}.
	\newblock Princeton University Press, 2006.
	
	\bibitem[Giusti et~al.(2016{\natexlab{a}})Giusti, Ghrist, and
	Bassett]{giusti2016twos}
	C~Giusti, R~Ghrist, and D~S Bassett.
	\newblock Two's company, three (or more) is a simplex: {A}lgebraic-topological
	tools for understanding higher-order structure in neural data.
	\newblock \emph{J Comput Neurosci}, 41\penalty0 (1):\penalty0 1--14,
	2016{\natexlab{a}}.
	
	\bibitem[Porter and Gleeson(2016)]{porter2016dynamical}
	M~A Porter and J~P Gleeson.
	\newblock Dynamical systems on networks: {A} tutorial.
	\newblock In \emph{Frontiers in Applied Dynamical Systems: {R}eviews and
		Tutorials}, volume~4. Springer-Verlag, 2016.
	
	\bibitem[Liben-Nowell and Kleinberg(2008)]{liben2008tracing}
	D~Liben-Nowell and J~Kleinberg.
	\newblock Tracing information flow on a global scale using internet
	chain-letter data.
	\newblock \emph{Proc Natl Acad Sci}, 105\penalty0 (12):\penalty0 4633--4638,
	2008.
	
	\bibitem[Scott(2012)]{Scott:2012a}
	J~Scott.
	\newblock \emph{Social Network Analysis}.
	\newblock Sage Publications, 2012.
	
	\bibitem[Hurd et~al.(2017)Hurd, Gleeson, and Melnik]{hurd2017framework}
	T~R Hurd, J~P Gleeson, and S~Melnik.
	\newblock A framework for analyzing contagion in assortative banking networks.
	\newblock \emph{PLoS One}, 12\penalty0 (2):\penalty0 e0170579, 2017.
	
	\bibitem[Sporns(2013)]{Sporns:2013aa}
	O~Sporns.
	\newblock Structure and function of complex brain networks.
	\newblock \emph{Dialogues Clin Neurosci}, 15\penalty0 (3):\penalty0 247--262,
	2013.
	
	\bibitem[Bassett and Sporns(2017)]{Bassett:2017a}
	D~S Bassett and O~Sporns.
	\newblock Network neuroscience.
	\newblock \emph{Nat Neurosci}, 20\penalty0 (3):\penalty0 353--364, 2017.
	
	\bibitem[Walker and Tordesillas(2010)]{Walker:2010aa}
	D~M Walker and A~Tordesillas.
	\newblock Topological evolution in dense granular materials: {A} complex
	networks perspective.
	\newblock \emph{Int J Solids Struct}, 47\penalty0 (5):\penalty0 624--639, 2010.
	
	\bibitem[Barrat et~al.(2004)Barrat, Barth{\'e}lemy, Pastor-Satorras, and
	Vespignani]{Barrat:2004aa}
	A~Barrat, M~Barth{\'e}lemy, R~Pastor-Satorras, and A~Vespignani.
	\newblock The architecture of complex weighted networks.
	\newblock \emph{Proc Natl Acad Sci}, 101\penalty0 (11):\penalty0 3747--3752,
	2004.
	
	\bibitem[Newman(2004)]{Newman:2004aa}
	M~E~J Newman.
	\newblock Analysis of weighted networks.
	\newblock \emph{Phys Rev E}, 70\penalty0 (5):\penalty0 056131, 2004.
	
	\bibitem[Alexander(1998)]{Alexander:1998aa}
	S~Alexander.
	\newblock Amorphous solids: {T}heir structure, lattice dynamics and elasticity.
	\newblock \emph{Phys Rep}, 296\penalty0 (2--4):\penalty0 65--236, 1998.
	
	\bibitem[Wyart(2005)]{Wyart:2005a}
	M~Wyart.
	\newblock On the rigidity of amorphous solids.
	\newblock \emph{Annales De Physique}, 30\penalty0 (3):\penalty0 1, 2005.
	
	\bibitem[Liu and Nagel(2010)]{Liu:2010aa}
	A~J Liu and S~R Nagel.
	\newblock The jamming transition and the marginally jammed solid.
	\newblock \emph{Annu Rev Condens Matter Phys}, 1\penalty0 (1):\penalty0
	347--369, 2010.
	
	\bibitem[van Hecke(2010)]{VanHecke2010}
	M~van Hecke.
	\newblock Jamming of soft particles: {G}eometry, mechanics, scaling and
	isostaticity.
	\newblock \emph{J Phys Condens Matter}, 22\penalty0 (3):\penalty0 033101, 2010.
	
	\bibitem[Masuda et~al.(2017)Masuda, Porter, and Lambiotte]{masuda2016}
	N~Masuda, M~A Porter, and R~Lambiotte.
	\newblock Random walks and diffusion on networks.
	\newblock \emph{Phys Rep}, 716-717:\penalty0 1--58, 2017.
	
	\bibitem[Krioukov et~al.(2010)Krioukov, Papadopoulos, Kitsak, Vahdat, and
	Bogu{\~{n}}{\'a}]{boguna2010}
	D~Krioukov, F~Papadopoulos, M~Kitsak, A~Vahdat, and M~Bogu{\~{n}}{\'a}.
	\newblock Hyperbolic geometry of complex networks.
	\newblock \emph{Phys Rev E}, 82\penalty0 (3):\penalty0 036106, 2010.
	
	\bibitem[Skiena(2008)]{skiena2008algorithm}
	S~Skiena.
	\newblock \emph{The Algorithm Design Manual}.
	\newblock Springer-Verlag, 2008.
	
	\bibitem[Watts and Strogatz(1998)]{Watts:1998aa}
	D~J Watts and S~H Strogatz.
	\newblock Collective dynamics of `small-world' networks.
	\newblock \emph{Nature}, 393\penalty0 (6684):\penalty0 440--442, 1998.
	
	\bibitem[Latora and Marchiori(2001)]{latora2001efficient}
	V~Latora and M~Marchiori.
	\newblock Efficient behavior of small-world networks.
	\newblock \emph{Phys Rev Lett}, 87\penalty0 (19):\penalty0 198701, 2001.
	
	\bibitem[Rubinov and Sporns(2009)]{Rubinov2009}
	M~Rubinov and O~Sporns.
	\newblock Complex network measures of brain connectivity: {U}ses and
	interpretations.
	\newblock \emph{NeuroImage}, 52\penalty0 (3):\penalty0 1059--1069, 2009.
	
	\bibitem[Latora and Marchiori(2003)]{Latora2003}
	V~Latora and M~Marchiori.
	\newblock Economic small-world behavior in weighted networks.
	\newblock \emph{Eur Phys J B}, 32:\penalty0 249--263, 2003.
	
	\bibitem[Estrada and Hatano(2016)]{estrada2016}
	E~Estrada and N~Hatano.
	\newblock Communicability angle and the spatial efficiency of networks.
	\newblock \emph{SIAM Review}, 58\penalty0 (4):\penalty0 692--715, 2016.
	
	\bibitem[Ar\'evalo et~al.(2010{\natexlab{a}})Ar\'evalo, Zuriguel, and
	Maza]{Arevalo:2010aa}
	R~Ar\'evalo, I~Zuriguel, and D~Maza.
	\newblock Topology of the force network in the jamming transition of an
	isotropically compressed granular packing.
	\newblock \emph{Phys Rev E}, 81\penalty0 (4):\penalty0 041302,
	2010{\natexlab{a}}.
	
	\bibitem[Gross and Yellen(2005)]{Gross:2005a}
	J~L Gross and J~Yellen.
	\newblock \emph{Graph theory and its Applications}.
	\newblock CRC Press, 2005.
	
	\bibitem[Kavitha et~al.(2009)Kavitha, Liebchen, Mehlhorn, Michail, Rizzi,
	Ueckerdt, and Zweig]{Kavitha:2009a}
	T~Kavitha, C~Liebchen, K~Mehlhorn, D~Michail, R~Rizzi, T~Ueckerdt, and K~A
	Zweig.
	\newblock Cycle bases in graphs characterization, algorithms, complexity, and
	applications.
	\newblock \emph{Comp Sci Rev}, 3\penalty0 (4):\penalty0 199--243, 2009.
	
	\bibitem[Griffin(2017)]{Griffin:2017a}
	C~Griffin.
	\newblock Graph theory: {P}enn {S}tate {Math 485} lecture notes.
	\newblock Webpage, 2017.
	\newblock URL \url{http://www.personal.psu.edu/cxg286/Math485.pdf}.
	\newblock (This manuscript includes contributions by S Shekhar.).
	
	\bibitem[Horton(1987)]{Horton:1987a}
	J~D Horton.
	\newblock A polynomial-time algorithm to find the shortest cycle basis of a
	graph.
	\newblock \emph{SIAM Journal on Computing}, 16\penalty0 (2):\penalty0 358--366,
	1987.
	
	\bibitem[Mehlhorn and Michail(2007)]{Mehlhorn:2006a}
	K~Mehlhorn and D~Michail.
	\newblock Implementing minimum cycle basis algorithms.
	\newblock \emph{J Exp Algorithmics}, 11, 2007.
	
	\bibitem[Walker et~al.(2014{\natexlab{a}})Walker, Tordesillas, and
	Froyland]{Tordesillas:2014a}
	D~M Walker, A~Tordesillas, and G~Froyland.
	\newblock Mesoscale and macroscale kinetic energy fluxes from granular fabric
	evolution.
	\newblock \emph{Phys Rev E}, 89\penalty0 (3):\penalty0 032205,
	2014{\natexlab{a}}.
	
	\bibitem[Walker et~al.(2015{\natexlab{a}})Walker, Tordesillas, Brodu, Dijksman,
	Behringer, and Froyland]{Walker:2015b}
	D~M Walker, A~Tordesillas, N~Brodu, J~A Dijksman, R~P Behringer, and
	G~Froyland.
	\newblock Self-assembly in a near-frictionless granular material:
	{C}onformational structures and transitions in uniaxial cyclic compression of
	hydrogel spheres.
	\newblock \emph{Soft Matter}, 11:\penalty0 2157--2173, 2015{\natexlab{a}}.
	
	\bibitem[Smart and Ottino(2008{\natexlab{b}})]{Smart:2008aa}
	A~G Smart and J~M Ottino.
	\newblock Evolving loop structure in gradually tilted two-dimensional granular
	packings.
	\newblock \emph{Phys Rev E}, 77\penalty0 (4):\penalty0 041307,
	2008{\natexlab{b}}.
	
	\bibitem[Ar\'evalo et~al.(2009)Ar\'evalo, Zuriguel, and Maza]{Arevalo:2009aa}
	R~Ar\'evalo, I~Zuriguel, and D~Maza.
	\newblock Topological properties of the contact network of granular materials.
	\newblock \emph{Int J Bifurc Chaos}, 19\penalty0 (02):\penalty0 695--702, 2009.
	
	\bibitem[Ar\'evalo et~al.(2010{\natexlab{b}})Ar\'evalo, Zuriguel, Trevijano,
	and Maza]{Arevalo:2010ba}
	R~Ar\'evalo, I~Zuriguel, S~A Trevijano, and D~Maza.
	\newblock Third order loops of contacts in a granular force network.
	\newblock \emph{Int J Bifurc Chaos}, 20\penalty0 (03):\penalty0 897--903,
	2010{\natexlab{b}}.
	
	\bibitem[Tordesillas et~al.(2010{\natexlab{a}})Tordesillas, Walker, and
	Lin]{Tordesillas:2010aa}
	A~Tordesillas, D~M Walker, and Q~Lin.
	\newblock Force cycles and force chains.
	\newblock \emph{Phys Rev E}, 81\penalty0 (1):\penalty0 011302,
	2010{\natexlab{a}}.
	
	\bibitem[Rivier(2006)]{rivier2006extended}
	N~Rivier.
	\newblock Extended constraints, arches and soft modes in granular materials.
	\newblock \emph{J Non-Cryst Solids}, 352\penalty0 (42--49):\penalty0
	4505--4508, 2006.
	
	\bibitem[Newman(2003)]{Newman:2003aa}
	M~E~J Newman.
	\newblock The structure and function of complex networks.
	\newblock \emph{SIAM Review}, 45\penalty0 (2):\penalty0 167--256, 2003.
	
	\bibitem[Barrat and Weigt(2000)]{Barrat:2000aa}
	A~Barrat and M~Weigt.
	\newblock On the properties of small-world network models.
	\newblock \emph{Eur Phys J B}, 13\penalty0 (3):\penalty0 547--560, 2000.
	
	\bibitem[Saram\"aki et~al.(2007)Saram\"aki, Kivel\"a, Onnela, Kaski, and
	Kert\'esz]{jari2007}
	J~Saram\"aki, M~Kivel\"a, J-P Onnela, K~Kaski, and J~Kert\'esz.
	\newblock Generalizations of the clustering coefficient to weighted complex
	networks.
	\newblock \emph{Phys Rev E}, 75\penalty0 (2):\penalty0 027105, 2007.
	
	\bibitem[Onnela et~al.(2005)Onnela, Saram\"{a}ki, Kert\'{e}sz, and
	Kaski]{onnela2005}
	J-P Onnela, J~Saram\"{a}ki, J~Kert\'{e}sz, and K~Kaski.
	\newblock Intensity and coherence of motifs in weighted complex networks.
	\newblock \emph{Phys Rev E}, 71\penalty0 (6):\penalty0 065103, 2005.
	
	\bibitem[Zhang and Horvath(2005)]{zhang2005general}
	B~Zhang and S~Horvath.
	\newblock A general framework for weighted gene co-expression network analysis.
	\newblock \emph{Stat Appl Genet Mol Biol}, 4\penalty0 (1):\penalty0 17, 2005.
	
	\bibitem[Freeman(1977)]{Freeman:1977aa}
	L~C Freeman.
	\newblock A set of measures of centrality based on betweenness.
	\newblock \emph{Sociometry}, 40\penalty0 (1):\penalty0 35--41, 1977.
	
	\bibitem[Girvan and Newman(2002)]{Girvan2002}
	M~Girvan and M~E~J Newman.
	\newblock Community structure in social and biological networks.
	\newblock \emph{Proc Natl Acad Sci}, 99\penalty0 (12):\penalty0 7821--7826,
	2002.
	
	\bibitem[Estrada and Rodr\'{i}guez-Vel\'azquez(2005)]{Estrada:2005aa}
	E~Estrada and J~A Rodr\'{i}guez-Vel\'azquez.
	\newblock Subgraph centrality in complex networks.
	\newblock \emph{Phys Rev E}, 71\penalty0 (5):\penalty0 056103, 2005.
	
	\bibitem[Estrada et~al.(2012)Estrada, Hatano, and Benzi]{estrada2012}
	E~Estrada, N~Hatano, and M~Benzi.
	\newblock The physics of communicability in complex networks.
	\newblock \emph{Phys Rep}, 514\penalty0 (3):\penalty0 89--119, 2012.
	
	\bibitem[Estrada and Rodr\'{\i}guez-Vel\'azquez(2005)]{Estrada:2005ab}
	E~Estrada and J~A Rodr\'{\i}guez-Vel\'azquez.
	\newblock Spectral measures of bipartivity in complex networks.
	\newblock \emph{Phys Rev E}, 72\penalty0 (4):\penalty0 046105, 2005.
	
	\bibitem[Milo et~al.(2002)Milo, Shen-Orr, Itzkovitz, Kashtan, Chklovskii, and
	Alon]{Milo:2002a}
	R~Milo, S~S Shen-Orr, S~Itzkovitz, N~Kashtan, D~Chklovskii, and U~Alon.
	\newblock Network motifs: {S}imple building blocks of complex networks.
	\newblock \emph{Science}, 298\penalty0 (5594):\penalty0 824--827, 2002.
	
	\bibitem[Shen-Orr et~al.(2002)Shen-Orr, Milo, Mangan, and
	Alon]{shenorr2002network}
	S~S Shen-Orr, R~Milo, S~Mangan, and U~Alon.
	\newblock Network motifs in the transcriptional regulation network of
	{E}scherichia coli.
	\newblock \emph{Nat Genet}, 31\penalty0 (1):\penalty0 64--68, 2002.
	
	\bibitem[Milo et~al.(2004)Milo, Itzkovitz, Kashtan, Levitt, Shen-Orr,
	Ayzenshtat, Sheffer, and Alon]{milo2004superfamilies}
	R~Milo, S~Itzkovitz, N~Kashtan, R~Levitt, S~Shen-Orr, I~Ayzenshtat, M~Sheffer,
	and U~Alon.
	\newblock Superfamilies of evolved and designed networks.
	\newblock \emph{Science}, 303\penalty0 (5663):\penalty0 1538--1542, 2004.
	
	\bibitem[Alon(2007)]{alon2007network}
	U~Alon.
	\newblock Network motifs: {T}heory and experimental approaches.
	\newblock \emph{Nat Rev Genet}, 8\penalty0 (6):\penalty0 450--461, 2007.
	
	\bibitem[Schreiber and Schwobbermeyer(2005)]{schreiber2005frequency}
	F~Schreiber and H~Schwobbermeyer.
	\newblock Frequency concepts and pattern detection for the analysis of motifs
	in networks.
	\newblock \emph{Transactions on Computational Systems Biology}, III:\penalty0
	89--104, 2005.
	
	\bibitem[Wernicke(2006)]{wernicke2006efficient}
	S~Wernicke.
	\newblock Efficient detection of network motifs.
	\newblock \emph{IEEE/ACM Trans Comput Biol Bioinform}, 3\penalty0 (4):\penalty0
	347--359, 2006.
	
	\bibitem[Grochow and Kellis(2007)]{grochow2007network}
	J~A Grochow and M~Kellis.
	\newblock Network motif discovery using sub-graph enumeration and
	symmetry-breaking.
	\newblock \emph{RECOMB}, pages 92--106, 2007.
	
	\bibitem[Omidi et~al.(2009)Omidi, Schreiber, and Masoudi-Nejad]{omidi2009moda}
	S~Omidi, F~Schreiber, and A~Masoudi-Nejad.
	\newblock {MODA}: {A}n efficient algorithm for network motif discovery in
	biological networks.
	\newblock \emph{Genes Genet Syst}, 84\penalty0 (5):\penalty0 385--395, 2009.
	
	\bibitem[Kashani et~al.(2009)Kashani, Ahrabian, Elahi, Nowzari-Dalini, Ansari,
	Asadi, Mohammadi, Schreiber, and Masoudi-Nejad]{kashani2009kavosh}
	Z~R Kashani, H~Ahrabian, E~Elahi, A~Nowzari-Dalini, E~S Ansari, S~Asadi,
	S~Mohammadi, F~Schreiber, and A~Masoudi-Nejad.
	\newblock Kavosh: {A} new algorithm for finding network motifs.
	\newblock \emph{BMC Bioinformatics}, 10\penalty0 (318):\penalty0 318, 2009.
	
	\bibitem[Paulau et~al.(2015)Paulau, Feenders, and Blasius]{paulau2015motif}
	P~V Paulau, C~Feenders, and B~Blasius.
	\newblock Motif analysis in directed ordered networks and applications to food
	webs.
	\newblock \emph{Sci Rep}, 5:\penalty0 11926, 2015.
	
	\bibitem[Sporns and Kotter(2004)]{sporns2004motifs}
	O~Sporns and R~Kotter.
	\newblock Motifs in brain networks.
	\newblock \emph{PLoS Biol}, 2\penalty0 (11):\penalty0 e369, 2004.
	
	\bibitem[Xu et~al.(2008)Xu, Zhang, and Small]{Xu:2008aa}
	X~Xu, J~Zhang, and M~Small.
	\newblock Superfamily phenomena and motifs of networks induced from time
	series.
	\newblock \emph{Proc Natl Acad Sci}, 105\penalty0 (50):\penalty0 19601--19605,
	2008.
	
	\bibitem[Walker et~al.(2014{\natexlab{b}})Walker, Tordesillas, Small,
	Behringer, and Tse]{Walker:2014aa}
	D~M Walker, A~Tordesillas, M~Small, R~P Behringer, and C~K Tse.
	\newblock A complex systems analysis of stick-slip dynamics of a laboratory
	fault.
	\newblock \emph{Chaos}, 24\penalty0 (1):\penalty0 013132, 2014{\natexlab{b}}.
	
	\bibitem[Walker et~al.(2015{\natexlab{b}})Walker, Tordesillas, Zhang,
	Behringer, And{\`o}, Viggiani, Druckrey, and Alshibli]{Walker:2015c}
	D~M Walker, A~Tordesillas, J~Zhang, R~P Behringer, E~And{\`o}, G~Viggiani,
	A~Druckrey, and K~Alshibli.
	\newblock Structural templates of disordered granular media.
	\newblock \emph{Int J Solids Struct}, 54:\penalty0 20--30, 2015{\natexlab{b}}.
	
	\bibitem[Tordesillas et~al.(2012{\natexlab{a}})Tordesillas, Walker, Froyland,
	Zhang, and Behringer]{Tordesillas:2012aa}
	A~Tordesillas, D~M Walker, G~Froyland, J~Zhang, and R~P Behringer.
	\newblock Transition dynamics and magic-number-like behavior of frictional
	granular clusters.
	\newblock \emph{Phys Rev E}, 86\penalty0 (1):\penalty0 011306,
	2012{\natexlab{a}}.
	
	\bibitem[Peixoto(2017)]{peixoto2017}
	T~P Peixoto.
	\newblock Bayesian stochastic blockmodeling.
	\newblock \emph{arXiv}, arXiv:1705.10225 [stat.ML], 2017.
	
	\bibitem[Giusti et~al.(2016{\natexlab{b}})Giusti, Papadopoulos, Owens, Daniels,
	and Bassett]{Giusti:2016a}
	C~Giusti, L~Papadopoulos, E~T Owens, K~E Daniels, and D~S Bassett.
	\newblock Topological and geometric measurements of force-chain structure.
	\newblock \emph{Phys Rev E}, 94\penalty0 (3):\penalty0 032909,
	2016{\natexlab{b}}.
	
	\bibitem[Papadopoulos et~al.(2016)Papadopoulos, Puckett, Daniels, and
	Bassett]{papadopoulos2016evolution}
	L~Papadopoulos, J~G Puckett, K~E Daniels, and D~S Bassett.
	\newblock Evolution of network architecture in a granular material under
	compression.
	\newblock \emph{Phys Rev E}, 94\penalty0 (3):\penalty0 032908, 2016.
	
	\bibitem[Walker and Tordesillas(2012)]{Walker:2012aa}
	D~M Walker and A~Tordesillas.
	\newblock Taxonomy of granular rheology from grain property networks.
	\newblock \emph{Phys Rev E}, 85\penalty0 (1):\penalty0 011304, 2012.
	
	\bibitem[Tordesillas et~al.(2013{\natexlab{a}})Tordesillas, Walker, And{\`o},
	and Viggiani]{Tordesillas:2013a}
	A~Tordesillas, D~M Walker, E~And{\`o}, and G~Viggiani.
	\newblock Revisiting localized deformation in sand with complex systems.
	\newblock \emph{Proc Math Phys Eng Sci}, 469\penalty0 (2152),
	2013{\natexlab{a}}.
	
	\bibitem[Walker and Tordesillas(2014)]{Walker:2014ba}
	D~M Walker and A~Tordesillas.
	\newblock Examining overlapping community structures within grain property
	networks.
	\newblock In \emph{2014 IEEE International Symposium on Circuits and Systems
		(ISCAS)}, pages 1275--1278, 2014.
	
	\bibitem[Walker et~al.(2012)Walker, Tordesillas, Pucilowski, Lin, Rechenmacher,
	and Abedi]{Walker:2012a}
	D~M Walker, A~Tordesillas, S~Pucilowski, Q~Lin, A~L Rechenmacher, and S~Abedi.
	\newblock Analysis of grain-scale measurements of sand using kinematical
	complex networks.
	\newblock \emph{Int J Bifurc Chaos}, 22\penalty0 (12):\penalty0 1230042, 2012.
	
	\bibitem[Newman and Girvan(2004)]{NG2004}
	M~E~J Newman and M~Girvan.
	\newblock Finding and evaluating community structure in networks.
	\newblock \emph{Phys Rev E}, 69\penalty0 (2):\penalty0 026113, 2004.
	
	\bibitem[Newman(2006{\natexlab{a}})]{Newman2006b}
	M~E~J Newman.
	\newblock Finding community structure in networks using the eigenvectors of
	matrices.
	\newblock \emph{Phys Rev E}, 74\penalty0 (3):\penalty0 036104,
	2006{\natexlab{a}}.
	
	\bibitem[Rosvall and Bergstrom(2008)]{rosvall2008maps}
	M~Rosvall and C~T Bergstrom.
	\newblock Maps of random walks on complex networks reveal community structure.
	\newblock \emph{Proc Natl Acad Sci}, 105\penalty0 (4):\penalty0 1118--1123,
	2008.
	
	\bibitem[Clauset(2005)]{clauset2005finding}
	A~Clauset.
	\newblock Finding local community structure in networks.
	\newblock \emph{Phys Rev E}, 72\penalty0 (2):\penalty0 026132, 2005.
	
	\bibitem[Jeub et~al.(2015)Jeub, Balachandran, Porter, Mucha, and
	Mahoney]{jeub2015}
	L~G~S Jeub, P~Balachandran, M~A Porter, P~J Mucha, and W~M Mahoney.
	\newblock Think locally, act locally: {T}he detection of small, medium-sized,
	and large communities in large networks.
	\newblock \emph{Phys Rev E}, 91\penalty0 (1):\penalty0 012821, 2015.
	
	\bibitem[Ahn et~al.(2010)Ahn, Bagrow, and Lehmann]{ahn2010link}
	Y~Y Ahn, J~P Bagrow, and S~Lehmann.
	\newblock Link communities reveal multiscale complexity in networks.
	\newblock \emph{Nature}, 466\penalty0 (7307):\penalty0 761--764, 2010.
	
	\bibitem[Good et~al.(2010)Good, de~Montjoye, and Clauset]{Good2010}
	B~H Good, Y~A de~Montjoye, and A~Clauset.
	\newblock Performance of modularity maximization in practical contexts.
	\newblock \emph{Phys Rev E}, 81\penalty0 (4):\penalty0 046106, 2010.
	
	\bibitem[Fortunato and Barth\'{e}lemy(2007)]{Fortunato2007}
	S.~Fortunato and M.~Barth\'{e}lemy.
	\newblock Resolution limit in community detection.
	\newblock \emph{Proc Natl Acad Sci}, 104\penalty0 (1):\penalty0 36--41, 2007.
	
	\bibitem[Bassett et~al.(2013)Bassett, Porter, Wymbs, Grafton, Carlson, and
	Mucha]{bassett2013robust}
	D~S Bassett, M~A Porter, N~F Wymbs, S~T Grafton, J~M Carlson, and P~J Mucha.
	\newblock Robust detection of dynamic community structure in networks.
	\newblock \emph{Chaos}, 23\penalty0 (1):\penalty0 013142, 2013.
	
	\bibitem[Newman(2006{\natexlab{b}})]{Newman2006}
	M~E~J Newman.
	\newblock Modularity and community structure in networks.
	\newblock \emph{Proc Natl Acad Sci}, 103\penalty0 (23):\penalty0 8577--8582,
	2006{\natexlab{b}}.
	
	\bibitem[Bazzi et~al.(2016)Bazzi, Porter, Williams, McDonald, Fenn, and
	Howison]{Bazzi2016}
	M~Bazzi, M~A Porter, S~Williams, M~McDonald, D~J Fenn, and S~D Howison.
	\newblock Community detection in temporal multilayer networks, with an
	application to correlation networks.
	\newblock \emph{Multiscale Model Simul}, 14\penalty0 (1):\penalty0 1--41, 2016.
	
	\bibitem[Sarzynska et~al.(2016)Sarzynska, Leicht, Chowell, and
	Porter]{Sarzynska:2015aa}
	M~Sarzynska, E~A Leicht, G~Chowell, and M~A Porter.
	\newblock Null models for community detection in spatially embedded, temporal
	networks.
	\newblock \emph{Journal of Complex Networks}, 4:\penalty0 363--406, 2016.
	
	\bibitem[Brandes et~al.(2008)Brandes, Delling, Gaertler, G\"{o}rke, Hoefer,
	Nikoloski, and Wagner]{Brandes2008}
	U~Brandes, D~Delling, M~Gaertler, R~G\"{o}rke, M~Hoefer, Z~Nikoloski, and
	D~Wagner.
	\newblock On modularity clustering.
	\newblock \emph{IEEE Trans on Knowl Data Eng}, 20\penalty0 (2):\penalty0
	172--188, 2008.
	
	\bibitem[Blondel et~al.(2008)Blondel, Guillaume, Lambiotte, and
	Lefebvre]{Blondel2008}
	V~D Blondel, J~L Guillaume, R~Lambiotte, and E~Lefebvre.
	\newblock Fast unfolding of community hierarchies in large networks.
	\newblock \emph{Journal of Statistical Mechanics: Theory and Experiment},
	2008\penalty0 (10):\penalty0 P10008, 2008.
	
	\bibitem[Jeub et~al.(2011--2016)Jeub, Bazzi, Jutla, and Mucha]{genlouvain2016}
	L~G~S Jeub, M~Bazzi, I~S Jutla, and P~J Mucha.
	\newblock A generalized {L}ouvain method for community detection implemented in
	{\sc matlab}, 2011--2016.
	\newblock URL \url{https://github.com/GenLouvain/GenLouvain}.
	
	\bibitem[Lancichinetti and Fortunato(2012)]{Lancichinetti2012}
	A~Lancichinetti and S~Fortunato.
	\newblock Consensus clustering in complex networks.
	\newblock \emph{Sci Rep}, 2:\penalty0 336, 2012.
	
	\bibitem[Jeub et~al.(2017)Jeub, Sporns, and Fortunato]{jeub-santo2017}
	L~G~S Jeub, O~Sporns, and S~Fortunato.
	\newblock Multiresolution consensus clustering in networks.
	\newblock \emph{arXiv}, arXiv:1710.02249 [cs.SI], 2017.
	
	\bibitem[Boccaletti et~al.(2014)Boccaletti, Bianconi, Criado, {Del Genio},
	G\'omez-Garde{\~n}es, Romance, Sendi{\~n}a-Nadal, Wang, and
	Zanin]{Boccaletti2014}
	S~Boccaletti, G~Bianconi, R~Criado, C~I {Del Genio}, J~G\'omez-Garde{\~n}es,
	M~Romance, I~Sendi{\~n}a-Nadal, Z~Wang, and M~Zanin.
	\newblock The structure and dynamics of multilayer networks.
	\newblock \emph{Phys Rep}, 544\penalty0 (1):\penalty0 1--122, 2014.
	
	\bibitem[Mucha et~al.(2010)Mucha, Richardson, Macon, Porter, and
	Onnela]{Mucha2010}
	P~J Mucha, T~Richardson, K~Macon, M~A Porter, and J-P Onnela.
	\newblock Community structure in time-dependent, multiscale, and multiplex
	networks.
	\newblock \emph{Science}, 328\penalty0 (5980):\penalty0 876--878, 2010.
	
	\bibitem[De~Domenico et~al.(2013)De~Domenico, Sol\'e-Ribalta, Cozzo, Kivel\"a,
	Moreno, Porter, Gomez, and Arenas]{domenico2013mathematical}
	M~De~Domenico, A~Sol\'e-Ribalta, E~Cozzo, M~Kivel\"a, Y~Moreno, M~A Porter,
	S~Gomez, and A~Arenas.
	\newblock Mathematical formulation of multi-layer networks.
	\newblock \emph{Phys Rev X}, 3\penalty0 (4):\penalty0 041022, 2013.
	
	\bibitem[Kolda and Bader(2009)]{Kolda2009Tensor}
	T~G Kolda and B~W Bader.
	\newblock Tensor decompositions and applications.
	\newblock \emph{SIAM Rev}, 51\penalty0 (3):\penalty0 455--500, 2009.
	
	\bibitem[Cranmer et~al.(2015)Cranmer, Menninga, and Mucha]{Cranmer:2014ut}
	S~J Cranmer, E~J Menninga, and P~J Mucha.
	\newblock Kantian fractionalization predicts the conflict propensity of the
	international system.
	\newblock \emph{Proc Nat Acad Sci}, 112\penalty0 (38):\penalty0 11812--11816,
	2015.
	
	\bibitem[Danchev and Porter(2017)]{danchev2016}
	V~Danchev and M~A Porter.
	\newblock Neither global nor local: {H}eterogeneous connectivity in spatial
	network structures of world migration.
	\newblock \emph{Social Networks}, in press;
	\url{https://doi.org/10.1016/j.socnet.2017.06.003}, 2017.
	
	\bibitem[Bassett et~al.(2011)Bassett, Wymbs, Porter, Mucha, Carlson, and
	Grafton]{Bassett2011b}
	D~S Bassett, N~F Wymbs, M~A Porter, P~J Mucha, J~M Carlson, and S~T Grafton.
	\newblock Dynamic reconfiguration of human brain networks during learning.
	\newblock \emph{Proc Natl Acad Sci}, 108\penalty0 (18):\penalty0 7641--7646,
	2011.
	
	\bibitem[Bassett et~al.(2015{\natexlab{b}})Bassett, Yang, Wymbs, and
	Grafton]{bassett2015learning}
	D~S Bassett, M~Yang, N~F Wymbs, and S~T Grafton.
	\newblock Learning-induced autonomy of sensorimotor systems.
	\newblock \emph{Nat Neurosci}, 18\penalty0 (5):\penalty0 744--751,
	2015{\natexlab{b}}.
	
	\bibitem[Braun et~al.(2015)Braun, Schafer, Walter, Erk, Romanczuk-Seiferth,
	Haddad, Schweiger, Grimm, Heinz, Tost, Meyer-Lindenberg, and
	Bassett]{braun2015dynamic}
	U~Braun, A~Schafer, H~Walter, S~Erk, N~Romanczuk-Seiferth, L~Haddad, J~I
	Schweiger, O~Grimm, A~Heinz, H~Tost, A~Meyer-Lindenberg, and D~S Bassett.
	\newblock Dynamic reconfiguration of frontal brain networks during executive
	cognition in humans.
	\newblock \emph{Proc Natl Acad Sci}, 112\penalty0 (37):\penalty0 11678--11683,
	2015.
	
	\bibitem[Blinder et~al.(2013)Blinder, Tsai, Kaufhold, Knutsen, Suhl, and
	Kleinfeld]{Blinder:2013aa}
	P~Blinder, P~S Tsai, J~P Kaufhold, P~M Knutsen, H~Suhl, and D~Kleinfeld.
	\newblock The cortical angiome: {A}n interconnected vascular network with
	noncolumnar patterns of blood flow.
	\newblock \emph{Nat Neurosci}, 16\penalty0 (7):\penalty0 889--897, 2013.
	
	\bibitem[Katifori et~al.(2010)Katifori, Sz\"oll\ifmmode~\mbox{\H{o}}\else
	\H{o}\fi{}si, and Magnasco]{Katifori:2010a}
	E~Katifori, G~J Sz\"oll\ifmmode~\mbox{\H{o}}\else \H{o}\fi{}si, and M~O
	Magnasco.
	\newblock Damage and fluctuations induce loops in optimal transport networks.
	\newblock \emph{Phys Rev Lett}, 104\penalty0 (4):\penalty0 048704, 2010.
	
	\bibitem[Lee et~al.(2017)Lee, Fricker, and Porter]{shl2017}
	S~H Lee, M~D Fricker, and M~A Porter.
	\newblock Mesoscale analyses of fungal networks as an approach for quantifying
	phenotypic traits.
	\newblock \emph{Journal of Complex Networks}, 5\penalty0 (1):\penalty0
	145--159, 2017.
	
	\bibitem[Bebber et~al.(2007)Bebber, Hynes, Darrah, Boddy, and
	Fricker]{Bebber:2007a}
	D~P Bebber, J~Hynes, P~R Darrah, L~Boddy, and M~D Fricker.
	\newblock Biological solutions to transport network design.
	\newblock \emph{Proc R Soc Lond B Biol Sci}, 274\penalty0 (1623):\penalty0
	2307--2315, 2007.
	
	\bibitem[Banavar et~al.(2000)Banavar, Colaiori, Flammini, Maritan, and
	Rinaldo]{Banavar:2000a}
	J~R Banavar, F~Colaiori, A~Flammini, A~Maritan, and A~Rinaldo.
	\newblock Topology of the fittest transportation network.
	\newblock \emph{Phys Rev Lett}, 84\penalty0 (20):\penalty0 4745--4748, 2000.
	
	\bibitem[Gastner and Newman(2006)]{Newman:2006a}
	M~T Gastner and M~E~J Newman.
	\newblock Optimal design of spatial distribution networks.
	\newblock \emph{Phys Rev E}, 74\penalty0 (1):\penalty0 016117, 2006.
	
	\bibitem[Kurant and Thiran(2006)]{Kurant:2006a}
	M~Kurant and P~Thiran.
	\newblock Extraction and analysis of traffic and topologies of transportation
	networks.
	\newblock \emph{Phys Rev E}, 74\penalty0 (3):\penalty0 036114, 2006.
	
	\bibitem[Bertsekas(1998)]{Bertsekas:1998}
	D~P Bertsekas.
	\newblock \emph{Network Optimization: {C}ontinuous and Discrete Models}.
	\newblock Athena Scientific, 1998.
	
	\bibitem[Ahuja et~al.(1993)Ahuja, Magnanti, and Orlin]{Ahuja:flow}
	R~K Ahuja, T~L Magnanti, and J~B Orlin.
	\newblock \emph{Network Flows: {T}heory, Algorithms, and Applications}.
	\newblock Prentice-Hall, Inc., 1993.
	
	\bibitem[Kaczynski et~al.(2004)Kaczynski, Mischaikow, and
	Mrozek]{Kaczynski:2004aa}
	T~Kaczynski, K~Mischaikow, and M~Mrozek.
	\newblock \emph{Computational Homology}.
	\newblock Springer-Verlag, 2004.
	
	\bibitem[Kesten(2006)]{kesten-whatis}
	H~Kesten.
	\newblock What is ... percolation?
	\newblock \emph{Not Am Math Soc}, 53\penalty0 (5):\penalty0 572--573, 2006.
	
	\bibitem[Stauffer and Aharony(1994)]{Stauffer:1994a}
	D~Stauffer and A~Aharony.
	\newblock \emph{Introduction to Percolation Theory}.
	\newblock CRC Press, 1994.
	
	\bibitem[Saberi(2015)]{saberi2015}
	A~A Saberi.
	\newblock Recent advances in percolation theory and its applications.
	\newblock \emph{Phys Rep}, 578:\penalty0 1--32, 2015.
	
	\bibitem[Broadbent and Hammersley(1957)]{broadbent1957percolation}
	S~Broadbent and J~Hammersley.
	\newblock Percolation processes {I}. {C}rystals and mazes.
	\newblock \emph{Proc Camb Philos Soc}, 53:\penalty0 629--641, 1957.
	
	\bibitem[Albert and Barab\'{a}si(2002)]{Albert2002}
	R~Albert and A-L Barab\'{a}si.
	\newblock Statistical mechanics of complex networks.
	\newblock \emph{Rev Mod Phys}, 74\penalty0 (1):\penalty0 47--98, 2002.
	
	\bibitem[Erd{\H o}s and Re\'nyi(1959)]{Erdos:1959aa}
	P~Erd{\H o}s and A~Re\'nyi.
	\newblock On random graphs {I}.
	\newblock \emph{Publicationes Mathematicae Debrecen}, 6:\penalty0 290--297,
	1959.
	
	\bibitem[Erd{\H o}s and Re\'nyi(1960)]{Erdos:1960aa}
	P~Erd{\H o}s and A~Re\'nyi.
	\newblock On the evolution of random graphs.
	\newblock \emph{Publications of the Mathematical Institute of the Hungarian
		Academy of Sciences}, 5:\penalty0 17--61, 1960.
	
	\bibitem[Stauffer(1979)]{Stauffer1979a}
	D.~Stauffer.
	\newblock Scaling theory of percolation clusters.
	\newblock \emph{Phys Rep}, 54\penalty0 (1):\penalty0 1--74, 1979.
	
	\bibitem[Slotterback et~al.(2012)Slotterback, Mailman, Ronaszegi, van Hecke,
	Girvan, and Losert]{Slotterback:2012aa}
	S~Slotterback, M~Mailman, K~Ronaszegi, M~van Hecke, M~Girvan, and W~Losert.
	\newblock Onset of irreversibility in cyclic shear of granular packings.
	\newblock \emph{Phys Rev E}, 85\penalty0 (2):\penalty0 021309, 2012.
	
	\bibitem[Kondic et~al.(2012)Kondic, Goullet, O'Hern, Kram\'ar, Mischaikow, and
	Behringer]{Kondic:2012aa}
	L~Kondic, A~Goullet, C~S O'Hern, M~Kram\'ar, K~Mischaikow, and R~P Behringer.
	\newblock Topology of force networks in compressed granular media.
	\newblock \emph{EPL (Europhysics Letters)}, 97\penalty0 (5):\penalty0 54001,
	2012.
	
	\bibitem[Kram\'ar et~al.(2013)Kram\'ar, Goullet, Kondic, and
	Mischaikow]{Kramar:2013aa}
	M~Kram\'ar, A~Goullet, L~Kondic, and K~Mischaikow.
	\newblock Persistence of force networks in compressed granular media.
	\newblock \emph{Phys Rev E}, 87\penalty0 (4):\penalty0 042207, 2013.
	
	\bibitem[Kram\'ar et~al.(2014{\natexlab{a}})Kram\'ar, Goullet, Kondic, and
	Mischaikow]{Kramar:2014b}
	M~Kram\'ar, A~Goullet, L~Kondic, and K~Mischaikow.
	\newblock Quantifying force networks in particulate systems.
	\newblock \emph{Physica D}, 283:\penalty0 37--55, 2014{\natexlab{a}}.
	
	\bibitem[Kram\'ar et~al.(2014{\natexlab{b}})Kram\'ar, Goullet, Kondic, and
	Mischaikow]{Kramar:2014aa}
	M~Kram\'ar, A~Goullet, L~Kondic, and K~Mischaikow.
	\newblock Evolution of force networks in dense particulate media.
	\newblock \emph{Phys Rev E}, 90\penalty0 (5):\penalty0 052203,
	2014{\natexlab{b}}.
	
	\bibitem[Ardanza-Trevijano et~al.(2014)Ardanza-Trevijano, Zuriguel, Ar\'evalo,
	and Maza]{Ardanza-Trevijano:2014aa}
	S~Ardanza-Trevijano, I~Zuriguel, R~Ar\'evalo, and D~Maza.
	\newblock Topological analysis of tapped granular media using persistent
	homology.
	\newblock \emph{Phys Rev E}, 89\penalty0 (5):\penalty0 052212, 2014.
	
	\bibitem[Kondic et~al.(2016)Kondic, Kram\'ar, Pugnaloni, Carlevaro, and
	Mischaikow]{Kondic:2016a}
	L~Kondic, M~Kram\'ar, L~A Pugnaloni, C~M Carlevaro, and K~Mischaikow.
	\newblock Structure of force networks in tapped particulate systems of disks
	and pentagons. {II}. {P}ersistence analysis.
	\newblock \emph{Phys Rev E}, 93\penalty0 (6):\penalty0 062903, 2016.
	
	\bibitem[Pugnaloni et~al.(2016)Pugnaloni, Carlevaro, Kram\'ar, Mischaikow, and
	Kondic]{Pugnaloni:2016a}
	L~A Pugnaloni, C~M Carlevaro, M~Kram\'ar, K~Mischaikow, and L~Kondic.
	\newblock Structure of force networks in tapped particulate systems of disks
	and pentagons. {I}. {C}lusters and loops.
	\newblock \emph{Phys Rev E}, 93\penalty0 (6):\penalty0 062902, 2016.
	
	\bibitem[Feng(1985)]{Feng1985}
	S~Feng.
	\newblock Percolation properties of granular elastic networks in two
	dimensions.
	\newblock \emph{Phys Rev B}, 32\penalty0 (1):\penalty0 510--513, 1985.
	
	\bibitem[Moukarzel and Duxbury(1995)]{Moukarzel:1995a}
	C~Moukarzel and P~M Duxbury.
	\newblock Stressed backbone and elasticity of random central-force systems.
	\newblock \emph{Phys Rev Lett}, 75\penalty0 (22):\penalty0 4055--4058, 1995.
	
	\bibitem[Jacobs and Thorpe(1995)]{Jacobs1995}
	D~Jacobs and M~F Thorpe.
	\newblock Generic rigidity percolation: {T}he pebble game.
	\newblock \emph{Phys Rev Lett}, 75\penalty0 (22):\penalty0 4051--4054, 1995.
	
	\bibitem[Aharonov and Sparks(1999)]{Aharonov:1999a}
	E~Aharonov and D~Sparks.
	\newblock Rigidity phase transition in granular packings.
	\newblock \emph{Phys Rev E}, 60\penalty0 (6):\penalty0 6890--6896, 1999.
	
	\bibitem[Lois et~al.(2008)Lois, Blawzdziewicz, and O'Hern]{Lois:2008a}
	G~Lois, J~Blawzdziewicz, and C~S O'Hern.
	\newblock Jamming transition and new percolation universality classes in
	particulate systems with attraction.
	\newblock \emph{Phys Rev Lett}, 100\penalty0 (2):\penalty0 028001, 2008.
	
	\bibitem[Shen et~al.(2012)Shen, O'Hern, and Shattuck]{Shen:2012a}
	T~Shen, C~S O'Hern, and M~D Shattuck.
	\newblock Contact percolation transition in athermal particulate systems.
	\newblock \emph{Phys Rev E}, 85\penalty0 (1):\penalty0 011308, 2012.
	
	\bibitem[Kovalcinova et~al.(2015)Kovalcinova, Goullet, and
	Kondic]{Kovalcinova:2015a}
	L~Kovalcinova, A~Goullet, and L~Kondic.
	\newblock Percolation and jamming transitions in particulate systems with and
	without cohesion.
	\newblock \emph{Phys Rev E}, 92\penalty0 (3):\penalty0 032204, 2015.
	
	\bibitem[Henkes et~al.(2016)Henkes, Quint, Fily, and Schwarz]{Henkes2016}
	S~Henkes, D~A Quint, Y~Fily, and J~M Schwarz.
	\newblock {Rigid Cluster Decomposition Reveals Criticality in Frictional
		Jamming}.
	\newblock \emph{Phys Rev Lett}, 116\penalty0 (2):\penalty0 028301, 2016.
	
	\bibitem[Thorpe(1985)]{Thorpe1985a}
	M~F Thorpe.
	\newblock \emph{Rigidity Percolation}.
	\newblock Springer-Verlag, 1985.
	
	\bibitem[Thorpe and Duxbury(1999)]{Thorpe:1999aa}
	M~F Thorpe and P~M Duxbury, editors.
	\newblock \emph{Rigidity Theory and Applications}.
	\newblock Springer-Verlag, 1999.
	
	\bibitem[Kovalcinova et~al.(2016)Kovalcinova, Goullet, and
	Kondic]{Kovalcinova:2016a}
	L~Kovalcinova, A~Goullet, and L~Kondic.
	\newblock Scaling properties of force networks for compressed particulate
	systems.
	\newblock \emph{Phys Rev E}, 93\penalty0 (4):\penalty0 042903, 2016.
	
	\bibitem[Pastor-Satorras and Miguel(2012)]{PastorSatorras2012}
	R~Pastor-Satorras and M-C Miguel.
	\newblock Percolation analysis of force networks in anisotropic granular
	matter.
	\newblock \emph{Journal of Statistical Mechanics: Theory and Experiment},
	2012\penalty0 (02):\penalty0 P02008, 2012.
	
	\bibitem[Pathak et~al.(2017)Pathak, Esposito, Coniglio, and
	Ciamarra]{pathak2017force}
	S~N Pathak, V~Esposito, A~Coniglio, and M~P Ciamarra.
	\newblock Force percolation transition of jammed granular systems.
	\newblock \emph{Phys Rev E}, 96\penalty0 (4):\penalty0 042901, 2017.
	
	\bibitem[Edelsbrunner(2010)]{Edelsbrunner:2010a}
	H~Edelsbrunner.
	\newblock \emph{Computational Topology: {A}n Introduction}.
	\newblock American Mathematical Society, 2010.
	
	\bibitem[Carlsson(2009)]{Carlsson:2009a}
	G~Carlsson.
	\newblock Topology and data.
	\newblock \emph{Bull Am Math Soc}, 46\penalty0 (2):\penalty0 255--308, 2009.
	
	\bibitem[Ghrist(2014)]{ghrist2014elementary}
	R~Ghrist.
	\newblock \emph{Elementary Applied Topology}.
	\newblock Createspace, 2014.
	\newblock Available at \url{https://www.math.upenn.edu/~ghrist/notes.html}.
	
	\bibitem[Dlotko et~al.(2012)Dlotko, Juda, Mrozek, and
	Ghrist]{dlotko2012distributed}
	P~Dlotko, M~Juda, M~Mrozek, and R~Ghrist.
	\newblock Distributed computation of coverage in sensor networks by homological
	methods.
	\newblock \emph{Appl Algebr Eng Comm}, 23:\penalty0 29--58, 2012.
	
	\bibitem[Taylor et~al.(2015)Taylor, Klimm, Harrington, Kram\'ar, Mischaikow,
	Porter, and Mucha]{taylor2015topological}
	D~Taylor, F~Klimm, H~A Harrington, M~Kram\'ar, K~Mischaikow, M~A Porter, and
	P~J Mucha.
	\newblock Topological data analysis of contagion maps for examining spreading
	processes on networks.
	\newblock \emph{Nat Commun}, 6:\penalty0 7723, 2015.
	
	\bibitem[Sizemore et~al.(2017{\natexlab{a}})Sizemore, Giusti, and
	Bassett]{sizemore2016classification}
	A~Sizemore, C~Giusti, and D~S Bassett.
	\newblock Classification of weighted networks through mesoscale homological
	features.
	\newblock \emph{Journal of Complex Networks}, 5\penalty0 (2):\penalty0 245,
	2017{\natexlab{a}}.
	
	\bibitem[Sizemore et~al.(2017{\natexlab{b}})Sizemore, Giusti, Kahn, Betzel, and
	Bassett]{sizemore2016cliques}
	A~Sizemore, C~Giusti, A~E Kahn, R~F Betzel, and D~S Bassett.
	\newblock Cliques and cavities in the human connectome.
	\newblock \emph{Journal of Computational Neuroscience:
		https://doi.org/10.1007/s10827-017-0672-6}, 2017{\natexlab{b}}.
	
	\bibitem[Otter et~al.(2017)Otter, Porter, Tullmann, Grindrod, and
	Harrington]{otter2015}
	N~Otter, M~A Porter, U~Tullmann, P~Grindrod, and H~A Harrington.
	\newblock A roadmap for the computation of persistent homology.
	\newblock \emph{EPJ Data Science}, 6\penalty0 (1):\penalty0 17, 2017.
	
	\bibitem[Patania et~al.(2017)Patania, Vaccarino, and Petri]{Patania2017}
	A~Patania, F~Vaccarino, and G~Petri.
	\newblock Topological analysis of data.
	\newblock \emph{EPJ Data Science}, 6\penalty0 (1):\penalty0 7, 2017.
	
	\bibitem[Stolz et~al.(2017)Stolz, Harrington, and Porter]{Stolz2016}
	B~J Stolz, H~A Harrington, and M~A Porter.
	\newblock Persistent homology of time-dependent functional networks constructed
	from coupled time series.
	\newblock \emph{Chaos}, 27\penalty0 (4):\penalty0 047410, 2017.
	
	\bibitem[Kozlov(2007)]{kozlov2007combinatorial}
	D~Kozlov.
	\newblock \emph{Combinatorial Algebraic Topology}, volume~21.
	\newblock Springer-Verlag, 2007.
	
	\bibitem[Nanda and Sazdanovi{\'{c}}(2014)]{nanda2014simplicial}
	V~Nanda and R~Sazdanovi{\'{c}}.
	\newblock \emph{Simplicial Models and Topological Inference in Biological
		Systems}, pages 109--141.
	\newblock Springer-Verlag, 2014.
	
	\bibitem[Petri et~al.(2013)Petri, Scolamiero, Donato, and
	Vaccarino]{Petri2013a}
	G~Petri, M~Scolamiero, I~Donato, and F~Vaccarino.
	\newblock Topological strata of weighted complex networks.
	\newblock \emph{PLoS One}, 8\penalty0 (6):\penalty0 1--8, 2013.
	
	\bibitem[Dantu(1957)]{Dantu1957}
	P~Dantu.
	\newblock {Contribution l'{\'{e}}tude m{\'{e}}chanique et
		g{\'{e}}om{\'{e}}trique des milieux pulv{\'{e}}rulents}.
	\newblock In \emph{Proceedings of the fourth International Conference on Soil
		Mechanics and Foundation Engineering, London}, pages 144--148, 1957.
	
	\bibitem[Drescher and {de Josselin de Jong}(1972)]{Drescher1972}
	A~Drescher and G~{de Josselin de Jong}.
	\newblock {Photoelastic Verification Of A Mechanical Model For Flow Of A
		Granular Material}.
	\newblock \emph{Journal Of The Mechanics And Physics Of Solids}, 20\penalty0
	(5):\penalty0 337--340, 1972.
	
	\bibitem[Luding(1997)]{luding1997stress}
	S~Luding.
	\newblock Stress distribution in static two-dimensional granular model media in
	the absence of friction.
	\newblock \emph{Phys Rev E}, 55\penalty0 (4):\penalty0 4720, 1997.
	
	\bibitem[Silbert et~al.(2002)Silbert, Grest, and Landry]{silbert2002statistics}
	L~E Silbert, G~S Grest, and J~W Landry.
	\newblock Statistics of the contact network in frictional and frictionless
	granular packings.
	\newblock \emph{Phys Rev E}, 66\penalty0 (6):\penalty0 061303, 2002.
	
	\bibitem[Tordesillas(2007)]{Tordesillas:2007aa}
	A~Tordesillas.
	\newblock Force chain buckling, unjamming transitions and shear banding in
	dense granular assemblies.
	\newblock \emph{Philos Mag}, 87\penalty0 (32):\penalty0 4987--5016, 2007.
	
	\bibitem[Majmudar et~al.(2007)Majmudar, Sperl, Luding, and
	Behringer]{Majmudar:2007aa}
	T~S Majmudar, M~Sperl, S~Luding, and R~P Behringer.
	\newblock Jamming transition in granular systems.
	\newblock \emph{Phys Rev Lett}, 98\penalty0 (5):\penalty0 058001, 2007.
	
	\bibitem[Snoeijer et~al.(2004{\natexlab{a}})Snoeijer, Vlugt, van Hecke, and van
	Saarloos]{Snoeijer-2004-FNE}
	J~H Snoeijer, T~J~H Vlugt, M~van Hecke, and W~van Saarloos.
	\newblock Force network ensemble: A new approach to static granular matter.
	\newblock \emph{Phys Rev Lett}, 92\penalty0 (5):\penalty0 54302,
	2004{\natexlab{a}}.
	
	\bibitem[Snoeijer et~al.(2004{\natexlab{b}})Snoeijer, Vlugt, Ellenbroek,
	Van~Hecke, and Van~Leeuwen]{Snoeijer-2004-ETF}
	J~H Snoeijer, T~J~H Vlugt, W~G Ellenbroek, M~Van~Hecke, and J~M~J Van~Leeuwen.
	\newblock Ensemble theory for force networks in hyperstatic granular matter.
	\newblock \emph{Phys Rev E}, 70\penalty0 (6):\penalty0 61306,
	2004{\natexlab{b}}.
	
	\bibitem[Tighe et~al.(2010)Tighe, Snoeijer, Vlugt, and van Hecke]{Tighe2010a}
	B~P Tighe, J~H Snoeijer, T~J~H Vlugt, and M~van Hecke.
	\newblock The force network ensemble for granular packings.
	\newblock \emph{Soft Matter}, 6\penalty0 (13):\penalty0 2908--2917, 2010.
	
	\bibitem[Kollmer and Daniels(2017)]{Kollmer2017}
	J~E Kollmer and K~E Daniels.
	\newblock An experimental investigation of the force network ensemble.
	\newblock In \emph{Powders and Grains 2017}, volume 140, page 02024, 2017.
	
	\bibitem[Liu et~al.(2011)Liu, Nagel, Van~Saarloos, and Wyart]{Liu:2011aa}
	A~J Liu, S~R Nagel, W~Van~Saarloos, and M~Wyart.
	\newblock \emph{The jamming scenario --- An introduction and outlook}.
	\newblock Oxford University Press, 2011.
	
	\bibitem[Henkes et~al.(2010)Henkes, {van Hecke}, and {van
		Saarloos}]{Henkes:2010aa}
	S~Henkes, M~{van Hecke}, and W~{van Saarloos}.
	\newblock Critical jamming of frictional grains in the generalized isostaticity
	picture.
	\newblock \emph{EPL (Europhysics Letters)}, 90\penalty0 (1):\penalty0 14003,
	2010.
	
	\bibitem[Shundyak et~al.(2007)Shundyak, van Hecke, and van
	Saarloos]{Shundyak:2007aa}
	K~Shundyak, M~van Hecke, and W~van Saarloos.
	\newblock Force mobilization and generalized isostaticity in jammed packings of
	frictional grains.
	\newblock \emph{Phys Rev E}, 75\penalty0 (1):\penalty0 010301, 2007.
	
	\bibitem[Stumpf and Porter(2012)]{stumpf2012critical}
	M~P Stumpf and M~A Porter.
	\newblock Mathematics. critical truths about power laws.
	\newblock \emph{Science}, 335\penalty0 (6069):\penalty0 665--666, 2012.
	
	\bibitem[Tordesillas et~al.(2010{\natexlab{b}})Tordesillas, O'Sullivan, Walker,
	and Paramitha]{Tordesillas:2010a}
	A~Tordesillas, P~O'Sullivan, D~M Walker, and Paramitha.
	\newblock Evolution of functional connectivity in contact and force chain
	networks: {F}eature vectors, $k$-cores and minimal cycles.
	\newblock \emph{Comptes Rendus M{\'e}canique}, 338\penalty0 (10):\penalty0
	556--569, 2010{\natexlab{b}}.
	
	\bibitem[Duxbury et~al.(1999)Duxbury, Jacobs, Thorpe, and
	Moukarzel]{Duxbury1999}
	P~Duxbury, D~Jacobs, M~Thorpe, and C~Moukarzel.
	\newblock {Floppy modes and the free energy: {R}igidity and connectivity
		percolation on {B}ethe lattices}.
	\newblock \emph{Phys Rev E}, 59\penalty0 (2):\penalty0 2084--2092, 1999.
	
	\bibitem[Maxwell(1864)]{Maxwell:1864aa}
	J~C Maxwell.
	\newblock On the calculation of the equilibrium and stiffness of frames.
	\newblock \emph{Philosophical Magazine Series 4}, 27\penalty0 (182):\penalty0
	294--299, 1864.
	
	\bibitem[Laman(1970)]{Laman:1970aa}
	G~Laman.
	\newblock On graphs and rigidity of plane skeletal structures.
	\newblock \emph{J Eng Math}, 4\penalty0 (4):\penalty0 331--340, 1970.
	
	\bibitem[Asimow and Roth(1978)]{Asimow:1978aa}
	L~Asimow and B~Roth.
	\newblock The rigidity of graphs.
	\newblock \emph{Trans Am Math Soc}, 245:\penalty0 279--289, 1978.
	
	\bibitem[Crapo(1979)]{Crapo:1979aa}
	H~Crapo.
	\newblock Structural rigidity.
	\newblock \emph{Structural Topology}, 1:\penalty0 26--45, 1979.
	
	\bibitem[Guyon et~al.(1990)Guyon, Roux, Hansen, Bideau, Troadec, and
	Crapo]{Guyon:1990aa}
	E~Guyon, S~Roux, A~Hansen, D~Bideau, J-P Troadec, and H~Crapo.
	\newblock Non-local and non-linear problems in the mechanics of disordered
	systems: {A}pplication to granular media and rigidity problems.
	\newblock \emph{Rep Prog Phys}, 53\penalty0 (4):\penalty0 373, 1990.
	
	\bibitem[Moukarzel(1998)]{Moukarzel:1998aa}
	C~F Moukarzel.
	\newblock Isostatic phase transition and instability in stiff granular
	materials.
	\newblock \emph{Phys Rev Lett}, 81\penalty0 (8):\penalty0 1634--1637, 1998.
	
	\bibitem[Tordesillas et~al.(2011)Tordesillas, Lin, Zhang, Behringer, and
	Shi]{Tordesillas:2011aa}
	A~Tordesillas, Q~Lin, J~Zhang, R~P Behringer, and J~Shi.
	\newblock Structural stability and jamming of self-organized cluster
	conformations in dense granular materials.
	\newblock \emph{J Mech Phys Solids}, 59\penalty0 (2):\penalty0 265--296, 2011.
	
	\bibitem[Tordesillas et~al.(2012{\natexlab{b}})Tordesillas, Pucilowski, Walker,
	Peters, and Hopkins]{Tordesillas:2010b}
	A~Tordesillas, S~Pucilowski, D~M Walker, J~Peters, and M~Hopkins.
	\newblock A complex network analysis of granular fabric evolution in
	three-dimensions.
	\newblock \emph{Dynam Cont Dis Ser B}, 19\penalty0 (4--5):\penalty0 417--495,
	2012{\natexlab{b}}.
	
	\bibitem[Walker et~al.(2014{\natexlab{c}})Walker, Tordesillas, Ren, Dijksman,
	and Behringer]{Walker:2014c}
	D~M Walker, A~Tordesillas, J~Ren, J~A Dijksman, and R~P Behringer.
	\newblock Uncovering temporal transitions and self-organization during slow
	aging of dense granular media in the absence of shear bands.
	\newblock \emph{EPL (Europhysics Letters)}, 107\penalty0 (1):\penalty0 18005,
	2014{\natexlab{c}}.
	
	\bibitem[Jeng and Schwarz(2008)]{Jeng2008}
	M~Jeng and J~M Schwarz.
	\newblock {On the Study of Jamming Percolation}.
	\newblock \emph{J Stat Phys}, 131\penalty0 (4):\penalty0 575--595, 2008.
	
	\bibitem[Jeng and Schwarz(2010)]{Jeng2010}
	M~Jeng and J~M Schwarz.
	\newblock Force-balance percolation.
	\newblock \emph{Phys Rev E}, 81\penalty0 (1):\penalty0 011134, 2010.
	
	\bibitem[Cao and Schwarz(2012)]{Cao2012}
	L~Cao and J~M Schwarz.
	\newblock {Correlated percolation and tricriticality}.
	\newblock \emph{Phys Rev E}, 86\penalty0 (6):\penalty0 061131, 2012.
	
	\bibitem[Lopez et~al.(2013)Lopez, Cao, and Schwarz]{Lopez:2013aa}
	J~H Lopez, L~Cao, and J~M Schwarz.
	\newblock Jamming graphs: {A} local approach to global mechanical rigidity.
	\newblock \emph{Phys Rev E}, 88\penalty0 (6):\penalty0 062130, 2013.
	
	\bibitem[Heroy et~al.(2017)Heroy, Taylor, Shi, Forest, and Mucha]{heroy2017}
	S~Heroy, D~Taylor, F~B Shi, M~G Forest, and P~J Mucha.
	\newblock Rigid graph compression: {M}otif-based rigidity analysis for
	disordered fiber networks.
	\newblock \emph{arXiv}, arXiv:1711.05790 [cond-mat.dis-nn], 2017.
	
	\bibitem[Oda and Kazama(1998)]{Oda:1998a}
	M~Oda and H~Kazama.
	\newblock Microstructure of shear bands and its relation to the mechanisms of
	dilatancy and failure of dense granular soils.
	\newblock \emph{G{\'e}otechnique}, 48\penalty0 (4):\penalty0 465--481, 1998.
	
	\bibitem[Tordesillas et~al.(2009)Tordesillas, Zhang, and
	Behringer]{Tordesillas:2009b}
	A~Tordesillas, J~Zhang, and R~Behringer.
	\newblock Buckling force chains in dense granular assemblies: {P}hysical and
	numerical experiments.
	\newblock \emph{Geomech Geoeng}, 4\penalty0 (1):\penalty0 3--16, 2009.
	
	\bibitem[Bagi(2007)]{Bagi:2007a}
	K~Bagi.
	\newblock On the concept of jammed configurations from a structural mechanics
	perspective.
	\newblock \emph{Granular Matter}, 9\penalty0 (1):\penalty0 109--134, 2007.
	
	\bibitem[Tordesillas and Muthuswamy(2009)]{Tordesillas:2009aa}
	A~Tordesillas and M~Muthuswamy.
	\newblock On the modeling of confined buckling of force chains.
	\newblock \emph{J Mech Phys Solids}, 57\penalty0 (4):\penalty0 706--727, 2009.
	
	\bibitem[Cates et~al.(1998)Cates, Wittmer, Bouchaud, and Claudin]{Cates:1998aa}
	M~E Cates, J~P Wittmer, J-P Bouchaud, and P~Claudin.
	\newblock Jamming, force chains, and fragile matter.
	\newblock \emph{Phys Rev Lett}, 81\penalty0 (9):\penalty0 1841--1844, 1998.
	
	\bibitem[Muthuswamy and Tordesillas(2006)]{Muthuswamy2006a}
	M~Muthuswamy and A~Tordesillas.
	\newblock How do interparticle contact friction, packing density and degree of
	polydispersity affect force propagation in particulate assemblies?
	\newblock \emph{Journal of Statistical Mechanics: Theory and Experiment},
	2006\penalty0 (09):\penalty0 P09003, 2006.
	
	\bibitem[Kob and Barrat(1997)]{kob1997aging}
	W~Kob and J~L Barrat.
	\newblock Aging effects in a {L}ennard-{J}ones glass.
	\newblock \emph{Phys Rev Lett}, 78\penalty0 (24):\penalty0 4581--4584, 1997.
	
	\bibitem[Kabla and Debregeas(2004)]{kabla2004contact}
	A~Kabla and G~Debregeas.
	\newblock Contact dynamics in a gently vibrated granular pile.
	\newblock \emph{Phys Rev Lett}, 92\penalty0 (3):\penalty0 35501, 2004.
	
	\bibitem[Steinhardt et~al.(1983)Steinhardt, Nelson, and
	Ronchetti]{Steinhardt:1983a}
	P~J Steinhardt, D~R Nelson, and M~Ronchetti.
	\newblock Bond-orientational order in liquids and glasses.
	\newblock \emph{Phys Rev B}, 28\penalty0 (2):\penalty0 784--805, 1983.
	
	\bibitem[Ar\'evalo et~al.(2013)Ar\'evalo, Pugnaloni, Zuriguel, and
	Maza]{Arevalo:2013aa}
	R~Ar\'evalo, L~A Pugnaloni, I~Zuriguel, and D~Maza.
	\newblock Contact network topology in tapped granular media.
	\newblock \emph{Phys Rev E}, 87\penalty0 (2):\penalty0 022203, 2013.
	
	\bibitem[Nowak et~al.(1998)Nowak, Knight, Ben-Naim, Jaeger, and
	Nagel]{Nowak:1998a}
	E~R Nowak, J~B Knight, E~Ben-Naim, H~M Jaeger, and S~R Nagel.
	\newblock Density fluctuations in vibrated granular materials.
	\newblock \emph{Phys Rev E}, 57\penalty0 (2):\penalty0 1971--1982, 1998.
	
	\bibitem[Pugnaloni et~al.(2010)Pugnaloni, S\'anchez, Gago, Damas, Zuriguel, and
	Maza]{Pugnaloni:2010aa}
	L~A Pugnaloni, I~S\'anchez, P~A Gago, J~Damas, I~Zuriguel, and D~Maza.
	\newblock Towards a relevant set of state variables to describe static granular
	packings.
	\newblock \emph{Phys Rev E}, 82\penalty0 (5):\penalty0 050301, 2010.
	
	\bibitem[Pugnaloni et~al.(2008)Pugnaloni, Mizrahi, Carlevaro, and
	Vericat]{Pugnaloni:2008aa}
	L~A Pugnaloni, M~Mizrahi, C~M Carlevaro, and F~Vericat.
	\newblock Nonmonotonic reversible branch in four model granular beds subjected
	to vertical vibration.
	\newblock \emph{Phys Rev E}, 78\penalty0 (5):\penalty0 051305, 2008.
	
	\bibitem[Gago et~al.(2009)Gago, Bueno, and Pugnaloni]{Gago:2009aa}
	P~A Gago, N~E Bueno, and L~A Pugnaloni.
	\newblock High intensity tapping regime in a frustrated lattice gas model of
	granular compaction.
	\newblock \emph{Granular Matter}, 11\penalty0 (6):\penalty0 365--369, 2009.
	
	\bibitem[Carlevaro and Pugnaloni(2011)]{Carlevaro:2011a}
	C~M Carlevaro and L~A Pugnaloni.
	\newblock Steady state of tapped granular polygons.
	\newblock \emph{Journal of Statistical Mechanics: Theory and Experiment},
	2011\penalty0 (01):\penalty0 P01007, 2011.
	
	\bibitem[Ar{\'e}valo et~al.(2013)Ar{\'e}valo, Pugnaloni, Maza, and
	Zuriguel]{Arevalo:2013ba}
	R~Ar{\'e}valo, L~A Pugnaloni, D~Maza, and I~Zuriguel.
	\newblock Tapped granular packings described as complex networks.
	\newblock \emph{Philosophical Magazine}, 93\penalty0 (31-33):\penalty0
	4078--4089, 2013.
	
	\bibitem[Itzkovitz and Alon(2005)]{itzkovitz2005subgraphs}
	S~Itzkovitz and U~Alon.
	\newblock Subgraphs and network motifs in geometric networks.
	\newblock \emph{Phys Rev E}, 71\penalty0 (2):\penalty0 026117, 2005.
	
	\bibitem[Shoval and Alon(2010)]{shoval2010snap}
	O~Shoval and U~Alon.
	\newblock Snap{S}hot: {N}etwork motifs.
	\newblock \emph{Cell}, 143\penalty0 (2):\penalty0 326--326.e1, 2010.
	
	\bibitem[Brodu et~al.(2015)Brodu, Dijksman, and Behringer]{Brodu2015}
	N~Brodu, J~A Dijksman, and R~P Behringer.
	\newblock Spanning the scales of granular materials through microscopic force
	imaging.
	\newblock \emph{Nat Commun}, 6:\penalty0 6361, 2015.
	
	\bibitem[Dijksman et~al.(2017)Dijksman, Brodu, and Behringer]{Dijksman2017}
	J~A Dijksman, N~Brodu, and R~P Behringer.
	\newblock Refractive index matched scanning and detection of soft particles.
	\newblock \emph{Rev Sci Instrum}, 88\penalty0 (5):\penalty0 051807, 2017.
	
	\bibitem[Sepiani and Ghazavi(2009)]{sepiani2009thermo}
	H~A Sepiani and A~Ghazavi.
	\newblock A thermo-micro-mechanical modeling for smart shape memory alloy woven
	composite under in-plane biaxial deformation.
	\newblock \emph{International Journal of Mechanics and Materials in Design},
	5\penalty0 (2):\penalty0 111, 2009.
	
	\bibitem[Tighe and Vlugt(2011)]{Tighe:2011a}
	B~P Tighe and T~J~H Vlugt.
	\newblock Stress fluctuations in granular force networks.
	\newblock \emph{Journal of Statistical Mechanics: Theory and Experiment},
	2011\penalty0 (04):\penalty0 P04002, 2011.
	
	\bibitem[Daniels et~al.(2017)Daniels, Kollmer, and Puckett]{Daniels2017}
	K~E Daniels, J~E Kollmer, and J~G Puckett.
	\newblock Photoelastic force measurements in granular materials.
	\newblock \emph{Rev Sci Instrum}, 88\penalty0 (5):\penalty0 051808, 2017.
	
	\bibitem[Hurley et~al.(2016)Hurley, Hall, Andrade, and Wright]{Hurley:2016a}
	R~C Hurley, S~A Hall, J~E Andrade, and J~Wright.
	\newblock Quantifying interparticle forces and heterogeneity in 3d granular
	materials.
	\newblock \emph{Phys Rev Lett}, 117\penalty0 (9):\penalty0 098005, 2016.
	
	\bibitem[Mukhopadhyay and Peixinho(2011)]{Mukhopadhyay2011}
	S~Mukhopadhyay and J~Peixinho.
	\newblock Packings of deformable spheres.
	\newblock \emph{Phys Rev E}, 84\penalty0 (1):\penalty0 011302, 2011.
	
	\bibitem[Saadatfar et~al.(2012)Saadatfar, Sheppard, Senden, and
	Kabla]{Saadatfar2012}
	M~Saadatfar, A~P Sheppard, T~J Senden, and A~J Kabla.
	\newblock Mapping forces in a {3D} elastic assembly of grains.
	\newblock \emph{J Mech Phys Solids}, 60\penalty0 (1):\penalty0 55--66, 2012.
	
	\bibitem[P{\"{o}}schel and Schwager(2005)]{Poschel2005}
	T~P{\"{o}}schel and T~Schwager.
	\newblock \emph{Computational Granular Dynamics: {M}odels and Algorithms}.
	\newblock Springer-Verlag, 2005.
	
	\bibitem[Weis and Schr\"oter(2017)]{weis:17}
	S~Weis and M~Schr\"oter.
	\newblock Analyzing {X}-ray tomographies of granular packings.
	\newblock \emph{Rev Sci Instrum}, 88\penalty0 (5):\penalty0 051809, 2017.
	
	\bibitem[Tordesillas and Muthuswamy(2008)]{Tordesillas:2008a}
	A~Tordesillas and M~Muthuswamy.
	\newblock A thermomicromechanical approach to multiscale continuum modeling of
	dense granular materials.
	\newblock \emph{Acta Geotechnica}, 3\penalty0 (3):\penalty0 225--240, 2008.
	
	\bibitem[Huang and Daniels(2016)]{Huang:2016a}
	Y~Huang and K~E Daniels.
	\newblock Friction and pressure-dependence of force chain communities in
	granular materials.
	\newblock \emph{Granular Matter}, 18\penalty0 (4):\penalty0 85, 2016.
	
	\bibitem[Navakas et~al.(2014)Navakas, D{\v z}iugys, and Peters]{Navakas:2014aa}
	R~Navakas, A~D{\v z}iugys, and B~Peters.
	\newblock A community-detection based approach to identification of
	inhomogeneities in granular matter.
	\newblock \emph{Physica A}, 407:\penalty0 312--331, 2014.
	
	\bibitem[Walker et~al.(2011)Walker, Tordesillas, Einav, and
	Small]{Walker:2011a}
	D~M Walker, A~Tordesillas, I~Einav, and M~Small.
	\newblock Complex networks in confined comminution.
	\newblock \emph{Phys Rev E}, 84\penalty0 (2):\penalty0 021301, 2011.
	
	\bibitem[Radjai et~al.(1999)Radjai, Roux, and Moreau]{Radjai:1999aa}
	F~Radjai, S~Roux, and J~J Moreau.
	\newblock Contact forces in a granular packing.
	\newblock \emph{Chaos}, 9\penalty0 (4):\penalty0 544--550, 1999.
	
	\bibitem[Pe{\~n}a et~al.(2009)Pe{\~n}a, Herrmann, and Lind]{Pena:2009aa}
	A~A Pe{\~n}a, H~J Herrmann, and P~G Lind.
	\newblock Force chains in sheared granular media of irregular particles.
	\newblock \emph{AIP Conference Proceedings}, 1145\penalty0 (1):\penalty0
	321--324, 2009.
	
	\bibitem[Ostojic et~al.(2007)Ostojic, Vlugt, and Nienhuis]{Ostojic:2007a}
	S~Ostojic, T~J~H Vlugt, and B~Nienhuis.
	\newblock Universal anisotropy in force networks under shear.
	\newblock \emph{Phys Rev E}, 75\penalty0 (3):\penalty0 030301, 2007.
	
	\bibitem[Kondic et~al.(2017)Kondic, Kram{\'a}r, Koval{\v c}inov{\'a}, and
	Mischaikow]{Kondic:2017a}
	L~Kondic, M~Kram{\'a}r, L~Koval{\v c}inov{\'a}, and K~Mischaikow.
	\newblock Evolution of force networks in dense granular matter close to
	jamming.
	\newblock \emph{EPJ Web Conf}, 140:\penalty0 15014, 2017.
	
	\bibitem[Tadanaga et~al.(2017)Tadanaga, Clark, Majmudar, and
	Kondic]{kondic-impact2017}
	T~Tadanaga, A~H Clark, T~Majmudar, and L~Kondic.
	\newblock Granular response to impact: Topology of the force networks.
	\newblock \emph{arXiv}, arXiv:1709.06957 [cond-mat.soft], 2017.
	
	\bibitem[Lim and Behringer(2017)]{lim-impact2017}
	M~X Lim and R~P Behringer.
	\newblock Topology of force networks in granular media under impact.
	\newblock \emph{arXiv}, arXiv:1709.01884 [cond-mat.soft], 2017.
	
	\bibitem[Tordesillas et~al.(2015{\natexlab{a}})Tordesillas, Tobin, Cil,
	Alshibli, and Behringer]{Tordesillas:2015a}
	A~Tordesillas, S~T Tobin, M~Cil, K~Alshibli, and R~P Behringer.
	\newblock Network flow model of force transmission in unbonded and bonded
	granular media.
	\newblock \emph{Phys Rev E}, 91\penalty0 (6):\penalty0 062204,
	2015{\natexlab{a}}.
	
	\bibitem[Tordesillas et~al.(2015{\natexlab{b}})Tordesillas, Pucilowski, Tobin,
	Kuhn, And{\`o}, Viggiani, Druckrey, and Alshibli]{Tordesillas:2015b}
	A~Tordesillas, S~Pucilowski, S~Tobin, M~R Kuhn, E~And{\`o}, G~Viggiani,
	A~Druckrey, and K~Alshibli.
	\newblock Shear bands as bottlenecks in force transmission.
	\newblock \emph{EPL (Europhysics Letters)}, 110\penalty0 (5):\penalty0 58005,
	2015{\natexlab{b}}.
	
	\bibitem[Tordesillas et~al.(2013{\natexlab{b}})Tordesillas, Cramer, and
	Walker]{Tordesillas:2013b}
	A~Tordesillas, A~Cramer, and D~M Walker.
	\newblock Minimum cut and shear bands.
	\newblock \emph{AIP Conference Proceedings}, 1542\penalty0 (1):\penalty0
	507--510, 2013{\natexlab{b}}.
	
	\bibitem[Lin and Tordesillas(2013)]{Lin:2013a}
	Q~Lin and A~Tordesillas.
	\newblock Constrained optimisation in granular network flows: {G}ames with a
	loaded dice.
	\newblock \emph{AIP Conference Proceedings}, 1542\penalty0 (1):\penalty0
	547--550, 2013.
	
	\bibitem[Lin and Tordesillas(2014)]{Lin:2014a}
	Q~Lin and A~Tordesillas.
	\newblock Towards an optimization theory for deforming dense granular
	materials: {M}inimum cost maximum flow solutions.
	\newblock \emph{J Ind Manag Optim}, 10\penalty0 (1):\penalty0 337--362, 2014.
	
	\bibitem[Walker and Tordesillas(2013)]{Walker:2013b}
	D~M Walker and A~Tordesillas.
	\newblock Understanding multi-scale structural evolution in granular systems
	through gmems.
	\newblock \emph{AIP Conference Proceedings}, 1542\penalty0 (1):\penalty0
	145--148, 2013.
	
	\bibitem[Zhang and Small(2006)]{Zhang:2006aa}
	J~Zhang and M~Small.
	\newblock Complex network from pseudoperiodic time series: {T}opology versus
	dynamics.
	\newblock \emph{Phys Rev Lett}, 96\penalty0 (23):\penalty0 238701, 2006.
	
	\bibitem[Yang and Yang(2008)]{Yang:2008a}
	Y~Yang and H~Yang.
	\newblock Complex network-based time series analysis.
	\newblock \emph{Physica A}, 387\penalty0 (5--6):\penalty0 1381--1386, 2008.
	
	\bibitem[Lacasa et~al.(2008)Lacasa, Luque, Ballesteros, Luque, and
	Nu{\~n}o]{Lacasa:2008a}
	L~Lacasa, B~Luque, F~Ballesteros, J~Luque, and J~C Nu{\~n}o.
	\newblock From time series to complex networks: The visibility graph.
	\newblock \emph{Proc Natl Acad Sci}, 105\penalty0 (13):\penalty0 4972--4975,
	2008.
	
	\bibitem[Gao and Jin(2009)]{Gao:2009a}
	Z~Gao and N~Jin.
	\newblock Complex network from time series based on phase space reconstruction.
	\newblock \emph{Chaos}, 19\penalty0 (3):\penalty0 033137, 2009.
	
	\bibitem[Marwan et~al.(2009)Marwan, Donges, Zou, Donner, and
	Kurths]{Marwan:2009a}
	N~Marwan, J~F Donges, Y~Zou, R~V Donner, and J~Kurths.
	\newblock Complex network approach for recurrence analysis of time series.
	\newblock \emph{Phys Lett A}, 373\penalty0 (46):\penalty0 4246--4254, 2009.
	
	\bibitem[MacKay(2003)]{MacKay:2003}
	D~J~C MacKay.
	\newblock \emph{Information Theory, Inference and Learning Algorithms}.
	\newblock Cambridge University Press, 2003.
	
	\bibitem[Rechenmacher(2006)]{Rechenmacher:2006a}
	A~L Rechenmacher.
	\newblock Grain-scale processes governing shear band initiation and evolution
	in sands.
	\newblock \emph{J Mech Phys Solids}, 54\penalty0 (1):\penalty0 22--45, 2006.
	
	\bibitem[Rechenmacher et~al.(2010)Rechenmacher, Abedi, and
	Chupin]{Rechenmacher:2010a}
	A~L Rechenmacher, S~Abedi, and O~Chupin.
	\newblock Evolution of force chains in shear bands in sands.
	\newblock \emph{G{\'e}otechnique}, 60\penalty0 (5):\penalty0 343--351, 2010.
	
	\bibitem[And{\`o} et~al.(2012)And{\`o}, Hall, Viggiani, Desrues, and
	B{\'e}suelle]{Ando:2012a}
	E~And{\`o}, S~A Hall, G~Viggiani, J~Desrues, and P~B{\'e}suelle.
	\newblock Grain-scale experimental investigation of localised deformation in
	sand: {A} discrete particle tracking approach.
	\newblock \emph{Acta Geotechnica}, 7\penalty0 (1):\penalty0 1--13, 2012.
	
	\bibitem[Amon et~al.(2017)Amon, Born, Daniels, Dijksman, Huang, Parker,
	Schr{\"o}ter, Stannarius, and Wierschem]{amon_focus_2017}
	A~Amon, P~Born, K~E Daniels, J~A Dijksman, K~Huang, D~Parker, M~Schr{\"o}ter,
	R~Stannarius, and A~Wierschem.
	\newblock Preface: {F}ocus on imaging methods in granular physics.
	\newblock \emph{Rev Sci Instrum}, 88\penalty0 (5):\penalty0 051701, 2017.
	
	\bibitem[Stannarius(2017)]{stannarius:17}
	R~Stannarius.
	\newblock Magnetic resonance imaging of granular materials.
	\newblock \emph{Rev Sci Instrum}, 88\penalty0 (5):\penalty0 051806, 2017.
	
	\bibitem[Porter et~al.(2015)Porter, Kevrekidis, and Daraio]{PT2015}
	M~A Porter, P~G Kevrekidis, and C~Daraio.
	\newblock Granular crystals: {N}onlinear dynamics meets materials engineering.
	\newblock \emph{Physics Today}, 68\penalty0 (11):\penalty0 44, 2015.
	
	\bibitem[Cundall and Strack(1979)]{Cundall1979}
	P~A Cundall and O~D~L Strack.
	\newblock {Discrete Numerical-model For Granular Assemblies}.
	\newblock \emph{Geotechnique}, 29\penalty0 (1):\penalty0 47--65, 1979.
	
	\bibitem[Papanikolaou et~al.(2013)Papanikolaou, O'Hern, and
	Shattuck]{Papanikolaou2013}
	S~Papanikolaou, C~S O'Hern, and M~D Shattuck.
	\newblock Isostaticity at frictional jamming.
	\newblock \emph{Phys Rev Lett}, 110\penalty0 (19):\penalty0 198002, 2013.
	
	\bibitem[Somfai et~al.(2005)Somfai, Roux, Snoeijer, van Hecke, and van
	Saarloos]{Somfai2005}
	E~Somfai, J~N Roux, J~H Snoeijer, M~van Hecke, and W~van Saarloos.
	\newblock Elastic wave propagation in confined granular systems.
	\newblock \emph{Phys Rev E}, 72\penalty0 (2):\penalty0 21301, 2005.
	
	\bibitem[Rosvall et~al.(2014)Rosvall, Esquivel, Lancichinetti, West, and
	Lambiotte]{Rosvall2014NatComm}
	M~Rosvall, A~V Esquivel, A~Lancichinetti, J~D West, and R~Lambiotte.
	\newblock {Memory in network flows and its effects on spreading dynamics and
		community detection}.
	\newblock \emph{Nat Commun}, 5:\penalty0 4630, 2014.
	
	\bibitem[Holme(2015)]{Holme2015EurPhysJB}
	P~Holme.
	\newblock Modern temporal network theory: A colloquium.
	\newblock \emph{Eur Phys J B}, 88\penalty0 (9):\penalty0 234, 2015.
	
	\bibitem[Butts(2009)]{butts2009}
	C~T Butts.
	\newblock Revisiting the foundations of network analysis.
	\newblock \emph{Science}, 325\penalty0 (5939):\penalty0 414--6, 2009.
	
	\bibitem[Gravish et~al.(2012)Gravish, Franklin, Hu, and
	Goldman]{gravish2012entangled}
	N~Gravish, S~V Franklin, D~L Hu, and D~I Goldman.
	\newblock Entangled granular media.
	\newblock \emph{Phys Rev Lett}, 108\penalty0 (20):\penalty0 208001, 2012.
	
	\bibitem[Murphy et~al.(2016)Murphy, Reiser, Choksy, Singer, and
	Jaeger]{murphy2015freestanding}
	K~A Murphy, N~Reiser, D~Choksy, C~E Singer, and H~M Jaeger.
	\newblock Freestanding loadbearing structures with {Z}-shaped particles.
	\newblock \emph{Granular Matter}, 18\penalty0 (2):\penalty0 26, 2016.
	
	\bibitem[Hidalgo et~al.(2009)Hidalgo, Zuriguel, Maza, and
	Pagonabarraga]{Hidalgo:2009aa}
	R~Cruz Hidalgo, I~Zuriguel, D~Maza, and I~Pagonabarraga.
	\newblock Role of particle shape on the stress propagation in granular
	packings.
	\newblock \emph{Phys. Rev. Lett.}, 103\penalty0 (11):\penalty0 118001, 2009.
	
	\bibitem[Trepanier and Franklin(2010)]{trepanier2010column}
	M~Trepanier and S~V Franklin.
	\newblock Column collapse of granular rods.
	\newblock \emph{Phys Rev E}, 82\penalty0 (1):\penalty0 011308, 2010.
	
	\bibitem[Schreck et~al.(2010)Schreck, Xu, and O{'}Hern]{Schreck:2010a}
	C~F Schreck, N~Xu, and C~S O{'}Hern.
	\newblock A comparison of jamming behavior in systems composed of dimer- and
	ellipse-shaped particles.
	\newblock \emph{Soft Matter}, 6\penalty0 (13):\penalty0 2960--2969, 2010.
	
	\bibitem[Athanassiadis et~al.(2014)Athanassiadis, Miskin, Kaplan, Rodenberg,
	Lee, Merritt, Brown, Amend, Lipson, and Jaeger]{Athanassiadis:2014a}
	A~G. Athanassiadis, M~Z. Miskin, P~Kaplan, N~Rodenberg, S~H Lee, J~Merritt,
	E~Brown, J~Amend, H~Lipson, and H~M Jaeger.
	\newblock Particle shape effects on the stress response of granular packings.
	\newblock \emph{Soft Matter}, 10\penalty0 (1):\penalty0 48--59, 2014.
	
	\bibitem[Harrington and Durian(2017)]{durian2017}
	M~Harrington and D~J Durian.
	\newblock Anisotropic particles strengthen granular pillars under compression.
	\newblock \emph{arXiv}, arXiv:1709.09511 [cond-mat.soft], 2017.
	
	\bibitem[Az\'ema et~al.(2013)Az\'ema, Radjai, and Dubois]{Azema:2013aa}
	E~Az\'ema, F~Radjai, and F~Dubois.
	\newblock Packings of irregular polyhedral particles: {S}trength, structure,
	and effects of angularity.
	\newblock \emph{Phys Rev E}, 87\penalty0 (6):\penalty0 062203, 2013.
	
	\bibitem[G\'{o}mez et~al.(2009)G\'{o}mez, Jensen, and Arenas]{Gomez:2009a}
	S~G\'{o}mez, P~Jensen, and A~Arenas.
	\newblock Analysis of community structure in networks of correlated data.
	\newblock \emph{Phys Rev E}, 80\penalty0 (1):\penalty0 016114, 2009.
	
	\bibitem[Traag and Bruggeman(2009)]{Traag:2009a}
	V~A Traag and J~Bruggeman.
	\newblock Community detection in networks with positive and negative links.
	\newblock \emph{Phys Rev E}, 80\penalty0 (3):\penalty0 036115, 2009.
	
	\bibitem[Zhang et~al.(2016)Zhang, Bassett, and Winkelstein]{Zhang:2015}
	S~Zhang, D~S Bassett, and B~A Winkelstein.
	\newblock Stretch-induced network reconfiguration of collagen fibers in the
	human facet capsular ligament.
	\newblock \emph{J R Soc Interface}, 13\penalty0 (114):\penalty0 20150883, 2016.
	
	\bibitem[Puckett and Daniels(2013)]{Puckett2013}
	J~G Puckett and K~E Daniels.
	\newblock Equilibrating temperaturelike variables in jammed granular
	subsystems.
	\newblock \emph{Phys Rev Lett}, 110\penalty0 (5):\penalty0 058001, 2013.
	
	\bibitem[Shaebani et~al.(2012)Shaebani, Madadi, Luding, and
	Wolf]{shaebani2012influence}
	M~R Shaebani, M~Madadi, S~Luding, and D~E Wolf.
	\newblock Influence of polydispersity on micromechanics of granular materials.
	\newblock \emph{Phys Rev E}, 85\penalty0 (1):\penalty0 011301, 2012.
	
	\bibitem[Kumar et~al.(2016)Kumar, Magnanimo, Ramaioli, and Luding]{Kumar:2016a}
	N~Kumar, V~Magnanimo, M~Ramaioli, and S~Luding.
	\newblock Tuning the bulk properties of bidisperse granular mixtures by small
	amount of fines.
	\newblock \emph{Powder Technol}, 293:\penalty0 94--112, 2016.
	
	\bibitem[Slanina(2017)]{slanina2017}
	F~Slanina.
	\newblock Localization in random bipartite graphs: Numerical and empirical
	study.
	\newblock \emph{Phys Rev E}, 95\penalty0 (5):\penalty0 052149, 2017.
	
	\bibitem[Harary(1972)]{harary1972}
	F~Harary.
	\newblock \emph{Graph Theory}.
	\newblock Addison-Wesley, 1972.
	
	\bibitem[Shi et~al.(2013)Shi, Wang, Forest, and Mucha]{Shi:2013aa}
	F~Shi, S~Wang, M~G Forest, and P~J Mucha.
	\newblock Percolation-induced exponential scaling in the large current tails of
	random resistor networks.
	\newblock \emph{Multiscale Model Sim}, 11\penalty0 (4):\penalty0 1298--1310,
	2013.
	
	\bibitem[Shi et~al.(2014)Shi, Wang, Forest, Mucha, and Zhou]{Shi:2014aa}
	F~Shi, S~Wang, M~G Forest, P~J Mucha, and R~Zhou.
	\newblock Network-based assessments of percolation-induced current
	distributions in sheared rod macromolecular dispersions.
	\newblock \emph{Multiscale Model Sim}, 12\penalty0 (1):\penalty0 249--264,
	2014.
	
	\bibitem[Abhilash et~al.(2014)Abhilash, Baker, Trappmann, Chen, and
	Shenoy]{abhilash2014remodeling}
	A~S Abhilash, B~M Baker, B~Trappmann, C~S Chen, and V~B Shenoy.
	\newblock Remodeling of fibrous extracellular matrices by contractile cells:
	{P}redictions from discrete fiber network simulations.
	\newblock \emph{Biophys J}, 107\penalty0 (8):\penalty0 1829--1840, 2014.
	
	\bibitem[Purohit et~al.(2011)Purohit, Litvinov, Brown, Discher, and
	Weisel]{purohit2011protein}
	P~K Purohit, R~I Litvinov, A~E Brown, D~E Discher, and J~W Weisel.
	\newblock Protein unfolding accounts for the unusual mechanical behavior of
	fibrin networks.
	\newblock \emph{Acta Biomater}, 7\penalty0 (6):\penalty0 2374--2783, 2011.
	
	\bibitem[Bullmore and Sporns(2009)]{Bullmore2009}
	E~Bullmore and O~Sporns.
	\newblock Complex brain networks: {G}raph theoretical analysis of structural
	and functional systems.
	\newblock \emph{Nat Rev Neurosci}, 10\penalty0 (3):\penalty0 186--198, 2009.
	
	\bibitem[Blunt(2001)]{Blunt:2001a}
	M~J Blunt.
	\newblock Flow in porous media --- pore-network models and multiphase flow.
	\newblock \emph{Curr Opin Colloid Interface Sci}, 6\penalty0 (3):\penalty0
	197--207, 2001.
	
	\bibitem[Al-Raoush et~al.(2003)Al-Raoush, Thompson, and
	Willson]{Al-Raoush:2003a}
	R~Al-Raoush, K~Thompson, and C~S Willson.
	\newblock Comparison of network generation techniques for unconsolidated porous
	media.
	\newblock \emph{Soil Sci Soc Am J}, 67\penalty0 (6):\penalty0 1687--1700, 2003.
	
	\bibitem[Vo et~al.(2013)Vo, Walker, and Tordesillas]{Vo:2013}
	K~Vo, D~M Walker, and A~Tordesillas.
	\newblock Transport pathways within percolating pore space networks of granular
	materials.
	\newblock \emph{AIP Conf Proc}, 1542\penalty0 (1):\penalty0 551--554, 2013.
	
	\bibitem[Walker et~al.(2013)Walker, Vo, and Tordesillas]{Walker:2013a}
	D~M Walker, K~Vo, and A~Tordesillas.
	\newblock On reynolds' dilatancy and shear band evolution: {A} new perspective.
	\newblock \emph{Int J Bifurc Chaos}, 23\penalty0 (09):\penalty0 1330034, 2013.
	
	\bibitem[van~der Linden et~al.(2016)van~der Linden, Narsilio, and
	Tordesillas]{Linden:2016a}
	J~H van~der Linden, G~A Narsilio, and A~Tordesillas.
	\newblock Machine learning framework for analysis of transport through complex
	networks in porous, granular media: {A} focus on permeability.
	\newblock \emph{Phys Rev E}, 94\penalty0 (2):\penalty0 022904, 2016.
	
	\bibitem[Russell et~al.(2016)Russell, Walker, and Tordesillas]{Russell:2016a}
	S~Russell, D~M Walker, and A~Tordesillas.
	\newblock A characterization of the coupled evolution of grain fabric and pore
	space using complex networks: {P}ore connectivity and optimized flows in the
	presence of shear bands.
	\newblock \emph{J Mech Phys Solids}, 88:\penalty0 227--251, 2016.
	
	\bibitem[Jimenez-Martinez and Negre(2017)]{Martinez:2017a}
	J~Jimenez-Martinez and C~F~A Negre.
	\newblock Eigenvector centrality for geometric and topological characterization
	of porous media.
	\newblock \emph{Phys Rev E}, 96\penalty0 (1):\penalty0 013310, 2017.
	
	\bibitem[Laubie et~al.(2017)Laubie, Radjai, Pellenq, and Ulm]{porous2017}
	H~Laubie, F~Radjai, R~Pellenq, and F~J Ulm.
	\newblock Stress transmission and failure in disordered porous media.
	\newblock \emph{Phys Rev Lett}, 119\penalty0 (7):\penalty0 075501, 2017.
	
	\bibitem[Newman and Clauset(2016)]{newman2016structure}
	M~E~J Newman and A~Clauset.
	\newblock Structure and inference in annotated networks.
	\newblock \emph{Nat Commun}, 7:\penalty0 11863, 2016.
	
	\bibitem[Hric et~al.(2016)Hric, Peixoto, and Fortunato]{tiago2016}
	D~Hric, T~P Peixoto, and S~Fortunato.
	\newblock Network structure, metadata, and the prediction of missing nodes and
	annotations.
	\newblock \emph{Phys Rev X}, 6\penalty0 (3):\penalty0 031038, 2016.
	
	\bibitem[Palla et~al.(2008)Palla, Farkas, Pollner, Derenyi, and
	Vicsek]{palla2008fundamental}
	G~Palla, I~J Farkas, P~Pollner, I~Derenyi, and T~Vicsek.
	\newblock Fundamental statistical features and self-similar properties of
	tagged networks.
	\newblock \emph{New J Phys}, 10\penalty0 (12):\penalty0 123026, 2008.
	
	\bibitem[Edelsbrunner and Harer(2010)]{edels2010}
	H~Edelsbrunner and J~Harer.
	\newblock \emph{Computational Topology: {A}n Introduction}.
	\newblock American Mathematical Society, 2010.
	
	\bibitem[Ramola and Chakraborty(2017)]{Ramola:2017a}
	K~Ramola and B~Chakraborty.
	\newblock Stress response of granular systems.
	\newblock \emph{J Stat Phys}, 169\penalty0 (1):\penalty0 1--17, 2017.
	
	\bibitem[Taylor-King et~al.(2017)Taylor-King, Basanta, Chapman, and
	Porter]{jkt2017}
	J~P Taylor-King, D~Basanta, S~J Chapman, and M~A Porter.
	\newblock Mean-field approach to evolving spatial networks, with an application
	to osteocyte network formation.
	\newblock \emph{Phys Rev E}, 96\penalty0 (1):\penalty0 012301, 2017.
	
	\bibitem[Beguerisse-Diaz et~al.(2014)Beguerisse-Diaz, Garduno-Hernandez,
	Vangelov, Yaliraki, and Barahona]{beguerisse2014interest}
	M~Beguerisse-Diaz, G~Garduno-Hernandez, B~Vangelov, S~N Yaliraki, and
	M~Barahona.
	\newblock Interest communities and flow roles in directed networks: {T}he
	{T}witter network of the {UK} riots.
	\newblock \emph{J R Soc Interface}, 11\penalty0 (101):\penalty0 20140940, 2014.
	
	\bibitem[Bassett et~al.(2010)Bassett, Greenfield, Meyer-Lindenberg, Weinberger,
	Moore, and Bullmore]{Bassett2010}
	D~S Bassett, D~L Greenfield, A~Meyer-Lindenberg, D~R Weinberger, S~Moore, and
	E~Bullmore.
	\newblock Efficient physical embedding of topologically complex information
	processing networks in brains and computer circuits.
	\newblock \emph{PLoS Comput Biol}, 6\penalty0 (4):\penalty0 e1000748, 2010.
	
	\bibitem[Modes et~al.(2016)Modes, Magnasco, and Katifori]{Modes:2016a}
	C~D Modes, M~O Magnasco, and E~Katifori.
	\newblock Extracting hidden hierarchies in {3D} distribution networks.
	\newblock \emph{Phys Rev X}, 6\penalty0 (3):\penalty0 031009, 2016.
	
	\bibitem[Tighe et~al.(2008)Tighe, van Eerd, and Vlugt]{Tighe2008}
	B~P Tighe, A~R~T van Eerd, and T~J~H Vlugt.
	\newblock Entropy maximization in the force network ensemble for granular
	solids.
	\newblock \emph{Phys Rev Lett}, 100\penalty0 (23):\penalty0 238001, 2008.
	
	\bibitem[Ronhovde et~al.(2012)Ronhovde, Chakrabarty, Sahu, Sahu, Kelton, Mauro,
	and Nussinov]{ronhovde2011detection}
	P~Ronhovde, S~Chakrabarty, M~Sahu, K~K Sahu, K~F Kelton, N~Mauro, and
	Z~Nussinov.
	\newblock Detection of hidden structures for arbitrary scales in complex
	physical systems.
	\newblock \emph{Sci Rep}, 2:\penalty0 329, 2012.
	
	\bibitem[Agarwala and Shenoy(2017)]{amorphous2017}
	A~Agarwala and V~B Shenoy.
	\newblock Topological insulators in amorphous systems.
	\newblock \emph{Phys Rev Lett}, 118\penalty0 (23):\penalty0 236402, 2017.
	
	\bibitem[Muller and Luding(2009)]{muller2009homogeneous}
	M~K Muller and S~Luding.
	\newblock Homogeneous cooling with repulsive and attractive long‚Äêrange
	interactions.
	\newblock \emph{AIP Conf Proc}, 1145\penalty0 (1):\penalty0 697--700, 2009.
	
	\bibitem[Mitarai and Nori(2006)]{Mitarai:2006a}
	N~Mitarai and F~Nori.
	\newblock Wet granular materials.
	\newblock \emph{Adv Phys}, 55\penalty0 (1-2):\penalty0 1--45, 2006.
	
	\bibitem[Wrobel and Sundararaghavan(2014)]{wrobel2014directed}
	M~R Wrobel and H~G Sundararaghavan.
	\newblock Directed migration in neural tissue engineering.
	\newblock \emph{Tissue Eng Part B Rev}, 20\penalty0 (2):\penalty0 93--105,
	2014.
	
	\bibitem[Huttenlocher and Poznansky(2008)]{huttenlocher2008reverse}
	A~Huttenlocher and M~C Poznansky.
	\newblock Reverse leukocyte migration can be attractive or repulsive.
	\newblock \emph{Trends in Cell Biol}, 18\penalty0 (6):\penalty0 298--306, 2008.
	
	\bibitem[Hartveit and Veruki(2012)]{hartveit2012electrical}
	E~Hartveit and M~L Veruki.
	\newblock Electrical synapses between aii amacrine cells in the retina:
	Function and modulation.
	\newblock \emph{Brain Res}, 1487:\penalty0 160--172, 2012.
	
	\bibitem[Nualart-Marti et~al.(2013)Nualart-Marti, Solsona, and
	Fields]{nualart2013biochim}
	A~Nualart-Marti, C~Solsona, and R~D Fields.
	\newblock Gap junction communication in myelinating glia.
	\newblock \emph{Biochim Biophys Acta}, 1828\penalty0 (1):\penalty0 69--78,
	2013.
	
	\bibitem[Pahtz et~al.(2010)Pahtz, Herrmann, and Shinbrot]{Pahtz:2010a}
	T~Pahtz, H~J Herrmann, and T~Shinbrot.
	\newblock Why do particle clouds generate electric charges?
	\newblock \emph{Nat Phys}, 6\penalty0 (5):\penalty0 364--368, 2010.
	
	\bibitem[Ladoux et~al.(2015)Ladoux, Nelson, Yan, and
	Mege]{ladoux2015mechanotransduction}
	B~Ladoux, W~J Nelson, J~Yan, and R~M Mege.
	\newblock The mechanotransduction machinery at work at adherens junctions.
	\newblock \emph{Integr Biol (Camb)}, 7\penalty0 (10):\penalty0 1109--1119,
	2015.
	
	\bibitem[Bausch and Kroy(2006)]{Bausch:2006aa}
	A~R Bausch and K~Kroy.
	\newblock A bottom-up approach to cell mechanics.
	\newblock \emph{Nat Phys}, 2\penalty0 (4):\penalty0 231--238, 2006.
	
	\bibitem[Broedersz and MacKintosh(2014)]{Broedersz:2014aa}
	C~P Broedersz and F~C MacKintosh.
	\newblock Modeling semiflexible polymer networks.
	\newblock \emph{Rev Mod Phys}, 86\penalty0 (3):\penalty0 995--1036, 2014.
	
	\bibitem[Lieleg et~al.(2009)Lieleg, Schmoller, Claessens, and
	Bausch]{Lieleg:2009a}
	O~Lieleg, K~M Schmoller, M~M Claessens, and A~R Bausch.
	\newblock Cytoskeletal polymer networks: {V}iscoelastic properties are
	determined by the microscopic interaction potential of cross-links.
	\newblock \emph{Biophys J}, 96\penalty0 (11):\penalty0 4725--4732, 2009.
	
	\bibitem[Fletcher and Mullins(2010)]{Fletcher:2010aa}
	D~A Fletcher and R~D Mullins.
	\newblock Cell mechanics and the cytoskeleton.
	\newblock \emph{Nature}, 463\penalty0 (7280):\penalty0 485--492, 2010.
	
	\bibitem[Mizuno et~al.(2007)Mizuno, Tardin, Schmidt, and
	MacKintosh]{Mizuno2007}
	D~Mizuno, C~Tardin, C~F Schmidt, and F~C MacKintosh.
	\newblock Nonequilibrium mechanics of active cytoskeletal networks.
	\newblock \emph{Science}, 315\penalty0 (5810):\penalty0 370--373, 2007.
	
	\bibitem[Gardel et~al.(2008)Gardel, Kasza, Brangwynne, Liu, and
	Weitz]{Gardel:2008aa}
	M~L Gardel, K~E Kasza, C~P Brangwynne, J~Liu, and D~A Weitz.
	\newblock Mechanical response of cytoskeletal networks.
	\newblock \emph{Methods in Cell Biol}, 89:\penalty0 487--519, 2008.
	
	\bibitem[Majumdar et~al.(2017)Majumdar, Foucard, Levine, and Gardel]{actin2017}
	S~Majumdar, L~C Foucard, A~J Levine, and M~L Gardel.
	\newblock Encoding mechano-memories in actin networks.
	\newblock \emph{arXiv}, arXiv:1706.05336 [cond-mat.soft], 2017.
	
	\bibitem[Billen et~al.(2009)Billen, Wilson, Rabinovitch, and
	Baljon]{Billen2009}
	J~Billen, M~Wilson, A~Rabinovitch, and A~R~C Baljon.
	\newblock Topological changes at the gel transition of a reversible polymeric
	network.
	\newblock \emph{EPL (Europhysics Letters)}, 87\penalty0 (6):\penalty0 68003,
	2009.
	
	\bibitem[Kim et~al.(2014)Kim, Litvinov, Weisel, and Alber]{Kim2014}
	O~V Kim, R~I Litvinov, J~W Weisel, and M~S Alber.
	\newblock Structural basis for the nonlinear mechanics of fibrin networks under
	compression.
	\newblock \emph{Biomaterials}, 35\penalty0 (25):\penalty0 6739--6749, 2014.
	
	\bibitem[Gavrilov et~al.(2015)Gavrilov, Komarov, and Khalatur]{Gavrilov2015}
	A~A Gavrilov, P~V Komarov, and P~G Khalatur.
	\newblock Thermal properties and topology of epoxy networks: {A} multiscale
	simulation methodology.
	\newblock \emph{Macromolecules}, 48\penalty0 (1):\penalty0 206--212, 2015.
	
	\bibitem[Liang et~al.(2016)Liang, Jones, Chen, Sun, and Jiao]{Liang:2016}
	L~Liang, C~Jones, S~Chen, B~Sun, and Y~Jiao.
	\newblock Heterogeneous force network in {3D} cellularized collagen networks.
	\newblock \emph{Phys Biol}, 13\penalty0 (6):\penalty0 066001, 2016.
	
	\bibitem[Venkatesan et~al.(2017)Venkatesan, Vivek-Ananth, Sreejith,
	Mangalapandi, Hassanali, and Samal]{samal2017}
	S~Venkatesan, R~P Vivek-Ananth, R~P Sreejith, P~Mangalapandi, A~A Hassanali,
	and A~Samal.
	\newblock Network approach towards understanding the crazing in glassy
	amorphous polymers.
	\newblock \emph{arXiv}, arXiv:1710.01996 [cond-mat.soft], 2017.
	
	\bibitem[Bouzid and Del~Gado(2017)]{Bouzid2017}
	M~Bouzid and E~Del~Gado.
	\newblock Network topology in soft gels: Hardening and softening materials
	network topology in soft gels: Hardening and softening materials network
	topology in soft gels: Hardening and softening materials.
	\newblock \emph{Langmuir}, 2017.
	\newblock URL \url{http://dx.doi.org/10.1021/acs.langmuir.7b02944}.
	
	\bibitem[Ahnert et~al.(2017)Ahnert, Grant, and Pickard]{Ahnert:2017aa}
	S~E Ahnert, W~P Grant, and C~J Pickard.
	\newblock Revealing and exploiting hierarchical material structure through
	complex atomic networks.
	\newblock \emph{npj Computational Materials}, 3\penalty0 (1):\penalty0 35,
	2017.
	
	\bibitem[Setford(2014)]{setford2014}
	J~Setford.
	\newblock Models of granular networks in two and three dimensions.
	\newblock Undergraduate Thesis, Department of Physics, University of Oxford
	(available at
	\url{http://www.math.ucla.edu/~mason/research/setford-final.pdf}), 2014.
	
	\bibitem[M{\"u}lken and Blumen(2011)]{Mulken:2011aa}
	O~M{\"u}lken and A~Blumen.
	\newblock Continuous-time quantum walks: {M}odels for coherent transport on
	complex networks.
	\newblock \emph{Phys Rep}, 502\penalty0 (2--3):\penalty0 37--87, 2011.
	
	\bibitem[Bianconi(2015)]{Bianconi:2015a}
	G~Bianconi.
	\newblock Interdisciplinary and physics challenges of network theory.
	\newblock \emph{EPL (Europhysics Letters)}, 111\penalty0 (5):\penalty0 56001,
	2015.
	
	\bibitem[Biamonte et~al.(2017)Biamonte, Faccin, and
	De~Domenico]{Biamonte:2017a}
	J~Biamonte, M~Faccin, and M~De~Domenico.
	\newblock Complex networks: From classical to quantum.
	\newblock \emph{arXiv}, arXiv:1702.08459 [quant-ph], 2017.
	
	\bibitem[Boccaletti et~al.(2006)Boccaletti, Latora, Moreno, Chavez, and
	Hwang]{bocca2006}
	S~Boccaletti, V~Latora, Y~Moreno, M~Chavez, and D-U Hwang.
	\newblock Complex networks: Structure and dynamics.
	\newblock \emph{Phys Rep}, 424\penalty0 (4):\penalty0 175--308, 2006.
	
	\bibitem[Liu and Zhang(2011)]{Liu:2011}
	Y~Liu and X~Zhang.
	\newblock Metamaterials: {A} new frontier of science and technology.
	\newblock \emph{Chem Soc Rev}, 40:\penalty0 2494--2507, 2011.
	
	\bibitem[Turpin et~al.(2014)Turpin, Bossard, Morgan, Werner, and
	Werner]{Turpin:2014}
	J~P Turpin, J~A Bossard, K~L Morgan, D~H Werner, and P~L Werner.
	\newblock Reconfigurable and tunable metamaterials: {A} review of the theory
	and applications.
	\newblock \emph{International Journal of Antennas and Propagation},
	2014\penalty0 (429837), 2014.
	
	\bibitem[Lee et~al.(2012)Lee, Singer, and Thomas]{Lee2012}
	J~H Lee, J~P Singer, and E~L Thomas.
	\newblock Micro-/nanostructured mechanical metamaterials.
	\newblock \emph{Adv Mater}, 24\penalty0 (36):\penalty0 4782--4810, 2012.
	
	\bibitem[Greaves et~al.(2011)Greaves, Greer, Lakes, and Rouxel]{Greaves:2011a}
	G~N Greaves, A~L Greer, R~S Lakes, and T~Rouxel.
	\newblock Poisson's ratio and modern materials.
	\newblock \emph{Nat Mater}, 10\penalty0 (11):\penalty0 823--837, 2011.
	
	\bibitem[Rocklin et~al.(2017)Rocklin, Zhou, Sun, and Mao]{Rocklin:2015}
	D~Z Rocklin, S~Zhou, K~Sun, and X~Mao.
	\newblock Transformable topological mechanical metamaterials.
	\newblock \emph{Nat Commun}, 8:\penalty0 14201, 2017.
	
	\bibitem[Fang et~al.(2006)Fang, Xi, Xu, Ambati, Srituravanich, Sun, and
	Zhang]{Fang:2006}
	N~Fang, D~Xi, J~Xu, M~Ambati, W~Srituravanich, C~Sun, and X~Zhang.
	\newblock Ultrasonic metamaterials with negative modulus.
	\newblock \emph{Nat Mater}, 5\penalty0 (6):\penalty0 452--456, 2006.
	
	\bibitem[Nicolaou and Motter(2012)]{Nicolaou:2012aa}
	Z~G Nicolaou and A~E Motter.
	\newblock Mechanical metamaterials with negative compressibility transitions.
	\newblock \emph{Nat Mater}, 11\penalty0 (7):\penalty0 608--613, 2012.
	
	\bibitem[Simovski et~al.(2012)Simovski, Belov, Atrashchenko, and
	Kivshar]{Simovski2012}
	C~R Simovski, P~A Belov, A~V Atrashchenko, and Y~S Kivshar.
	\newblock Wire metamaterials: {P}hysics and applications.
	\newblock \emph{Adv Mater}, 24\penalty0 (31):\penalty0 4229--4248, 2012.
	
	\bibitem[Smith et~al.(2004)Smith, Pendry, and Wiltshire]{Smith:2004aa}
	D.~R. Smith, J.~B. Pendry, and M.~C.~K. Wiltshire.
	\newblock Metamaterials and negative refractive index.
	\newblock \emph{Science}, 305\penalty0 (5685):\penalty0 788--792, 2004.
	
	\bibitem[Eiben and Smith(2015)]{eiben2015from}
	A~E Eiben and J~Smith.
	\newblock From evolutionary computation to the evolution of things.
	\newblock \emph{Nature}, 521\penalty0 (7553):\penalty0 476--482, 2015.
	
	\bibitem[D\'iaz-Manr\'iquez et~al.(2016)D\'iaz-Manr\'iquez, Toscano,
	Barron-Zambrano, and Tello-Leal]{diaz2016review}
	A~D\'iaz-Manr\'iquez, G~Toscano, J~H Barron-Zambrano, and E~Tello-Leal.
	\newblock A review of surrogate assisted multiobjective evolutionary
	algorithms.
	\newblock \emph{Comput Intell Neurosci}, 2016\penalty0 (9420460), 2016.
	
	\bibitem[Papadimitriou(2014)]{papadimitriou2014algorithms}
	C~Papadimitriou.
	\newblock Algorithms, complexity, and the sciences.
	\newblock \emph{Proc Natl Acad Sci}, 111\penalty0 (45):\penalty0 15881--15887,
	2014.
	
	\bibitem[Goldberg(1989)]{Goldberg:1989}
	D~E Goldberg.
	\newblock \emph{Genetic Algorithms in Search, Optimization and Machine
		Learning}.
	\newblock Addison-Wesley Longman Publishing Co., Inc., 1989.
	
	\bibitem[McGhee(1999)]{McGhee:1997}
	G~R McGhee.
	\newblock \emph{Theoretical Morphology: {T}he Concept and its Applications}.
	\newblock Columbia University Press, 1999.
	
	\bibitem[Valera et~al.(2017)Valera, Guo, Kelly, Matz, Cantu, Percus, Hyman,
	Srinivasan, and Viswanathan]{valera2017machine}
	M~Valera, Z~Guo, P~Kelly, S~Matz, A~Cantu, A~G Percus, J~D Hyman, G~Srinivasan,
	and H~S Viswanathan.
	\newblock Machine learning for graph-based representations of three-dimensional
	discrete fracture networks.
	\newblock \emph{arXiv}, arXiv:1705.09866 [physics.geo-ph], 2017.
	
	\bibitem[Avena-Koenigsberger et~al.(2014{\natexlab{a}})Avena-Koenigsberger,
	Go{\~n}i, Sol{\'e}, and Sporns]{Avena-Koenigsberger:2014}
	A~Avena-Koenigsberger, J~Go{\~n}i, R~Sol{\'e}, and O~Sporns.
	\newblock Network morphospace.
	\newblock \emph{J R Soc Interface}, 12\penalty0 (103), 2014{\natexlab{a}}.
	
	\bibitem[Avena-Koenigsberger et~al.(2014{\natexlab{b}})Avena-Koenigsberger,
	Go{\~n}i, Betzel, van~den Heuvel, Griffa, Hagmann, Thiran, and
	Sporns]{Avena-Koenigsberger:2013}
	A~Avena-Koenigsberger, J~Go{\~n}i, R~F Betzel, M~P van~den Heuvel, A~Griffa,
	P~Hagmann, J-P Thiran, and O~Sporns.
	\newblock Using pareto optimality to explore the topology and dynamics of the
	human connectome.
	\newblock \emph{Philos Trans R Soc Lond B Biol Sci}, 369\penalty0 (1653),
	2014{\natexlab{b}}.
	
	\bibitem[Go{\~n}i et~al.(2013)Go{\~n}i, Avena-Koenigsberger, {de Mendizabal},
	{van den Heuvel}, Betzel, and Sporns]{Goni:2013}
	J~Go{\~n}i, A~Avena-Koenigsberger, NV~{de Mendizabal}, M~P {van den Heuvel},
	R~F Betzel, and O~Sporns.
	\newblock Exploring the morphospace of communication efficiency in complex
	networks.
	\newblock \emph{PLoS One}, 8\penalty0 (3):\penalty0 e58070, 2013.
	
	\bibitem[Jaeger and de~Pablo(2016)]{jaeger2016evolutionary}
	H~M Jaeger and J~J de~Pablo.
	\newblock Perspective: {E}volutionary design of granular media and block
	copolymer patterns.
	\newblock \emph{APL Materials}, 4\penalty0 (5):\penalty0 053209, 2016.
	
	\bibitem[Miskin and Jaeger(2013)]{miskin2013adapting}
	M~Z Miskin and H~M Jaeger.
	\newblock Adapting granular materials through artificial evolution.
	\newblock \emph{Nat Mater}, 12:\penalty0 326--331, 2013.
	
	\bibitem[Miskin and Jaeger(2014)]{miskin2014evolving}
	M~Z Miskin and H~M Jaeger.
	\newblock Evolving design rules for the inverse granular packing problem.
	\newblock \emph{Soft Matter}, 10:\penalty0 3708--3715, 2014.
	
	\bibitem[Roth and Jaeger(2016)]{roth2016optimizing}
	L~K Roth and H~M Jaeger.
	\newblock Optimizing packing fraction in granular media composed of overlapping
	spheres.
	\newblock \emph{Soft Matter}, 12:\penalty0 1107--1115, 2016.
	
	\bibitem[Yan et~al.(2017)Yan, Ravasio, Brito, and Wyart]{Yan:2017a}
	L~Yan, R~Ravasio, C~Brito, and M~Wyart.
	\newblock Architecture and coevolution of allosteric materials.
	\newblock \emph{Proc Natl Acad Sci}, 114\penalty0 (10):\penalty0 2526--2531,
	2017.
	
	\bibitem[Ellenbroek et~al.(2015)Ellenbroek, Hagh, Kumar, Thorpe, and van
	Hecke]{Ellenbroek:2015aa}
	W~G Ellenbroek, V~F Hagh, A~Kumar, M~F Thorpe, and M~van Hecke.
	\newblock Rigidity loss in disordered systems: {T}hree scenarios.
	\newblock \emph{Phys Rev Lett}, 114\penalty0 (13):\penalty0 135501, 2015.
	
	\bibitem[Goodrich et~al.(2015)Goodrich, Liu, and Nagel]{Goodrich:2015a}
	C~P Goodrich, A~J Liu, and S~R Nagel.
	\newblock The principle of independent bond-level response: {T}uning by pruning
	to exploit disorder for global behavior.
	\newblock \emph{Phys Rev Lett}, 114\penalty0 (22):\penalty0 225501, 2015.
	
	\bibitem[Rocks et~al.(2017)Rocks, Pashine, Bischofberger, Goodrich, Liu, and
	Nagel]{Rocks:2017a}
	J~W Rocks, N~Pashine, I~Bischofberger, C~P Goodrich, A~J Liu, and S~R Nagel.
	\newblock Designing allostery-inspired response in mechanical networks.
	\newblock \emph{Proc Natl Acad Sci}, 114\penalty0 (10):\penalty0 2520--2525,
	2017.
	
	\bibitem[Driscoll et~al.(2016)Driscoll, Chen, Beuman, Ulrich, Nagel, and
	Vitelli]{Driscoll2016}
	M~M Driscoll, B~G-G Chen, T~H Beuman, S~Ulrich, S~R Nagel, and V~Vitelli.
	\newblock The role of rigidity in controlling material failure.
	\newblock \emph{Proc Natl Acad Sci U S A}, 113\penalty0 (39):\penalty0
	10813--10817, 2016.
	
	\bibitem[Shekhawat et~al.(2013)Shekhawat, Zapperi, and
	Sethna]{shekhawat2013damage}
	A~Shekhawat, S~Zapperi, and J~P Sethna.
	\newblock From damage percolation to crack nucleation through finite size
	criticality.
	\newblock \emph{Phys Rev Lett}, 110\penalty0 (18):\penalty0 185505, 2013.
	
	\bibitem[Reid et~al.(2017)Reid, Pashine, Jaeger, Liu, Nagel, and
	de~Pablo]{ReidAuxeticMetamaterials:2017}
	D~R Reid, N~Pashine, H~M Jaeger, A~J Liu, S~R Nagel, and J~J de~Pablo.
	\newblock Auxetic metamaterials from disordered networks.
	\newblock \emph{arXiv}, arXiv:1710.02493 [cond-mat.soft], 2017.
	
	\bibitem[Quinn and Winkelstein(2011)]{Quinn:2011}
	K~P Quinn and B~A Winkelstein.
	\newblock Preconditioning is correlated with altered collagen fiber alignment
	in ligament.
	\newblock \emph{J Biomech Eng}, 133\penalty0 (6):\penalty0 064506--064506,
	2011.
	
	\bibitem[Zhao et~al.(2014)Zhao, Chen, and Reich]{Zhao:2014}
	R~Zhao, C~S Chen, and D~H Reich.
	\newblock Force-driven evolution of mesoscale structure in engineered {3D}
	microtissues and the modulation of tissue stiffening.
	\newblock \emph{Biomaterials}, 35\penalty0 (19):\penalty0 5056--5064, 2014.
	
	\bibitem[Han et~al.(2013)Han, Heo, Driscoll, Smith, Mauck, and
	Elliott]{Han:2013}
	W~M Han, S-J Heo, T~P Driscoll, L~J Smith, R~L Mauck, and D~M Elliott.
	\newblock Macro- to microscale strain transfer in fibrous tissues is
	heterogeneous and tissue-specific.
	\newblock \emph{Biophys J}, 105\penalty0 (3):\penalty0 807--817, 2013.
	
	\bibitem[Pong et~al.(2011)Pong, Adams, Bray, Feinberg, Sheehy, Werdich, and
	Parker]{Pong:2011}
	T~Pong, W~J Adams, M-A Bray, A~W Feinberg, S~P Sheehy, A~A Werdich, and K~K
	Parker.
	\newblock Hierarchical architecture influences calcium dynamics in engineered
	cardiac muscle.
	\newblock \emph{Exp Biol Med}, 236\penalty0 (3):\penalty0 366--373, 2011.
	
	\bibitem[Sporns(2014)]{Sporns:2014aa}
	O~Sporns.
	\newblock Towards network substrates of brain disorders.
	\newblock \emph{Brain}, 137\penalty0 (8):\penalty0 2117--2118, 2014.
	
\end{thebibliography}
\end{document}